%% file: Ashutosh_Thesis.tex
\documentclass{dmathesis}
\usepackage{amsmath}   
\usepackage{amssymb}   
\usepackage{mathtools}
\usepackage{MnSymbol, graphicx}
\usepackage{fancyhdr}
\usepackage{subfigure} 
\usepackage{float}     
\usepackage{enumitem}  
\usepackage{blindtext} 
\usepackage[english]{babel}		
\usepackage{float}		     	
\usepackage{tabularx,colortbl}	
\usepackage{multirow}           
\newcolumntype{C}{>{\centering\arraybackslash}X} 
\usepackage{physics}			
\usepackage{braket}             
\usepackage{tikz}               
\usepackage{tikz-feynman}       
\tikzfeynmanset{compat=1.1.0}   
\usepackage{adjustbox}          
\usepackage{makecell}           
\usepackage{scalerel}          
\usetikzlibrary{arrows,decorations.markings,plotmarks} 
\usepackage{gensymb}  
\usepackage[toc,page]{appendix} 
\definecolor{darkblue}{rgb}{0.0, 0.0, 0.55} 
\usepackage{empheq}
\definecolor{myblue}{rgb}{.8, .8, 1}   
\usepackage{hyperref} 
\hypersetup{colorlinks=true,linkcolor=darkblue,filecolor=magenta,
urlcolor=darkblue,} 
\usepackage[T1]{fontenc}
\usepackage[utf8]{inputenc}

\input RoyalIn.fd

\input ArtNouvc.fd

\input{format}

\begin{document}
\date{}

\input{frontpage}

\pagenumbering{arabic}
\setcounter{page}{1}

\include{chapter1}

\include{chapter2}

\include{chapter3}
\include{chapter4}

\include{chapter5}

\include{chapter6}

\include{chapter7}


\begin{appendices}
\chapter{Availability of analysis code files and experimental data}
Mathematical analysis files, source codes, and experimental data of the experiments reported in this thesis can be shared with interested readers upon reasonable request. You may send your request for the same to me at \textcolor{blue}{asinghrri@gmail.com} with a cc to my supervisor Prof. Urbasi Sinha at \textcolor{blue}{usinha@rri.res.in}.

\chapter{List of Abbreviations}
$\begin{array}{ll}
\text{ADC} & \text{Amplitude Damping Channel} \\
\text{ADE} & \text{Asymptotic Decay of Entanglement} \\
\text{AL} & \text{Aspheric lens} \\
\text{BPF} & \text{Band Pass Filter} \\
\text{CW} & \text {Continuous Wave} \\
\text{BS} & \text {Beam Splitter} \\
\text{DSI} & \text {Displaced Sagnac Interferometer} \\
\text{EM} & \text{Entanglement Measure} \\
\text{EOF} & \text{Entanglement of Formation} \\
\text{ESD} & \text{Entanglement Sudden Death} \\
\text{HWP} & \text{Half Wave Plate} \\ 
\text{IF} & \text{Interference Filter} \\
\text{LN}  & \text{Log Negativity} \\
\text{LOCC} & \text{Local Operations and Classical Communication} \\
\text{LUO} & \text{Local Unitary Operation} \\
\text{MLE} & \text{Maximum Likelihood Estimation} \\
\text{N} & \text{Negativity} \\
\text{NOT} & \text{Pauli Operator} (\sigma_x) \\
\text{NPT} & \text{Negative Partial Transpose} \\
\text{NSD} & \text{Negativity Sudden Death} \\
\end{array}$

\newpage
$\begin{array}{ll}
\text{PBS} & \text{Polarizing Beam Splitter} \\
\text{PCC} & \text{Pearson Correlation Coefficient} \\
\text{PPT} & \text{Positive Partial Transpose} \\
\text{PT} & \text{Partial Transpose} \\
\text{QIP} & \text{Quantum Information Processing} \\
\text{QST} & \text{Quantum State Tomography} \\  
\text{QWP} & \text{Quarter Wave Plate} \\  
\text{SMF} & \text{Single Mode Fiber} \\
\text{SPAD} & \text{Single-Photon Avalanche Diode} \\
\text{SPDC} & \text{Spontaneous Parametric Down Conversion} \\
\end{array}$
\end{appendices}

\end{document}

%% file: format.tex
\usepackage{fancyhdr}
\usepackage{epsfig}
\usepackage{cite}
\usepackage{graphicx}
\usepackage{amsmath}
\usepackage{theorem}
\usepackage{amssymb}
\usepackage{latexsym}
\usepackage{epic}
\usepackage{palatino}
\usepackage{xcolor}
\definecolor{maroon}{RGB}{128,0,0}

\renewcommand{\baselinestretch}{1.5}
\newcounter{ind}

{\theorembodyfont{\rmfamily}}
{\theorembodyfont{\rmfamily}}
{\theorembodyfont{\rmfamily}}
{\theorembodyfont{\rmfamily}}
{\theorembodyfont{\rmfamily}}
{\theorembodyfont{\rmfamily}}

\date{}

%% file: frontpage.tex
\pagenumbering{roman}

\setcounter{page}{1}

\newpage

\thispagestyle{empty}
\begin{center}
  \vspace*{0.75cm}
  {\Large\bf \color{maroon} Creation, Characterization, and Manipulation of Quantum Entanglement in a Photonic System}\\
  
  \vspace*{0.6cm}
  {\LARGE\bf Ashutosh Singh}  \\
   \vspace*{0.6cm}
 {\Large \color{maroon} RAMAN RESEARCH INSTITUTE - BANGALORE}\\
          
          \vspace{0.5cm}
          
           \begin{center}
  \includegraphics[scale=0.7]{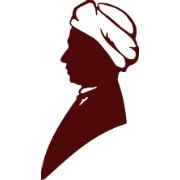}
   \end{center}

 \vspace{0.5cm}

  {\large A thesis submitted for the degree of\\
         [1mm]\Large{\color{maroon}DOCTOR OF PHILOSOPHY}\\
       [1mm] \large{ to} \\
      [2mm] \Large \color{maroon}JAWAHARLAL NEHRU UNIVERSITY - NEW DELHI}
       
       \vspace{0.5cm}
        \begin{center}
  \includegraphics[scale=0.05]{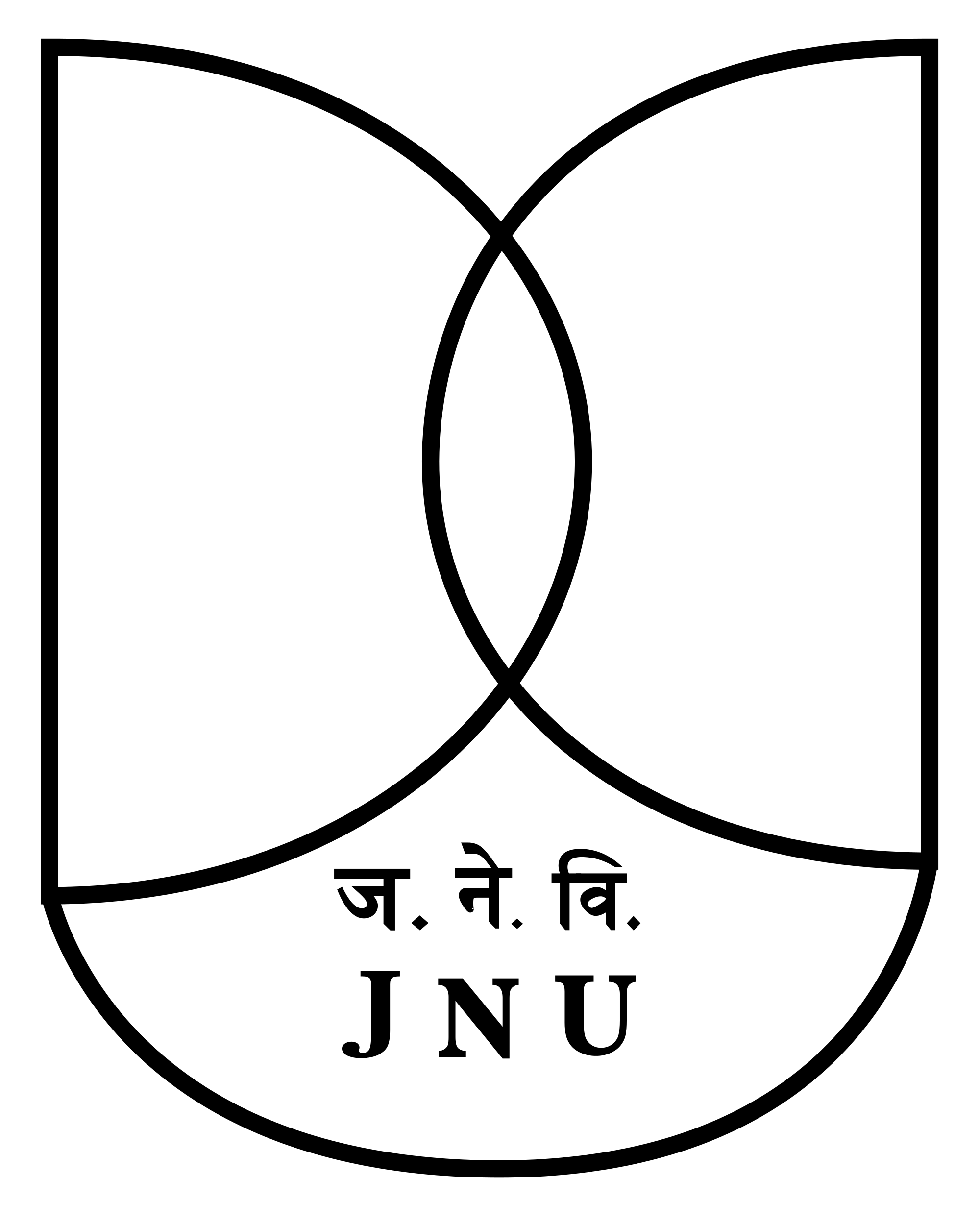}
   \end{center}         
  \vspace{-0.1 cm}
  \large{Thesis submitted: 22 Jan, 2020}\\
Copyright \copyright  Ashutosh Singh (2020). All rights reserved.\\[1pc]
\end{center}

\newpage
\thispagestyle{empty}
\begin{center}
 \vspace*{2cm}
  \textit{\Huge {\textcolor{maroon}{Dedication}}}\\ 
  \vspace{1cm}
\textit{\Large{\textcolor{darkblue}{This thesis is dedicated to my parents Shri Om Prakash Singh and Smt. Lalsa Singh for their endless love, unwavering support, encouragement, and ardent faith in me.  I can hardly describe my thanks and appreciation for you in words. You have been my source of inspiration and motivation. Thank you so much for everything!}}}\\
\end{center}

\chapter*{Declaration}
\addcontentsline{toc}{chapter}{\numberline{}Declaration}
The scientific work reported in this thesis is based on the research carried out by me at Quantum Information and Computing lab, Light and Matter Physics Group, Raman Research Institute-Bangalore, India, with the able support of my supervisor and collaborators from India and abroad. No part of this thesis has been submitted elsewhere for any other degree or qualification and it is all my own work unless referenced to the contrary in the text. This thesis has been run through the plagiarism checker software Turnitin.

\vspace{50pt}
\begin{center}
\begin{tabular*}{\textwidth}{c @{\extracolsep{\fill}} cccc}
     & & & \\
     & & &\\
     & & & \\
\makecell{Prof. Urbasi Sinha
\\
(Supervisor)}
& & &
\makecell{Ashutosh Singh
\\
(Student)}
\end{tabular*}
\end{center}



\chapter*{Certificate}
\addcontentsline{toc}{chapter}{\numberline{}Certificate}
This is to certify that the thesis titled ``\textbf{Creation, characterization and manipulation of quantum entanglement in a photonic system}" submitted by \textbf{Ashutosh Singh} for the award of the degree of Doctor of Philosophy of Jawaharlal Nehru University is his original work. This work has not been submitted to any other University for the award of any other degree or diploma.

\vspace{50 pt}
\begin{center}
\begin{tabular*}{\textwidth}{c @{\extracolsep{\fill}} cccc}
     & & & \\
     & & &\\
     & & & \\
\makecell{Prof. Ravi Subrahmanyan\\
Director and Center Chairperson\\
Raman Research Institute,\\
Bengaluru-560080, India.}
& & &
\makecell{Prof. Urbasi Sinha\\
Supervisor}
\end{tabular*}
\end{center}

\chapter*{Acknowledgements}
\addcontentsline{toc}{chapter}{\numberline{}Acknowledgements}

The research work presented in this thesis would not have been possible without the help and support of many individuals, whom I would like to thank here.

First and foremost, I would like to thank and convey my sincere gratitude to Prof. Urbasi Sinha for being my Ph.D. supervisor. I thank her for giving me a start in this field, for her guidance and for providing state-of-the-art lab facilities at RRI. Our long insightful discussions in the weekly group meetings and individual project meetings helped me shape my project and Ph.D. Due credits to her mentorship, immense support whether financial or otherwise, and the freedom that she gave me to learn and explore different avenues of research in quantum information science and complete the thesis.

I would like to thank Prof. Ravi Subramanian (Director, RRI) and Prof. Reji Philip for being part of my thesis advisory committee. Their comments and suggestions in the annual reviews helped me keep my research on track and complete it in a stipulated time. I would like to thank Prof. Dipankar Home (Bose Institute, Kolkata) for being my external examiner in the Ph.D. comprehensive exam. His cheerful disposition always made discussions very pleasant. I have learned a lot from him through the collaborative work that we did together and insightful discussions on the foundations of quantum mechanics and quantum information during his visits to RRI. Thanks to all the LAMP faculties who taught me various quantum optics and quantum information courses in the first year of Ph.D. coursework. My heartfelt thanks to Prof. R. Srikanth for stimulating and enlightening discussions.

My special thanks and sincere gratitude to Prof. A.R.P. Rau (Louisiana State University, USA) for collaborating with us and having stimulating and insightful discussions in the formulation of this project as well as on quantum information during his annual visits at RRI. Thanks to him for being an amazing host, arranging the lab visits, and showing me the places around in Louisiana during my visit to LSU, USA in March 2019. My sincere thanks to Prof. Aephraim M. Steinberg (University of Toronto, Canada) for sharing his maximum likelihood estimation code for quantum state tomography. 

I would like to thank Siva Pradyumna, visiting student fellow at RRI, for the discussions we had during the formulation of the project and the collaborative work we did together. Thanks to Mr. Ijaz Ahamad (IISER-Mohali) for the collaborative work on the entanglement measures for two-qubit pure states. Thanks to my labmates and friends - Debadrita, Animesh, Pradeep, Karthik, Siva, Sudhi, Siman, Surya, Sourav, Kaushik, Prathwiraj, Nagalakshi, Rakshita,  Rishabh, Sanchari, Abhishek, Pradosh, and Karthik H S - who made my life at RRI enjoyable. My special thanks to my labmate Surya N. Sahoo for his generous help on various occasions and for stimulating discussions.

A token of thanks is due to V S Naresh - administrative officer, accounts section, purchase section, Er. Mohamed Ibrahim and his team from mechanical engineering workshop,  Er. G. B. Suresh - in charge estate and buildings, Er. Munees from the electrical dept, Jacob Rajan - IT dept,  Savitha M. D. - secretary to the LAMP dept, Vidya and Radha from RRI administration, library staff, canteen staff, Shiva and Manju for keeping my office desk spick and span, and everyone who helped me in some or the other ways during the course of my Ph.D. journey. I would like to thank the vibrant sports community at RRI which helped me de-stress and relax when felt exhausted after work and allowed me to forget the problems and socialize with other students. 

My special thanks to my uncle Prof. O.N. Singh (IIT-BHU) for his constant push and support during my Ph.D. and my fianc\'ee Dr. Monika Singh for her unwavering support and encouragement during highs and lows of my life. My heartfelt thanks to my parents and family members without whose encouragement and support this would not have been possible. In the end, I would like to thank the almighty for this wonderful life full of happiness and opportunities.

\newpage
\thispagestyle{empty}
\addcontentsline{toc}{chapter}{\numberline{}Synopsis}
\begin{center}
\textbf{\Large\color{maroon}{``Creation, characterization, and manipulation of quantum entanglement in a photonic system"}}

  \vspace*{1cm}
  \textbf{\large Ashutosh Singh}

  \vspace*{0.5cm}
  {\large A thesis submitted for the degree of Doctor of Philosophy\\
   to \\
Jawaharlal Nehru University}

  \vspace*{1cm}
  \textbf{\large Synopsis}
\end{center}

Quantum superposition and entanglement are the quintessential characteristics of quantum mechanics and play an invaluable role in most quantum information processing protocols. In the existing literature, they have often been the subject of intense debate due to their counter-intuitive characteristics. While the former refers to the possibility of a quantum system simultaneously being in two (or more) different states, the latter can be seen as the superposition of the states of two (or more) different systems (or two properties of the same system) in such a correlated manner that their joint state cannot be factorised into the state of individual systems. Quantum entanglement is a non-classical correlation shared among quantum systems, which is now seen as an indispensable resource in quantum computation and communication.

Technological advancement over the last few decades has witnessed a rapid growth in research explorations in the field of quantum information science. This is fuelled by its direct impact on the upcoming quantum technologies with commercial applications. Exponential speed up of certain tasks in quantum computation as well as enabling some other tasks in quantum information  processing such as teleportation, superdense coding, ensuring unconditional security through quantum key distribution protocols, etc., are some of the attractive advantages offered by quantum information science which are either impossible or less efficient in the realm of classical information. Most of these quantum advantages are manifestations of the superposition principle and quantum entanglement which appear only in the framework of quantum mechanics.


The goal of quantum information processing is to prepare a set of qubits; quantum mechanical two-level systems (or qudits, d-level systems), in a known coherent superposition state and to perform arbitrary computation with them. This requires the implementation of an arbitrary unitary matrix on the Hilbert space of the qubits and finally qubit states are readout to get the result of computation. In a physical system, we don't get hold of the unitary matrix directly, rather it is constructed through the Hamiltonian in the Hilbert space of the qubits. If $H(t)$ is the Hamiltonian in the system Hilbert space then corresponding unitary matrix is generated as
\begin{equation}
U=\exp\left(-\frac{\iota}{\hbar} \int_0^t dt' H(t')\right),
\end{equation} 
where $H\in H_S$; $H_S$ being the Hilbert space of the system.

Evolution of the system $\rho_s(0)$ is given by
\begin{equation}
\rho_s(t)=U\rho_s(0)U^\dag
\end{equation}

In a real physical system, however, the Hamiltonian that we engineer need not be acting only on the system Hilbert space but also on the environment and in that case, we get
\begin{equation}
U=\exp\left(-\frac{\iota}{\hbar} \int_0^t dt'[H(t')+\delta H(t')]\right),
\end{equation}
where $H\in H_S, ~\delta H(t')\in H_S\otimes H_E$, $H_E$ is the Hilbert space of the environment.

When Hamiltonian acts on both, the system as well as the environment, it generates correlation between them. If we look at the state of the system alone by tracing out the environment degrees of freedom, the correlations are lost and it appears as decoherence on the quantum system. Thus, quantum decoherence is a ubiquitous and unavoidable phenomenon arising because of entanglement between quantum systems and their environment. 

Like classical correlations, entanglement also decays with time in the presence of noise in the ambient environment. More precisely, the presence of decoherence in the computing devices and communication channels, due to the unavoidable and irreversible interaction between the system and environment, causes the degradation in entanglement present in the system as the computation evolves or particles propagate. The decay of entanglement depends on the initial state as well as the amount and type of noise (amplitude damping, phase damping, etc.) acting on the system. In the presence of multiple stochastic noises, decoherence of individual qubits follows the additive law of relaxation rates whereas the decay of entanglement does not. In fact, entanglement may not decay asymptotically at all, and disentanglement can happen in finite time. Entanglement Sudden Death (ESD) is the phenomenon wherein a multipartite entangled state disentangles in finite time even when individual qubits decohere only asymptotically in time due to noise. ESD has been experimentally demonstrated in atomic and photonic systems.

Since quantum entanglement is an indispensable resource in quantum
information processing, manipulation that prolongs entanglement will help realize protocols that would otherwise suffer due to short entanglement decay times. Entanglement purification and distillation schemes could possibly recover the initial correlation from the ensemble of noise-degraded correlated states so long as the system has not completely disentangled. Therefore, delay or avoidance of ESD is important. For this purpose, local NOT operation in the computational basis on one or both qubits ($\sigma_x\otimes \mathbf{I}$,~$\mathbf{I}\otimes\sigma_x$,~or $\sigma_x\otimes\sigma_x$) has been proposed to combat the amplitude damping noise. Here, we report an all-optical implementation of the NOT operations; a theoretical proposal as well as the experimental demonstration, that can hasten, delay, or avert ESD, all depending on when it is applied during the process of decoherence for the polarization entangled photonic qubit system. We give analytical expressions for the probabilities corresponding to the hastening, delay, and avoidance of ESD which depend only on the parameters of the initial state density matrix of the system. These results are then experimentally verified in a suitably planned experiment.

Our proposal has an advantage over other known decoherence suppression schemes such as decoherence suppression using weak measurement and quantum measurement reversal, and delayed-choice decoherence suppression. There, as the strength of weak interaction increases, the success probability of decoherence suppression decreases. In our scheme, however, we can manipulate the ESD, in principle, with unit success probability as long as we perform the NOT operation at the appropriate time. Delay and avoidance of ESD, in particular, will find applications in the practical realization of quantum information and computation protocols that might otherwise suffer due to a short lifetime of entanglement. Also, it will have implications towards such control in other physical systems. A major advantage of the manipulation of ESD in a photonic system is that one has complete control over the damping parameters, unlike in most atomic systems, and hence provides a perfect test bed for such a study. The experiment that I report in this thesis is important for practical noise engineering in quantum information processing. Now, I will outline the structure of my thesis.

\textbf{In this thesis}, I will present our work on creation and characterization of quantum entanglement in a photonic system, and manipulation of its dynamics in the presence of an amplitude damping channel. When a two-qubit entangled system passes through an amplitude damping channel, some initial entangled states follow the asymptotic decay of entanglement whereas others undergo the sudden death of entanglement. The class of states which undergo ESD are then manipulated by local NOT operations ($\sigma_x$) on one or both qubits. This thesis is divided into seven chapters as given below.

\textbf{In chapter one}, I discuss the notations and basic concepts of quantum information that shall be helpful in understanding the rest of the thesis. An introduction to quantum entanglement along with the entanglement measures is given. Next, evolution equation of the open quantum system is discussed using the Kraus operator formalism. Then the basics of non-linear optics such as birefringence, second order non-linearity, phase matching condition, and spontaneous parametric  down-conversion are presented. 

\textbf{In chapter two}, I discuss the theoretical framework for, and experimental implementation of a type-I polarization entangled photon source based on spontaneous parametric down conversion. The entangled state is characterized by Quantum State Tomography (QST) and Maximum Likelihood Estimation (MLE). The concept of QST is discussed and it is motivated as to how a density matrix reconstructed by linear inversion of the QST data may not satisfy the requirements of a physical state; i.e., normalization, positivity, and  hermiticity. The process of obtaining a physical state from the above density matrix using MLE, is discussed.

\textbf{In chapter three}, I address the issue of quantitative non-equivalence of different entanglement measures for two-qubit pure states. Given a non-maximally entangled state, an operationally significant question is to quantitatively assess as to what extent the state is away from the maximally entangled state, which is of importance in evaluating the efficacy of the state for its various uses as a resource. It is this question which is examined in this paper for two-qubit pure entangled states in terms of different entanglement measures like Negativity (N), Logarithmic Negativity (LN), and Entanglement of Formation (EOF). Although these entanglement measures are defined differently, to what extent they differ in quantitatively addressing the earlier mentioned question has remained uninvestigated. Theoretical estimate in this paper shows that an appropriately defined parameter characterizing the fractional deviation of any given entangled state from the maximally entangled state in terms of N is quite different from that computed in terms of EOF with their values differing up to $\sim 15\%$ for states further away from the maximally entangled state. Similarly, the values of such fractional deviation parameters estimated using the entanglement measures LN and EOF, respectively, also strikingly differ among themselves with the maximum value of this difference being around $23\%$. This analysis is complemented by illustration of these differences in terms of empirical results obtained from a suitably planned experimental study. Thus, such appreciable amount of quantitative non-equivalence between the entanglement measures in addressing the experimentally relevant question considered in the present paper highlights the requirement of an appropriate quantifier for such intent. We indicate directions of study that can be explored towards finding such a quantifier.

\textbf{In chapter four}, I present the theoretical proposal for the manipulation of entanglement sudden death in an all optical set up. Evolution of a two-qubit entangled state is studied in the presence of an Amplitude Damping Channel (ADC). Two levels of the qubits are realized by the H- and V-polarization states of the photon and ADC is mimicked by the action of a half-wave plate (HWP) acting only on the V-polarization component of the photon in a displaced Sagnac interefrometer (DSI). As a result, polarization and momentum modes of the photon get entangled. When momentum modes of the photon are traced out, it acts as ADC. We see that the entangled state $|\psi\rangle=\alpha|HH\rangle+\beta|VV\rangle$ undergoes ESD for $|\alpha|<|\beta|$ and asymptotic decay for $|\alpha|>|\beta|$ in the presence of an ADC. To manipulate the ESD, we again use a $\text{HWP}@45^o$ but acting on both H- as well as V-polarization components and hence mimics NOT ($\sigma_x$) operation. The complete theoretical analysis of the experiment is given and results of the manipulation are compared  with the ESD results. We see the phenomenon of hastening, delay, and avoidance of ESD for either cases when NOT operation is applied on only one or both the qubits.

\textbf{In chapter five}, I present our experimental results on ESD followed by its manipulation in the photonic system. This requires preparation and characterization of a non-maximally entangled state. These entangled photons are then passed through the DSI. ADC is artificially introduced using a HWP acting only on  the vertical polarization component of the entangled state in the DSI. Initially, ESD experiment is performed by setting the ADC HWPs at non-zero angles and then reconstructing the state by QST and MLE for each setting and entanglement is computed from the density matrix. Next, ESD manipulation experiment is set up and for each setting of the ADC HWP, NOT operation is performed and output modes are appropriately traced out. Resulting state is reconstructed using QST and MLE; and then entanglement is computed from the density matrix. The results of ESD and manipulation experiments are compared with the simulation results.


\textbf{In chapter six}, I extend the theoretical proposal for manipulation of ESD in a $2\otimes 2$ system to $2\otimes 3$ and $3\otimes 3$ entangled system. We find that these higher dimensional systems can also be manipulated similar to the $2\otimes 2$ systems and the phenomenon of hastening, delay, and avoidance of ESD also occur as seen in $2\otimes 2$ systems. Our study, thus, indicates the possibility of controlled manipulation of higher dimensional entangled quantum systems evolving in a noisy environment.

\textbf{In chapter seven}, I conclude with the summary of the thesis followed by future scope of work.


\vspace{50 pt}
\begin{center}
\begin{tabular*}{\textwidth}{c @{\extracolsep{\fill}} cccc}
     & & & \\
     & & &\\
     & & & \\
\makecell{Prof. Urbasi Sinha
\\
(Supervisor)}
& & &
\makecell{Ashutosh Singh
\\
(Student)}
\end{tabular*}
\end{center}

\chapter*{List of publications}
\begin{enumerate}
\item {\textbf{Ashutosh Singh}, Siva Pradyumna, A. R. P. Rau, and Urbasi Sinha, ``Manipulation of entanglement sudden death in an all-optical setup," \href{https://doi.org/10.1364/JOSAB.34.000681}{J. Opt. Soc. Am. B \textbf{34}, 681-690 (2017)}}.
\item {Urbasi Sinha, Surya Narayan Sahoo, \textbf{Ashutosh Singh}, Kaushik Joarder, Rishab Chatterjee, Sanchari Chakraborti, Invited review article: ``Single-Photon Sources", \href{https://www.osa-opn.org/home/articles/volume_30/september_2019/features/single-photon_sources/}{Opt. Photon. News \textbf{30} (9), 32-39 (2019)}.\\ 
\textbf{ArXiv version}: ``Single photon sources: ubiquitous tools in quantum information processing," \href{https://arxiv.org/abs/1906.09565}{https://arxiv.org/abs/1906.09565 (2019).}} 
\item {\textbf{Ashutosh Singh}, Ijaz Ahamed, Dipankar Home, and Urbasi Sinha, ``Revisiting comparison between entanglement measures for two-qubit pure states," \href{https://doi.org/10.1364/JOSAB.37.000157}{J. Opt. Soc. Am. B \textbf{37}, 157-166 (2020)}}.
\item \textbf{Ashutosh Singh} and Urbasi Sinha, ``Entanglement protection in higher-dimensional systems",  Phys. Scr. \href{https://doi.org/10.1088/1402-4896/ac8200}{\textbf{97}, 085104 (2022).}
\end{enumerate}

\vspace{10pt}
\begin{center}
\begin{tabular*}{\textwidth}{c @{\extracolsep{\fill}} cccc}
     & & & \\
     & & &\\
     & & & \\
\makecell{Prof. Urbasi Sinha
\\
(Supervisor)}
& & &
\makecell{Ashutosh Singh
\\
(Student)}
\end{tabular*}
\end{center}

\tableofcontents
\listoffigures
\listoftables
\clearpage


%% file: chapter1.tex
\setcounter{equation}{0}
\chapter{Introduction and basics}

\section{Motivation}

Quantum Information Processing (QIP) exploits the fundamentals laws of quantum mechanics for information processing, communication, computation, metrology, and sensing [\ref{1.01}, \ref{1.03}]. The fundamental difference which makes QIP more efficient, attractive, and fascinating is its ability to use entangled states and the information processing of the more general input states in the form of superposition of different possible eigenstates of a quantum system as opposed to the information processing via classical systems which can handle only bits. Moreover, two distant parties which may be space-like separated can share arbitrary and unknown quantum information without physically sending the quantum system in which the information is stored using a prior shared entanglement between them [\ref{1.01}, \ref{1.03}]. Further, no-cloning theorem and uncertainty principle ensures that arbitrary and unknown quantum information can neither be copied nor eavesdropped without leaving behind a trace for the same. 

The possibility of preparation and controlled manipulation of multi-partite superposition states with exquisite precision has opened up the doors for quantum computation which is fundamentally equivalent to doing massive parallel computation with classical systems. All the aforementioned quantum advantages require maintenance of quantum coherence (well defined phase relationship) [\ref{1.04}] among different parts of the multi-partite systems for large time scales as well as over long distances. It means that the interaction of the quantum systems with their environment need to be minimized. Ideally, system should be isolated from the environment to avoid loss of quantum coherence due to interaction of these systems with the environment. 

However, practical realization of an ideal isolated quantum system may not be possible and systems will inevitably interact with its environment leading to decoherence [\ref{1.05}, \ref{1.06}] in the system. The term decoherence refers to a dynamical process through which a quantum mechanical system looses the phase relationship between its different parts. Decoherence is an unavoidable and irreversible phenomenon arising because of entanglement between system and environment due to interaction. It is the biggest obstacle towards practical realization of exponential speedup promised for certain tasks is quantum computation. In the presence of local stochastic noises in the environment, system doesn't only loose the single particle coherence, even the non-local coherence (entanglement) degrades in the multi-particle system [\ref{1.06}, \ref{1.07}]. Since entanglement is the quintessential resource for QIP protocols, protecting  entanglement against the detrimental effects of the noise is important. There are several schemes to protect entanglement from decoeherence such as weak measurement and reversal, dynamical decoupling, quantum bang-bang protocol, delayed choice decoherence suppression [\ref{1.06}], etc.

In this thesis, we propose another scheme [\ref{1.08}] to protect quantum entanglement in the presence of an amplitude damping channel (ADC). We know that under the action of an ADC, excited state of a two-level atomic system decays down to ground state emitting the energy to the environment. Thus, as time passes, excited state population decreases and ground state population increases. Therefore, any process which flips the population between ground and excited states of the quantum system is expected to give some interesting results on the system dynamics. Motivated by this idea, we propose a local unitary operation (NOT operation, $\sigma_x$) on the individual qubit of a two-qubit entangled state during the process of decoherence and study its impact on entanglement preservation in a noisy environment. It turns out that the action of NOT operation on one or both qubits preserves entanglement for longer duration [\ref{1.08}] for states which would have otherwise become separable in finite time due to the action of ADC. Here, we present the results based on simulation as well as taken some steps to verify them experimentally in a two-qubit entangled state evolving in the presence of an ADC in a photonic system. 

Consider a practical scenario where Charlie prepares a bipartite entangled state for some QIP task and he has to send the entangled particles to Alice and Bob through a quantum channel which is noisy and can potentially cause disentanglement before the particles reach the two parties. In this scenario, we ask the following question: given a bipartite entangled state which would undergo finite time disentanglement (also known as Entanglement Sudden Death or ESD) in the presence of an ADC, can we alter the time of disentanglement by some suitable local unitary operations during the process of decoherence? To answer this question, we explore not only qubit-qubit system but higher-dimensional systems such as qubit-qutrit and qutrit-qutrit system as well. If such systems undergo ESD, we propose a set of local unitary operations such that when applied on the subsystems during  the  process   of  decoherence, these operations can manipulate the disentanglement dynamics, in particular,  delay the  time at which ESD occurs. Depending on the combination of local unitary operations, and the time  of their  application, this  method is shown  to be  able to hasten, delay or completely avoid the ESD even in higher-dimensional systems. Such an entanglement protection scheme will not only facilitate the aforementioned task but will also find application where two parties, say Alice and Bob, share an entangled pair for some quantum information processing task and they know a-priori that ADC is present in the environment, and therefore they can decide whether they are faced with the prospect of ESD. Then, they can locally apply suitable local unitary operations at appropriate time to delay or avoid the ESD. Our proposed scheme for preserving entanglement longer will also find application in entanglement distillation protocols.

\section{Outline of the thesis}

This thesis deals with creation, characterization, and manipulation of quantum entanglement in a  photonic system. It has two main parts which deal with static and dynamic properties of entanglement, respectively, in bipartite systems. It is organized in a way that the current introductory chapter deals with several basic and technical concepts which are required for understanding the research work reported in this thesis.

In our exploration of the static properties of entanglement, we experimentally prepare a polarization entangled photon source using Spontaneous Parametric Down-Conversion (SPDC) and  reconstruct the state (density matrix) via Quantum State Tomography (QST) and Maximum Likelihood Estimation (MLE) [\ref{1.09}]. For characterization of the experimentally prepared states and quantification of entanglement, we use different computable measures of entanglement. In this process, we make some non-trivial observations on the comparison of entanglement measures for two-qubit pure states and indicate directions for further study in higher-dimensional systems [\ref{1.10}]. In order to comprehend this aspect of the thesis, a basic understanding of the concepts such as SPDC based single and entangled photon sources, quantification of entanglement using different entanglement measures, as well as QST and MLE are essential.

In our investigation of entanglement dynamics, we study the evolution of entanglement in a noisy environment. The noise under consideration is a local, identical but independent Markovian noise acting on each qubit. It is found that under the action of an Amplitude Damping Channel (ADC), an initially entangled state can either undergo Asymptotic Decay of Entanglement (ADE) or finite time disappearance of entanglement, also known as Entanglement Sudden Death (ESD) [\ref{1.07}]. For the state undergoing ESD, we propose a local unitary operation (more specifically NOT operation, also known as $\sigma_x$) on one or both qubits such that depending on the time of application of NOT operation, ESD can be hastened, delayed or avoided [\ref{1.08}]. The phenomenon of ESD and ADE are then experimentally demonstrated for polarization entangled photonic qubit system in a Displaced Sagnac Interferometer (DSI). This is followed by a discussion on our attempts towards experimental demonstration of manipulation of ESD in such a system. In order to appreciate this portion of the thesis, we require to understand the evolution of open quantum systems, in general, and entangled states, in particular. The overarching theme of this thesis being entanglement, it is important to have a basic understanding of the same; both theoretically as well as experimentally and some basic concepts of quantum information. 

In the end, we theoretically extend the idea of manipulation of ESD to higher-dimensional systems such as $2\otimes 3$ and $3 \otimes 3$ systems. By choosing suitable measure(s) of entanglement for this purpose, we study the evolution of entanglement in the presence of ADC and phenomenon of ESD is observed. Like $2\otimes 2$ systems, we propose a set of local unitary operations for qubit as well as qutrit subsystems wherein we indicate more than one possible choices of local unitaries for qutrit. We then demonstrate the phenomenon of hastening, delay, and avoidance of ESD in these systems as well.

To start with, we will discuss the basic terminology, notation, and some basic concepts of quantum information to set the stage for the rest of the thesis. This chapter begins by motivating the subject of this thesis which is followed by a primer on the concept of state vector and density matrices, concept of pure and mixed states, followed by a detailed introduction to classical bit and quantum-bit (qubit). Then, we introduce the concept of quantum entanglement and its measures for pure and mixed states. Next, evolution of closed quantum systems is introduced which is followed by open quantum system dynamics using the Kraus operator formalism. The chapter ends with the discussion of the basics of birefringence, nonlinear optics, and spontaneous parametric down conversion (SPDC).

\begin{itemize}
\item Let us begin the introduction with some basic concepts of quantum mechanics and quantum information.
\end{itemize}
\section{Basics of quantum mechanics and quantum information}
\subsection{Concept of state vector and  density matrix}
In quantum mechanics, the state of an isolated physical system is described by a \textit{state vector}, a unit vector in the Hilbert space, which completely characterizes the physical properties of the system. The simplest quantum mechanical system is a qubit, which resides in a two-dimensional Hilbert space $\mathbb{C}^2$. If $|0\rangle$ and $|1\rangle$ are taken as the basis vectors for the Hilbert space, then the general state vector can be written [\ref{1.01}] as
\begin{equation}
|\Psi\rangle=\alpha|0\rangle+\beta|1\rangle ,
\label{eq1.01}
\end{equation}
 where $\alpha$ and $\beta$ are, in general, complex numbers satisfying the normalization condition for the state vectors: $|\alpha|^2+|\beta|^2=1$.


The state vectors for quantum systems are idealized descriptions which represent only pure states and not the statistical (incoherent) mixtures - ensemble comprising of many pure states with weighted probabilities which are frequently encountered in experiments. The \textit{density matrix} or density operator is an alternate and more general description of quantum systems for both pure as well as mixed states [\ref{1.01}, \ref{1.03}]. All the accessible information of a quantum system is encoded in the density matrix.

Consider an ensemble of systems all in the same state $|\Psi\rangle$, also known as pure state. The density matrix $\hat{\rho}$ for such a \textit{pure state} $|\Psi\rangle$ is defined as the outer product of the wave function and its conjugate as given below  [\ref{1.01}, \ref{1.03}].
\begin{equation}
\hat{\rho}:= |\Psi\rangle\langle\Psi|~.
\label{eq1.02}
\end{equation}
 
Let us consider an ensemble of n-systems where not all of them are in the same state but $n_i$ no. of systems are in the state $|\Psi_i\rangle$, etc., where $\sum_i n_i=n$. The probability of finding an individual system in the state $|\Psi_i\rangle$ of such a ensemble is given by $p_i=n_i/n$. In such a situation, when all the systems of the ensemble  $\{|\Psi_i\rangle\}$ are not in the same state, but a statistical mixture of the states  $\{|\Psi_i\rangle\}$ with respective probabilities $p_i$ then it is said to be in \textit{mixed state} [\ref{1.01}, \ref{1.03}]. The density matrix of such a ensemble is given by
\begin{equation}
\hat{\rho}:= \sum_i p_i |\Psi_i\rangle\langle\Psi_i|.
\label{eq1.03}
\end{equation}
 
Any physical density matrix satisfies the following three cardinal properties:
\begin{equation}
\begin{aligned}
\begin{split}
 \text{Tr}[\hat{\rho}] &=1     ~~~~ \text{(Normalization)} \\ 
 \hat{\rho}^{\dag} &=\hat{\rho}  ~~~~\text{(Hermiticity)} \\ 
\text{det}[\hat{\rho}] &\geq 0 ~~~~  \text{(Positive semidefiniteness)} \\
 \end{split}
 \end{aligned}
 \label{eq1.04}
\end{equation}

A pure state density matrix, in addition, satisfies the condition: $\hat{\rho}^2=\hat{\rho}$, i.e., it is also a projector.

\subsection{Pure and mixed states}
Consider an ensemble of particles in the states $\{|\Psi_i\rangle\}$. If all the particles are in the same state then the ensemble is represented by a \textit{pure state}. For a pure state, $\hat{\rho}^2=\hat{\rho}$, and thus
\begin{equation}
\text{Tr}[\hat{\rho}^2]=1~~~~\text{(For pure state)}.
\label{eq1.05}
\end{equation} 

When all the particles of the ensemble  $\{|\Psi_i\rangle\}$ are not in the same state, but a statistical mixture of the states  $\{|\Psi_i\rangle\}$ with respective probabilities $p_i$ then it is said to be in \textit{mixed state} and its density matrix is given by
\begin{equation}
\hat{\rho}=\sum_i p_i |\Psi_i\rangle\langle\Psi_i|.
\label{eq1.06}
\end{equation}

For a mixed state $\hat{\rho}^2\neq\hat{\rho}$ and thus
\begin{equation}
\text{Tr}[\hat{\rho}^2]< 1 ~~~~\text{(For mixed state)}.
\label{eq1.07}
\end{equation}

The term $\text{Tr}[\hat{\rho}^2]$ is defined as the \textit{ purity} of a state. It can take a minimum value of $1/d$ for maximally mixed state, where $d$ is the dimensionality of the system, and $1$ for pure state.

\subsection{Bit and Qubit}

Classical bit, also known as \textit{bit}, is the basic unit of digital information in classical information  theory, which can take logical values `0' or `1'. These two possibilities are mutually exclusive. Because of its binary nature, it can be encoded into the property of any physical system which has two stable states, e.g., ON or OFF  state of a transistor, direction of magnetic domain in a magnetic media, etc.  On the other hand, \textit{quantum bit} or \textit{qubit} is the quantum mechanical analogue of the classical bit which can exist in an arbitrary superposition of the two possible states, say $|0\rangle$ and $|1\rangle$. The term `\textit{qubit}' was first introduced by Schumacher in 1995 [\ref{1.11}]. The general state of a qubit is given by
\begin{equation}
|\Psi\rangle=\alpha |0\rangle + \beta |1\rangle ,
\label{eq1.08}
\end{equation}
 where $\alpha$ and $\beta$ are in general complex, and $|\alpha|^2+|\beta|^2=1$.

An example of a classical and quantum bit is illustrated using a switch [\ref{1.12}] is shown in Fig.~\ref{fig1.01} below. Quantum superposition is the key difference between the two. A classical bit always remains in one of the two possible orthogonal states (`0', or `1') whereas a qubit can exist in a coherent superposition of the two orthogonal states as well.

\begin{figure} [H]
\centering
\includegraphics[width=0.9\linewidth]{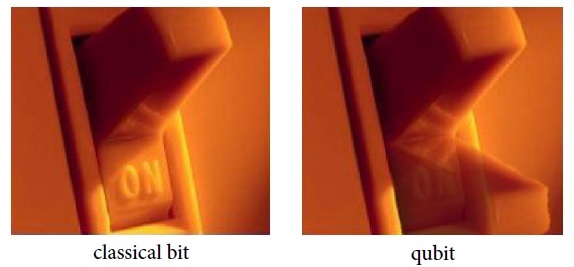}
\caption{\textit{Difference between a classical bit and qubit illustrated using a switch [Image credit: Ref. \ref{1.12}]. A classical bit always remains in one of the two possible orthogonal states (`0', or `1') whereas a qubit can exist in a coherent superposition of the two orthogonal states as well.}\label{fig1.01}}
\end{figure}

 The state of a qubit is an element of two-dimensional Hilbert space. The states $|0\rangle$ and $|1\rangle$ are the eigenstates of the Pauli matrix $\sigma_z$ which form the basis vectors of this space, and known as computational basis states.
 \begin{equation}
|0\rangle  = \left(\begin{array}{cc}
1\\0
\end{array} \right),~~\text{and}~~
~~|1\rangle =\left(\begin{array}{cc}
0\\1
\end{array} \right).
\label{eq1.09}
\end{equation}
 
If the normalization is written implicitly into the coefficients, and global phases are ignored, then Eq.~(\ref{eq1.08}) can be rewritten as 
\begin{equation}
|\Psi\rangle=\cos\left(\frac{\theta}{2}\right)|0\rangle+\exp(i\phi)\sin\left(\frac{\theta}{2}\right)|1\rangle ,
\label{eq1.10}
\end{equation}
where $0\leq\theta\leq\pi$ and $0\leq\phi<2\pi$. 

If we consider a point with ($\theta, \phi$) as defined above and $r=1$ as the spherical polar coordinates ($r, \theta, \phi$) on a sphere then  corresponding point on the Poincare sphere represents a state of the quantum system whose density matrix is given by
\begin{equation}
\hat{\rho}=\frac{1}{2}(\mathbb{I}_2+\vec{r}.\vec{\sigma}) ,
\label{eq1.11}
\end{equation}
where $\vec{r}$ is the Poincare vector given by\begin{equation}
\vec{r}=\left(\begin{array}{c}
r_D\\
r_R\\
r_H\end{array}\right)=\left(\begin{array}{c}
\sin(\theta)\cos(\phi)\\
\sin(\theta)\sin(\phi)\\
\cos(\theta)\end{array}\right),~\vec{\sigma}=\left(\begin{array}{c}
\sigma_x\\
\sigma_y\\
\sigma_z\end{array}\right),
\label{eq1.12}
\end{equation}
and, $\mathbb{I}_2$ is the identity matrix, and $\sigma_i$s are the Pauli matrices as given below.
\begin{equation}
\begin{aligned}
~\mathbb{I}_2=\left(\begin{array}{cc}
1 & 0\\
0 & 1\end{array}\right),~
\sigma_x=\left(\begin{array}{cc}
0 & 1\\
1 & 0\end{array}\right),~
~\sigma_y=\left(\begin{array}{cc}
0 & -i\\
i & 0\end{array}\right),~
\sigma_z=\left(\begin{array}{cc}
1 & 0\\
0 & -1\end{array}\right).
\end{aligned}
\label{eq1.13}
\end{equation}

\begin{figure} [H]
\begin{center}
\begin{tikzpicture}
\definecolor{bananamania}{rgb}{0.98, 0.91, 0.71}
\definecolor{antiflashwhite}{rgb}{0.95, 0.95, 0.96}
\definecolor{babyblueeyes}{rgb}{0.63, 0.79, 0.95}
\draw[line width=0.5mm, black, fill=antiflashwhite, opacity=0.3] (0,0) circle (4cm);
\draw[line width=0.5mm,gray , fill=babyblueeyes, opacity=0.25] (0,0) ellipse (4cm and 1cm);
\draw[line width=0.5mm,brown, fill=bananamania, opacity=0.3] (0,0) ellipse (1.5cm and 4cm);
\draw[dashed, line width=0.4mm,gray] (-4,0)-- (4,0);
\draw[dashed, line width=0.4mm, gray] (0,-4)-- (0,4);
\draw[dashed, line width=0.4mm, gray] (-1.5,-0.95)-- (1.5,0.95);
\draw[line width=0.3mm, red, arrows={-Triangle[angle=90:5pt,black,fill=red]}](0,0)--(1,2.2) ;
\draw[red] (0,1.5) arc (90:40:0.7);
\node at (0.4,1.7) {$\theta$};
\draw [dotted, line width=0.3mm, red](0,0)--(1,-0.5);
\draw[dotted, line width=0.25mm, red](0,2.2)--(1,2.2);
\draw[dotted, line width=0.25mm, red](1,2.2)--(1,-0.5);
\draw[red] (-0.2,-0.1) arc (-138:-45:0.3);
\node at (0.2,-0.35) {$\phi$};
\node at (0,4.4){$|H\rangle$};
\node at (0,-4.4){$|V\rangle$};
\node at (-1.9,-1.25){$|D\rangle$};
\node at (1.9,1.25){$|A\rangle$};
\node at (4.4,0){$|R\rangle$};
\node at (-4.4,0){$|L\rangle$};
\node at (0.83,2.45){$|\Psi\rangle$};
\begin{scope}[shift={(5.5,-4)}]
 \draw[line width=0.3mm, black, arrows={-Triangle[angle=90:5pt,black,fill=red]}](-5.5,4)--(-4.1,4) ;
 \draw[line width=0.3mm, black, arrows={-Triangle[angle=90:5pt,black,fill=red]}](-5.5,4)--(-5.5,5.4) ;
  \draw[line width=0.3mm, black, arrows={-Triangle[angle=90:5pt,black,fill=red]}](-5.5,4)--(-6.3,3.5) ; 
\node at (-6.4,3.7){$x$}; 
\node at (-4.2,3.7){$y$}; 
\node at (-5.8,5.3){$z$};
\end{scope}
\end{tikzpicture}
\caption{\textit{Poincare sphere representation of a qubit. All points at the suraface of Poincare sphere represent pure states whereas those lying inside the spehere represent mixed state. Point at the center of the sphere is maximally mixed.}\label{fig1.02}}
\end{center}
\end{figure}
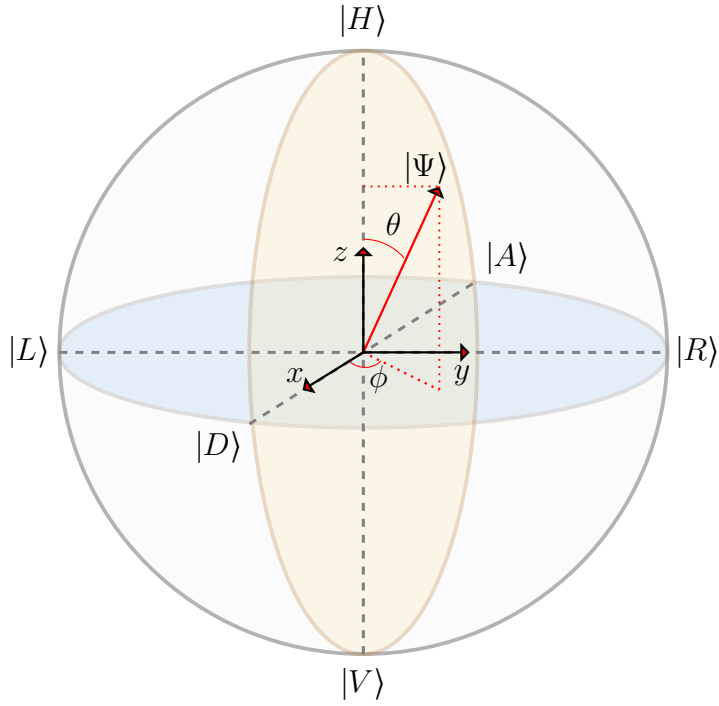

Three components of $\vec{r}$ ($r_D$, $r_R$, $r_H$) can be viewed as the coordinates in the three dimensional polarization space and give the polarization component of the photon in diagonal/anti-diagonal (D/A), right-/left-circular (R/L), and horizontal/vertical (H/V),  bases, respectively. A spin-1/2 particle such as electron, polarization state of a photon, etc., are good examples of a qubit. If we represent the basis states $|H\rangle\equiv |0\rangle$ and $|V\rangle\equiv |1\rangle$, then the general polarization state of a photon can be written as Eq.~(\ref{eq1.10}) or (\ref{eq1.11}) and it can be represented on the Poincare sphere as shown in the Fig.~\ref{fig1.02}. 
 
Note that Eq.~(\ref{eq1.11}) with $|\vec{r}|=1$ captures all the pure states which lie on the surface of the Poincare sphere. All the states in Eq.~(\ref{eq1.11}) with $|\vec{r}|<1$ are mixed states and they lie inside the sphere, and maximally mixed state ($\mathbb{I}/2$) lies at the center of the sphere. It is important to bear in mind that the orthogonal states lie $180^\circ$ apart, i.e., on the diametrically opposite sides of the sphere. The $|H\rangle$ and $|V\rangle$ polarization states lie on the opposite sides ($\theta=180^\circ$ apart) of the pole, and  $|D\rangle$, $|R\rangle$, $|A\rangle$, and $|L\rangle$ lie on the equator at $\phi=90^\circ$ apart from the adjacent ones. These polarization states can be expressed in terms of the bases states $|H\rangle$ and $|V\rangle$ as follows:
\begin{equation}
\begin{aligned}
|D\rangle = \frac{|H\rangle +|V\rangle}{\sqrt{2}},~~|A\rangle = \frac{|H\rangle -|V\rangle}{\sqrt{2}},~~\\
|R\rangle = \frac{|H\rangle + i |V\rangle}{\sqrt{2}},~~|L\rangle = \frac{|H\rangle - i |V\rangle}{\sqrt{2}}.
\end{aligned}
\label{eq1.14}
\end{equation}
Such a visualization of a polarization qubit on the Poincare sphere will be very convenient and intuitive to understand the unitary transformations on the Hilbert space of a qubit and projective measurements on different bases set for quantum state tomography in next chapter.

\subsection{Partial trace and reduced density matrix}
Consider a composite quantum system characterized by the density matrix $\hat{\rho}^{(AB)}$ comprising of two subsystems A and B. Suppose we are interested in only one of the subsystems; say A, then this situation is captured by the reduced density matrix. For subsystem A, it is obtained by tracing out [\ref{1.03}] the subsystem B as follows:
\begin{equation}
\hat{\rho}^{(A)}=\text{Tr}_B[\hat{\rho}^{(AB)}].
\end{equation}
\begin{equation}
\text{Tr}_B(|a_i\rangle\langle a_j|\otimes |b_k\rangle\langle b_l|)\equiv |a_i\rangle\langle a_j \text{Tr}(|b_k\rangle\langle b_l|)= |a_i\rangle\langle a_j|\langle b_k|b_l\rangle.
\end{equation} 

where $\{|a_i\rangle,|a_j\rangle\}$, and $\{|b_k\rangle,|b_l\rangle\}$ are the basis vectors of the state space A and B, respectively. Physically, partial trace is the unique operation for obtaining the correct measurement statistics of a subsystem of the composite quantum system.
 
\subsection{Partial transpose}

Let us consider a bipartite joint system density matrix $\hat{\rho}^{(AB)}$ with the density matrices of its individual subsystems given by $\hat{\rho}^{(A)}=\text{Tr}_B[\hat{\rho}^{(AB)}]$ and  $\hat{\rho}^{(B)}=\text{Tr}_A[\hat{\rho}^{(AB)}]$. The elements of the density matrix $\hat{\rho}^{(AB)}$ can be written as
\begin{equation}
\rho_{m\mu,n\nu}=\langle a_m,b_\mu|\hat{\rho}^{(AB)}|a_n,b_\nu\rangle
\end{equation}
where \{$a_m$\} and \{$b_\mu$\} are the orthonormal bases sets in the Hilbert space of subsystems A and B, respectively. 

As the name suggests, partial transpose [\ref{1.13}] is the transpose taken with respect to one of the subsystems of a  bipartite quantum system. Using the above representation, partial transposition with respect to subsystem A can be written as follows:
\begin{equation}
\text{PT}^A(\rho_{m\mu,n\nu})=\rho_{n\mu,m\nu}
\end{equation}

Likewise, partial transposition with respect to subsystem B can be written as follows:
\begin{equation}
\text{PT}^B(\rho_{m\mu,n\nu})=\rho_{m\nu,n\mu}
\end{equation}

This can be further illustrated by taking the explicit example of the density matrix of a $2\otimes 2$ dimensional system and dividing it into $2\times 2$ blocks and doing transpose in each block to get the partially transposed density matrix with respect to subsystem B.
\begin{equation}
\begin{split}
 \hat{\rho}^{(AB)} & =\left(\begin{array}{cc|cc}
\rho_{11,11} & \color{magenta}{\rho_{11,12}} & \rho_{11,21} & \color{magenta}{\rho_{11,22}}\\
\color{blue}{\rho_{12,11}} & \rho_{12,12} & \color{blue}{\rho_{12,21}} & \rho_{12,22}\\
\hline
\rho_{21,11} & \color{magenta}{\rho_{21,12}} & \rho_{21,21} & \color{magenta}{\rho_{21,22}}\\
\color{blue}{\rho_{22,11}} & \rho_{22,12} & \color{magenta}{\rho_{22,21}} & \rho_{22,22}\\
\end{array}\right)\\
\Rightarrow  \text{PT}^B(\hat{\rho}^{(AB)})& =\left(\begin{array}{cc|cc}
\rho_{11,11} & \color{blue}{\rho_{12,11}} & \rho_{11,21} & \color{blue}{\rho_{12,21}}\\
\color{magenta}{\rho_{11,12}} & \rho_{12,12} & \color{magenta}{\rho_{11,22}} & \rho_{12,22}\\
\hline
\rho_{21,11} & \color{blue}{\rho_{22,11}} & \rho_{21,21} & \color{blue}{\rho_{22,21}} \\
\color{magenta}{\rho_{21,12}} & \rho_{22,12} & \color{magenta}{\rho_{21,22}} & \rho_{22,22}\\
\end{array}\right)
\end{split}
\end{equation}

It is evident that the partial transpose operation interchanges the entries in blue and magenta in each block. Partial transpose operation finds application in deciding the (non)separability of a bipartite state using PPT criterion (discussed in Sec. \ref{sec1.4.3}).

\subsection{Fidelity} Fidelity is a distance measure of two quantum states.  Fidelity $F(\hat{\rho},\hat{\sigma})$ of the states $\hat{\rho}$ and $\hat{\sigma}$ measures the ``closeness" of the states and it is defined as [\ref{1.14}] 
\begin{equation}
F(\hat{\rho},\hat{\sigma}):= \text{Tr}\left[\sqrt{\sqrt{\hat{\rho}}~\hat{\sigma}~\sqrt{\hat{\rho}}}~~ \right] .
\end{equation}
Fidelity of a pure state $|\psi\rangle$ and an arbitrary state $\hat{\rho}$ is defined as
\begin{equation}
F(|\Psi\rangle,\hat{\rho})=\text{Tr}\left[\sqrt{\langle\Psi|\hat{\rho}|\Psi\rangle|\Psi\rangle\langle\Psi|}\right]=\sqrt{\langle\Psi|\hat{\rho}|\Psi\rangle} ,
\end{equation}
i.e., fidelity equals the square root of the overlap between $\Psi$ and $\hat{\rho}$. \\
Fidelity between two pure states $|\phi\rangle$ and $|\psi\rangle$ is defined as
\begin{equation}
F(|\phi\rangle, |\psi\rangle)=|\langle\phi|\psi\rangle| .
\end{equation}

It is symmetric in its inputs and a good distance measure between the quantum states. Some useful properties of the fidelity are:
\begin{enumerate}[label=(\roman*)]
\item $0 \leq F(\rho,\sigma)\leq 1$
\item $F(\rho,\sigma)=F(\sigma,\rho)$
\item $F(\rho,\rho)=1$
\item $F(\rho,\sigma)=1~\Rightarrow \rho=\sigma$
\item $F(\rho,\sigma)=0~\Rightarrow \rho \perp \sigma$
\end{enumerate}
In literature, some physicists choose to define Fidelity as the overlap squared. So, one should be careful and consistent with the fidelity definition.
 
\begin{itemize}
\item Given a bipartite entangled state $\hat{\rho}^{(AB)}$, how do we know whether it is entangled or not (separable)? How to quantify entanglement? What are the basic requirements for the entanglement measure? We will learn all these things in the next section.
\end{itemize}

\section{Quantum entanglement}
 To quote E. Schr\"{o}dinger who coined the term \textit{entanglement} to describe the strange connection between the quantum systems. 
 \begin{itemize}
 \item ``\textit{I would not call that one but rather the characteristic trait of quantum mechanics, the one that enforces its entire departure from classical lines of thought.}"
 \item ``\textit{The best possible knowledge of the whole does not include the best possible knowledge of its parts. By the interaction the two representatives [the quantum states] have become entangled.}"  
 \end{itemize}
 
\subsection{Separable and entangled states}

A bipartite state $\hat{\rho}^{(AB)}$ (whether pure or mixed) is said to be \textit{separable}, classically correlated or unentangled, if it can be prepared ``classically" by two parties; say Alice and Bob, by local operation and classical communication (LOCC) alone. The state created in this manner can contain only classical correlations. States which cannot be prepared by LOCC alone are said to be entangled.

Mathematically speaking, a bipartite \textit{pure state} $|\Psi\rangle_{AB}$ is said to be \textit{separable} if it can be written as the direct product of the states from the subsystem Hilbert spaces, i.e.,
\begin{equation}
|\Psi\rangle_{AB}=|\Psi\rangle_A\otimes|\Psi\rangle_B~,
\end{equation}
 where $|\psi\rangle_{AB}\in\mathcal{H}^{AB}=\mathcal{H}^A\otimes\mathcal{H}^B$, $|\psi\rangle_A\in\mathcal{H}^A$, and $|\psi\rangle_B\in\mathcal{H}^B$. The joint state $|\Psi\rangle_{AB}$ is said to be \textit{entangled} [\ref{1.15}] if it is not separable.


On the other hand, a bipartite \textit{mixed state} $\hat{\rho}^{(AB)}$ is said to be \textit{separable} if it can be written as the convex combination of the  pure product states from the subsystem Hilbert spaces, i.e.,
\begin{equation}
\hat{\rho}^{(AB)}=\sum_i p_i~\hat{\rho}_i^{(A)}\otimes\hat{\rho}_i^{(B)}.
\label{eq1.25}
\end{equation}
where $\hat{\rho}^{(AB)}\in\mathcal{H}^{AB}=\mathcal{H}^A\otimes\mathcal{H}^B$, $\hat{\rho}^{(A)}_i=|\psi_i\rangle_A\langle\psi_i|\in \mathcal{H}^A$, $\hat{\rho}^{(B)}_i=|\phi_i\rangle_B\langle\phi_i|\in \mathcal{H}^B$, $\sum_i p_i=1$ ($p_i$ are non-negative coefficients), and $|\psi_i\rangle_A$ and $|\phi_i\rangle_B$ denote the normalized states of subsystems A and B, respectively.

The above separability criterion is not operational, as it is very difficult to find a composition of the form (\ref{eq1.25}) or to prove that it doesn't exist. Therefore, we need operational criterion and measures of entanglement, discussed below.

\subsection{Basic requirements for an entanglement measure}

A measure of entanglement $E(\hat{\rho})$ must assign a non-negative value for each state $\hat{\rho}$. A good measure of entanglement $E(\hat{\rho}_{AB})$ is required to satisfy the following criterion [\ref{1.15}-\ref{1.17}]:

\begin{enumerate}
\item\textit{Separability:} For separable states, the measure of entanglement must vanish.
\begin{equation}
E(\hat{\rho}_{AB})=0~\forall ~\text{separable}~\hat{\rho}_{AB}~.
\end{equation}
\item \textit{Normalization:} The entanglement of a maximally entangled qudit-qudit system is given by
\begin{equation}
E(\hat{\rho}_{AB})=\log_2d.
\end{equation}
\item \textit{Invariant under unitary transformation:}  $E(\hat{\rho}_{AB})$ is invariant under all local unitary transformations $U_A\otimes U_B$. i.e., 
\begin{equation}
 E\left(\hat{\rho}_{AB}\right)=E\left((U_A\otimes U_B)\hat{\rho}_{AB}(U_A\otimes U_B)^\dag \right). 
\end{equation}
It should not depend on the choice of basis.
\item \textit{Non increasing under LOCC:} On an average, $E(\hat{\rho}_{AB})$ should not increase under local operation and classical communication (LOCC). \begin{equation}
E(\hat{\rho}_{AB})\geq E(\Theta(\hat{\rho}_{AB})),
\end{equation}
 where  $\Theta(\hat{\rho}_{AB})$ is completely positive trace preserving (CPTP) map.
\item \textit{Continuity:} If the distance between two density matrices is vanishingly small, the difference between their entanglement should vanish, i.e., 
\begin{equation}
E(\hat{\rho})-E(\hat{\sigma}) \rightarrow 0~,~\text{if}~||\hat{\rho}-\hat{\sigma}||\rightarrow 0.
\end{equation}
\item\textit{Additivity:} The entanglement contained in $n$ identical copies of a state $\rho$ equals $n$ times the entanglement of one copy.
\begin{equation}
E(\hat{\rho}^{\otimes n})=n E(\hat{\rho}).
\end{equation}
\item\textit{Sub-additivity:} The entanglement of the tensor product of the two states  $\hat{\rho}$ and $\hat{\sigma}$ should not exceed the sum of entanglement of each of the states.
\begin{equation}
E(\hat{\rho}\otimes\hat{\sigma})\leq E(\hat{\rho})+E(\hat{\sigma}).
\end{equation}
\item\textit{Convexity:} The entanglement measure should satisfy the convexity property, i.e. 
\begin{equation}
E(\lambda\hat{\rho}+(1-\lambda)\hat{\sigma})\leq \lambda E(\hat{\rho})+(1-\lambda)E(\hat{\sigma}).
\end{equation} 
\end{enumerate}

From the definition of entanglement, first condition is obvious: separable states are not entangled. Second and third conditions are essential for entanglement to be considered global property of the joint system and they render it impossible to create and distribute the entanglement via LOCC alone. It should be noted that whether all these requirements are necessary or not, is still an open question.

\subsection{Measures of entanglement}
Some measures that fulfil most of the above requirements are discussed below.

\subsubsection{(i) Schmidt decomposition}

Consider a $n\otimes m$ dimensional Hilbert space $\mathcal{H}_{AB}$ defined by Kronecker product of the two Hilbert spaces $\mathcal{H}_{A}$ and $\mathcal{H}_{B}$. For a \textit{pure state} $|\Psi\rangle_{AB}$ from $\mathcal{H}_{AB}$, there exists an orthonormal bases $\{|a_i\rangle,~i=1,2...n\}$ and $\{|b_i\rangle,~i=1,2...m\}$ in the spaces $\mathcal{H}_{A}$ and $\mathcal{H}_{B}$, respectively, such that
\begin{equation}
|\Psi\rangle_{AB}=\sum_{i=1}^d \sqrt{\lambda_i}|a_i\rangle\otimes|b_i\rangle~; ~d\leq \text{min}(n,m).
\end{equation}
where $\lambda_i$'s are non-negative coefficients known as \textit{Schmidt coefficients} satisfying $\sum_{i=1}^d \lambda_i=1$, and above decomposition is known as \textit{Schmidt decomposition} [\ref{1.01}]. The bases $|a_i\rangle$ and $|b_i\rangle$ are called Schmidt bases for A and B, respectively, and  the number of non-zero values of $\lambda_i$ is called Schmidt number for the state $|\psi\rangle_{AB}$. The values $\lambda_i$'s are precisely the features of the state $|\Psi\rangle_{AB}$ that remains intact when the subsystems are subjected to local unitary transformation.  For a pure state $|\Psi\rangle_{AB}$ the $\lambda_i$'s are equal to the eigenvalues of the density matrix of subsystem A, given by $\hat{\rho}_A=\text{Tr}_B[|\Psi_{AB}\rangle\langle\Psi_{AB}|]$. The state $|\Psi\rangle_{AB}$ is \textit{entangled} [\ref{1.01}] if it has more than one Schmidt coefficients and \textit{separable} iff there is only one Schmidt coefficient, i.e., $\lambda_i=1$.

\subsubsection{(ii) Entropy of entanglement}

Consider a bipartite pure state $|\Psi_{AB}\rangle$ with density matrix $\hat{\rho}_{AB}$ ($=|\Psi_{AB}\rangle\langle\Psi_{AB}|$) in the Hilbert space $H_{AB}$ of dimension $d_A\otimes d_B$. The quantum state of the individual subsystem is obtained by tracing out the other subsystems such that $\hat{\rho}_{i}=\text{Tr}_{j}[\hat{\rho}_{ij}]~;~i,j=A,B$. If the joint state $|\Psi_{AB}\rangle$ is entangled, then the marginal density operators will be mixed, implying the presence of quantum entanglement. Entropy of entanglement [\ref{1.18}] uses the Von Neumann entropy of density matrix, defined by
\begin{equation}
S(\hat{\rho})=-\text{Tr}[\hat{\rho}\log_2(\hat{\rho})]~=-\sum_i \lambda_i\log\lambda_i.
\end{equation}
where $\lambda_i$'s are the eigenvalues of the density matrix $\hat{\rho}$. For pure bipartite states, the entropy of entanglement is defined as the Von Neumann entropy of the reduced density matrices.
\begin{equation}
E(\hat{\rho}_{AB})=S(\hat{\rho}_A)=S(\hat{\rho}_B)~,
\end{equation}
where $\hat{\rho}_A=\text{Tr}_B[\hat{\rho}_{AB}]$, and $\hat{\rho}_B=\text{Tr}_A[\hat{\rho}_{AB}]$.

Note that von Neumann entropy is zero for a pure entangled state and non-zero for its reduced states which makes it a measure of entanglement. 

\subsubsection{(iii) Entanglement of formation}

Entanglement of formation [\ref{1.19}] is an information theoretic measure of entanglement for bipartite mixed states which quantifies the resource required to create the given state. Consider a bipartite mixed state density matrix $\hat{\rho}$ as given below.  
\begin{equation}
\hat{\rho}=\sum_i p_i |\psi_i\rangle\langle\psi_i|.
\end{equation}
Note that it can have infinitely many pure state decompositions.

Entanglement of formation ($\mathcal{E}$) of the state $\hat{\rho}$ is defined as the minimum average entanglement over all possible pure-state decompositions of $\hat{\rho}$.
\begin{equation}
\mathcal{E}(\hat{\rho})= \text{inf} \sum_i p_i ~\mathcal{E}(|\psi_i\rangle)~.
\end{equation}
For a maximally mixed state, entanglement of formation is trivially zero.

\subsubsection{(iv) Concurrence and Tangle}

Concurrence ($C$) [\ref{1.20}] is a measure of entanglement defined for $2\otimes 2$ and $2\otimes 3$ dimensional systems. For a pure state $|\Psi\rangle_{AB}$, concurrence is defined as 
\begin{equation}
C(|\Psi\rangle_{AB})=|\langle\Psi_{AB}|\tilde{\Psi}_{AB}\rangle| ,
\end{equation}
where $\tilde{\Psi}_{AB}=\sigma_y\otimes \sigma_y|\Psi\rangle_{AB}$ is referred as spin flipped state vector and $\sigma_y$ is the second Pauli matrix. 

The concurrence of a mixed entangled state $\hat{\rho}_{AB}$ is defined as 
\begin{equation}
C(\hat{\rho}_{AB})= \text{Max}\{0,\Lambda\},
\end{equation}
where $\Lambda=\sqrt{\lambda_1}-\sqrt{\lambda_2}-\sqrt{\lambda_3}-\sqrt{\lambda_4}$, and $\lambda_i's$ are the positive eigenvalues of the auxiliary matrix $\tilde{\rho}_{AB}$ defined by
\begin{equation}
\tilde{\rho}_{AB}=\hat{\rho}_{AB}(\sigma_y\otimes\sigma_y)\hat{\rho}^*_{AB}(\sigma_y\otimes\sigma_y).
\end{equation}

The complex conjugation is done in the computational basis: $|00\rangle,|01\rangle,|10\rangle,|11\rangle$. Tangle ($\tau$) is defined as the square of concurrence, i.e.,
\begin{equation}
\tau= [C(\hat{\rho}_{AB})]^2
\end{equation}

Initially, interest in Concurrence started due to its connection to the entanglement of formation ($\mathcal{E}$) as given below
\begin{equation}
\mathcal{E(C(\rho))}:=h\left( \frac{1+\sqrt{1+C^2}}{2} \right),
\end{equation}
where $h(x)=-x\log_2(x)-(1-x)\log_2(1-x)$.

\subsubsection{(v) Peres-Horodecki criterion and Negativity} \label{sec1.4.3}

 The Peres-Horodecki criterion for separability is also known as positive partial transpose (PPT) criterion [\ref{1.21}]. It is necessary condition for separability of pure as well as mixed states. It turns out that for $2\otimes2$ and $2\otimes3$ dimensional systems, the PPT criterion is also sufficient condition for separability. 
 
 When a positive map is applied to one of the subsystems of a composite quantum system in a separable state, it always maps to a valid quantum state. However, when the composite state is entangled, the same positive map, in general, does not results in a valid density operator. This happens because a positive operator does not behave like a completely positive operator in the presence of quantum entanglement in the system. The partial transposition operator is such an example of a positive but not completely positive operator. 

Let the composite system density matrix be written as
\begin{equation}
\rho_{AB}=\sum_{ijkl}p_{ij}^{kl}|i_Ak_B\rangle\langle j_Al_B| ,
\end{equation}
where $\{|i_A\rangle\}$, and $\{|k_B\rangle\}$ form the orthonormal basis set for the subsystem `A' and `B', respectively.
Then the partial transposed density matrix, with respect to the subsystem $`A'$, is given by
\begin{equation}
\rho^{T_A}_{AB}=\sum_{ijkl}p_{ij}^{kl}|j_Ak_B\rangle\langle i_Al_B| .
\end{equation}
If the state $\hat{\rho}_{AB}$ is \textit{separable} then above Eq.~(1.44) can be rewritten as
\begin{equation}
\hat{\rho}_{AB}^{T_A}=\sum_{ijkl} p_{ij}^{kl}|j_A\rangle\langle i_A| \otimes |k_B\rangle\langle l_B|=\sum_{ijkl} p_{ij}^{kl}(|i_A\rangle\langle j_A|)^{\text{T}} \otimes |k_B\rangle\langle l_B|.
\end{equation}

Since $(|i_A\rangle\langle j_A|)^{\text{T}}$ are again valid density matrices for Alice, it implies that  $\hat{\rho}_{AB}^{T_A}\geq 0$. Same argument can be given for partial transposition with respect to Bob's subsystem. In short, the partial transpose of a bipartite separable state 
$\hat{\rho}_{AB}$ with respect to any subsystem is positive. \\

\textit{\textit{Negativity} is a measure of entanglement based on the PPT criteria of separability and it is defined as the sum of absolute values of all the negative eigenvalues of the partially transposed density matrix with respect to one of its subsystems.}

Note that PPT criterion is necessary but not sufficient condition for separability in higher-dimensional ($d\otimes D$, $d, D\geq 3$) systems.  If a higher-dimensional system is found to be PPT, we can't say whether state is entangled or separable, and in that case we can use matrix realignment method (discussed below) for detecting some of the bound entangled states which may not be detected by PPT criterion.

\subsubsection{(vi) Matrix realignment method for detecting entanglement}
This criterion for inseparability of bipartite systems of arbitrary dimensions is motivated by Kronecker product approximation technique for density matrices. It is based on a realigned matrix [\ref{1.22}] obtained from the density matrix via realignment method as discussed below.

Consider a matrix $A=[a_{ij}]$ having dimension $m\otimes n$, where $a_{ij}$ is the $i$th row and $j$th column entry of this matrix. Let us define $vec(A)$ as
\begin{equation}
vec(A)=[a_{11},.....a_{m1},.....a_{m2},......a_{1n},......a_{mn}]^T~,
\end{equation}
where symbol $T$ denotes transpose. 

Now, consider a block matrix B of dimension $m\times m$ with each block of size $n\times n$. Let us define the realigned matrix $\tilde{B}$ of dimension $m^2\times n^2$ that contains the same elements as in matrix as $B$ but in a rearranged position as given below.
\begin{equation}
\tilde{B}=\left[\begin{array}{c}
vec(B_{1,1})^T\\
\vdots\\
vec(B_{m,1})^T\\
\vdots\\
vec(B_{1,n})^T\\
\vdots\\
vec(B_{m,n})^T\\
\end{array}\right]~,
\end{equation}
where $B_{i,j}$ denotes the block in the $i$th row and $j$th column.

The Singular value decomposition of B is given by
\begin{equation}
\tilde{B}=U D V^\dag =\sum_{i=1}^q d_i u_i v_i^\dag
\end{equation}
where D is a diagonal matrix with elements $d_1\geq d_2\geq d_2\geq.....d_q\geq 0$ with $q=\text{min}(m^2,~n^2)$, and $U={u_1,~u_2.......u_{m^2}}\in \mathcal{C}^{m^2\times m^2}$ \& $V={v_1,~v_2.......v_{m^2}}\in \mathcal{C}^{n^2\times n^2}$ are unitary matrices. 

where $d_is$ are given by non-negative square roots of the eigenvalues of the matrix $\tilde{B}\tilde{B}^\dag$ or $\tilde{B}^\dag\tilde{B}$.
In fact, rank of the matrix $\tilde{B}$ is equal to number of non-zero singular values $d_i$.  Inspired by this construction, Loan and Pitsianis gave the following decomposition for $B$.
\begin{equation}
B=\sum_{i=1}^r X_i\otimes Y_i~,
\end{equation}
where $vec(X_i)=\sqrt{d_i}u_i$ and $vec(Y_i)=\sqrt{d_i}v_i^*$.

For a given density matrix $\rho_{AB}$, we can write the realigned matrix $\tilde{\rho}_{AB}$ according to the transformation given in Eq.~(1.48). As an example, a bipartite $2\otimes 2$ density matrix $\rho_{AB}$ can be transformed as follows:
\begin{equation}
\rho_{AB}=\left(\begin{array}{cc|cc}
\rho_{11} & \rho_{12} & \rho_{13} & \rho_{14}\\
\rho_{21} & \rho_{22} & \rho_{23} & \rho_{24}\\
\hline
\rho_{31} & \rho_{32} & \rho_{33} & \rho_{34}\\
\rho_{41} & \rho_{42} & \rho_{43} & \rho_{44}\\
\end{array}\right)
\Rightarrow \tilde{\rho}_{AB} =\left(\begin{array}{cccc}
\rho_{11} & \rho_{21} & \rho_{12} & \rho_{22}\\
\hline
\rho_{31} & \rho_{41} & \rho_{32} & \rho_{42}\\
\hline
\rho_{13} & \rho_{23} & \rho_{14} & \rho_{24}\\
\hline
\rho_{33} & \rho_{43} & \rho_{34} & \rho_{44}\\
\end{array}\right)
\end{equation}

Realigned Negativity is then defined as
\begin{equation}
R(\rho_{AB})= \text{max}(0,~||\tilde{\rho}_{AB}||-1)
\end{equation} 
where symbol $||.||$ denotes trace norm.

This criteria is motivated by the fact that for any separable state, trace norm of the realigned matrix will not be greater than one. Hence, If $R(\rho)$ turns out to be greater than zero, state must be entangled. This criterion can detect some of the bound entangled states in higher-dimensional system which may not be detected by PPT criterion.

\begin{itemize}
\item Given a quantum system $\rho(0)$ at time $t=0$. How does it evolve with time in the presence of different environments? We will learn this in the next section.
\end{itemize}
 
\section{Time evolution of the quantum systems}
In this section, we introduce the minimum basic formalism required to understand and mathematically describe decoherence in the framework of quantum theory. A \textit{closed quantum system} is one which neither interchanges energy nor matter  with its environment. On the other hand, an \textit{open quantum system} interchanges  energy or matter or both with its environment.
 
\subsection{Evolution of the closed quantum systems}
 Evolution of a closed quantum system [\ref{1.01}] in the pure state $|\psi(t)\rangle$ is given by Schr$\ddot{o}$dinger equation:
 \begin{equation}
 \frac{d}{dt}|\psi(t)\rangle=-\frac{i}{\hbar}H(t)|\psi(t)\rangle ,
 \label{eq1.53}
\end{equation}  
where $H(t)$ is the Hamiltonian of the system.

Evolution of the states under Schr$\ddot{o}$dinger equation is unitary and governed by a unitary time-evolution operator $U(t,t')$ such that it takes the state $|\psi(t')\rangle$ to $|\psi(t)\rangle$ as 
\begin{equation}
|\psi(t)\rangle=U(t,t')|\psi(t')\rangle ,
\end{equation}
where $U(t',t')=\mathbb{I}$.

Substituting $|\psi(t)\rangle$ into Eq.~(\ref{eq1.53}), we get
\begin{equation}
\frac{\partial}{\partial t}U(t,t') =-\frac{i}{\hbar}H(t)U(t,t').
\end{equation}

For an isolated system (which is not driven by any external agency) Hamiltonian becomes time independent and Unitary operator can be written as
\begin{equation}
U(t,t')=exp[-\frac{i}{\hbar}H(t-t')].
\label{eq1.56}
\end{equation}

If the initial state of system is a statistical mixture then it is characterized by a density matrix $\rho$ instead of a state vector. Let the initial state be given by
\begin{equation}
\rho(t')=\sum_i p_i |\psi(t')\rangle\langle\psi(t')| .
\end{equation}

Each state vector of the ensemble evolves as per Eq.~(\ref{eq1.56}) and thus the state of the system at a later time 't' is given by
\begin{equation}
\begin{split}
\rho(t)&=\sum_i p_i U(t,t')|\psi(t')\rangle\langle\psi(t')|U^\dag (t,t'),\\
& =  U(t,t')\rho(t') U^\dag (t,t').
\end{split}
\label{eq1.58}
\end{equation}

Upon differentiating the above eqn. wrt. time, we get the eqn. of motion as
\begin{equation}
\frac{d}{dt}\rho(t)=-\frac{i}{\hbar}[H(t),\rho(t)].
\end{equation}
This is known as \textit{Liouville - von Neumann equation}. It gives the evolution of a closed quantum system $\rho$ with time.

\subsection{Evolution of the open quantum systems}
\begin{figure}[htb]
\centering
\begin{tikzpicture}
\draw[red,fill=gray,opacity=0.35] (0,0) ellipse (4cm and 3cm);
\draw[blue,fill=cyan,opacity=0.35] (0,0) ellipse (2.5cm and 1.5cm);
\draw[draw=red!60!black,line width=6pt,stealth-stealth] (1.5,0) -- (3.5,0) node[midway,text=white,font=\footnotesize\bfseries,sloped]{};
\draw[draw=red!60!black,line width=6pt,stealth-stealth] (-1.5,0) -- (-3.5,0) node[midway,text=white,font=\footnotesize\bfseries,sloped]{};
\node at (0,0.25) {($S,\mathcal{H}_S,\rho_S$)};
\node at (0,-0.5) {System};
\node at (0,1.9) {($E,\mathcal{H}_E,\rho_E$)};
\node at (0,2.4) {Environment};
\node at (0,3.4) {($S+E,\mathcal{H}_S\otimes \mathcal{H}_E,\rho$)};
\end{tikzpicture}
\caption{\textit{Schematic diagram of an open quantum system.}}\label{fig1.03}
\end{figure}
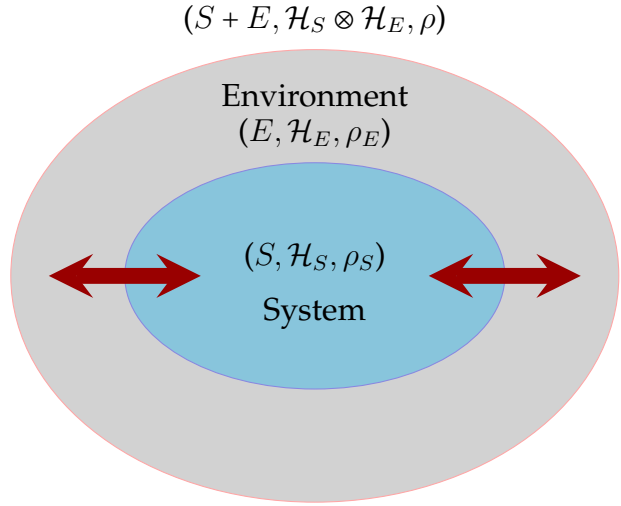

Let us consider a quantum system `S' denoted by density matrix $\rho_S$ in the Hilbert space $\mathcal{H}_S$ interacting with another quantum system `E', called environment, having density matrix $\rho_E$ in the Hilbert space $\mathcal{H}_E$. The state of the joint system; System + Environment (S+E), is given by the tensor product $\rho_S\otimes \rho_E$ which resides in the joint Hilbert space $\mathcal{H}_S\otimes \mathcal{H}_E$. The schematic of a quantum system interacting with its environment is shown in the Fig.~\ref{fig1.03}.

The joint system (S+E) behaves as a closed quantum system and follows the Hamiltonian dynamics, as discussed above. During the evolution, system and environment develop some kind of correlation and consequently, if we consider the evolution of the system alone, it becomes non-unitary and behaves as an open quantum system [\ref{1.01}, \ref{1.02}]. The Hamiltonian of the joint quantum system can be written as
\begin{equation}
H(t)=H_S\otimes \mathbb{I}_E+\mathbb{I}_S\otimes H_E + H_{int}(t),
\end{equation}
where $H_S$ and $H_E$ are the system and environment Hamiltonian, respectively, and $H_{int}(t)$ is the interaction Hamiltonian acting on the Hilbert space  $\mathcal{H}_S\otimes \mathcal{H}_E$. The $\mathbb{I}_S$ and $\mathbb{I}_E$ are the identity matrices in $\mathcal{H}_S$ and $\mathcal{H}_E$, respectively.

The reduced density matrix of the system ($\rho_s$) can be found by tracing over the degrees of the freedom of the environment from the joint system density matrix [\ref{1.01}, \ref{1.02}] as follows:
\begin{equation}
\rho_s = \text{Tr}_E(\rho).
\end{equation}

Substituting $\rho(t)$ from Eq.~(\ref{eq1.58}), we get
\begin{equation}
\rho_S(t)=\text{Tr}_E\left(U(t,t')\rho(t') U^\dag (t,t')\right).
\end{equation}
Above equation gives the evolution of an open quantum system. To illustrate this further, we consider a specific noise model of Amplitude Damping Channel (ADC) and discuss the time evolution of a two-level system using Kraus operator formalism.

\subsubsection{Amplitude damping channel}
Amplitude damping channel [\ref{1.01}, \ref{1.02}] is a noise model that captures the dissipative interaction of a quantum system with its environment. For example, spontaneous decay of the excited state of a two-level atomic system resulting in emission of a photon. Consider a two-level systems `S' in a cavity interacting with the vacuum mode of the electromagnetic field at absolute zero temperature. Initially, electromagnetic field (which is treated as environment `E') is assumed to be in the vacuum state. Let us label the ground and excited states of the system by $|g\rangle_S$ and $|e\rangle_S$, respectively, and label $|0\rangle_E$ and $|1\rangle_E$ as the vacuum state and the one photon state of the environment, respectively. 

Evolution of the system and environment is unitary and represented by following quantum map:
\begin{equation}
\begin{aligned}
\begin{split}
|g\rangle_S|0\rangle_E & \rightarrow |g\rangle_S|0\rangle_E\\ 
|e\rangle_S|0\rangle_E & \rightarrow \sqrt{1-p}|e\rangle_S|0\rangle_E+\sqrt{p}|g\rangle_S|1\rangle_E
\end{split}
\end{aligned}
\end{equation}

For an atom initially in excited state, there is a probability $(1-p)$ that the system remains in the excited state environment in vacuum state and there is probability $p$ that system de-excites to ground state with an excitation in the environment. Under the Born-Markov approximation; $p=1-\exp(-\Gamma t)$, where $t$ denotes time and $\Gamma$ is the decay rate of the  two-level atomic system.

Thus, a superposition of ground and excited states of the two-level atomic system evolves with the environment as follows:
\begin{equation}
(\alpha |g\rangle_S+\beta |e\rangle_S)|0\rangle_E\xrightarrow {U_{SE}} \alpha |g\rangle_S|0\rangle_E+\beta\left[\sqrt{1-p}|e\rangle_S|0\rangle_E+\sqrt{p}|g\rangle_S|1\rangle_E\right]
\end{equation}

By tracing out the environment in the $\{|0\rangle_E,~|1\rangle_E\}$ bases, we get the Kraus operators as follows:
\begin{equation}
K_\mu =~_E\langle\mu|U_{SE}|0\rangle_E,~~~\mu=0,~1.
\end{equation}
\begin{equation}
\Rightarrow ~ K_0=\left(\begin{array}{cc}
1 & 0\\
0 & \sqrt{1-p}
\end{array} \right),~
K_1=\left(\begin{array}{cc}
0 & \sqrt{p}\\
0 & 0
\end{array} \right).
\end{equation}

The Kraus operator $K_1$ governs the decay of the system from $|e\rangle_S$ to $|g\rangle_S$ and $K_0$ governs  the evolution when there is no decay. These operators are non-unitary operators and satisfy the completeness condition
\begin{equation}
\sum_i K^\dag_i K_i=\mathbb{I}
\end{equation}

Evolution of the system $\rho_S(0)$ is given by 
\begin{equation}
\rho_S(p)=\sum_{i=0}^1 K_i \rho_S(0) K^\dag_i,
\label{eq1.68}
\end{equation}

The Eq.~(\ref{eq1.68}) governs the evolution of a system $\rho_S(0)$ with time in the presence of an ADC.

\begin{itemize}
\item Let us learn some basics of birefringence and non-linearity.
Birefringence plays a very important role in most of the polarization dependent optical components such as half- and quarter-wave plates, polarizer, polarizing beam splitters, and in the phase matching of non-linear crystals used for spontaneous parametric down-conversion, etc.
\end{itemize}

\section{Basics of birefringence and nonlinear optics}
\subsection{Anisotropy and birefringence}
When an electric field $\textbf{E}$ is applied to a dielectric media, it causes the separation between positive and negative charge centres creating electric dipoles. The net dipole moment $\textit{P}$, electric field $\textit{E}$, and electric displacement vector $\textbf{D}$ are related as
\begin{equation}
\textbf{D}=\epsilon_0 \textbf{E} + \textbf{P}.
\end{equation}
When dipoles are always oriented along the direction of applied electric field, displacement vector $\textbf{D}$ is parallel to the applied electric field $\textbf{E}$, medium is called \textit{isotropic} [\ref{1.23}]. This is the case with the media which has same properties in all the directions.

In an isotropic media, we have
\begin{equation}
\textbf{P}=\epsilon_0\chi \textbf{E} \Rightarrow \textbf{P} \parallel \textbf{E}.
\end{equation} 
Thus,
 \begin{equation} 
 \textbf{D}=\epsilon \textbf{E} \Rightarrow \textbf{D} \parallel \textbf{E}.
 \end{equation}

where $\epsilon=\epsilon_0(1+\chi)$ is a scalar quantity, called dielectric permittivity, and $\chi$ is electric susceptibility.

 But, if the structure is not symmetric then dipoles may orient themselves in a direction other than field $\textbf{E}$ due to restoring forces from all the directions. As a result $\textbf{P} \nparallel \textbf{E}$. Such a medium is called \textit{anisotropic}. In such a medium, we have
 
\begin{equation}
\begin{split}
D_x=\epsilon_{xx}E_x+\epsilon_{xy}E_y+\epsilon_{xz}E_z,\\
D_y=\epsilon_{yx}E_x+\epsilon_{yy}E_y+\epsilon_{yz}E_z,\\
D_z=\epsilon_{zx}E_x+\epsilon_{zy}E_y+\epsilon_{zz}E_z.
\end{split}
\end{equation}
Or,
\begin{equation}
\left(\begin{array}{ccc}
 D_x\\D_y\\D_z \end{array} \right) 
 = \left(\begin{array}{ccc}
 \epsilon_{xx} & \epsilon_{xy} & \epsilon_{xz}\\
 \epsilon_{yx} & \epsilon_{yy} & \epsilon_{yz}\\
 \epsilon_{zx} & \epsilon_{zy} & \epsilon_{zz}  \end{array} \right)
  \left(\begin{array}{ccc}
 E_x\\E_y\\E_z \end{array} \right).
\end{equation}
Or, 
\begin{equation}
\textbf{D}=\bar{\bar{\epsilon}} \textbf{ E}.
\end{equation}
 In component form,
 \begin{equation}
 D_i=\sum_j \epsilon_{ij}E_j~;~i,j=1,2,3.
 \end{equation}
It can be shown that the $\bar{\bar{\epsilon}}$ tensor is symmetric; i.e,
\begin{equation}
\epsilon_{xy}=\epsilon_{yx},~\epsilon_{yz}=\epsilon_{zy},~\epsilon_{xz}=\epsilon_{zx}.
\end{equation}
One can rotate the coordinate system such that the $\epsilon$ matrix becomes diagonal, this coordinate system is known as \textit{principal axis coordinate system} [\ref{1.23}] and $\bar{\bar{\epsilon}}$ takes the following form: 
\begin{equation}
\bar{\bar{\epsilon}}=\left( \begin{array}{ccc}
\epsilon_{xx} & 0 & 0\\
0 & \epsilon_{yy} & 0\\
0 & 0 & \epsilon_{zz} 
\end{array} \right).
\end{equation}

In the principal axis system, if we rewrite the diagonal elements of $\bar{\bar{\epsilon}}$ tensor as
\begin{equation}
\begin{aligned}
\begin{split}
& \epsilon_x, ~&& \epsilon_y, ~&& \epsilon_z  ~~ && \text{:  Principal permittivity}\\
\textrm{then}\\
& k_x=\frac{\epsilon_x}{\epsilon_0}, ~&& k_y=\frac{\epsilon_y}{\epsilon_0}, ~&& k_z=\frac{\epsilon_z}{\epsilon_0} ~~ &&  \text{: Dielectric constant} \\
& n_x=\sqrt{k_x},~&& n_y=\sqrt{k_y},~&& n_z=\sqrt{k_z} ~~ &&  \text{: Refractive indices}.
\end{split}
\end{aligned}
\end{equation}

\textbf{Case(I):} If $\epsilon_{x}=\epsilon_{y}=\epsilon_{z}$, then medium is said to be \textit{isotropic}, for example: glass.

\textbf{Case(II):} If $\epsilon_{x}=\epsilon_{y}\neq \epsilon_{z}$, then medium is said to be \textit{uniaxial anisotropic medium}, for example: quartz.

\textbf{Case(III):} If $\epsilon_{x}\neq \epsilon_{y}\neq \epsilon_{z}$, then medium is said to be \textit{biaxial anisotropic medium}, for example: mica.

\subsection{Non-linearity}
In the linear optics, the response of a medium to an applied electric field $\textbf{E}$ is given by
\begin{equation}
\textbf{P}=\epsilon_0 \bar{\bar{\chi}}\textbf{E}~.
\end{equation}

This is known as \textit{linear approximation} [\ref{1.23}]. When the applied $\textbf{E}$ field becomes very large compared to the inter-atomic fields in the medium then linear approximation fails. The oscillators are no more harmonic, and induced polarization is given by
\begin{equation}
P_i=\epsilon_0\chi^{(1)}_{ij}E_j+\epsilon_0\chi^{(2)}_{ijk}E_jE_k+\epsilon_0\chi^{(3)}_{ijkl}E_jE_kE_l+.....
\end{equation}

where $\chi^{(1)}_{ij}$, $\chi^{(2)}_{ijk}$, and $\chi^{(3)}_{ijkl}$ are linear, second-, and third-order nonlinear optical susceptibilities, respectively. The $\chi^{(2)}_{ijk}$ plays a very important role in many nonlinear optical phenomena, e.g. second harmonic generation, parametric generation, etc. It has 27 elements and if a medium possesses center of inversion symmetry then all the 27 elements vanish and such a crystal doesn't exhibit nonlinear effects.

\begin{itemize}
\item We will now discuss some basic concepts related to Spontaneous Parametric Down-Conversion and how phase matching is achieved in a birefringent medium. This process is an important tool for the generation of single and entangled photons.
\end{itemize}

\section{Spontaneous Parametric Down-Conversion}

Generally speaking, Spontaneous Parametric Down-Conversion (SPDC) [\ref{1.24}, \ref{1.25}] is the process by which a parent photon decays into arbitrary number of lower energy daughter photons. For our purposes, we identify SPDC as a second order non-linear optical process wherein a high energy \textit{pump} photon ($\omega_p$), in the presence of a non-linear medium, spontaneously splits into two lower energy photons; historically known as \textit{signal} (photon with higher frequency, $\omega_s$), and \textit{idler} (photon with lower frequency, $\omega_i$), due to scattering by the zero point fluctuation of vacuum. This process is known as SPDC because it is generated by quantum vacuum fields (spontaneous), initial and final quantum mechanical states of the medium are identical and photon energy is always conserved (parametric), and signal and idler frequencies are lower than the pump frequency (down-conversion). It is also known as spontaneous parametric scattering (SPS), and optical parametric generation (OPG). An example of type-II SPDC process (discussed below) resulting in intersecting cones is shown below.

The SPDC is a three-wave mixing process which utilizes the second-order non-linear susceptibility, $\chi^{(2)}$. It satisfies energy and momentum conservation, collectively known as phase matching condition (discussed below). If the energies of the signal and idler photons are equal (different), the process is known as degenerate (non-degenerate) SPDC. The daughter photons may be correlated in energy, momentum, time and spatial degrees of freedom. Generally, these photons posses only classical correlations and special techniques need to be employed to make them entangled.

\subsection{Phase matching technique}

Consider a non-centro-symmetric uniaxial birefringent crystal. For efficient SPDC process, it is required that the phase velocities of all the interacting waves (pump, signal, and idler) are equal, so that the intensity of down-converted signal can coherently build up throughout the crystal at the expense of pump beam. If the phase velocities of the pump, signal, and idler were different then the energies would be exchanged back and forth between them, resulting in inefficient SPDC.

The frequencies and wave vectors of the three-wave interaction is governed by the following energy and momentum conservation laws, known as \textit{phase matching condition} [\ref{1.24}-\ref{1.26}]:
\begin{subequations}
\begin{equation}
\hbar \omega_p = \hbar \omega_s+\hbar \omega_i~,
\end{equation}
\begin{equation}
\textbf{k}_p = \textbf{k}_s+\textbf{k}_i + \Delta \textbf{k}.
\label{eq1.81b}
\end{equation}
\end{subequations}

where $\textbf{k}_p$, $\textbf{k}_s$, and $\textbf{k}_i$ are the $\textbf{k}$ vectors corresponding the frequencies $\omega_p$, $\omega_s$, and $\omega_i$ of the pump, signal, and idler photons respectively, and $\Delta \textbf{k}$ is the phase mismatch term.

\begin{figure}[h!]
\centering
\subfigure[Feynman diagram] 
{ 
\begin{tikzpicture}
\begin{feynman}
\vertex (a);
\vertex [right=of a] (b);
\vertex [above right=of b] (f1);
\vertex [below right=of b] (c);
\diagram* {
(a) -- [boson,blue,thick] (b) -- [boson,red,thick,edge label=\(\gamma_s(\omega_s)\)] (f1),
(b) -- [boson, red,thick, edge label'=\(\gamma_i(\omega_i)\)] (c),};
\end{feynman}
\draw [->,>=stealth] (-0.25,-1.6) -- (2.75,-1.6);
\node at (1.25,-1.4) {\scriptsize{Time}};
\node at (0,0.3) {\color{blue}{$\gamma_p(\omega_p)$}};
\end{tikzpicture}}
\hspace{1.5cm}
\subfigure[Energy conservation]
{
 \begin{tikzpicture}
\draw[line width=0.5mm, black] (10,0)-- (14,0);
\draw[line width=0.5mm, blue] (10,3)-- (14,3);
\draw[line width=0.5mm, blue, dashed] (12,1.5)-- (14,1.5);
\draw[line width=0.4mm, blue, arrows={-Triangle[angle=90:5pt,blue,fill=gray]}](11,0)-- (11,3);
\draw[line width=0.4mm, red, arrows={-Triangle[angle=90:5pt,blue,fill=gray]}](13,3)-- (13,1.5);
\draw[line width=0.4mm, red, arrows={-Triangle[angle=90:5pt,blue,fill=gray]}](13,1.5)-- (13,0);
\node at (10.5,1.5) {\color{blue}$\omega_p$};
\node at (12.5,2.25) {\color{red}$\omega_s$};
\node at (12.5,0.75) {\color{red}$\omega_i$};
\end{tikzpicture}}
\subfigure[Momentum conservation] 
{ 
\begin{tikzpicture}[scale=1]
 \draw[line width=0.4mm, blue, arrows={-Triangle[angle=90:5pt,blue,fill=gray]}](10,0)-- (18,0);
 \draw[line width=0.4mm, red, arrows={-Triangle[angle=90:5pt,red,fill=gray]}](10,0)-- (14,1.5);
 \draw[line width=0.4mm, red, arrows={-Triangle[angle=90:5pt,red,fill=gray]}](14,1.495)-- (17,0.4);
   \draw[line width=0.4mm, black, arrows={-Triangle[angle=90:5pt,black,fill=gray]}](17,0.4)--(18,0) ;
   
\node at (14,-0.5) {\color{blue}$\textbf{k}_p$};
\node at (12,1.2) {\color{red}$\textbf{k}_s$};
\node at (15.5,1.3) {\color{red}$\textbf{k}_i$};
\node at (17.7,0.5) {$\Delta\textbf{k}$};
\end{tikzpicture}}
\caption{\textit{Non-collinear phase matching condition for the  SPDC process where a pump photon having energy $\hbar\omega_p$ gets absorbed and two lower energy photons $\hbar\omega_s$ (signal) and $\hbar\omega_i$ (idler) get emitted. The $\mathbf{k_p,~k_s}$ and $\mathbf{k_i}$ are the momentum vectors of the pump, signal, and idler, respectively.}\label{fig1.04}}
\end{figure}
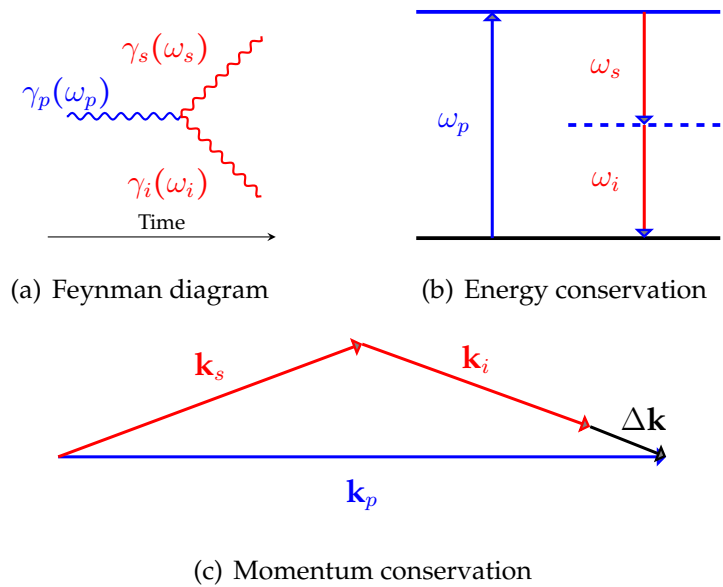

Phase matching is achieved whenever  $\Delta \textbf{k}=0$. When all $\textbf{k}$ vectors are parallel, it is called \textit{collinear phase matching} and when the $\textbf{k}$ vectors are non-parallel, it is called non-collinear phase matching. An example of \textit{non-collinear phase matching} is shown in Fig.~\ref{fig1.04}.

Now, 
\begin{equation}
\textbf{k}=\frac{ n \omega}{c} ~\hat{i}~.
\label{eq1.82}
\end{equation}

where, n is the refractive index at frequency $\omega$, $c$ is the velocity of light in vacuum, and $~\hat{i}$ is the unit vector parallel to $\textbf{k}$.

Using Eq.~(\ref{eq1.82}), under phase matching condition, Eq.~(\ref{eq1.81b}) can be rewritten as
\begin{equation}
n_p \omega_p \hat{i}_p = n_s \omega_s \hat{i}_s + n_i \omega_i \hat{i}_i . \label{eq1.83}
\end{equation}

If all the three waves propagate with parallel wave normals in the crystal, i.e. $\hat{i}_p=\hat{i}_s=\hat{i}_i$, then Eq.~(\ref{eq1.83}) reduces to
\begin{equation}
n_p=\frac{\omega_s}{\omega_p} n_s + \frac{\omega_i}{\omega_p} n_i~.
\label{eq1.84}
\end{equation}

 There are two different ways in which phase matching condition can be achieved: type-I and type-II phase matching.

\textbf{1) Type-I phase matching:} The signal and idler, both have same polarization, i.e. either both are $o$-rays or both $e$-rays, but orthogonal to pump polarization. In the case of a \textit{negative uniaxial crystal ($n_o>n_e$)}, an extra-ordinary polarized high energy pump photon down-converts into two ordinary polarized low energy photons. 
\begin{equation}
\textbf{k}^e_p(\theta) \rightarrow \textbf{k}^o_s + \textbf{k}^o_i .
\end{equation} 

where $\theta$ is the angle between the pump propagation direction and the  optic axis of the crystal, inside it. Whereas, in the case of a \textit{positive uniaxial crystal ($n_o<n_e$)}, an ordinary polarized high energy pump photon down-converts into two extra-ordinary polarized low energy photons. 

\begin{equation}
\textbf{k}^o_p \rightarrow \textbf{k}^e_s(\theta) + \textbf{k}^e_i(\theta) .
\end{equation} 

\begin{figure} [H]
\centering
\includegraphics[width=0.9\textwidth]{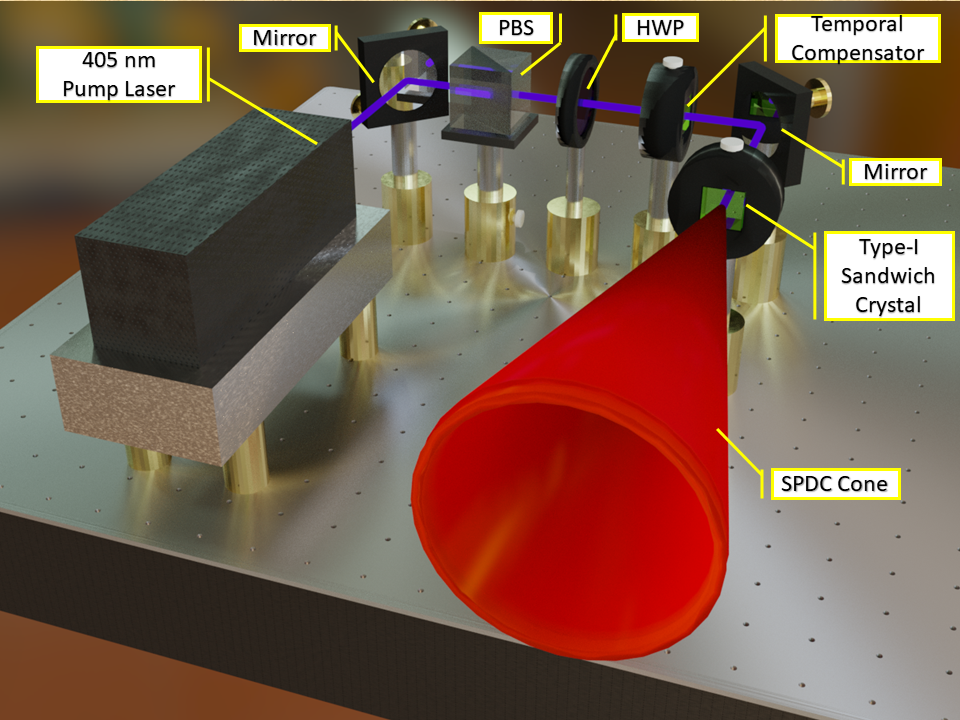}
\caption{\textit{Schemtaic of a type-I non-collinear SPDC process resulting in concentric  cones. A pair of photons collected from diamtrically opposite sides of the ring will be polarizatiion entangled. (Image Credit: S. N. Sahoo)}\label{fig1.05}}
\end{figure}

\begin{figure}[H]
\begin{center}
\includegraphics[width=0.9\textwidth]{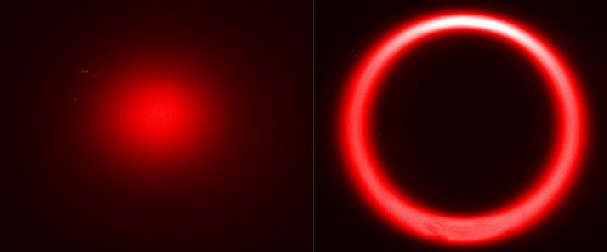}
\caption{\textit{Type-I SPDC ring in collinear (left) and non-collinear (right) configurations imaged by Andor CMOS camera.}\label{fig1.06}}
\end{center}
\end{figure}

As an example, schematic of a type-I SPDC source in paired-crystal geometry (more details can be found in the next chapter [\textcolor{darkblue}{2}]) resulting in a pair of concentric overlapping cones is shown in Fig.~\ref{fig1.05}. The cross section of such a SPDC cone is a ring in the non-collinear configuration and a Gaussian spot in the collinear configuration as shown in Fig.~\ref{fig1.06}. These images were captured from SPDC sources in our lab by Andor CMOS camera.

\textbf{2) Type-II phase matching:} The signal and idler have orthogonal polarizations, i.e., if one is $e$-ray, other is $o$-ray. In the case of a \textit{negative uniaxial crystal}, an extra ordinary polarized high energy pump photon down-converts into one ordinary and one extra-ordinary polarized low energy photons. 
\begin{equation}
\textbf{k}^e_p(\theta) \rightarrow \textbf{k}^e_s(\theta) + \textbf{k}^o_i .
\end{equation} 

Whereas, in the case of a \textit{positive uniaxial crystal}, an ordinary polarized high energy pump photon down-converts into one ordinary and one extra-ordinary polarized low energy photons. 
\begin{equation}
\textbf{k}^e_p(\theta) \rightarrow \textbf{k}^o_s + \textbf{k}^e_i(\theta) .
\end{equation} 

\begin{figure} [H]
\centering
\includegraphics[width=0.9\textwidth]{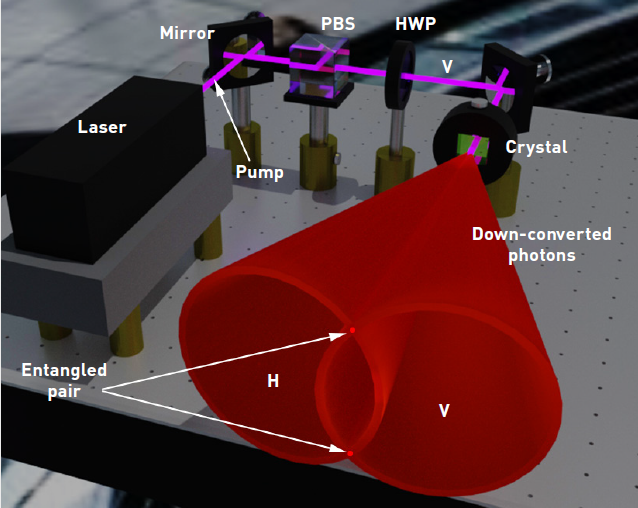}
\caption{\textit{Schematic of type-II non-collinear SPDC process resulting in intersecting cones [\ref{1.27}].}\label{fig1.07}}
\end{figure}

As an example, schematic of a type-II SPDC source resulting in intersecting cones is shown in Fig.~\ref{fig1.07}. The cross section of such a SPDC cone is  a pair of  intersecting rings in the non-collinear configuration and rings just touch each other in the collinear configuration as shown in Fig.~\ref{1.08}. These images were captured from SPDC sources in our lab by Andor CMOS camera.

\begin{figure}[H]
\begin{center}
\includegraphics[width=0.9\textwidth]{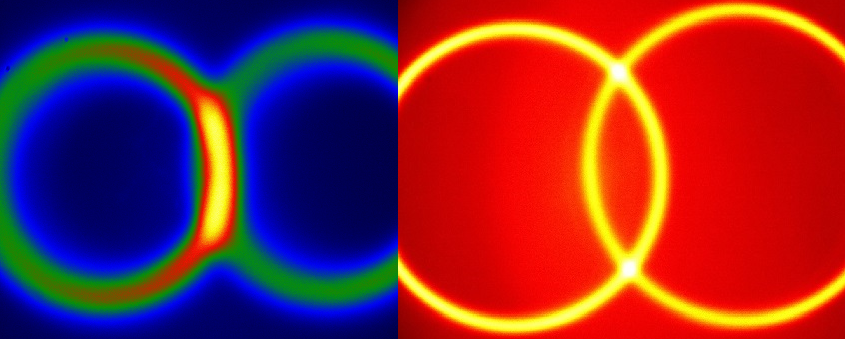}
\caption{\textit{Type-II SPDC ring in collinear (left) and non-collinear (right) configurations imaged by Andor CMOS camera. In the overlapping region of the two rings, one cannot comment whether which ring the photon belongs to, hence a pair of photons collected from these regions will be polarization entangled.}\label{fig1.08}}
\end{center}
\end{figure}

It should be noted that phase matching condition may not be satisfied in all uniaxial crystals. The desirable optical properties of a crystal are high non-linearity, large birefringence, and small dispersion. Depending on the relative orientation of the \textbf{k} vectors of pump, signal, and idler, phase matching is called collinear when all \textbf{k} vectors are parallel to each other, and non-collinear when the \textbf{k} vectors are non-parallel.

The refractive indices of the $e$-ray  ($n_e$) and $o$-ray ($n_o$) propagating in a uniaxial with a common wave normal at an angle $\theta$ to the optic axis are given by,
\begin{subequations}
\begin{equation}
n^{(o)}=n_o~,
\end{equation}
\begin{equation}
n^{(e)}(\theta)=\frac{n_o n_e}{[n_e^2\sin^2(\theta)+n_o^2\cos^2(\theta)]^{1/2}}~.
\end{equation}
\label{eq1.89}
\end{subequations}
where $n_0$ and $n_e$ are ordinary and extraordinary refractive indices of the crystal.

As an example, we choose BBO which is a negative uniaxial crystal with point symmetry 3. The effective non-linear optical constants [\ref{1.23}] of BBO for type-I phase matching are given by

\begin{equation}
d_\text{eff}=[d_{11} \cos(3\phi) - d_{22}\sin(3\phi)]\cos(\theta)+d_{31}\sin(\theta),
\end{equation}
where $\theta$ and $\phi$ are usual polar and azimuthal angles.

The Sellmier's equation for BBO are given by
\begin{equation}
\begin{split}
n^2_o=2.7359+\frac{0.01878}{\lambda^2-0.01822}-0.01354\lambda^2,\\
n^2_e=2.3753+\frac{0.01224}{\lambda^2-0.01667}-0.01516\lambda^2.
\end{split}
\end{equation}
where $\lambda$ is in microns. For $\lambda=0.405~\mu m$, we get $n_o=1.611$ and $n_e=1.544$.

\begin{figure}[H]
\begin{center}
\begin{tikzpicture}
\draw[red] (0,0) circle (3cm);
\draw[red] (0,0) ellipse (4cm and 3cm);
\draw[blue] (0,0) circle (3.5cm);
\draw[blue] (0,0) ellipse (4.5cm and 3.5cm);
 \draw[black, arrows={-Triangle[angle=90:5pt,black,fill=gray]}](0,0)--(0,4) ;
\draw[black, arrows={-Triangle[angle=90:5pt,black,fill=gray]}](0,0)--(6,0) ;
\draw[blue, arrows={-Triangle[angle=90:5pt,black,fill=gray]}](0,0)--(4,3.23) ; 
\draw[black] (0,0.6) arc (90:40:0.6);
\node at (0,4.2) {Z};
\node at (6.5,0) {Y};
\node at (4.5,3.5) {$\hat{i}$};
\node at (0.4,0.7) {\tiny{$\theta_{pm}$}};
\node at (-0.2,-0.2) {O};
\node at (0,4.7) {Optics axis};
\end{tikzpicture}
\caption{\textit{Wave surfaces for a negative uniaxial crystal. The blue curves show the wave surfaces at fundamental frequency and red curves show the  wave surfaces for down-converted frequency. The unit vector $\hat{i}$ points in the direction of index matched wave normals, and $\theta_{pm}$ is the phase matching angle.}\label{fig1.09}}
\end{center}
\end{figure}
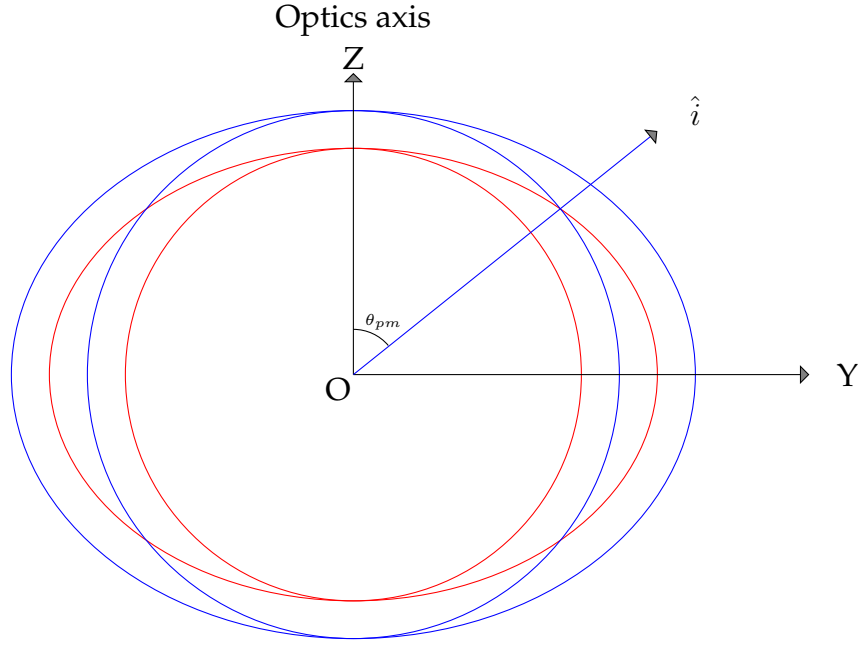

If $n_o$ and $n_e$ are known at each of the frequencies $\omega_p$, $\omega_s$, and $\omega_i$ then the phase matching angle $\theta_{pm}$ can be calculated for type-I or type-II phase matching using Eq.~(\ref{eq1.84}) and appropriate condition as given in Eq.~(\ref{eq1.89}). An example of phase matching for negative uniaxial crystal is shown in Fig.~\ref{fig1.09}.

Although SPDC process produces the brightest photon sources, the efficiency of SPDC is about 1 in $10^9$. Hence, we use ``Non-depleted pump approximation" for SPDC process. In the semi-classical analysis of SPDC,  the pump beam is treated classically, and signal and  idler fields are quantized [\ref{1.25}]. The interaction Hamiltonian is then written as
\begin{equation}
\begin{split}
\hat{H}&=\int_v d^3r~\chi^{(2)}\mathbf{E}_p(\mathbf{r},t)\hat{\mathbf{E}}^{(-)}_s(\mathbf{r},t)\hat{\mathbf{E}}^{(-)}_i(\mathbf{r},t) + h.c.\\
&=\frac{1}{v}\int_v d^3r~ \chi^{(2)}\mathbf{E}_p e^{i(\mathbf{k}_p. \mathbf{r}-\omega t)}\sum_{\mathbf{k}_s,m}\hat{\mathbf{P}}_{\mathbf{k}_s,m}\sqrt{\frac{\hbar\omega\mathbf{k}_s}{2\epsilon_0}}\hat{a}^\dag_{\mathbf{k}_s,m}e^{i(\mathbf{k}_s. \mathbf{r})}\sum_{\mathbf{k}_i,n}\hat{\mathbf{P}}_{\mathbf{k}_i,n}\sqrt{\frac{\hbar\omega\mathbf{k}_i}{2\epsilon_0}}\\
&~~~~~~~~~~~~~~~~~~~~~~~~~~~~~~~~~~~~~~~~~~~~~~~~~~~~~~~~~~~~~~~~~~~~~ \times \hat{a}^\dag_{\mathbf{k}_i,n}e^{i(\mathbf{k}_i. \mathbf{r})}
+ h.c.
\end{split}
\end{equation}
where $\mathbf{k}_p,\mathbf{k}_s, \mathbf{k}_i$ are the pump, signal and idler wave vectors, and $\hat{\mathbf{P}}_{\mathbf{k}_p}, \hat{\mathbf{P}}_{\mathbf{k}_s},\hat{\mathbf{P}}_{\mathbf{k}_i}$ are corresponding polarization vectors.

As soon as pump beam enters the non-linear crystal, SPDC process begins. The phase matching condition (discussed above) ensures that the down-conversion probability amplitudes add constructively throughout the crystal.

\section{References}
\begin{enumerate}
\item \textit{Michael A. Nielsen and Isaac L. Chuang, ``Quantum Computation and Quantum Information", \href{https://doi.org/10.1017/CBO9780511976667}{Cambridge University Press (2000).}} \label{1.01}
\item \textit{J. Preskill, \href{http://www.theory.caltech.
edu/~preskill/ph219/ph219_2004.html}{Lecture notes for physics 219: quantum computation (2004)}}.\label{1.02}
\item \textit{M. Wilde, ``Quantum Information Theory", \href{https://doi.org/10.1017/CBO9781139525343}{Cambridge University Press (2013)}}.\label{1.03}
\item \textit{Alexander Streltsov, Gerardo Adesso, Martin B. Plenio, ``Quantum Coherence as a Resource", \href{https://doi.org/10.1103/RevModPhys.89.041003}{Rev. Mod. Phys. \textbf{89}, 041003 (2017)}}.\label{1.04} 
\item \textit{W. H. Zurek, ``Decoherence, einselection, and the quantum origins of the classical", \href{https://doi.org/10.1103/RevModPhys.75.715}{Rev. Mod. Phys. \textbf{75}, 715 (2003)}}. \label{1.05}
\item \textit{Decoherence, Entanglement and Information Protection
in Complex Quantum Systems, edited by V. Akulin, A. Sarfati, G. Kurizki, and S. Pellegrin, \href{https://www.springer.com/gp/book/9781402032813}{Springer, New York, (2005)}}. \label{1.06}
\item \textit{Ting Yu, J. H. Eberly, ``Sudden Death of Entanglement", \href{https://doi.org/10.1126/science.1167343}{Science \textbf{323}, 5914, pp. 598-601 (2009)}}. \label{1.07}
\item \textit{Ashutosh Singh, Siva Pradyumna, A. R. P. Rau, and Urbasi Sinha, ``Manipulation of entanglement sudden death in an all-optical setup," \href{https://doi.org/10.1364/JOSAB.34.000681}{J. Opt. Soc. Am. B \textbf{34}, 681-690 (2017)}}.\label{1.08}
\item \textit{Daniel F. V. James, Paul G. Kwiat, William J. Munro, and Andrew G. White, ``Measurement of qubits", \href{https://doi.org/10.1103/PhysRevA.64.052312}{Phys. Rev. A \textbf{64}, 052312 (2001)}}.\label{1.09}
\item \textit{Ashutosh Singh, Ijaz Ahamed, Dipankar Home, and Urbasi Sinha, ``Revisiting comparison between entanglement measures for two-qubit pure states," \href{https://doi.org/10.1364/JOSAB.37.000157}{J. Opt. Soc. Am. B \textbf{37}, 157-166 (2020)}}. \label{1.10}
\item \textit{Schumacher B., ``Quantum coding", \href{https://doi.org/10.1103/PhysRevA.51.2738}{Phys. Rev. A \textbf{51}, 2738 (1995).}} \label{1.11}
\item \textit{M. Krenn, M. Malik, T. Scheidl, R. Ursin, and A. Zeilinger,
``Quantum Communication with Photons", \href{https://doi.org/10.1007/978-3-319-31903-2_18}{Springer Intl. Publishing, Cham, p. 455 (2016)}}.\label{1.12}
\item \textit{D. BruB and C. Macchiavello, ``How the first partial transpose was written", \href{https://doi.org/10.1007/s10701-005-7357-0}{Found Phys \textbf{35}, 1921–1926 (2005)}}. \label{1.13}
\item \textit{R. Jozsa, ``Fidelity for mixed quantum states," \href{https://doi.org/10.1080/09500349414552171}{J. Mod. Opt. \textbf{ 41}, 2315–2323 (1994)}}.\label{1.14}
\item \textit{Ryszard Horodecki, Paweł Horodecki, Michał Horodecki, and Karol Horodecki, ``Quantum entanglement", \href{https://doi.org/10.1103/RevModPhys.81.865}{Rev. Mod. Phys. \textbf{ 81}, 865 (2009)}}. \label{1.15}
\item\textit{ Michał Horodecki, ``Entanglement measures", Quantum Inf. Comput. \textbf{1}, 3 (2001)}.\label{1.16}
\item \textit{Martin B. Plbnio, Shashank  Virmani, ``An introduction to entanglement measures",  Quantum Inf. Comput. \textbf{7}, 3 (2007)}. \label{1.17}
\item \textit{J. Eisert, M. Cramer, and M. B. Plenio, ``Colloquium: Area laws for the entanglement entropy",  \href{https://doi.org/10.1103/RevModPhys.82.277}{Rev. Mod. Phys. \textbf{82}, 277 (2010)}}. \label{1.18}
\item \textit{W. K. Wootters, ``Entanglement of formation of an arbitrary state of two qubits", \href{https://doi.org/10.1103/PhysRevLett.80.2245}{Phys. Rev. Lett. \textbf{80}, 2245 - 2248 (1998)}}.\label{1.19}
\item \textit{W. K. Wootters, ``Entanglement of formation and concurrence", Quantum Inf. Comput. \textbf{1}, 27 - 44 (2001)}. \label{1.20}
\item \textit{G. Vidal and R. F. Werner, ``Computable measure of entanglement," \href{https://doi.org/10.1103/PhysRevA.65.032314}{Phys. Rev. A \textbf{65}, 032314 (2002)}}. \label{1.21}
\item \textit{Kai Chen, Ling-An Wu, ``A matrix realignment method for recognizing entanglement", \href{https://dl.acm.org/doi/10.5555/2011534.2011535}{Quantum Inf. Comput. 3, 3, pp.193-202 (2003)}}.\label{1.22}
\item \textit{Robert W. Boyd, ``Nonlinear Optics", Academic, San Diego (1992)}.\label{1.23}
\item \textit{ David C. Burnham and Donald L. Weinberg, ``Observation of simultaneity in parametric production of optical photon pairs", \href{https://doi.org/10.1103/PhysRevLett.25.84}{Phys. Rev. Lett. \textbf{25}, 84(1970)}}.\label{1.24}
\item \textit{Christophe Couteau, ``Spontaneous parametric down-conversion", \href{https://doi.org/10.1080/00107514.2018.1488463}{Contemp. Phys. \textbf{59}, 3, pp:291–304 (2018)}}.\label{1.25}
\item \textit{J. E. Midwinter and J. Warner,``The effects of phase matching method and of uniaxial crystal symmetry on the polar distribution of second-order non-linear optical polarization", \href{https://doi.org/10.1088/0508-3443/16/8/312}{Br. J. Appl. Phys. \textbf{16}, 1135}}.\label{1.26}
\item \textit{Urbasi Sinha, Surya Narayan Sahoo, Ashutosh Singh, Kaushik Joarder, Rishab Chatterjee, Sanchari Chakraborti, ``Single-Photon Sources," \href{https://www.osa-opn.org/home/articles/volume_30/september_2019/features/single-photon_sources/}{Optics and Photonics News, \textbf{30}, 9, pp. 32-39 (2019)}}.\label{1.27}
\end{enumerate}

%% file: chapter2.tex
\setcounter{equation}{0}
\chapter{Preparation and characterization of polarization entangled photon source}

In this chapter, we discuss the experimental techniques for preparation and characterization of a polarization Entangled Photon Source (EPS). We give a theoretical framework for and experimental realization of a Spontaneous Parametric Down-Conversion (SPDC) based type-I polarization-entangled photon source. The two-crystal geometry, where optic axes of the two crystals are orthogonal to each other, is used in type-I phase matching to produce the polarization entangled photons by pumping it with a diagonally polarized laser beam. This scheme causes the spatial and temporal walk-offs leading to decoherence in such sources when Continuous Wave (CW) diode lasers with low coherence time are used as pump. We discuss the origin of decoherence mechanism and its compensation using specially designed crystals. The source is characterized by complete two-qubit Quantum State Tomography (QST) - a technique to reconstruct the state (density matrix) of an unknown ensemble of quantum systems through a series of measurements.

To begin with, we introduce the concept of tomography in classical and quantum systems and then give a brief introduction to Stokes parameter and how polarization state of light can be expressed in terms of Stokes parameters. This is followed by experimental setup for the measurement of Stokes parameters and its relation to polarization density-matrix of qubits. Then we give a detailed discussion on single- and two-qubit QST. The tomographic measurements are inevitably affected by statistical errors, imperfect measurement settings, and device imperfections and if we directly reconstruct the density matrix from such measurements, the state may not be physical. To avoid this issue, we use Maximum Likelihood Estimation (MLE) to ensure that the reconstructed density matrices be necessarily physical - i.e, Hermitian, unit trace, and positive, even with imperfect measurements. 

Degree of entanglement is quantified by computing Concurrence of the reconstructed density matrix and closeness of the reconstructed state with the ideal/fiducial state is quantified by fidelity which is a measure of overlap between the two states. Effect of spectral filtering on the purity  and Concurrence of two-qubit entanglement is discussed. In the end, we briefly comment on the two-qubit entanglement characterization by measuring visibility of entanglement and make a remark on how Bell-state fidelity can be inferred by a smaller set of measurements in Mutually Unbiased Bases (MUBs).

\section{Quantum state tomography (QST)}\label{S2.1}

The word tomography has originated from greek words `\textit{tomos}' meaning `\textit{slice or section}' and `\textit{graphein}' meaning `\textit{to write}'. Thus, the word `\textit{tomography}' can be loosely translated to `\textit{section imaging}'. A very popular injury diagnosing tool in medical science is what is known as computerized tomography scan (or CT scan) which uses a programmed computer and rotating X-ray machine to generate different cross-sectional images of the human body. From a set of images that contain reduced information about the object, the complete object is reconstructed. For example, the 3D content of a picture can be extracted from several 2D pictures taken from different directions. This process of generating images is known as \textit{tomography}, the device used is called \textit{tomograph}, and the image generated is called \textit{tomogram}. This is also the basic idea of quantum state tomography (QST) wherein a series of measurements (known as tomographic set) are performed in different bases onto the quantum system for complete characterization of the quantum state.

While characterizing the classical systems, it may be possible to perform a series of measurements on the same system, hence a single copy of the classical system would suffice for its complete characterization. But, in the case of quantum systems, from the measurement postulate of quantum mechanics we know that the very act of measurement disturbs the state of the quantum system being measured. Hence, further measurements on the same system wouldn't give any meaningful information. Due to this limitation, QST must be performed on an ensemble of identically prepared systems in several stages. Therefore, QST can be interpreted as a technique to reconstruct the state of unknown quantum system by doing measurements over an ensemble of identically prepared particles. We will first discuss the ideal tomography of a single qubit followed by two-qubit system. Further generalization to multipartite systems is straight forward. In order to understand the single-qubit tomography, let us first understand the stokes parameter representation of polarization state of light.

\subsection{Stokes parameter representation of polarized light}
Let us consider a light wave whose x- and y-components of the electric field are given by
\begin{subequations}
\begin{align}
\begin{split}
E_x(t) &=E_{xo}\cos(\omega t+\delta_x )~,
\label{eq2.01a}
\end{split}\\
\begin{split}
E_y(t)&=E_{yo}\cos(\omega t+\delta_y) ~,
\label{eq2.01b}
\end{split}
\end{align}
\end{subequations}
 where $E_{xo}$ and $E_{yo}$ are the electric field amplitudes corresponding to  x- and y-components of the field, respectively, $\omega$ is the instantaneous angular frequency, and $\delta_x$, $\delta_y$ are the phase factors.

Eqs.~(\ref{eq2.01a}) and~(\ref{eq2.01b}) can be combined and rewritten as
\begin{equation}
\frac{E^2_x(t)}{E^2_{xo}}+\frac{E^2_y(t)}{E_{yo}^2}-\frac{2E_x(t)E_y(t)}{E_{xo}E_{yo}}\cos(\delta)=\sin^2(\delta)~,
\label{eq2.02}
\end{equation}
where $\delta =\delta_y-\delta_x$ is the relative phase between and x- and y-components of the field. The Eq.~(\ref{eq2.02}) represents the equation of an ellipse, known as \textit{polarization ellipse}. 

In order to represent Eq.~(\ref{eq2.02}) in terms of experimentally observable quantity of the optical field, we take an average over time which is given by
\begin{equation}
\frac{\langle E^2_x(t)\rangle}{E^2_{xo}}+\frac{\langle E^2_y(t) \rangle}{E_{yo}^2}-\frac{2 \langle E_x(t)E_y(t)\rangle}{E_{xo}E_{yo}}\cos(\delta)=\sin^2(\delta)~,
\label{eq2.03}
\end{equation}
where 
\begin{equation}
\langle E_i(t)E_j(t) \rangle=\lim_{t\to T}\frac{1}{T}\int_0^T E_i(t)E_j(t) dt~;~i,j=x,y.
\label{eq2.04}
\end{equation}

On multiplying Eq.~(\ref{eq2.03}) by $4 E^2_{xo}E^2_{yo}$, we get
\begin{equation}
4 E^2_{yo}\langle E^2_x(t)\rangle+4 E^2_{xo}\langle E^2_y(t)\rangle-8 E_{xo} E_{yo}\langle E_{xo}(t) E_{yo}(t)\rangle \cos(\delta)= [2E_{xo} E_{yo}\sin(\delta)]^2
\label{eq2.05}
\end{equation}
Using Eq.~(\ref{eq2.04}), time average of Eq.~(\ref{eq2.01a}) and ~(\ref{eq2.01b}) can be written as
\begin{subequations}
\begin{align}
\begin{split}
\langle E^2_x(t)\rangle &= \frac{1}{2} E^2_{xo}~,
\end{split}\\
\begin{split}
\langle E^2_x(t)\rangle &= \frac{1}{2} E^2_{xo}~,
\end{split}\\
\begin{split}
\langle E_x(t) E_y(t)\rangle &= \frac{1}{2} E_{xo} E_{yo}\cos(\delta)~.
\end{split}
\end{align}
\label{eq2.06}
\end{subequations}
Substituting values from Eq.~(\ref{eq2.06}) into Eq.~(\ref{eq2.03}), we get
\begin{equation}
2 E^2_{yo}E^2_{xo}+2 E^2_{xo}E^2_{yo}-[2 E_{xo}E_{yo}\cos(\delta)]^2=[2 E_{xo}E_{yo}\sin(\delta)]^2~.
\label{eq2.07}
\end{equation}
Upon adding and subtracting $E^4_{xo}+E^4_{yo}$ in Eq.~(\ref{eq2.07}) and rewriting, we get
\begin{equation}
(E^2_{xo}+E^2_{yo})^2-(E^2_{xo}-E^2_{yo})^2- (2E_{xo}E_{yo}\cos(\delta))^2=(2E_{xo}E_{yo}\sin(\delta))^2
\label{eq2.08}
\end{equation}  
Let us express quantities in the parentheses in Eq.~(\ref{eq2.08}) as follows:
\begin{subequations}
\begin{align}
\begin{split}
S_0 &=E^2_{xo}+E^2_{yo}~,
\end{split}\\
\begin{split}
S_1 &=E^2_{xo}-E^2_{yo}~,
\end{split}\\
\begin{split}
S_2 &=2E_{xo}E_{yo}\cos(\delta)~,
\end{split}\\
\begin{split}
S_3 &=2E_{xo}E_{yo}\sin(\delta)~.
\end{split}
\end{align}
\label{eq2.09}
\end{subequations}
Using Eq.~(\ref{eq2.09}) in Eq.~(\ref{eq2.08}), we get
\begin{equation}
S_0^2=S_1^2+S_2^2+S_3^2~.
\label{eq2.10}
\end{equation}
The parameters $\{S_i\}$ as given by Eq.~(\ref{eq2.09}) are the un-normalized  Stokes parameters which were first introduced by Sir G. G. Stokes in 1852 [\ref{2.01}]. It is clear that the Stokes parameters are the observables of optical field (intensity) on the polarization ellipse and hence real quantities. The Stokes parameter $S_0$ represents the total intensity of light. The parameters $S_1,~S_2,~\text{and}~S_3$ represent the horizontal/vertical, diagonal/anti-diagonal and right-/left-circular polarization components, respectively.

Using Schwarz's inequality, we find that for an arbitrary state of polarized light, the Stokes parameters follow the relation:
\begin{equation}
S_0^2\geq S_1^2+S_2^2+S_3^2~,
\end{equation}
where equality holds for polarized light and inequality for partially polarized light.

For polarization ellipse represented by Eq.~(\ref{eq2.02}), the angle of orientation $\psi$, and ellipticity angle $\chi$ of the ellipse [\ref{2.02}, \ref{2.03}] are given by 

\begin{subequations}
\begin{align}
\begin{split}
\tan(2\psi)=\frac{2E_{xo}E_{yo}\cos(\delta)}{E^2_{xo}-E^2_{yo}}~,
\end{split}\\
\begin{split}
\sin(2\chi)=\frac{2E_{xo}E_{yo}\sin(\delta)}{E^2_{xo}+E^2_{yo}}~.
\end{split}
\end{align}
\label{eq2.12}
\end{subequations}

From Eqs.~(\ref{eq2.09}) and (\ref{eq2.12}), we get
\begin{subequations}
\begin{align}
\begin{split}
\tan(2\psi)=\frac{S_2}{S_1},
\label{eq2.13a}
\end{split}\\
\begin{split}
\sin(2\chi)=\frac{S_3}{S_0}.
\end{split}
\label{eq2.13b}
\end{align}
\end{subequations}

From Eqs.~(\ref{eq2.13a}) and (\ref{eq2.13b}), we get $S_2=S_1\tan (2\psi)$ and $S_3=S_0\sin (2\chi)$. Substituting these values in Eq.~(\ref{eq2.10}), we get
\begin{subequations}
\begin{align}
\begin{split}
S_1 =S_0\cos (2\chi)\cos (2\psi)~,
\end{split}\\
\begin{split}
S_2 =S_0\cos (2\chi)\sin (2\psi)~,
\end{split}\\
\begin{split}
S_3 =S_0\sin (2\chi)~.
\end{split}
\end{align}
\end{subequations}

Let us arrange these in the from of Stokes vector as follows:
\begin{equation}
S=S_0\left(\begin{array}{c}
1\\
\cos(2\chi)\cos(2\psi)\\
\cos(2\chi)\sin(2\psi)\\
\sin(2\chi)
\end{array}\right)~.
\end{equation}
We get normalized Stokes parameters ($s_i$) by dividing the un-normalized Stokes parameters $S_i$ by $S_0$.
 \begin{equation}
 s:=\frac{S}{S_0}=\left(\begin{array}{c}
1\\
\cos(2\chi)\cos(2\psi)\\
\cos(2\chi)\sin(2\psi)\\
\sin(2\chi)
\end{array}\right)=\left(\begin{array}{c}
s_0\\ s_1 \\ s_2\\ s_3.
\end{array}\right)~.
\label{eq2.16}
 \end{equation}
Let us recall that the Spherical polar coordinates $(r,\theta,\phi)$ are related to Cartesian coordinates ($x,y,z$) as
\begin{subequations}
\begin{align}
\begin{split}
x=r\sin(\theta)\cos(\phi)~,
\end{split}\\
\begin{split}
y=r\sin(\theta)\sin(\phi)~,
\end{split}\\
\begin{split}
z=r\cos(\theta)~.
\end{split}
\end{align}
\label{eq2.17}
\end{subequations}
On comparing Eq.~(\ref{eq2.17}) with ($s_1,s_2,s_3$) in Eq.~(\ref{eq2.16}), we get
\begin{subequations}
\begin{align}
\begin{split}
r=1~,
\end{split}\\
\begin{split}
\theta=90-2\chi~,
\end{split}\\
\begin{split}
\phi=2\psi~.
\end{split}
\end{align}
\label{eq2.18}
\end{subequations}
The polarization state of light can be represented by stokes parameter on the surface of unit radius sphere, known as Poincare sphere, as illustrated in Fig.~\ref{fig2.01} below.

\begin{figure} [H]
\begin{center}
\includegraphics[width=0.7\textwidth]{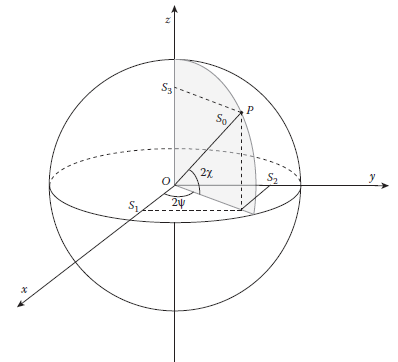}
\caption{\textit{Poincare sphere representation of polarization state of light depicting the stokes parameter [Image credit: Ref. \ref{2.02}].}\label{fig2.01}}
\end{center}
\end{figure}

\subsection{Single-qubit tomography}

A single-qubit's polarization density matrix  $\hat{\rho}$ can be uniquely represented by its Stokes parameters $\{s_0,s_1,s_2,s_3\}$ as follows [\ref{2.04}-\ref{2.07}]:
\begin{equation}
\hat{\rho}=\frac{1}{2}\sum_{i=0}^3 s_i\hat{\sigma}_i~,
\label{eq2.19}
\end{equation}
where $s_0$ is normalized to one (by definition, $s_0=\text{Tr}[\hat{\rho}]=1$), and $\hat{\sigma}_i$'s are the Pauli spin matrices given by
\begin{equation}
\begin{split}
\hat{\sigma}_0=\left(\begin{array}{cc}
1 & 0\\
0 & 1 \end{array}\right),~
\hat{\sigma}_1=\left(\begin{array}{cc}
0 & 1\\
1 & 0 \end{array}\right),~
\hat{\sigma}_2=\left(\begin{array}{cc}
0 & -i\\
i & 0 \end{array}\right),~
\hat{\sigma}_3=\left(\begin{array}{cc}
1 & 0\\
0 & -1 \end{array}\right)~.
\end{split}
\label{eq2.20}
\end{equation}

The Stokes parameters are, thus, the real coefficients of expansion of a single qubit polarization density matrix in terms of the Pauli matrices ($\sigma_i$). If we rewrite Eq.~(\ref{eq2.19}) using (\ref{eq2.20}), we get
\begin{equation}
\hat{\rho}=\frac{1}{2}\left(\begin{array}{cc}
s_0+s_3 & s_1-i s_2\\
s_1+i s_2 & s_0-s_3\end{array}\right)~.
\label{eq2.21}
\end{equation}
Invoking the condition of non-negativity for a density matrix  ($\text{det}(\rho)\geq 0$) on Eq.~(\ref{eq2.21}), we get
\begin{equation}
\begin{aligned}
\begin{split}
&s_0^2+s_1^2+s_2^2+s_3^2 \leq 2~,~\text{Or}\\
& s_1^2+s_2^2+s_3^2 \leq 1~;~(s_0=1).
\end{split}
\end{aligned}
\label{eq2.22}
\end{equation}

For all pure states $\sum_{i=1}^3 s^2_i=1$, for mixed states $\sum_{i=1}^3 s^2_i < 1$, and $\sum_{i=1}^3 s^2_i=0$ for maximally mixed states. Given the density matrix, the stokes parameters $s_i$ can be found by
\begin{equation}
s_i= \text{Tr}\left[\hat{\sigma}_i\hat{\rho}\right]~.
\label{eq2.23}
\end{equation}

One can define degree of polarization (DOP) of a single photon source using the stokes parameter as follows:
\begin{equation}
\text{DOP}=\sqrt{s_1^2+s_2^2+s_3^2}~.
\label{eq2.24}
\end{equation}
For a pure polarization state, $\text{DOP}=1$ and that for a maximally mixed state $\text{DOP}=0$, thus, the stokes parameters for maximally mixed state would be $\{1,0,0,0\}$. 

\subsubsection{Measurement of the single-qubit Stokes parameter}

The Stokes parameters are defined from a set of four intensity measurements [\ref{2.04}]: (i) with a filter that transmits
$50\%$ of the incident radiation, regardless of its polarization,
(ii) with a polarizer that transmits only horizontally polarized
light, (iii) with a polarizer that transmits only light polarized
at $45^\circ$ to the horizontal, and (iv) with a polarizer that transmits only right-circularly polarized light. Physically, each of these Stokes parameters correspond to a specific pair of projective measurements as given below [\ref{2.04}].
\begin{subequations}
\begin{align}
s_0 &=P_{|H\rangle}+P_{|V\rangle}=\frac{N_{|H\rangle}+N_{|V\rangle}}{N_{|H\rangle}+N_{|V\rangle}}~,\\
s_1 &=P_{|D\rangle}-P_{|A\rangle}=\frac{N_{|D\rangle}-N_{|A\rangle}}{N_{|D\rangle}+N_{|A\rangle}}~,\\
s_2 &=P_{|R\rangle}-P_{|L\rangle}=\frac{N_{|R\rangle}-N_{|L\rangle}}{N_{|R\rangle}+N_{|L\rangle}}~,\\
s_3 &=P_{|H\rangle}-P_{|V\rangle}=\frac{N_{|H\rangle}-N_{|V\rangle}}{N_{|H\rangle}+N_{|V\rangle}}~.
\end{align}
\label{eq2.25}
\end{subequations}
where $P_{|\phi\rangle}$ denotes the projection probability of the ensemble of photons in state $\hat{\rho}$ onto the bases state $|\phi\rangle$.

 If we use single photon detectors then the number of photons ($N_{|\phi\rangle}$) detected for a fixed acquisition time, say $\tau$, are related to the projection probability ($P_{|\phi\rangle}$) as given on the right side of Eq.~(\ref{eq2.25}). The three orthogonal bases sets: \{$|H\rangle,~|V\rangle$\}, \{$|D\rangle,~|A\rangle$\}, and \{$|R\rangle,~|L\rangle$\} are an example of mutually unbiased (MUB) set. The measurement bases set for determining the Stokes parameters are not unique and they are not required to be mutually unbiased either. However, the choice of bases set in Eq.~(\ref{eq2.25}) are natural as they are the eigen states of the Pauli spin matrices.

For single qubit tomography, we use quarter-wave plate (QWP), half-wave plate (HWP), and polarizing beam-splitter (PBS) combination as shown in Fig.~\ref{fig2.02} below, for projecting the input state onto different bases states. Note that a PBS transmits a horizontally polarized light and reflects vertically polarized light. The wave-plate settings for different projectors in the transmitted arm of the PBS are given in the table~[\ref{tab1.1}] below.

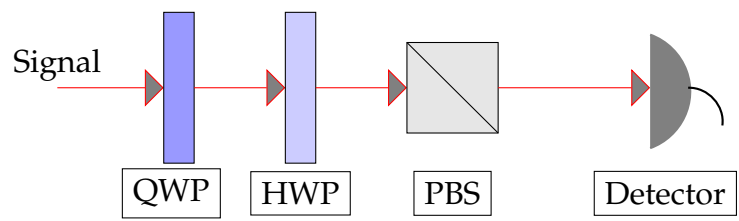
\begin{figure}[h!]
\centering
\begin{tikzpicture}
    \draw[red, arrows={-Triangle[angle=90:10pt,red,fill=gray]}](10,0)-- (11.4,0) ;
    \filldraw[fill=blue!40!white, draw=black!40!black] (11.4,-1) rectangle (11.8,1) ;
    \draw[red, arrows={-Triangle[angle=90:10pt,red,fill=gray]}] (11.8,0) -- (13,0) ;
    \filldraw[fill=blue!20!white, draw=black!40!black] (13,-1) rectangle (13.4,1) ;
    \draw[red, arrows={-Triangle[angle=90:10pt,red,fill=gray]}]  (13.4,0) -- (14.6,0) ;
    \filldraw[fill=gray!20!white, draw=black!40!black](14.6,-0.6) rectangle (15.8,0.6); 			\draw (14.6,0.6)--(15.8,-0.6);
    \draw[red, arrows={-Triangle[angle=90:10pt,red,fill=gray]}]  (15.8,0) -- (17.8,0) ;  
\draw[line width=0.3mm, color=gray, fill] (17.8,-0.8) arc (-70:70:0.8) ;  
\draw[-, thick] (18.3,0) edge[bend left=50] (18.75,-0.5);
\node[draw] at (18,-1.4) {Detector};
\node at (10,0.3) {Signal};
\node[draw] at (15.2,-1.4) {PBS};
\node[draw] at (13.2,-1.4) {HWP};
\node[draw] at (11.5,-1.4) {QWP};
 \end{tikzpicture}
\centering
\caption{\textit{Schematic of the experimental set up for quantum state tomography of a single qubit.}\label{fig2.02}}
\end{figure}

\begin{table}[H]
\begin{center}  
\resizebox{\textwidth}{!}{
\renewcommand{\arraystretch}{1.5}
\begin{tabular}{ |p{1cm}|p{4cm}|p{4cm}|p{4cm}|}
\hline
S.N.  & QWP orientation wrt fast axis (in Degree) & HWP orientation wrt fast axis (in Degree) & Projection in the transmitted arm \\
  \hline
   1. & 0 & 0 & $|H\rangle\langle H|$   \\
  \hline
   2. & 45 & 0 & $|R\rangle\langle R|$   \\
  \hline
   3. & 0 & 22.5 & $|L\rangle\langle L|$   \\
  \hline
   4. & 45 & 22.5 & $|D\rangle\langle D|$   \\
  \hline
   5. & 0 & 45 & $|V\rangle\langle V|$   \\
  \hline
   6. & 45 & 67.5 & $|A\rangle\langle A|$   \\
  \hline
\end{tabular}}
\caption{\textit{QST settings for projection onto different bases for single qubit.}\label{tab1.1}}
\end{center}
\end{table}
\subsubsection{Intuitive understanding of the single-qubit stokes parameter measurement}

Let us now intuitively understand how different settings of QWP, HWP, followed by a polarizer with transmission axis along horizontal direction ($|H\rangle\langle H|$ projection, or detection in the transmitted arm of the PBS) enables projection onto different bases in the QST setup (Fig.~\ref{fig2.02}). Let us consider a polarization qubit on the Poincare sphere. Using a QWP and a HWP, one can implement arbitrary unitary operation on the system. If we include a polarizer with transmission axis oriented along horizontal direction, then one can implement arbitrary projection operator on the polarization Hilbert space of a single qubit. Action of a QWP and a HWP whose fast axis makes an angle $q$ and $h$, respectively, with respect to vertical, on the polarization state of qubit are given by Jones matrix as given below.
\begin{subequations}
\begin{align}
\text{Q(q)} &=\left(\begin{array}{cc}
i-\cos(2q) & \sin(2q)\\
\sin(2q) & i+\cos(2q)
\end{array}\right)\\
\text{H(h)}& =\left(\begin{array}{cc}
\cos(2h) & -\sin(2h)\\
-\sin(2h) & -\cos(2h)
\end{array}\right)
\end{align}
\label{eq2.26}
\end{subequations}

Jones matrix of a linear polarizer with transmission axis along horizontal direction is given by
\begin{equation}
\text{LPH} =\left(\begin{array}{cc}
1 & 0\\
0 & 0
\end{array}\right)
\label{eq2.27}
\end{equation}

When these components are placed as in Fig.~\ref{fig2.02}, their general action for arbitrary settings of QWP, and HWP is given by
\begin{equation}
P_{|x\rangle\langle x|}=\text{LPH}.H(h).Q(q)
\label{eq2.28}
\end{equation}

\begin{figure} [H]
\begin{center}
\includegraphics[scale=1]{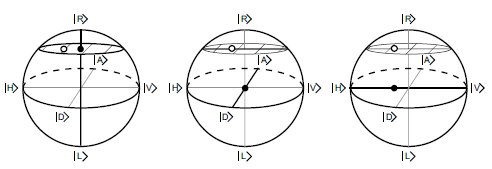}
\caption{\textit{Three digrams depict measurement along three linearly independent bases ($|R\rangle,~|D\rangle,~\text{and}~|H\rangle$ from left to right) which locate the state of single qubit  (represented by as an open circle in the Poincare sphere) in the Hilbert space [Image credit: Ref. \ref{2.05}].}\label{fig2.03}}.
\end{center}
\end{figure}

By choosing the appropriate settings of QWP and HWP angles, one can implement all the projections as given in Table~[\ref{tab1.1}]. Ideally, a single qubit QST requires a set of three linearly independent measurements. Each of these measurements specifies one degree of freedom, thus reduces the number of free parameters of the unknown state's Hilbert space by one. The first projective measurement locates the unknown state into a plane orthogonal to the measurement basis. The second measurement locates the unknown state along the line of intersection of the plane orthogonal to the current measurement basis and the plane located by first measurement. In the end, third measurement locates the state at the pinpoint on the Bloch sphere which is the point of intersection of the plane orthogonal to the current measurement basis and the line located by the second measurement. These steps can be understood from the Fig.~(\ref{fig2.03}) shown below [\ref{2.05}].

\begin{itemize}
\item Next, we will discuss a technique known as maximum likelihood estimation to get the physical state from the set of non-ideal measurements as discussed below.
\end{itemize}

\subsection{Maximum likelihood estimation}
As mentioned earlier, the density matrix representing a physical system must fulfil the following conditions:
\begin{subequations}
\begin{align}
\text{Tr}\left[\rho\right] = & 1~~ (\text{Normalization}),\\
\text{det}(\rho)\geq & 0~~ (\rho ~ \text{is non-negative}),
 \end{align}
 \label{eq2.29}
 \end{subequations}
note that positivity of the matrix $\hat{\rho}$ guarantees Hermiticity. 

An ideal QST requires exact measurements to be performed on an ensemble of infinite many systems to reveal the true probability distribution for different projection measurement, which are then used to reconstruct the density matrix of the system. Practically, any measurement has a finite sample size and each measurement is bound to be affected by experimental imperfections, and thus, reconstructed density matrix may not satisfy the above conditions. In such cases, a theoretical fit to the experimental data is required. This is achieved by constructing a density matrix in terms of the parameters to be fitted by an optimization procedure known as maximum likelihood estimation (MLE).\\

The basic approach for MLE is the following [\ref{2.04}].
\begin{enumerate}
\item A two-qubit density matrix has 15 independent parameters and one dependent parameter that ensures normalization. Therefore, the first step is to write a physical density matrix (i.e, normalized, hermitian, and positive) depending on 16 real parameters $\{t_1,t_2,.....,t_{16}   \}$. Let us denote this density matrix by $\hat{\rho}_p(t_1,t_2,....,t_{16})$.
\item We need to perform minimum 16 measurements to get hold of 16 independent parameters of the density matrix. Let us denote the experimental data set as $\{n_\nu,\nu=1,2...16\}$. Next, define a likelihood function that captures how good $\hat{\rho}_p(t_1,t_2,....,t_{16})$ is for the experimental data. Let us denote the likelihood function by  $\mathcal{L}(t_1,t_2,...,t_{16};n_1,n_2,...,n_{16})$.
\item The objective is to find the optimum values of the parameters\\ $\{t_1^{(opt)},~t_2^{(opt)},....t_{16}^{(opt)}\}$ using numerical optimization techniques for which the likelihood function is maximized or log-likelihood function is minimized.
\end{enumerate}

A detailed discussion on the execution of these steps is given in the following section below.
 
\subsubsection{T-matrix calculation}

Let us first understand the MLE of a single qubit. Consider a T-matrix of the form
\begin{equation}
T=\left(\begin{array}{cc}
t_1 & 0\\
t_3+i t_4 & t_2
\end{array}\right),
\label{eq2.30}
\end{equation}
where the real parameters $t_1,t_2,t_3,t_4$ are obtained by MLE fitting. Then the fitted density matrix of a single qubit system can be defined through the T-matrix as follows:
\begin{equation}
\hat{\rho} := \frac{T^\dag T}{\text{Tr}(T^\dag T)}.
\label{eq2.31}
\end{equation}
 Using Eq.(\ref{eq2.31}) into (\ref{eq2.30}) we get

\begin{equation}
\hat{\rho}=\frac{1}{t_1^2+t_2^2+t_3^2+t_4^2}\left(\begin{array}{cc}
t_1^2+t_3^2+t_4^2 & t_2(t_3-i t_4)\\
t_2(t_3+i t_4) & t_2^2
\end{array}\right).
\label{eq2.32}
\end{equation}

It should be noted here that by construction all the properties of the density matrix are satisfied by $\hat{\rho}$. The purpose of MLE process is to find the global minimum (not the local minimum) and to achieve this it is important to start with a good guess for the initial values of $t_i$'s.  To do so, we relate the $t_i$'s with the normalized stokes parameters $s_i$'s. Comparing Eq.(\ref{eq2.21}) and (\ref{eq2.33}), we get
\begin{subequations}
\begin{align}
s_1=\frac{2t_2t_3}{t_1^2+t_2^2+t_3^2+t_4^2},\\
s_2=\frac{2t_2t_4}{t_1^2+t_2^2+t_3^2+t_4^2},\\
s_3=\frac{t_1^2-t_2^2+t_3^2+t_4^2}{t_1^2+t_2^2+t_3^2+t_4^2}.
\end{align}
\label{eq2.33}
\end{subequations}

Upon solving for $t_i$'s, we get
\begin{subequations}
\begin{align}
t_1 &=\sqrt{\frac{(1-s_1^2-s_2^2-s_3^2}{(1-s_3)^2}}t_2\\
t_3 &=\frac{s_1}{1-s_3}t_2\\
t_4 &=\frac{s_2}{1-s_3}t^2
\end{align}
\label{eq2.34}
\end{subequations}

Thus we get the expressions for $t_1,t_3,t_4$  in the form of normalized stokes parameter and parameter $t_2$ for optimization.

\subsubsection{Numerical optimization}

For numerical optimization, we do Chi-squared test using `NMinimize' routine in Mathematica to find optimum set of t-parameters $\{t_1^{(opt)},~t_2^{(opt)},....t_{16}^{(opt)}\}$ which minimizes the Likelihood function. These optimum t-parameters are then plugged in to the Eq.(\ref{eq2.32}) to get the likelihood density matrix which is by definition physical.

\subsection{Two-qubit tomography}

\begin{figure}[H] 
\centering
\begin{tikzpicture}
    \draw[red, arrows={-Triangle[angle=90:10pt,red,fill=gray]}](10,0)-- (11.4,0) ;
    \filldraw[fill=blue!40!white, draw=black!40!black] (11.4,-1) rectangle (11.8,1) ;
    \draw[red, arrows={-Triangle[angle=90:10pt,red,fill=gray]}] (11.8,0) -- (13,0) ;
    \filldraw[fill=blue!20!white, draw=black!40!black] (13,-1) rectangle (13.4,1) ;
    \draw[red, arrows={-Triangle[angle=90:10pt,red,fill=gray]}]  (13.4,0) -- (14.6,0) ;
    \filldraw[fill=gray!20!white, draw=black!40!black](14.6,-0.6) rectangle (15.8,0.6); 			
    \draw (14.6,0.6)--(15.8,-0.6);
    \draw[red, arrows={-Triangle[angle=90:10pt,red,fill=gray]}]  (15.8,0) -- (17.8,0) ;    
\node[draw] at (15.2,-1.4) {PBS};
\node[draw] at (13.2,-1.4) {$H_1$};
\node[draw] at (11.5,-1.4) {$Q_1$};
\node at (10,0.3) {Signal};
    
\begin{scope}[shift={(0,-3)}]
 \draw[red, arrows={-Triangle[angle=90:10pt,red,fill=gray]}](10,0)-- (11.4,0) ;
    \filldraw[fill=blue!40!white, draw=black!40!black] (11.4,-1) rectangle (11.8,1) ;
    \draw[red, arrows={-Triangle[angle=90:10pt,red,fill=gray]}] (11.8,0) -- (13,0) ;
    \filldraw[fill=blue!20!white, draw=black!40!black] (13,-1) rectangle (13.4,1) ;
    \draw[red, arrows={-Triangle[angle=90:10pt,red,fill=gray]}]  (13.4,0) -- (14.6,0) ;
    \filldraw[fill=gray!20!white, draw=black!40!black](14.6,-0.6) rectangle (15.8,0.6); 			
    \draw (14.6,0.6)--(15.8,-0.6);
    \draw[red, arrows={-Triangle[angle=90:10pt,red,fill=gray]}]  (15.8,0) -- (17.8,0) ;
    
\draw[line width=0.3mm, color=gray, fill] (17.8,2.2) arc (-70:70:0.8) ;  
\draw[-, thick] (18.3,3) edge[bend left=50] (18.75,2.5);
\node[draw] at (18,1.6) {Detector};

\draw[line width=0.3mm, color=gray, fill] (17.8,-0.8) arc (-70:70:0.8) ;  
\draw[-, thick] (18.3,0) edge[bend left=50] (18.75,-0.5);
\node[draw] at (18,-1.4) {Detector};

\node[draw] at (15.2,-1.4) {PBS};
\node[draw] at (13.2,-1.4) {$H_2$};
\node[draw] at (11.5,-1.4) {$Q_2$};
\node at (10,0.3) {Idler};
\end{scope}
\end{tikzpicture}
\centering
\caption{\textit{Schematic of the experimental set up for quantum state tomography of two-qubit state.}\label{fig2.04}}
\end{figure}
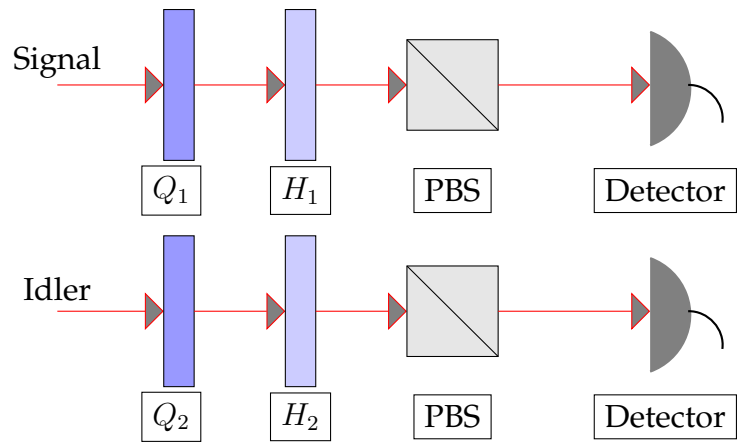

Like in single qubit case, the two-qubit polarization state is characterized by a density matrix
\begin{equation}
\hat{\rho}=\frac{1}{4}\sum_{i,j=0}^3 r_{ij}\hat{\sigma}_i\otimes{\hat{\sigma}}_j
\label{eq2.35}
\end{equation}  
where two photon stokes parameter $r_{ij}$ are the real coefficient of expansion of a two-qubit density matrix in terms of Pauli matrices $\hat{\sigma}_i\otimes\hat{\sigma}_j$. Normalization of the density matrix demands that $r_{00}=\text{Tr}[\hat{\rho}]=1$, and therefore the two-qubit density matrix is characterized by 15 real parameters. The experimental setup for characterizing two-qubit state is shown in the Fig.~\ref{fig2.04}.

The measurement settings for two-qubit QST are listed in Table~\ref{tab1.2} below.

\newcommand{\myelement}[3]{
 		&			&	0	&	0	&  \(\op{#3H}\)		\\ \cline{3-5}	
     	&			&	45	&	0	&  \(\op{#3R}\)		\\ \cline{3-5}	
   #1	& 	#2		&	0	&	22.5	& \(\op{#3L}\)		\\ \cline{3-5}	
     	&			&	45	&	22.5	& \(\op{#3D}\)		\\ \cline{3-5}	
     	&			&	0	&	45	& \(\op{#3V}\)		\\ \cline{3-5}	
     	&			&	45	&	67.5	& \(\op{#3A}\)		\\ \cline{1-5}}
     	
\renewcommand{\baselinestretch}{1.15}
\begin{table}[H] 
\begin{adjustbox}{width=\textwidth}
\noindent
\begin{tabularx}{\linewidth}{|c|c|c|c|c|}
    \cline{1-5}
    \(Q_1\) settings & \(H_1\) settings & \(Q_2\) settings & \(H_2\) settings & Projection in\\
   	( in degree) 	& ( in degree) 	&	( in degree) 	&	( in degree) 	& transmitted arms\\				\cline{1-5}	
    \myelement{0}{0}{H}
    \myelement{45}{0}{R}
    \myelement{0}{22.5}{L}
    \myelement{45}{22.5}{D}
    \myelement{0}{45}{V}
    \myelement{45}{67.5}{A}
\end{tabularx}
 \end{adjustbox}
 \caption{\textit{QST settings for projection onto different bases for two-qubit system.} \label{tab1.2}}
\end{table}

For MLE in a two-qubit system, the T-matrix takes the form
\begin{equation}
T=\left(\begin{array}{cccc}
t_1 & 0 & 0 & 0\\
t_5+i t_6 & t_2 & 0 & 0\\
t_{11}+i t_{12} & t_7+i t_8 & t_3 & 0\\
t_{15}+i t_{16} & t_{13}+i t_{14} & t_9+i t_{10} & t_4
\end{array}\right).
\label{eq2.36}
\end{equation}

The other expressions in the previous section and the procedure to get the physical density matrix remains exactly the same. 

\section{Preparation of the entangled photon source}

High fidelity polarization-entangled photons are produced by the non-linear optical process of SPDC in a two-crystal geometry [\ref{2.09}-\ref{2.14}]. Two identically cut beta-Barium Borate ($\beta-B_aB_2O_4$, also known as BBO) crystals with their optic axes aligned orthogonal to each other are oriented such that the pump beam propagation direction and optic axis of the first (second) crystal defines the horizontal (vertical) plane. If a horizontally (vertically) polarized pump beam is incident on such a crystal, then down-conversion occurs in the first (second) crystal, where pump beam is extraordinary polarized, due to type-I SPDC. This leads to a down-conversion light cone of vertically (horizontally) polarized photons from first (second) crystal. Under no pump depletion approximation, a diagonally polarized pump beam has equal probability of down-conversion in the either crystals.

If spatial overlap of the two down-conversion cones emanating from two crystals is high enough, then the SPDC polarization amplitudes from the two crystals add coherently leading to the generation of an entangled state of the kind: $|\Psi\rangle=[|HH\rangle+ \exp(i\phi)|VV\rangle]/\sqrt{2}$. The spatial overlap of the of the light cones is defined by the parameter $\theta_\text{dc}L/D$; where $\theta_\text{dc}$ is the cone opening angle, $L$ is the crystal thickness, and $D$ is the pump beam diameter [\ref{2.09}]. For coherent addition of the down-converted  polarization amplitudes from the two crystals, we must have: $\theta_{dc}L/D>>1$.

\begin{figure} [!htb]
\begin{center}
\includegraphics[clip, trim=4cm 15.5cm 3cm 2.5cm, width=0.9 \textwidth]{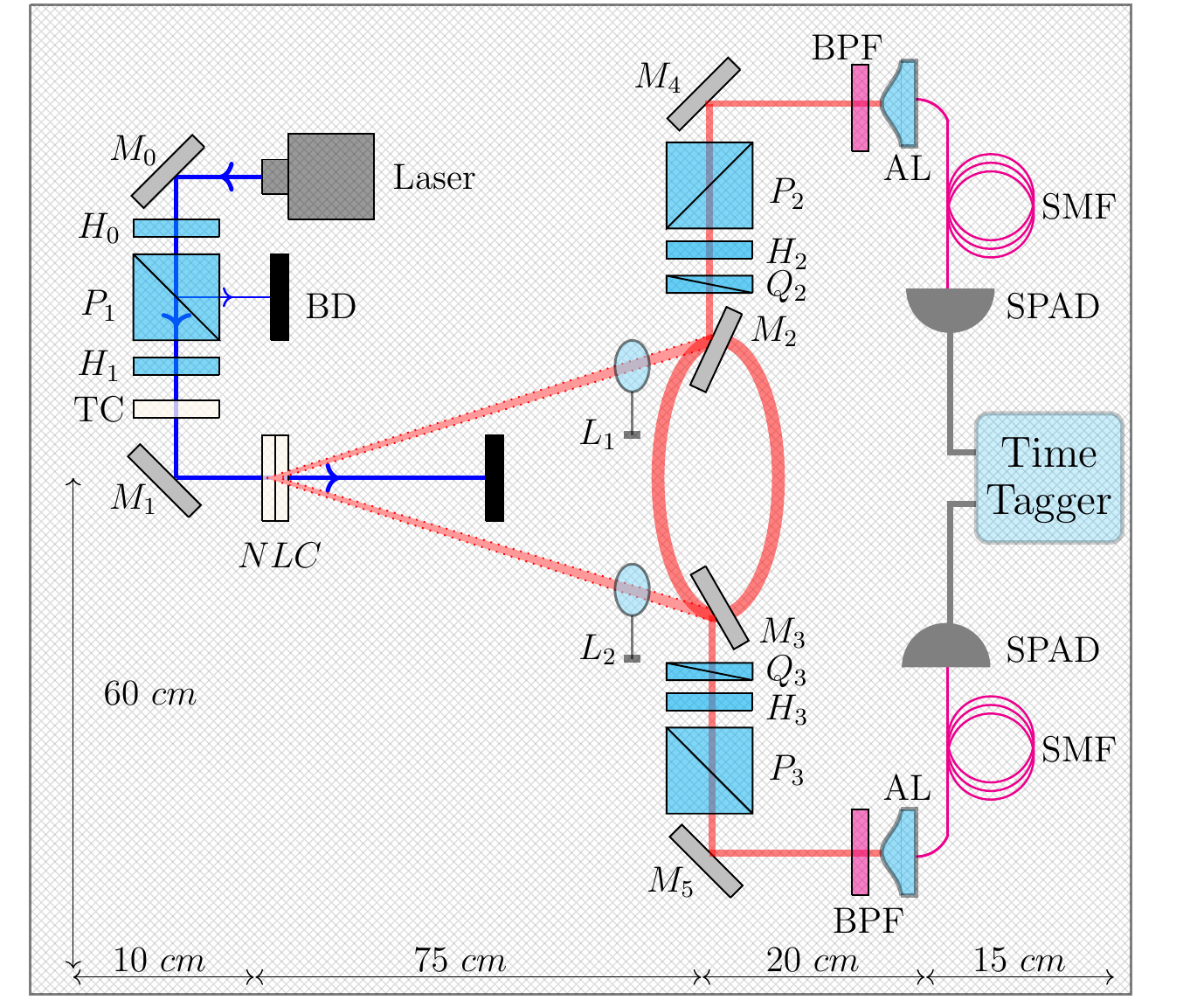}
\end{center}
\caption{\textit{Schematic of the experimental apparatus (not to scale) for preparation of SPDC based type-I polarization entangled photon source using two crystal geometry and characterization using quantum state tomography. Different symbols have the following meaning: P, polarizing beam splitter; Q, quarter wave plate; H, half wave plate; NLC, non-linear crystal; TC, temporal compensator; L, plano-convex lens; M, mirror; BPF, bandpass filter; AL, aspheric lens, SMF, single mod fiber; SPAD, single-photon avalanche diode; and TT, time tagger unit or coincidence module.}\label{fig2.05}}
\end{figure}

Schematic of the experimental setup to produce and characterize type-I polarization entangled photon source is shown in Fig.~\ref{fig2.05}. We used a 100 mW, CW diode laser having central wavelength at 405 nm and a bandwidth of 1.2 nm (Cobolt-06-01-Series) as pump. Two type-I BBO  crystals in sandwich configuration ($5\times 5\times 0.5~\text{mm}^3$ each, from Castech Inc. China) having their optic axis orthogonal to each other and phase matched at $\theta=28.9^\circ$ and $\phi=0^\circ$ with half opening angle of the cone equal to $3^\circ$ are used for generating entangled photons.
 
The polarization state of the pump beam is controlled  using a HWP acting on an input $|H\rangle$ polarized beam as follows:
\begin{equation}
|H\rangle\xrightarrow{\text{HWP at}~\theta} \sin(2\theta)|V\rangle + \cos(2\theta)|H\rangle.
\label{eq4.03}
\end{equation}

When a laser beam in the polarization state given by Eq.~(\ref{eq4.03}) is incident on the sandwich BBO crystal, it undergoes SPDC and resulting state is given by
\begin{equation}
\begin{aligned}
|\Psi\rangle &=\sin(2\theta)|HH\rangle+\cos(2\theta) \exp(i\phi)|VV\rangle,\\
 & =\alpha|HH\rangle+\beta \exp(i\phi)|VV\rangle .
\end{aligned}
\end{equation}
 where $\alpha=\sin(2\theta)$, $\beta=\cos(2\theta)$, and relative phase $\phi$ depends on the specifics of phase matching condition and crystal thickness. It can be controlled by tilting the BBO crystals (which in turn changes the opening angle of the cone) or by placing a tiltable quarter wave plate in the path of pump beam that introduces a relative phase between H- and V-polarization components of the laser beam polarization.
 
\begin{figure}[H]
\begin{center}
\includegraphics[scale=0.6]{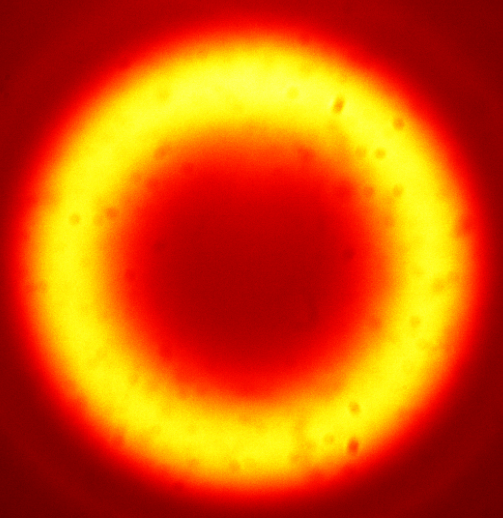}
\caption{\textit{SPDC ring of type-I polarization entangled photon source imaged by Andor CMOS camera. Photon at the diametrically opposite sides of the cone are entangled.} \label{fig2.06}}
\end{center}
\end{figure}

If the pump beam is spectrally filtered and a CMOS camera is placed in front of the crystal, one can image the SPDC cone in non-collinear configuration as shown in Fig.~\ref{fig2.06}. If certain conditions are satisfied (discussed in the next section), then the photons emitted along the diametrically opposite sides of the cone are found to be entangled.
 
\begin{figure}[H]
\begin{center}
\includegraphics[scale=0.33]{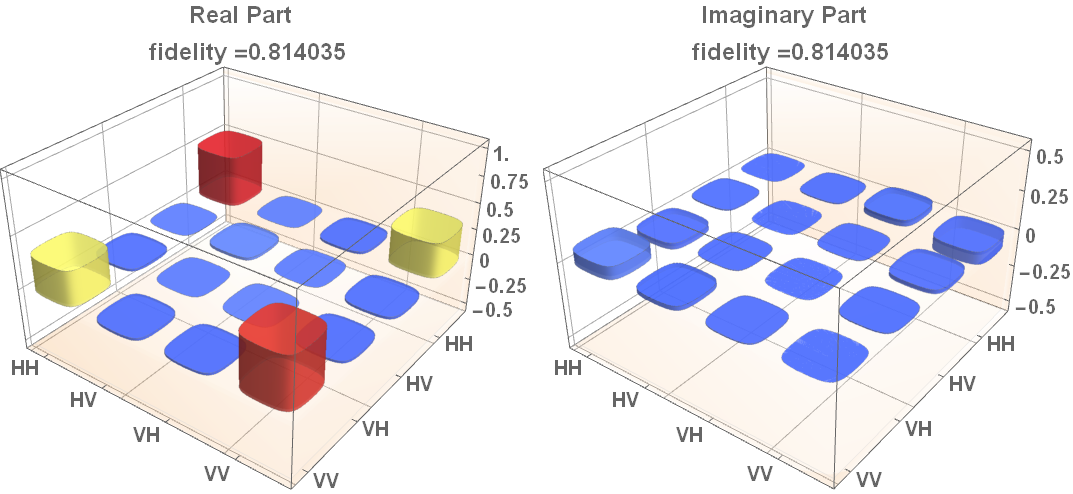}
\end{center}
\caption{\textit{The two-qubit entangled state reconstructed through QST. Left (right) figure shows the real (imaginary) part of the density matrix. The reconstructed state had  $81.4\%$ fidelity with the ideal state, concurrence = 0.669, and purity = 0.709.} \label{fig2.07}}
\end{figure}

For source characterization, we used QWP, HWP, followed by PBS on both the sides to implement arbitrary projectors in the two-qubit Hilbert space as discussed in Section [\ref{S2.1}]. A total 36-measurements are performed for complete two-qubit QST in the product bases. The SPDC photons are spectrally filtered using 810-10 nm band pass filters on either side and then coupled into single mode fibers and coincidence measurement is performed for a fixed acquisition time of  100 s for each setting. For pump HWP oriented at $23.1^\circ$, we did the complete two-qubit QST followed by maximum likelihood estimation to reconstruct the density matrix and found that the state has a fidelity of $81.4\%$ with the ideal state $|\psi\rangle=\sin(46.2^\circ)|HH\rangle+\cos(46.2^\circ)|VV\rangle$, concurrence = $0.669$, and purity of the state = 0.709. The real and imaginary parts of the density matrix are shown in Fig.~\ref{fig2.07}.
 
Ideally, we expect a maximally entangled state to be produced with this process but finite thickness of the BBO crystals and non-monochromatic pump beam causes decoherence. This leads to drop in entangled state purity, fidelity, and concurrence as discussed below.

\section{Decoherence due to walk-off}
There are two main processes of decoherence: Temporal walk-off and spatial walk-off as discussed below [\ref{2.11}-\ref{2.13}].

\subsection{Temporal walk-off}
``Spectral-temporal decoherence due to the timing information (Pump frequency dependent phase when using low coherence time pumps such as free-running diode lasers): Spectral decoherence [\ref{2.11}] arises due to frequency-dependence of the relative phase $\phi$. It can be intuitively understood (especially in the case of a pulsed pump) in the temporal domain (hence the term spectral-temporal decoherence) as follows: different propagation speeds of the pump and downconversion photons within the crystals leads to temporal which-crystal information. The $|HH\rangle$ photons emitted in the first crystal are delayed by $\Delta t$ compared to the $|VV\rangle$ photons generated in the second crystal. If this relative delay is comparable to, or greater than, the pump coherence time, entanglement quality will be reduced. This effect can be countered by ``precompensating" the pump by passing it through a birefringent crystal before the downconversion crystals, which reverses the effect of temporal walk-off as seen in the sandwich SPDC crystal."

Generation of entangled photons through SPDC process in a two-crystal geometry can be understood as follows: a V-polarized pump photon gets down-converted in the first crystal (whose optic axis is in vertical plane) into a pair of H-polarized photons, and a H-polarized pump photon gets down-converted in the second crystal (whose optic axis is in horizontal plane) into a pair of V-polarized photons. If these two processes occur in a coherent manner, i.e., they are indistinguishable, the two down-converted polarization amplitudes are coherently added and the resultant state becomes an entangled state as given below.
\begin{equation}
\begin{aligned}
\begin{split}
\alpha|H\rangle+\beta|V\rangle 
\xrightarrow {\text{SPDC}} \alpha |VV\rangle+\beta \exp(i\phi)|HH\rangle.
\end{split}
\end{aligned}
\end{equation}

The relative phase $\phi$ depends on the optical path difference/delay between the photons down-converted in the first and second crystals. It can be controlled by a tilting a quarter-wave plate in the pump beam (not shown in Fig.\ref{fig2.05}).

\begin{figure}[H]
\centering
\includegraphics[scale=0.5]{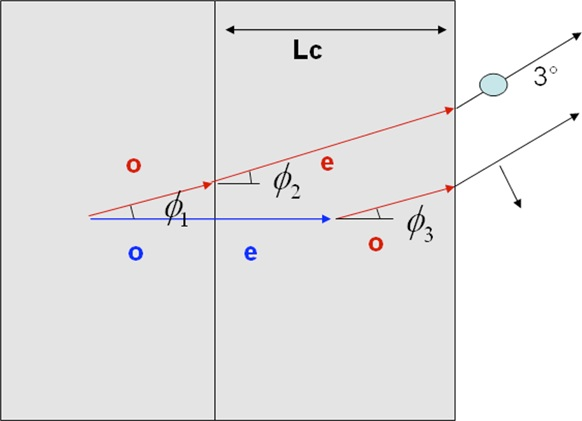}
\caption{\textit{Entangled photon generation and propagation inside the crystals. For clarity, only signal photon trajectories are drawn. The ``o" and ``e" indicate ordinary and extraordinary rays, respectively; $L_c$ is the length of each crystal [Image credit: Ref. \ref{2.14}]}.\label{fig2.08}}
\end{figure}

The propagation times for  V- and H-polarized photons are given by [\ref{2.14}]
\begin{subequations}
\begin{equation}
\tau_V=\frac{L_c}{2}\left[\frac{1}{\cos(\phi_1)V^o_{SPDC}}+\frac{1}{\cos(\phi_2)V^e_{SPDC}}\right]
\end{equation}
\begin{equation}
\tau_H=\frac{L_c}{2}\left[\frac{1}{V^o_{pump}}+\frac{1}{V^e_{pump}}+\frac{1}{\cos(\phi_3)V^o_{SPDC}}\right] 
\end{equation}
\end{subequations}

where $V^{(o,e)}_\text{(pump, SPDC)}$ are group velocities, and angles $\phi_1=1.807$, $\phi_2=1.84$, and $\phi_3=1.806$ are found by imposing the condition that signal and idler exit making an angle $3^\circ$ with the pump.

The propagation delay within the birefringent crystal (5 mm thick each) is given by
\begin{equation}
\Delta \tau = \tau_H-\tau_V=210 fs.
\end{equation}

The coherence time of the pump for a CW diode laser (405 nm Cobolt-06-01-Series) with 1.2 nm linewidth is given by
\begin{equation}
\tau_c=\frac{\lambda^2}{c\Delta\lambda}=455 fs.
\end{equation}

The temporal delay between the photons produced in the first an second crystal leads to distinguishability by providing which crystal information leading to decoherence. If $\Delta \tau$ is the delay between the $H$ and $V$ components produced in first and second crystals, and $\tau_c$ is the coherence time of pump beam then the off diagonal term of the two-qubit density matrix scales as $~\ \exp(-\Delta\tau/\tau_c)$ and resulting density matrix is given by

\begin{equation}
\renewcommand{\arraystretch}{1.5}
\hat{\rho}_\text{decohered}=\left(\begin{array}{cccc}
|\alpha|^2 & 0 & 0 & \alpha\beta^*\exp(-\Delta\tau/\tau_c) \\
0 & 0 & 0 & 0\\
0 & 0 & 0 & 0\\
\alpha^*\beta \exp(-\Delta\tau/\tau_c) & 0 & 0 & |\beta|^2 \end{array}\right).
\end{equation}

Thus, the coherence term of the two-qubit density matrix reduces to $0.5 \exp(-\Delta \tau/\tau_c)=0.3152$ as compared to 0.5, due to temporal walk-off alone.

\subsection{Spatial walk-off} 
``Spatial decoherence [\ref{2.11}] due to spatial mode (emission angle) dependent phase: Ordinary polarized downconversion photons from the first crystal acquire an additional phase in the second crystal, where they are extraordinarily polarized, both because of spatial walk-off and the fact that they have to traverse the additional (i.e., the second) crystal. Three phase terms — the extraordinary phase $\phi_e$, ordinary phase $\phi_o$, and external phase $\phi_\Delta$ -contribute to the total relative spatial phase. Thus, collecting the photons with moderate-size irises (e.g., 5-mm diameter) would greatly decohere the polarization entanglement."

Spatial decoherence can be eliminated by directing the downconversion photons through suitable birefringent compensating crystals that have the opposite phase characteristics as that of the downconversion crystals, a technique called spatial compensation. We have verified experimentally that in the case of single mode fiber optic collection of entangled photons, spatial decoherence does not occur.

\section{Temporal walk-off pre-compensation }

 The BBO crystals give rise to a delay of 210 fs which causes spectro-temporal decoherence leading to a mixed state having a fidelity of $~80\%$ with the Bell state $|\Psi\rangle=[|HH\rangle+|VV\rangle]/\sqrt{2}$. To compensate this temporal delay of $210$ fs, a BBO (Quartz) precompensator of the thickness 1.61 mm (5.65 mm) is required. Since we were using single-mode collection optics, we can ignore the spatial decoherence effects.

\subsection{Partial temporal compensation}

 To combat the spectral-temporal decoherence, initially, we used a 1.5mm thick suboptimal type-II BBO crystal cut for $\theta = 42^{\circ}$ phase matching ($5\times 5\times 1.5~\text{mm}^3$ from Castech Inc., China), available in our lab. It introduces a temporal delay of 360 fs between e- and o-rays at 405 nm. Thus, when this type-II crystal is used as pre-compensator, net delay would be 360 $\pm$ 210 = 570 fs, 150 fs. Thus, in one orientation, it would increase the temporal delay to 570 fs whereas in the orthogonal orientation, delay would be reduced to 150 fs as compared to 210 fs delay due to down-converting crystals.

For pump HWP oriented at $23.1^\circ$, full two-qubit QST was performed for zero- and ninety-degree orientations of the compensator followed by maximum likelihood estimation to reconstruct the density matrix. For zero-degree, it was found that the state fidelity, concurrence, and purity dropped to $53.83\%$, 0.115, and 0.4992, respectively, as compared to $81.4\%$, 0.669, and 0.709. This indicates that the temporal delays are getting added for this orientation leading to drop in the state properties. 

\begin{figure}[H]
\begin{center}
\includegraphics[scale=0.35]{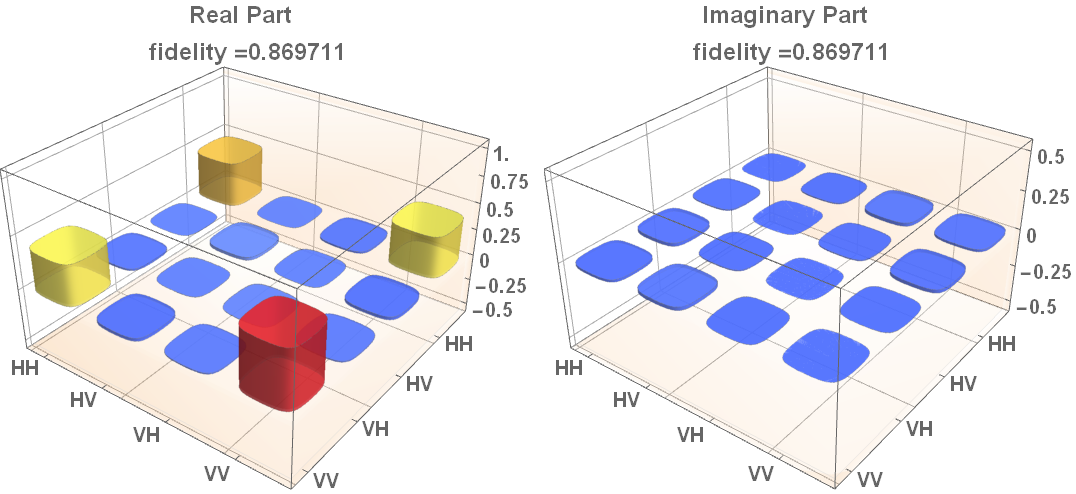}
\end{center}
\caption{\textit{The two-qubit entangled state reconstructed through QST with partial temporal compensation. The left (right) figure shows the real (imaginary) part of the density matrix. The reconstructed state has a fidelity of $ 86.97\%$ with the ideal state, concurrence = 0.764, and purity = 0.776.}\label{fig2.09}}
\end{figure}

Next, compensator was rotated by ninety-degrees and QST was performed. It was found that the state fidelity, concurrence, and purity improved to $86.97\%$, 0.764, and 0.776, respectively. This implies that the temporal delay is getting partially compensated for this orientation leading to improvement in the state properties. Real and imaginary parts of the density matrix is shown in Fig.~\ref{fig2.09}.

 \subsection{Complete temporal compensation}
 
A custom designed compensator with following specifications: $5~\text{mm}\times 5~\text{mm} \times (1.61\pm 0.05~\text{mm})$ type-I BBO crystal cut for $\theta=28.9^\circ,~~\phi=0^\circ$, was procured from (Castech Inc., China) for perfect temporal compensation. Again, for this compensator also, there exist two orientations; one in which delays get added and other for which delay, ideally, completely cancels out. After placing this pre-compensator in correct orientation, QST was performed for different pump HWP orientations. 

\begin{figure}[H]
\begin{center}
\includegraphics[scale=0.35]{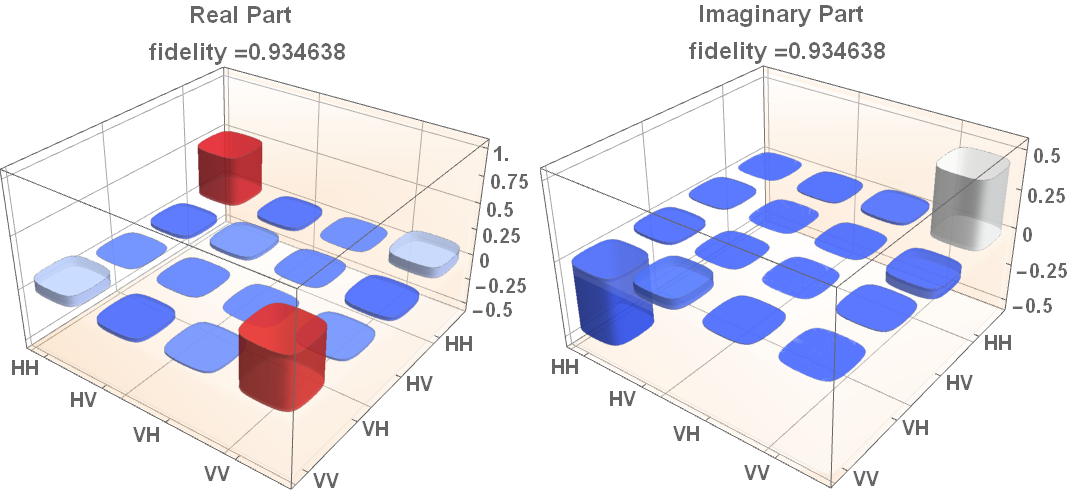}
\end{center}
\caption{\textit{The two-qubit entangled state reconstructed through QST with partial temporal compensation. The left (right) figure shows the real (imaginary) part of the density matrix. The reconstructed state has a fidelity of $ 93.46\%$ with the ideal state, concurrence = 0.909 and purity = 0.891.}\label{fig2.10}}
\end{figure}

For pump HWP at $22.5^{\circ}$ QST was performed and state was reconstructed using MLE. The reconstructed state had fidelity with ideal state: \\  
$|\psi\rangle=\sin(45^\circ)|HH\rangle+\exp(i \phi)\cos(45^\circ)|VV\rangle$, $\phi=283^\circ$ = $93.46\%$, concurrence = 0.909, purity = 0.891. The reconstructed state density matrix is given in Eq.~(\ref{eq2.44}) and 3D plot of the density matrix is shown in Fig.(\ref{fig2.10}).

\begin{equation} \resizebox{1\hsize}{!}{$
\renewcommand{\arraystretch}{1.5}
\rho_\text{exp}=\left(\begin{array}{cccc}
 0.486 & -0.0346-0.008 i & 0.003 +0.018 i & 0.101 +0.441i \\
 -0.035+0.008i & 0.033 & -0.005+0.001 i & -0.037-0.066 i \\
 0.003 -0.018 i & -0.005-0.001 i & 0.002 & 0.018 +0.006 i \\
 0.101 -0.441i & -0.0369+0.0659i & 0.018 -0.006i & 0.479 \\
\end{array}\right). $}
\label{eq2.44}
\end{equation}

\begin{figure}[h!]
\begin{center}
\includegraphics[scale=0.35]{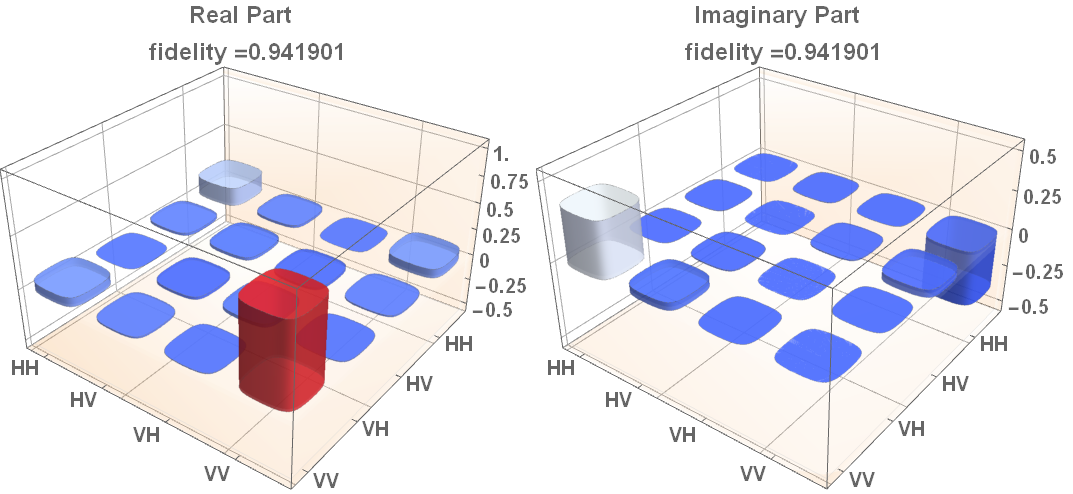}
\end{center}
\caption{\textit{The two-qubit entangled state reconstructed through QST with partial temporal compensation. The left (right) figure shows the real (imaginary) part of the density matrix. The reconstructed state has a fidelity of $ 95.03\%$ with the ideal state, concurrence = 0.717 and purity = 0.929.}\label{fig2.11}}
\end{figure}

Another QST was performed for pump HWP at $13.1^{\circ}$ and state was reconstructed using MLE. The reconstructed state had fidelity with ideal state ($|\psi\rangle=\sin(26.2^\circ)|HH\rangle+\exp(i \phi)\cos(26.2^\circ)|VV\rangle$, $\phi=133^\circ$) = $95.03\%$, concurrence = 0.717, purity = 0.929. The reconstructed state density matrix is given in Eq.~(\ref{eq2.45}) and 3D plot of the density matrix is shown in Fig.(\ref{fig2.11}).

\begin{equation}\resizebox{1\hsize}{!}{$\\
\renewcommand{\arraystretch}{1.5}
\rho_\text{exp}=\left(\begin{array}{cccc}
 0.184 & -0.028+0.0007 i & 0.0081 -0.0006 i & 0.0654 +0.349 i \\
 -0.0276-0.0007 i & 0.0272 & -0.0078+0.0012 i & -0.0317-0.0949 i \\
 0.0081 +0.00063 i & -0.0078-0.0012 i & 0.0023 & 0.0063 +0.0289 i \\
 0.0654 -0.349 i & -0.0317+0.0949 i & 0.0063 -0.0289i & 0.786 \\
\end{array}\right).$}
\label{eq2.45}
\end{equation}

Although after the near perfect temporal compensation, fidelity improved but it was not close to the ideal state and purity was just $\sim 93\%$, we decided to investigate the spectral filtering condition and its effect on state properties.

\section{Effect of spectral filtering on entanglement}

The phase matching, however, cannot be perfect in real experimental situations due to dispersion and the finite thickness of the nonlinear crystal. Therefore, the real phase-matching condition always contains the phase-mismatch term  and it is precisely the phase-mismatch term that gives rise to the spectral bandwidth of the SPDC process.

Therefore, we used a 810-10 nm BPF instead of 800-40 nm BPF used earlier for high collection efficiency. This lead to drop in the overall singles as well as coincidence rate. Next, coincidence was optimized with the new BPF and QST was performed for pump HWP at $23.1^\circ$ (near maximally entangled state) and $13.1^\circ$ (non-maximally entangled state required for the ESD expt).

The ideal two-qubit density matrix for pump HWP at $23.1^\circ$ is given by
\begin{equation}
\renewcommand{\arraystretch}{1.5}
\rho_\text{ideal}=\left(\begin{array}{cccc}
 0.520938 & 0 & 0 & 0.499561 \\
 0 & 0 & 0 & 0 \\
 0 & 0 & 0 & 0 \\
 0.499561 & 0 & 0 & 0.479062 \\
\end{array}\right).
\end{equation}

Experimentally reconstructed density matrix using QST and maximum likelihood estimation is given in Eq.~(\ref{eq2.47}) below. This can be represented using 3D plot of the real and imaginary parts of the density matrix as shown in Fig.~\ref{fig2.12} below.

\begin{equation}\resizebox{1\hsize}{!}{$
\renewcommand{\arraystretch}{1.5}
\rho_\text{exp}=\left(\begin{array}{cccc}
 0.510 & 0.0206+0.0233 i & -0.0209+0.0199 i & 0.483+0.0104 i \\
 0.0206 -0.0233 i & 0.0063 & 0.0012 +0.00047 i & 0.0266 -0.0279 i \\
 -0.0209-0.0199 i & 0.00115 -0.00047 i & 0.0037& -0.0141-0.0183 i \\
 0.483 -0.0104 i & 0.0266 +0.0279 i & -0.0141+0.0183 i & 0.480 \\
\end{array}\right).$}
\label{eq2.47}
\end{equation}

\begin{figure}[h!]
\begin{center}
\includegraphics[scale=0.35]{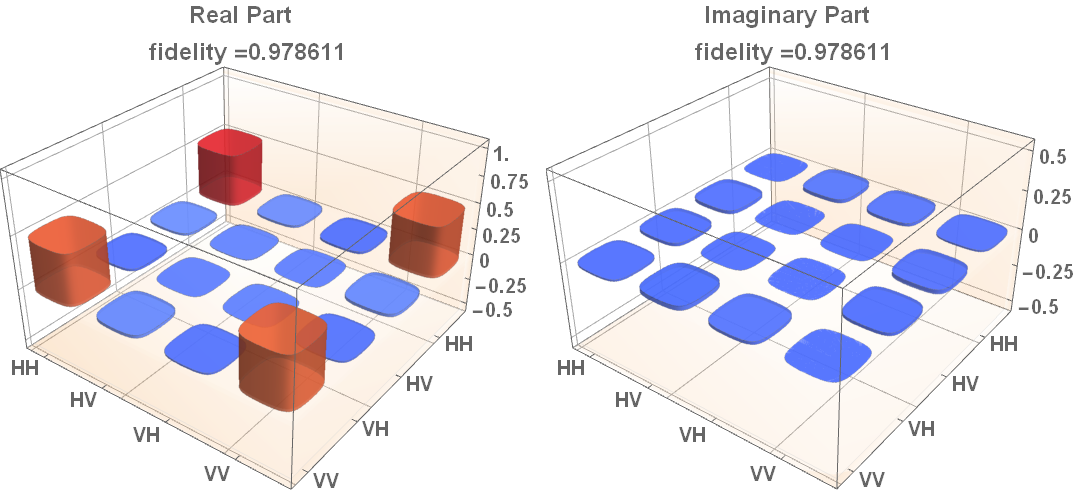}
\end{center}
\caption{\textit{The two-qubit entangled state reconstructed through QST with perfect temporal compensation and narrow spectral filtering using 810-10 nm band pass filters for pump HWP at $23.1^\circ$. Left (right) figure shows the real (imaginary) part of the density matrix. The reconstructed state has a fidelity of $ 97.86\%$ with the ideal state.}\label{fig2.12}}
\end{figure}

The reconstructed state has following properties: 
\begin{itemize}
\item Concurrence of the target/ideal state = 0.9991.
\item Concurrence of the reconstructed state = 0.9652.
\item Fidelity with the ideal state = 0.9785.
\item Purity of the state = 0.9657.
\end{itemize}

Next, QST was performed for pump HWP at $13.1^\circ$. The ideal/target state density matrix is given by

\begin{equation}\\
\renewcommand{\arraystretch}{1.5}
\rho_\text{ideal}=\left(\begin{array}{cccc}
 0.1949 & 0 & 0 & -0.1158-0.3788 i \\
 0 & 0 & 0 & 0 \\
0 & 0 & 0 & 0 \\
 -0.1158+0.3788 i & 0 & 0 & 0.8051 \\
\end{array}\right).
\end{equation}

Experimentally reconstructed density matrix using QST and maximum likelihood estimation is given in Eq.~(\ref{eq2.49}) below. This can be represented using 3D plot of the real and imaginary parts of the density matrix is shown in Fig.~\ref{fig2.13} below.

\begin{equation}\resizebox{1\hsize}{!}{$
\renewcommand{\arraystretch}{1.5}
\rho_\text{exp}=\left(\begin{array}{cccc}
 0.195 & 0.0148 -0.0044 i & 0.003 +0.0004 i & -0.2635-0.2835 i \\
 0.0148 +0.0044i & 0.0067 & -0.0054-0.0024 i & -0.0092-0.0366 i \\
 0.003-0.0004 i & -0.0054+0.0024 i & 0.0105 & -0.0045+0.0054 i \\
 -0.2635+0.2835i & -0.0092+0.0366i & -0.0045-0.0054i & 0.788 \\
\end{array}\right).$}
\label{eq2.49}
\end{equation}

\begin{figure}[!htb]
\begin{center}
\includegraphics[scale=0.35]{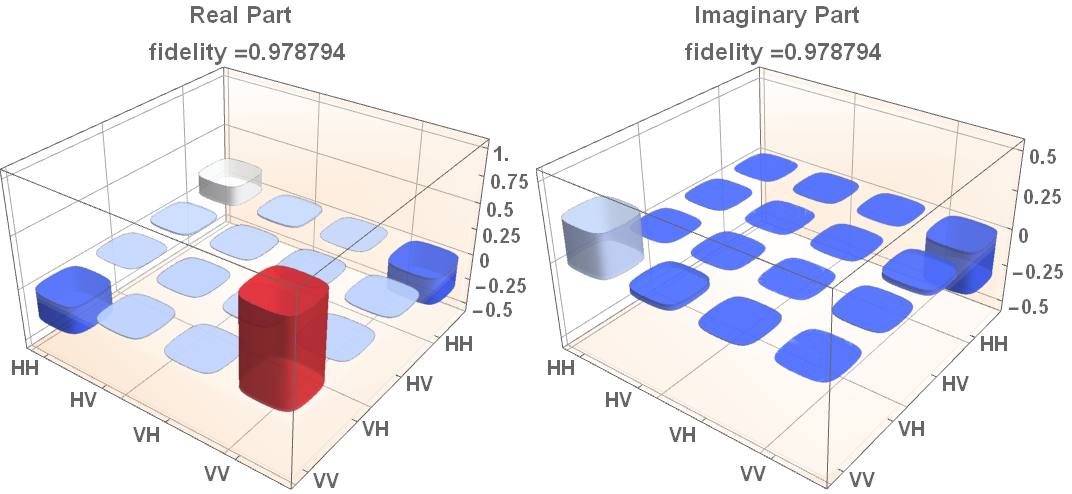}
\end{center}
\caption{\textit{The two-qubit entangled state reconstructed through QST with nearly perfect temporal precompensation and narrow spectral filtering using 810-10 nm band pass filters for pump HWP at $13.1^\circ$. Left (right) figure shows the real (imaginary) part of the density matrix. The reconstructed state has a fidelity of $ 97.88\%$ with the ideal state.}\label{fig2.13}}
\end{figure}

The reconstructed state has the following properties: 
\begin{itemize}
\item Concurrence of the ideal state = 0.7923.
\item Concurrence of the reconstructed state = 0.7659.
\item Fidelity with the ideal state = 0.9788.
\item Purity of the state: 0.9617.
\end{itemize}

\section{Visibility of entanglement}
The experimental setup to measure visibility of entanglement is same as shown in Fig.~\ref{fig2.01}. For measuring visibility in $|HH\rangle$ ($|VV\rangle$) bases, one one side QWP is set to zero degree and HWP is set to zero degree ($45^\circ$) so that it acts as $|H\rangle\langle H|$ ($|V\rangle\langle V|$) projector and on the other side HWP is rotated from $0^\circ$ to $140^\circ$ (keeping QWP to zero degree) in the steps of $2^\circ$ and coincidence was recorded for fixed acquisition time of 100 s each. Likewise, for measuring visibility in $|LL\rangle\langle LL|$ bases, one one side, HWP was set to zero degree and QWP was set to $22.5^\circ$ such that it  it acts as \text{or} $|L\rangle\langle L|$ projector and QWP on the other side is rotated from $0^\circ$ to $140^\circ$ and coincidence is recorded for a fixed acquisition time of 100 s each. Normalized coincidence is plotted as a function of the waveplate orientation for three different visibility measurements. 

\begin{figure}[H]
\begin{center}
\includegraphics[width=\textwidth]{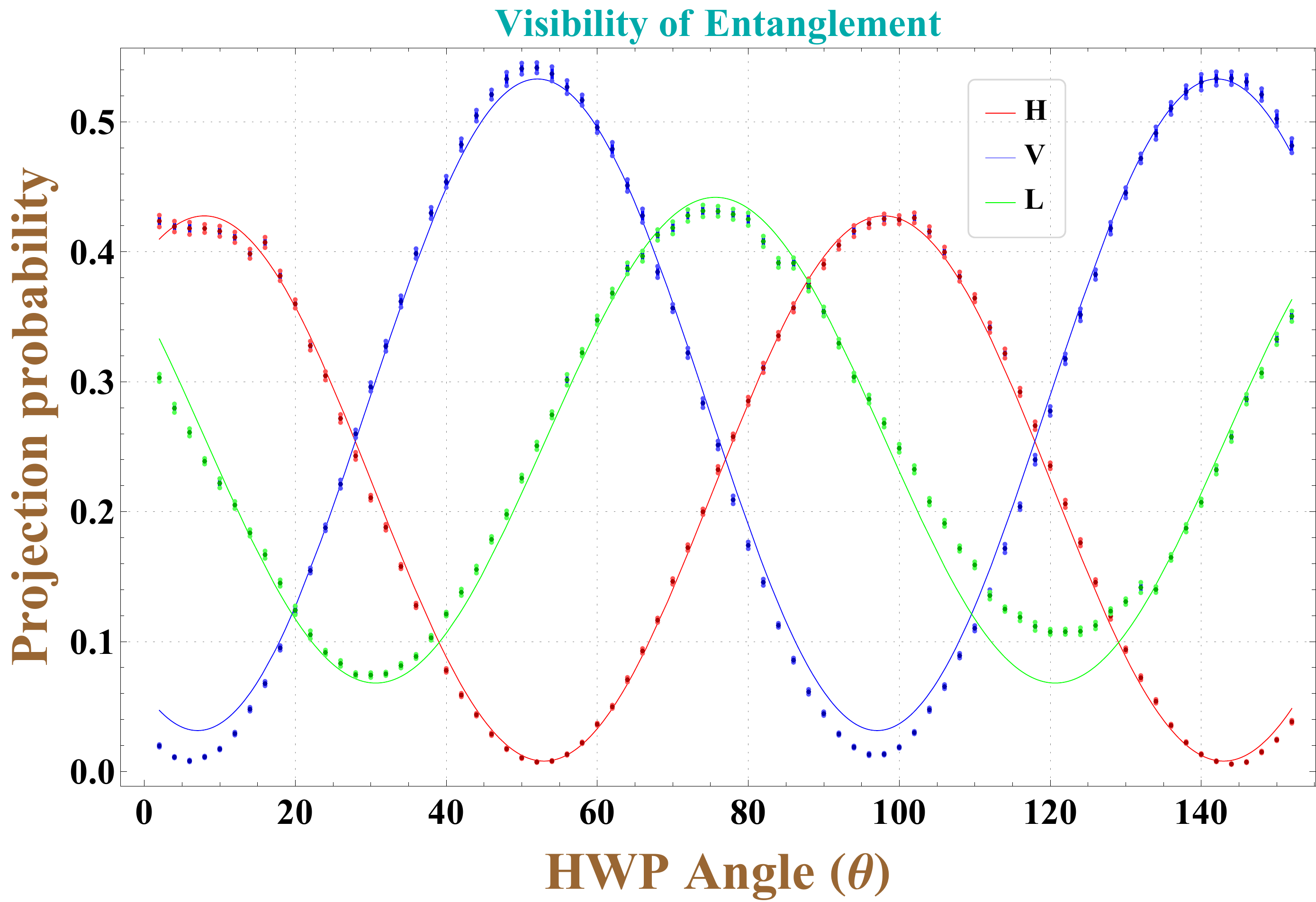}
\end{center}
\caption{\textit{Result of the entanglement visibility measurement in $|HH\rangle,~|VV\rangle,~\text{and}~|LL\rangle$ bases. Note that when this data was taken, soure fidelity was about $\sim 88\%$.}\label{fig2.14}}
\end{figure}

In the case of Bell states, one can give a lower bound on the fidelity of entanglement by performing Visibility measurement in two mutually unbiased bases, say $|HV\rangle$ and $|DA\rangle$. Fidelity of the experimental state from Bell state is then defined as the average value of visibility in the two-photon correlation function obtained in the two mutually unbiased bases.

\section{References}
\begin{enumerate}
\item \textit{G. G. Stokes, ``On the Composition and Resolution of Streams of Polarized Light from different Sources", Trans. Camb. Phil. Soc. 9, 399 (1852); Reprinted in \href{https:\\doi:10.1017/CBO9780511702266.010}{Mathematical and Physical Papers, Vol. \textbf{3}, 233-258 (1901)}}. \label{2.01}
\item \textit{Dennis H. Goldstein, ``Polarized Light", \href{https://doi.org/10.1201/b10436}{CRC Press (2011)}}. \label{2.02}
\item \textit{Arun Kumar, Ajoy Ghatak, ``Polarization of Light: With Applications in Optical Fibers", \href{https://doi.org/10.1117/3.861761}{Bellingham, WA: SPIE Press (2011)}}. \label{2.03}
\item \textit{Daniel F. V. James, Paul G. Kwiat, William J. Munro, and Andrew G. White, ``Measurement of qubits", \href{https://doi.org/10.1103/PhysRevA.64.052312}{Phys. Rev. A \textbf{64}, 052312 (2001)}}.\label{2.04} 
\item \textit{J.B.Altepeter, E.R.Jeffrey, P.G.Kwiat, ``Photonic State Tomography", \href{https://doi.org/10.1016/S1049-250X(05)52003-2}{Adv. At. Mol. Opt. Phys. \textbf{52},105-159 (2005)}}.\label{2.05}
\item\textit{Ermes Toninelli, Bienvenu Ndagano, Adam Vallés, Bereneice Sephton, Isaac Nape, Antonio Ambrosio, Federico Capasso, Miles J. Padgett, and Andrew Forbes, ``Concepts in quantum state tomography and classical implementation with intense light: a tutorial", \href{https://doi.org/10.1364/AOP.11.000067}{Adv. Opt. Photon. \textbf{11}, 67-134 (2019)}}. \label{2.06}
\item \textit{Nicholas Peters, Joseph Altepeter, Evan Jeffrey, David Branning, and Paul Kwiat, ``Precise creation, characterization, and manipulation of single optical qubits",  J. Quant. Inf. Comp. \textbf{3}, 503 (2003)}. \label{2.07}
\item \textit{E. Hecht and A. Zajac, Optics (Addision-Wesley, Reading,
MA, 1974)}.\label{2.08} 
\item \textit{Paul G. Kwiat, Edo Waks, Andrew G. White, Ian Appelbaum, and Philippe H. Eberhard, ``Ultrabright source of polarization-entangled photons," \href{https://doi.org/10.1103/PhysRevA.60.R773}{Phys. Rev. A \textbf{60}, 2, pp. R773-R776 (1999)}}.\label{2.09}
\item \textit{AG White, DFV James, PH Eberhard, PG Kwiat, ``Nonmaximally entangled states: Production, characterization, and utilization," \href{https://doi.org/10.1103/PhysRevLett.83.3103}{Phys. Rev. Lett. \textbf{83}, 16, pp. 3103-3107 (1999)}}.\label{2.10}
\item  \textit{Radhika Rangarajan, Michael Goggin, and Paul Kwiat, ``Optimizing type-I polarization-entangled photons," \href{https://doi.org/10.1364/OPEX.13.008951}{ Optics express \textbf{17}, 21, pp. 18920-18933 (2009)}}.\label{2.11}
\item \textit{J. B. Altepeter, E. R. Jeffrey, and P. G. Kwiat, ``Phase-compensated ultra-bright source of entangled photons," \href{https://doi.org/10.1364/OPEX.13.008951}{Opt. Express 13, 8951-8959 (2005)}}.\label{2.12}
\item \textit{G. M. Akselrod, J. B. Altepeter, E. R. Jeffrey, and P. G. Kwiat, ``Phase-compensated ultra-bright source of entangled photons: erratum," \href{https://doi.org/10.1364/OE.15.005260}{Opt. Express 15,pp. 5260-5261 (2007)}}.\label{2.13}
\item \textit{S Cialdi, F Castelli, I Boscolo, MG Paris, ``Generation of entangled photon pairs using small-coherence-time continuous wave pump lasers,"  \href{https://doi.org/10.1364/AO.47.001832}{Appl. Opt. \textbf{47}, 11, pp. 1832-1836 (2008)}}.\label{2.14}
\item \textit{So-Young Baek and Yoon-Ho Kim, ``Spectral properties of entangled photon pairs generated via frequency-degenerate type-I spontaneous parametric down-conversion", \href{https://doi.org/10.1103/PhysRevA.77.043807}{Phys. Rev. A \textbf{77}, 043807 (2008)}}.\label{2.15}
\end{enumerate}

%% file: chapter3.tex
\setcounter{equation}{0}
\chapter {Revisiting comparison between entanglement measures}

Given a non-maximally entangled state, an operationally significant question is to quantitatively assess as to what extent the state is away from the maximally entangled state, which is of importance in evaluating the efficacy of the state for its various uses as a resource. It is this question which is examined in this chapter for two-qubit pure entangled states in terms of different entanglement measures like Negativity (N), Logarithmic Negativity (LN), and Entanglement of Formation (EOF). Although these entanglement measures are defined differently, to what extent they differ in quantitatively addressing the earlier mentioned question has remained uninvestigated. Theoretical estimate in this chapter shows that an appropriately defined parameter characterizing the fractional deviation of any given entangled state from the maximally entangled state in terms of N is quite different from that computed in terms of EOF with their values differing up to $\sim 15\%$ for states further away from the maximally entangled state. Similarly, the values of such fractional deviation parameters estimated using the entanglement measures LN and EOF, respectively, also strikingly differ among themselves with the maximum value of this difference being around $23\%$. This analysis is complemented by illustration of these differences in terms of empirical results obtained from a suitably planned experimental study. Thus, such appreciable amount of quantitative non-equivalence between the entanglement measures in addressing the experimentally relevant question considered in the present chapter highlights the requirement of an appropriate quantifier for such intent. We indicate directions of study that can be explored towards finding such a quantifier.

\section{Introduction: Background and Motivation}

Entanglement lies at the core of Quantum Foundational studies leading to Information Theoretic applications and forms the bedrock of Quantum Computation. One of the key concepts used for studying entanglement is what is known as Entanglement Measure (EM) which is invoked for quantifying entanglement. For this purpose, different EMs have been proposed. It was argued by Bennett \textit{et al.} [\ref{3.01}, \ref{3.02}] that the Entanglement of Formation (EOF), intended to quantify the resources needed to create a given entangled state, satisfies the criterion of being nonincreasing under local operations and classical communication (LOCC); for bipartite pure states it is given by the von Neumann entropy of reduced density matrix relevant to either Alice or Bob, also known as Entanglement Entropy. Justification of the above EM from the thermodynamic considerations was given by Popescu \textit{et al.} [\ref{3.03}], followed by a comprehensive analysis due to Vedral \textit{et al.} [\ref{3.04},\ref{3.05}] and Vidal [\ref{3.06}] who argued that only one EM is not sufficient to completely quantify entanglement of pure states for bipartite systems. Subsequently, $\dot{Z}$yczkowski \textit{et al.} [\ref{3.07},\ref{3.08}] defined Negativity as a ``quantity capable of measuring a degree of entanglement". Later, Negativity was proved to be a valid EM [\ref{3.09}-{\ref{3.11}] by showing that it is an entanglement monotone, i.e., nonincreasing under LOCC.

In this chapter, we have used the particular expression of Negativity (N) given by Vidal and Werner [\ref{3.10}], who also defined another quantity called Logarithmic Negativity (LN = $\log_{2}(2\text{N}+1)$) as a valid EM which exhibits a form of monotonicity under LOCC (non-increasing under deterministic distillation protocols) and signifies an upper bound of distillable entanglement. In a separate line of work, for bipartite qubit states, Wootters [\ref{3.12}] expressed EOF as a monotonic function of a quantity called `Concurrence' and argued that Concurrence can also be regarded as a measure of entanglement. Note that, for bipartite pure qubit states, Concurrence is twice of Negativity [\ref{3.13}], thus implying that EOF is also a monotonic function of N and LN for such states. The Fig.~\ref{fig3.1} below shows the comparison of different EMs for  a two-qubit pure state: $|\psi\rangle=c_0|00\rangle+c_1|11\rangle$, where $c_0$ and $c_1$ are the Schmidt coefficients. It is worth noting that (a) Concurrence and twice Negativity, and (b) Entanglement of Formation and Entanglement Entropy match with each other. Therefore, in this work, while considering essentially two-qubit pure states,  we focus on N, LN and EOF as the relevant EMs as these are the ones which do not overlap.
	  
\begin{figure} [h!] 
\begin{center}
\includegraphics[clip, trim=0cm 0cm 0cm 0cm, width=0.8 \columnwidth]{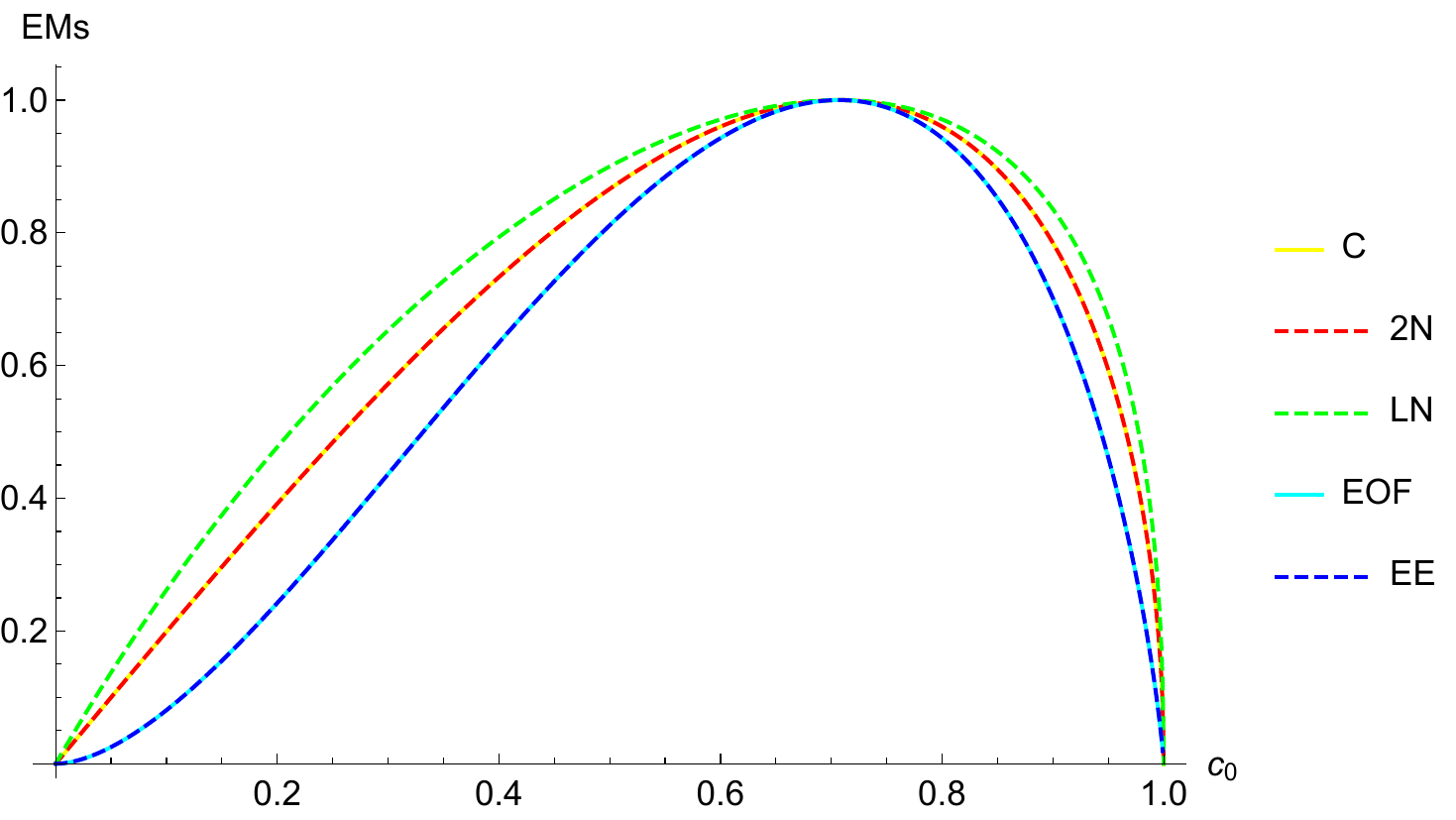}
\end{center}
\caption{\textit{A comparison of different entanglement measures with respect to the state parameter $c_0$ for two-qubit pure states. Here, C, N, LN, EOF, and EE denote Concurrence, Negativity, Logarithmic Negativity, Entanglement of Formation, and Entanglement Entropy, respectively.}\label{fig3.1}}
\end{figure}
	 
From the Fig.~\ref{fig3.1}, it can be seen that although for any given two-qubit pure state, the values of N, LN and EOF differ among themselves, these EMs are monotonic with respect to each other. Hence, for comparing the amount of entanglement between two-qubit pure states, N, LN and EOF are all equivalent in the sense that all these EMs give the same result in answering the question as to whether a given two-qubit pure state is more (less) entangled than any other state. On the other hand, in this chapter, an operationally relevant different question is addressed; i.e., of quantifying the percentage deviation of a given state from the maximally entangled state- it is in this context, we consider the issue of comparison between the various EMs. To this end, by appropriately defining the measures of such percentage deviations in terms of N, LN and EOF, respectively, we present in Section \ref{sec3.2} the theoretical estimates of these measures for the general class of two-qubit pure states by varying values of the Schmidt coefficients. This thorough study reveals a considerable amount of disagreement between the computed measures of percentage deviations from the maximally entangled state using N, LN and EOF, respectively. 

A complementary line of study of this issue is then presented in Section \ref{sec3.3} by considering a range of states produced in a relevant experimental study for which the quantities N, LN and EOF are determined from the density matrix reconstructed using quantum state tomography. The results obtained in this way confirm significant quantitative non-equivalence between these EMs in capturing the extent to which a given non-maximal entangled state is deviating from the maximally entangled state. This finding, therefore, underscores the need for identifying an appropriate quantifier for addressing such an empirically relevant question even in the simplest case of  $2\otimes 2$ composite systems. 

In Section~\ref{sec3.4}, we extend the current analysis to higher dimensional systems; bipartite pure qutrit states and tripartite pure qubit states, in particular. In Section~\ref{sec3.5}, we discuss possible directions of study for addressing this issue based on the various suggested ideas of `distance measure' between quantum states, as well as a line of study is outlined in terms of the deviation of the maximum value of the violation of Bell-CHSH inequality for a given state from that  corresponding to the maximally entangled state. Further, indications have been given about the way  the results obtained from such studies can be compared with those obtained from different EMs. This is followed by concluding remarks in Section~\ref{sec3.6}.
	 
\section{Theoretical study of the deviation of any given state from the MES using different EMs}
\label{sec3.2}

Consider a two-qubit pure state with Schmidt coefficients $c_0$ and $c_1$ as given below.
\begin{align}
\ket{\Psi} = c_{0}\ket{0}\ket{0}+c_1\ket{1}\ket{1}, 
\label{eq3.1}
\end{align}
where $c_{0}$ and $c_{1}$ satisfy the relation $0 \leq c_{0},c_{1} \leq 1 $ and $c_{0}^2 + c_{1}^2 = 1$. The state (\ref{eq3.1}) is maximally entangled for $c_{0},c_{1}=1/\sqrt{2}$, and  separable for $c_0, c_1 = 0,1$.
     
The three EMs discussed above for a two-qubit pure state are given by
\begin{subequations}
\begin{align}
\begin{split}
\text{N} &= c_{0}c_{1},
\end{split}\\
\begin{split}
\text{LN} &= \log_{2}(2c_{0}c_{1}+1),
\end{split}\\
\begin{split}
\text{EOF} &= - c_{0}^2\log_{2}c_{0}^2-c_{1}^2\log_{2}c_{1}^2.
\end{split}
\end{align}
\end{subequations}
    
In order to quantify the deviation of a given entangled state from the maximally entangled state, the following parameters are defined as measures of fractional deviations in terms of the quantities N, LN and EOF whose maximum values for the maximally entangled state are 0.5, 1, and 1, respectively.
\begin{subequations}
\begin{align}
\begin{split}
Q_{\text{N}} &= (0.5 - \text{N})/0.5~ ,  
\end{split}\\
\begin{split}
Q_{\text{L}} &= (1-\text{LN}),  
\end{split}\\
\begin{split}
Q_{\text{E}} &= (1 -\text{EOF}).
\end{split}
\end{align}
\end{subequations}

Note that all the above parameters range from 0 to 1, with 0 for the maximally entangled state and 1 for the separable state. For quantifying the extent to which these three parameters differ with each other, the following quantities are defined as absolute differences between the respective fractional deviations defined above.
\begin{subequations}
\begin{equation}
\Delta Q_{\text{NL}} = |Q_{\text{N}}- Q_{\text{L}}|, 
\end{equation}
\begin{equation}
\Delta Q_{\text{EL}} = |Q_{\text{E}}- Q_{\text{L}}|,
\end{equation}
\begin{equation}
\Delta Q_{\text{NE}} = |Q_{\text{N}}- Q_{\text{E}}|.
\end{equation}
\label{eq4}
\end{subequations}
    
Different values of the quantities $Q_{\text{N}}$, $Q_{\text{L}}$, $Q_{\text{E}}$ , $\Delta Q_{\text{NL}}$, $\Delta Q_{\text{EL}}$, and $\Delta Q_{\text{NE}}$ corresponding to different values of Schmidt coefficients have been incorporated in Table~\ref{tab3.1} as percentage values.
\begin{table}[htb]
\begin{center}
\resizebox{\textwidth}{!}{
\renewcommand{\arraystretch}{2}
\begin{tabular}{|l|c|c|c|c|c|c|c|c|c|r|}
\hline
$\mathbf{c_{0}}$ &  \textbf{N} & \textbf{LN} & \textbf{EOF} & $\mathbf{Q_{N}}(\%)$ & $\mathbf{Q_{L}}(\%)$ & $\mathbf{Q_{E}}(\%)$ &   $ \mathbf{\Delta Q_{NL}}(\%)$  & $\mathbf{\Delta Q_{EL}}(\%)$  & $\mathbf{\Delta Q_{NE}}(\%)$ \\
\hline \hline
0.1  &  0.099 & 0.262 &   0.081  & 80.10  & 73.82 & 91.92&  6.28 & 18.10  & 11.82 \\
\hline
0.2  &  0.196 & 0.477 &   0.242 & 60.81 & 52.29  & 75.77 & 8.52 & 23.48  & 14.96  \\
\hline
0.4  &  0.367 & 0.793 &   0.634 & 26.68  & 20.66  & 36.57   & 6.02  & 15.91  & 9.89  \\
\hline
0.7  &  0.499 & 0.999 &   0.999 & 0.02  & 0.01  &  0.03  &  0.01  & 0.01  & 0.01 \\
\hline
0.7071 &  0.5 & 1  & 1 & 0 & 0 & 0  &  0  & 0  & 0  \\
\hline
0.8  &  0.480 & 0.971 &   0.943 & 4.00  & 2.91  & 5.73  &  1.09  & 2.82  & 1.73 \\
\hline
0.9  &  0.392 & 0.836 &   0.701 & 21.54 & 16.44  & 29.85  & 5.10  & 13.41  & 8.31   \\
\hline
\end{tabular}}
\caption{\textit{Differences between the respective fractional deviation parameters for different EMs for two-qubit pure states given as percentage values.}\label{tab3.1}}
\end{center}
\end{table}

It is evident from Table~\ref{tab3.1} that for a given entangled state the percentage deviations are different for different EMs. For example, for a state with $c_{0}=0.4$ (where $c_{0}=0.7071$ corresponds to the maximally entangled state),  its percentage deviation from the maximally entangled state is $26.68\%$ when N is used to quantify entanglement; the percentage deviation is $20.66\%$ when LN is used to quantify entanglement, and is $36.57\%$ when one uses EOF. Thus, in this case, the differences in the percentage deviations are, respectively, given by  $\Delta Q_{\text{NL}}=6.02\%$ , $\Delta Q_{\text{EL}}=15.91\%$, and $\Delta Q_{\text{NE}}=9.89\%$.
      
Numerical study by optimization of the Schmidt coefficients to find the maximum deviations in $\Delta Q$ leads to the following results: 

\begin{itemize}
\item Maximum value of $\Delta Q_{\text{NL}}$ is 8.61\% corresponding to the states with $c_{0}$ = 0.227  and 0.974.
\item Maximum value of $\Delta Q_{\text{EL}}$  is 23.57\% corresponding to the states with  $c_{0}$ = 0.217 and 0.976.
\item Maximum value of $\Delta Q_{\text{NE}}$  is 14.99\% corresponding to the states with $c_{0}$ = 0.210 and 0.978.
\end{itemize}
   
Note that the disagreement between the values of $\Delta Q$s increases for the states further away from the maximally entangled state, reaching a maximum value, and then decreases for the states getting closer to the separable state. Now, in order to analyze the way the above mentioned differences between the deviation parameters occur, we obtain the following results by studying the derivatives of different EMs with respect to the Schmidt coefficient  $c_{0}$ characterizing the two-qubit pure state:
     
The derivative of N with respect to $c_{0}$ is given by
\begin{align}
\dfrac{\text{dN}}{dc_{0}} = \frac{1-2c_{0}^2}{\sqrt{1-c_{0}^2}}.
\label{eq3.05}
\end{align}    
The derivative of LN with respect to $c_{0}$ is given by
\begin{align}
\dfrac{d\text{LN}}{dc_{0}} =  \frac{2(1-2c_{0}^2)}{\sqrt{1-c_{0}^2}~[2c_{0}\sqrt{1-c_{0}^2}+1]\ln(2)}.
\label{eq3.06}
\end{align}  
The derivative of E with respect to $c_{0}$ is given by
\begin{align}
\dfrac{d\text{EOF}}{dc_{0}} = \frac{2c_{0}\log_{2}[(1-c_{0}^2)/c_{0}^2]}{\ln(2)}.
\label{eq3.07}
\end{align}  

It can then be seen that the values of 2N, LN and EOF increase with $c_{0}$ starting from 0 and reach their respective maximum values corresponding to the maximally entangled state when $c_{0}$ is $1/\sqrt{2}$ and then start decreasing with further increasing values of $c_{0}$. Note that although the values of 2N, LN and EOF start from 0, the quantity LN kicks off rapidly due to the higher value of its derivative with respect to the state parameter $c_{0}$ as compared to that of N and EOF. Hence, for any $c_{0}$, the quantity LN is always greater than 2N and EOF, and has least deviation from the maximally entangled state. On the other hand, for any value of $c_{0}$, EOF is always less than 2N and LN,  and  has the highest deviation from the maximally entangled state. These features are illustrated in Figs.~\ref{fig3.2}-\ref{fig3.5}. Monotonicity of N, EOF and LN with respect to each other can be further verified from Eqs.~(\ref{eq3.05}-\ref{eq3.07}) by computing the quantities $\frac{d\text{N}}{d\text{EOF}}$, etc.

\begin{figure}[H] 
\begin{center}
\includegraphics[width=0.6\columnwidth, keepaspectratio]{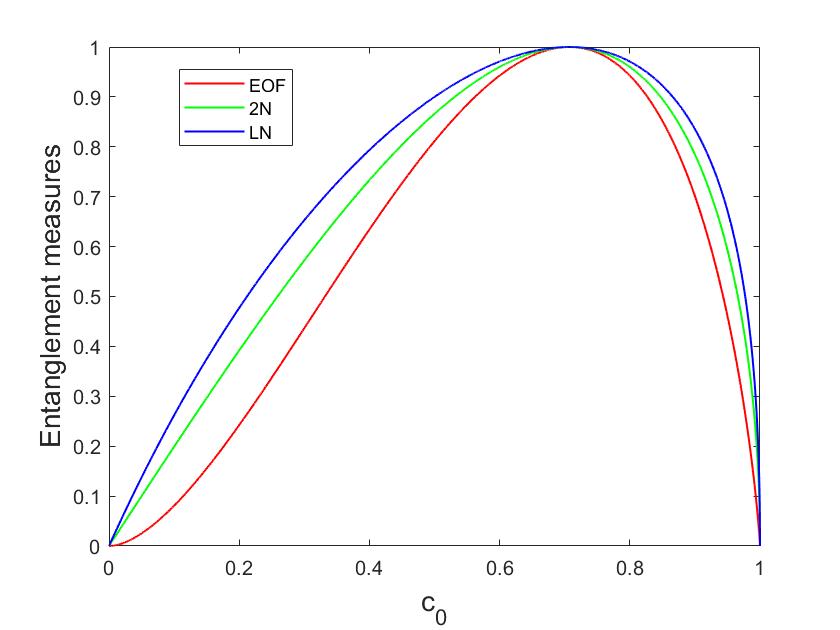}
\end{center}
\caption{\textit{This figure illustrates the variation of different entanglement measures with respect to the state parameter $c_{0}$.}\label{fig3.2}}
\end{figure}

\begin{figure}[H] 
\begin{center}
\includegraphics[width=0.6\columnwidth, keepaspectratio]{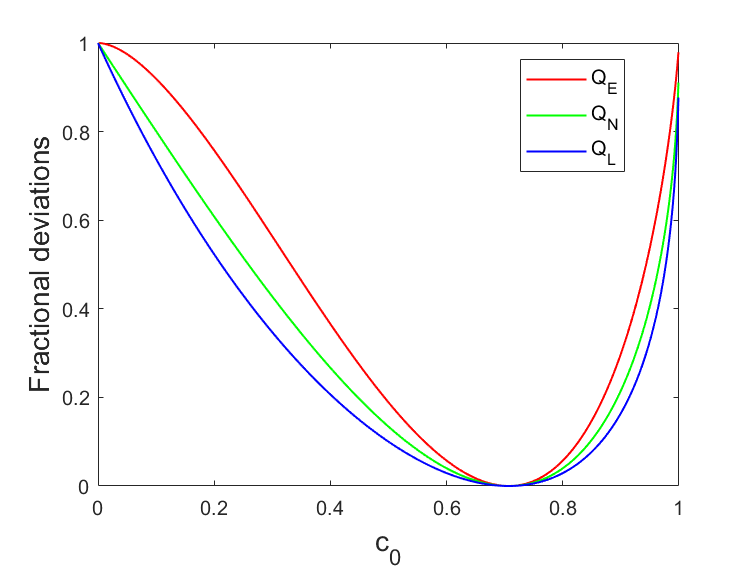}
\end{center}
\caption{\textit{This figure shows how the fractional deviations of a given state from the maximally entangled state calculated using different entanglement measures N, LN and EOF vary with respect to the state parameter $c_{0}$.}\label{fig3.3}}
\end{figure}

\begin{figure}[H] 
\centering
\includegraphics[width=0.6\columnwidth, keepaspectratio]{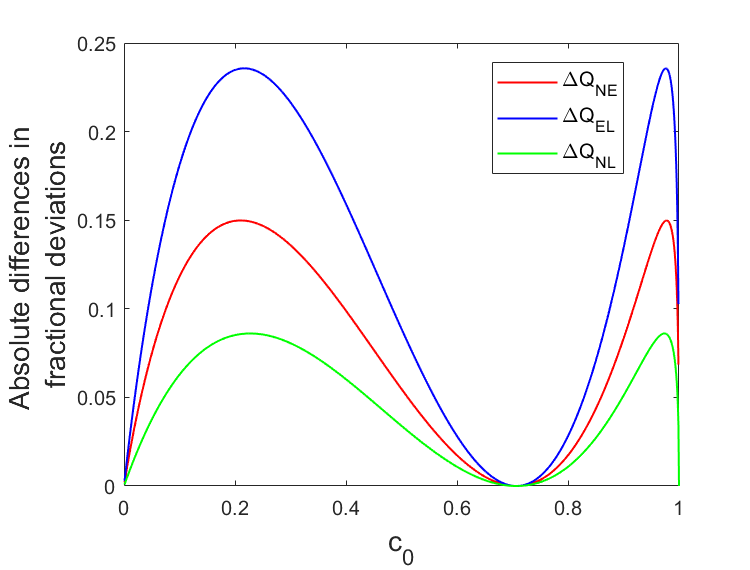}
\caption{\textit{ This figure illustrates the differences in the fractional deviations of the respective states from the maximally entangled state calculated using different entanglement measures N, LN and EOF by varying the state parameter $c_{0}$.}\label{fig3.4}}
\end{figure}

\begin{figure}[H] 
\centering
\includegraphics[width=0.7\columnwidth, keepaspectratio]{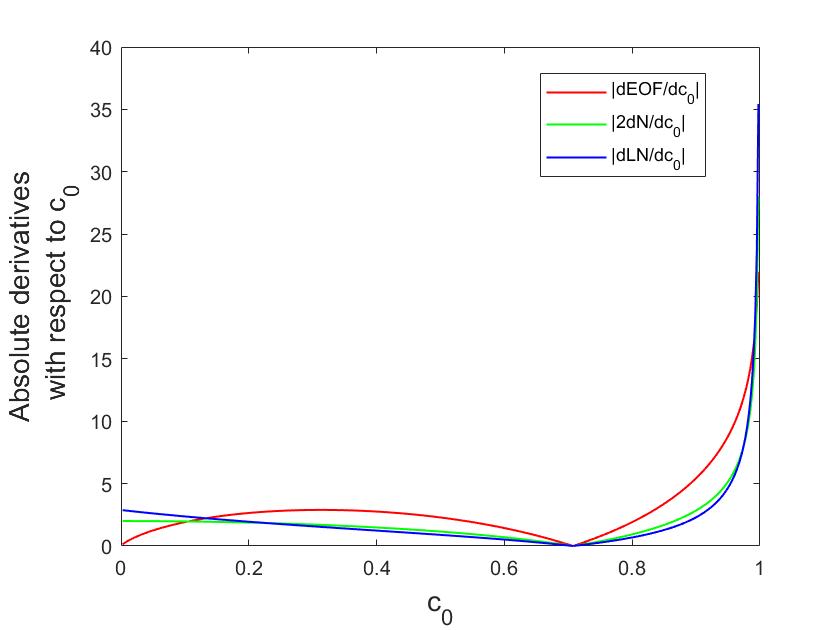}
\caption{\textit{This figure shows how the absolute values of the derivatives of different entanglement measures; 2N, LN and EOF with respect to state parameter $c_{0}$, vary with $c_{0}$.}\label{fig3.5}}
\end{figure}

\section{Deviation of the experimental states from MES using different EMs} \label{sec3.3}
 
Polarization entangled photon pairs are produced by the second order non-linear optical process of Spontaneous Parametric Down-Conversion (SPDC) in a two-crystal geometry [\ref{3.14}]. A 100 mW, Continuous Wave (CW) diode laser having central wavelength at 405 nm and a bandwidth of 1.2 nm (405 nm Cobolt-06-01-Series) was used as the pump laser. Two type-I BBO ($\beta-B_aB_2O_4$) crystals in sandwich configuration ($5\times 5\times 0.5 ~\text{mm}^3$ each from Castech Inc., China) having their optic axes orthogonal to each other and phase matched at $\theta=28.9^\circ$ and $\phi=0^\circ$ with half opening angle of the cone equal to $3^\circ$ was used for producing entangled photon pairs. Schematic of the experimental set up is shown in Fig.~\ref{fig4} below. 

 \begin{figure} [!htb]
\begin{center}
\includegraphics[clip, trim=4.5cm 16cm 3.95cm 2.5cm, width=0.7\linewidth]{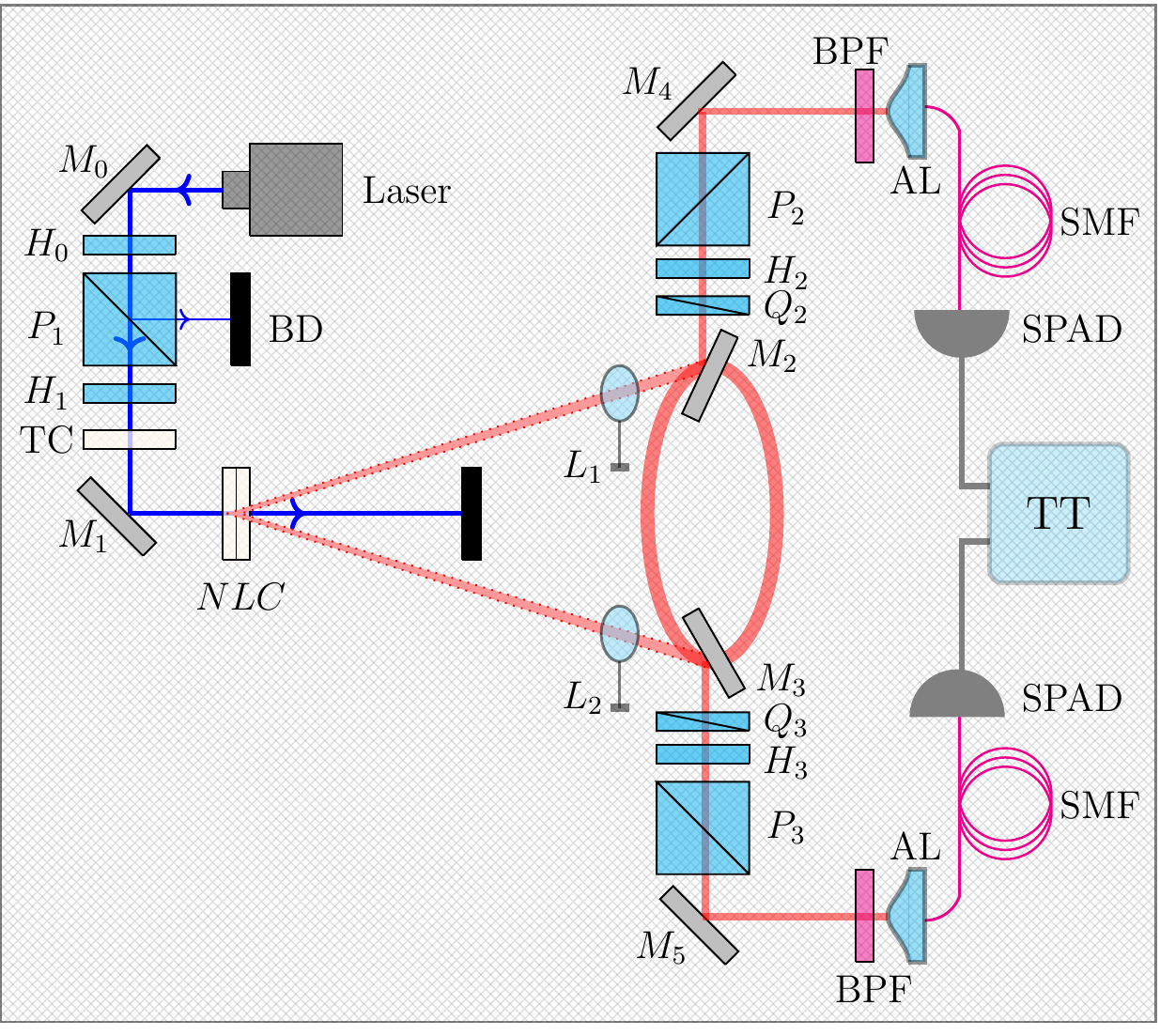}
\end{center}
\caption{\textit{Schematic of the experimental apparatus (not to scale) for preparation of SPDC based type-I polarization entangled photon source using two-crystal geometry and characterization using quantum state tomography. Different symbols have the following meaning: P, polarizing beam splitter; Q, quarter wave plate; H, half wave plate; NLC, non-linear crystal; TC, temporal compensator; L, plano-convex lens; M, mirror; BPF, bandpass filter; AL, aspheric lens; SMF, single mode fiber; SPAD, single-photon avalanche diode; and TT, time tagger unit or coincidence module.}}
\label{fig4}
\end{figure}

The pump beam is passed through a half-wave plate (HWP) and polarizing beam splitter (PBS) to get pure H-polarized laser beam. This laser beam is then passed through a HWP ($\text{H}_1$) with fast axis oriented at an angle $\theta$ with respect to vertical, which prepares the pump polarization state to be inputted to the BBO crystals for the preparation of different entangled states.

\begin{equation}
|H\rangle \xrightarrow{\text{HWP at} ~\theta} \sin(2\theta)|H\rangle+\cos(2\theta)|V\rangle =\alpha |H\rangle+\beta|V\rangle,
\end{equation}
where $\alpha=\sin(2\theta)$, and $\beta=\cos(2\theta)$.

Generation of entangled photons through SPDC process in a two-crystal geometry can be understood as follows: a H-polarized pump photon gets down-converted in the first crystal (whose optic axis is in horizontal plane) into a pair of V-polarized photons, and a V-polarized pump photon gets down-converted in the second crystal (whose optic axis is in vertical plane) into a pair of H-polarized photons. If these two processes occur in a coherent manner, i.e., they are indistinguishable, the two down-converted polarization amplitudes are coherently added and the resultant state becomes an entangled state as given below.
\begin{equation}
\begin{aligned}
\begin{split}
\alpha|H\rangle+\beta|V\rangle 
\xrightarrow {\text{SPDC}} \alpha |VV\rangle+\beta \exp(i\phi)|HH\rangle.
\end{split}
\end{aligned}
\end{equation}

The relative phase $\phi$ depends on the optical path difference/delay between the photons down-converted in the first and second crystals. It can be controlled by a tilting a quarter-wave plate in the pump beam (not shown in Fig.\ref{fig4}).

The SPDC photons created in the first crystal get delayed compared to those created in the second crystal, thus giving rise to temporal distinguishability leading to drop in the quality of entanglement. This temporal delay is pre-compensated [\ref{3.15}] using another type-I BBO crystal (TC) of thickness 1.6 mm. The SPDC photons are then passed through a Quantum State Tomography (QST) setup [\ref{3.16}]  consisting of quarter-wave plate, half-wave plate and PBS on either side and collected through single mode fiber using aspheric lens and 810-10 nm band pass filter used for spectral filtering. These photons are then detected by single photon detectors and 36-coincidence measurements are performed for acquisition time of 60 s each. These measurements correspond to the projections in different bases that are required in QST for the state reconstruction. The maximum-likelihood estimation (MLE) [\ref{3.16}] is used to get the physical state (density matrix)  from the QST data which is expected to have some experimental imperfections. 

Here, we have prepared three different two-qubit entangled states (three sets each) for pump HWP oriented at $23.1^\circ$ (State-I), $13.1^\circ$ (State-II) and $9.1^\circ$ (State-III). These states have average purity (where purity is denoted by P and defined as Tr[$\rho^2$], $\rho$ being the density matrix of the system) better than $95.7\%$. The representative 3D plots of the density matrices reconstructed through QST and MLE are shown in Figs.~\ref{fig3.7}-\ref{fig3.9} below.
\begin{figure}[H]
\centering
\includegraphics[width=0.7\columnwidth, keepaspectratio]{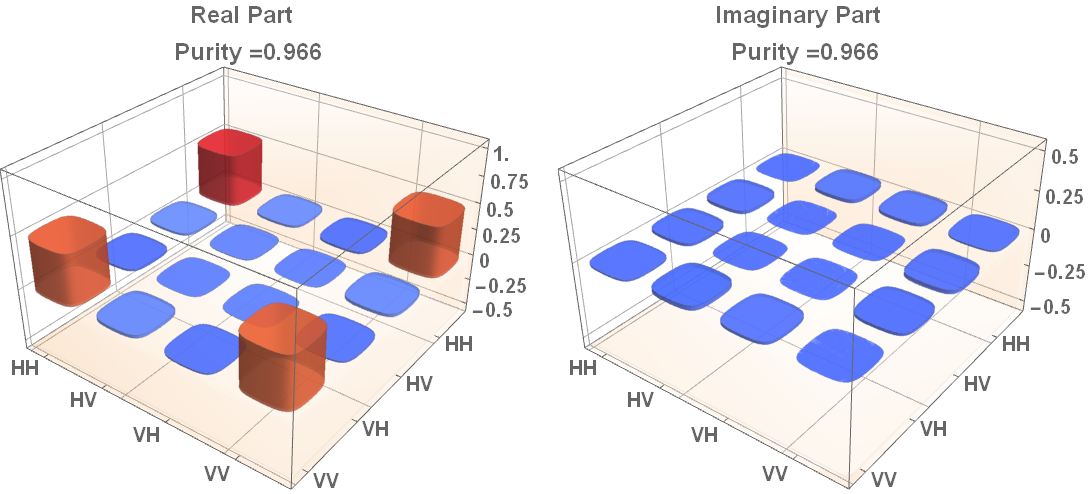}
\caption{\textit{Representative 3D plot of the experimentally reconstructed density matrix for state-I with P=0.966 and 2N=0.964.}\label{fig3.7}}
\end{figure}

\begin{figure}[H]
\centering
\includegraphics[width=0.7\columnwidth, keepaspectratio]{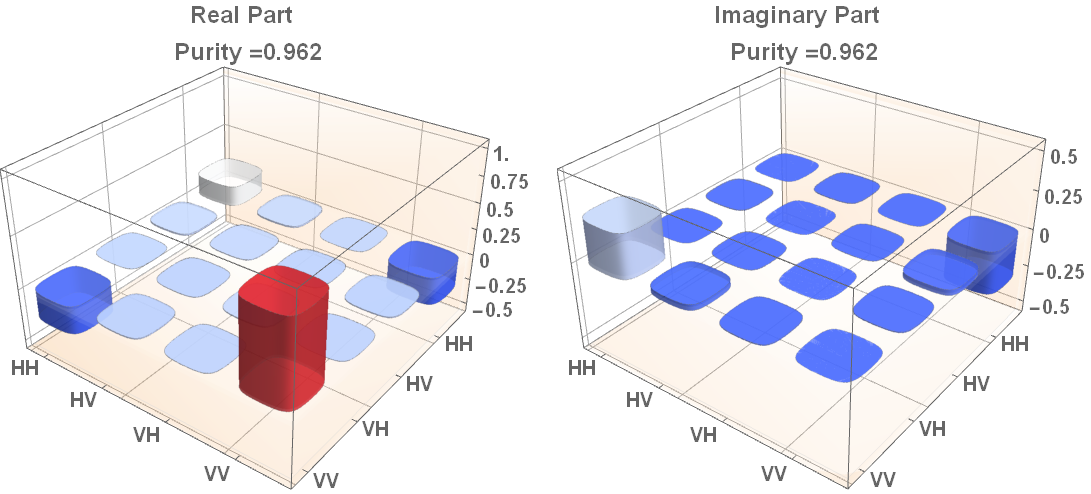}
\caption{\textit{Representative 3D plot of the experimentally reconstructed density matrix for state-II with P=0.962 and 2N=0.759.}\label{fig3.8}}
\end{figure}

\begin{figure}[H]
\centering
\includegraphics[width=0.7\columnwidth, keepaspectratio]{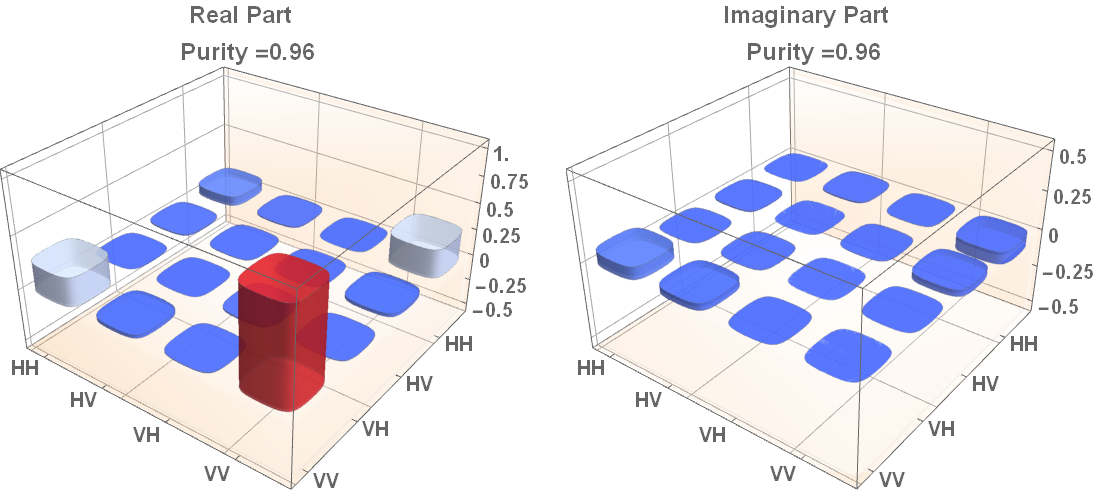}
\caption{\textit{Representative 3D plot of the experimentally reconstructed density matrix for state-III with P=0.960 and 2N=0.544.}\label{fig3.9}}
\end{figure}

Properties of different experimentally prepared entangled states such as purity, quantification of entanglement by different EMs, and their respective deviations from the maximally entangled state are summarized in the Table~\ref{tab3.2} below. The statistical error due to reconstruction occurs in the third decimal place (indicated in parentheses in the Table \ref{tab3.2}) for P, 2N, LN and EOF. Thus statistical errors in other derived quantities would also be of the same order.
\begin{table}[!htb] 
\begin{center}
\resizebox{\textwidth}{!}{
\renewcommand{\arraystretch}{2}
\begin{tabular}{|l|c|c|c|c|c|c|c|c|c|c|c|c|c|r|}
\hline
\textbf{States} & $\mathbf{2N_{ideal}}$ &   $\mathbf{P_{expt}}$ & $\mathbf{2N_{expt}}$ & $\mathbf{LN_{expt}}$ & $\mathbf{EOF_{expt}}$  &  $\mathbf{Q_L (\%)}$  &  $\mathbf{Q_N (\%)}$  &  $\mathbf{Q_E (\%)}$  & $\mathbf{\Delta Q_{NL}(\%)}$  & $\mathbf{\Delta Q_{NE}(\%)}$ & $\mathbf{\Delta Q_{EL}(\%)}$  \\
\hline
\hline
I  & 0.999 &   0.961(5) & 0 0.960(5) & 0.971(4) & 0.944(6) & 4.01 & 2.93 & 5.57 & 1.09  & 1.55 & 2.64\\
\hline
II  & 0.792 &  0.957(4) &  0.749(9) & 0.808(7) & 0.663(12) & 25.07 & 19.32 & 33.68 & 5.75 & 8.61 & 14.36 \\
\hline
III  & 0.593 & 0.958(2)   &  0.547(3) & 0.629(3) & 0.413(5) & 45.32 & 37.07 & 58.70 & 8.25  & 13.39 & 21.64\\
\hline
\end{tabular}}
\caption{\textit{Comparison tables for the properties of experimentally prepared two-qubit entangled states and its deviation from the intended maximally entangled state. Different Q-values are reported as percentage quantity.}\label{tab3.2}}
\end{center} 
\end{table}
It is evident from the experimentally prepared states considered here that $\Delta Q_{\text{EL}}$  $>$ $\Delta Q_{\text{NE}}$ $> $ $\Delta Q_{\text{NL}}$. Further, for the State-III, the values of $\Delta Q_{\text{NL}}$,  $\Delta Q_{\text{EL}}$, and  $\Delta Q_{\text{NE}}$ are very close to the maximum deviation obtained by numerical optimization, albeit with small impurity in the experimental states. These observations are in close agreement with the expectations from the analytical derivations that have been shown in the theory Section \ref{sec3.2}.

\section{Extending the analysis to higher dimensional systems}
\label{sec3.4}
In this section, we will discuss the way the present analysis can be extended to higher dimensional systems by considering bipartite pure qutrit states and multipartite states such as tripartite pure qubit states.
\subsection{Bipartite pure qutrit states}

Consider a two-qutrit pure state with Schmidt coefficients $c_{0}$, $c_{1}$, and $c_{2}$ as given below.
\begin{align}
\ket{\Phi} = c_{0}\ket{0}\ket{0} + c_{1}\ket{1}\ket{1} + c_{2}\ket{2}\ket{2},
\label{eq10}
\end{align}
where $0 \leq c_{0},~c_{1},~c_{2} \leq 1$ and $c_{0}^2$ + $c_{1}^2$  + $c_{2}^2$ = 1. The state (\ref{eq10}) is maximally entangled for $c_{0}$ = $c_{1}$ = $c_{2}$ = 1/$\sqrt{3}$ and separable when one of the Schmidt coefficients is one and others are zero. 

We will first recap some of the well known entanglement measures for higher dimensional systems. Negativity ($\mathcal{N}$) [\ref{3.17}], Entanglement of formation ($\mathcal{E}$) [\ref{3.18}], normalized generalized Concurrence ($\mathcal{C}$) [\ref{3.19}], and Linear entropy ($\mathcal{I}$) [\ref{3.20}] for two-qutrit  pure states are defined as follows:
 \begin{subequations}
\begin{align}
\begin{split}
\mathcal{N} &= c_{0}c_{1} + c_{1}c_{2} + c_{2}c_{0},
\end{split} \\
\begin{split}
\mathcal{E} &= -c_{0}^2(log_{2}c_{0}^2) - c_{1}^2(log_{2}c_{1}^2) - c_{2}^2(log_{2}c_{2}^2),
\end{split} \\
\begin{split}
\mathcal{C} &= \sqrt{3(c_{0}^2c_{1}^2+c_{1}^2c_{2}^2+c_{2}^2c_{0}^2)},
\label{eq3.11c}
\end{split} \\
\begin{split}
\mathcal{I} &= 3(c_{0}^2c_{1}^2+c_{1}^2c{2}^2+c_{2}^2c_{0}^2).
\label{eq3.11d}
\end{split} 
\end{align}
\end{subequations}

From Eq.~(\ref{eq3.11c}) and (\ref{eq3.11d}), it is clear that the Linear entropy is just the square of generalized Concurrence. Hence, we shall consider generalized Concurrence and compare it with other EMs. In order to quantify the percentage deviation of a given non-maximally entangled two-qutrit pure state from the maximally entangled state using different EMs, we define the Q-parameters as given below.
\begin{subequations}
\begin{align}
\begin{split}
Q_{\mathcal{E}} &= \frac{\log_{2}(3) - \mathcal{E}}{\log_{2}(3)},
\end{split}\\
\begin{split}
 Q_{\mathcal{N}} &= (1 - \mathcal{N}),
\end{split}\\
\begin{split}
Q_{\mathcal{C}} &= (1 - \mathcal{C}).
\end{split}
\end{align}
\end{subequations}

In this context, in a recent earlier work from our group [\ref{3.21}], extensive theoretical as well as experimental analysis on the certification and quantification of entanglement in spatial-bin bipartite photonic qutrits was done. Here, significant differences between the computed deviations of any pure non-maximally entangled bipartite qutrit state from the maximally entangled state in terms of the EMs such as $\mathcal{E}$ and $\mathcal{N}$ were reported for a different pair of Schmidt coefficients from the ones reported here. Further, it was shown that $\mathcal{E}$ and $\mathcal{N}$ are not monotonic with respect to each other. In this work, we consider, in addition to the comparison between $\mathcal{E}$ and $\mathcal{N}$, the entanglement measure $\mathcal{C}$ as well.

In order to quantify to what extent these three parameters differ with each other, the following quantities are appropriate measures.
\begin{subequations}
\begin{align}
\begin{split}
\Delta Q_{\mathcal{NE}} &= |Q_{\mathcal{N}}- Q_{\mathcal{E}}|~,
\end{split}\\
\begin{split}
\Delta Q_{\mathcal{EC}} &= |Q_{\mathcal{E}}- Q_{\mathcal{C}}|~,
\end{split}\\
\begin{split}
\Delta Q_{\mathcal{NC}} &= |Q_{\mathcal{N}}- Q_{\mathcal{C}}|~.
\end{split}
\end{align}
\end{subequations}

Now, we will present some specific cases exemplifying the non-monotonic nature of these EMs.
\begin{itemize}
\item Consider a state $\ket{\phi_{1}}$ with Schmidt coefficients $c_{0}$ = 0.9755 and $c_{1}$ = 0.0361, and another state $\ket{\phi_{2}}$ with Schmidt coefficients $c_{0}$=0.1403 and $c_{1}$ = 0.1346. 
    $$ \mathcal{E}(\phi_{1}) = 0.2878\  ~\text{and}~ \  \mathcal{N}(\phi_{1}) = 0.2546~,$$
    $$ \mathcal{E}(\phi_{2}) = 0.2698\  ~\text{and}~  \mathcal{N}(\phi_{2}) = 0.2885~.$$
Here, $\mathcal{E}(\phi_{1}) > \mathcal{E}(\phi_{2})$ but $\mathcal{N}(\phi_{1}) < \mathcal{N}(\phi_{2})$, showing that $\mathcal{E}$ and $\mathcal{N}$ are not monotonic with respect to each other.

\item Consider a state $\ket{\phi_{3}}$ with Schmidt coefficients $c_{0}$ = 0.4134 and $c_{1}$ = 0.8275, and another state $\ket{\phi_{4}}$ with Schmidt coefficients $c_{0}$ = 0.7452 and $c_{1}$ = 0.1143. 
    $$ \mathcal{N}(\phi_{3}) = 0.8136\  ~\text{and}~  \mathcal{C}(\phi_{3}) = 0.8495~, $$
    $$ \mathcal{N}(\phi_{4}) = 0.6498 \  ~\text{and}~      \mathcal{C}(\phi_{4}) = 0.8705~. $$
Here, $\mathcal{C}(\phi_{3}) < \mathcal{C}(\phi_{4})$ but $\mathcal{N}(\phi_{3}) > \mathcal{N}(\phi_{4})$, showing that $\mathcal{C}$ and $\mathcal{N}$ are not monotonic with respect to each other.

\item Consider a state $\ket{\phi_{5}}$ with Schmidt coefficients $c_{0}$ = 0.4134 and $c_{1}$ = 0.8275 and another state $\ket{\phi_{6}}$ with Schmidt coefficients $c_{0}$ = 0.2334 and $c_{1}$ = 0.8052. 
    $$ \mathcal{E}(\phi_{5}) = 1.2128\  ~\text{and}~  \mathcal{C}(\phi_{5}) = 0.8495~,$$
    $$ \mathcal{E}(\phi_{6}) = 1.1542 \  ~\text{and}~  \mathcal{C}(\phi_{6}) = 0.8559~.$$
Here, $\mathcal{C}(\phi_{5}) <\mathcal{C}(\phi_{6})$ but $\mathcal{E}(\phi_{5}) > \mathcal{E}(\phi_{6})$, showing that $\mathcal{C}$ and $\mathcal{E}$ are not monotonic with respect to each other.
\end{itemize}

Different values of the quantities $Q_{\mathcal{E}}$, $Q_{\mathcal{N}}$, $Q_{\mathcal{C}}$, $\Delta Q_{\mathcal{NE}}$, $\Delta Q_{\mathcal{EC}}$, and $\Delta Q_{\mathcal{NC}}$ corresponding to different values of Schmidt coefficients have been incorporated in Table \ref{tab3.3} as percentage values.
\begin{table}[!h]
\begin{center}
\resizebox{\textwidth}{!}{
\renewcommand{\arraystretch}{2}
\begin{tabular}{|l|c|c|c|c|c|c|c|c|c|r|}
\hline
\textbf{$c_{0}$} & \textbf{$c_{1}$} & $\mathcal{E}$ & $\mathcal{N}$ & $\mathcal{C}$ & $Q_\mathcal{E}$ & $Q_\mathcal{N}$  & $Q_\mathcal{C}$ & $\Delta Q_\mathcal{NE}$ & $\Delta Q_\mathcal{EC}$ & $\Delta Q_\mathcal{NC}$\\
\hline \hline
0.1 & 0.1 & 0.1614 & 0.2080 & 0.2431 & 89.81 & 79.20 & 75.69 & 10.61 & 14.12 & 3.51 \\
\hline
0.3 & 0.8 & 1.2347 & 0.8116 & 0.8741 &  22.10 & 18.84 & 12.59 & 3.25 & 9.51 & 6.26 \\
\hline
0.5774 & 0.5774 & 1.5850 & 1 & 1 & 0 & 0 & 0 & 0 & 0 & 0 \\
\hline
0.6 & 0.6 & 1.5755  & 0.9950 & 0.9968  & 0.60& 0.50 & 0.32 & 0.10 & 0.28 & 0.18 \\
\hline
0.9 & 0.3 & 0.8911 & 0.6495 & 0.6990 &  43.78 & 35.05 & 30.09 & 8.73 & 13.69 & 4.96\\
\hline
\end{tabular}}
\caption{\textit{Differences in the percentage deviations of entanglement measures of a given state from the value corresponding to the maximally entangled two-qutrit pure state.}\label{tab3.3}}
\end{center}
\end{table}

\vspace{4mm}
\textbf{Observations:}
\begin{itemize}
\item For non-maximally entangled two-qutrit pure states, a given EM is not always greater (or less) than any other EM for different values of the state parameters, unlike for the case of two-qubit pure states.

\item From numerical optimization, we find that the $\Delta Q_{\mathcal{NE}}$ takes a maximum value of 13.09\% when $c_{0} = 0.7071$ and  $c_{1} = 0.7071$, $\Delta Q_{\mathcal{EC}}$ takes a maximum value of 23.81\% when $c_{0} = 0.5$  and  $c_{1} = 0.8660$, and $\Delta Q_{\mathcal{NC}}$ takes a maximum value of 36.60\% when $c_{0} = 0.7071$ and  $c_{1} = 0.7071$.
\end{itemize}

We would like to emphasize an important and interesting point that although in the bipartite qubit case different EMs show different deviations of a given state from maximally entangled state, the EMs are monotonic with respect to each other. But in the bipartite qutrit case, different EMs not only provide different estimations of the deviation of any non-maximally entangled state from the maximally entangled state, the EMs can also be non-monotonic with respect to each other.

\subsection{Tripartite pure qubit states}
Here, we indicate the directions for extending the analysis presented in this chapter to multipartite systems focusing primarily on tripartite pure qubit states. Let us consider the usually discussed tripartite pure qubit states such as GHZ-type state [\ref{3.22}], W-state [\ref{3.23}] and Cluster state [\ref{3.24}]. In tripartite system, Cluster state is same as the GHZ state [\ref{3.24}]. Let us consider the following tripartite pure qubit state
\begin{align}
\ket{\xi_1} = c_{0}\ket{000} + c_{1}\ket{111},
\label{eq21}
\end{align}  
where $0 \leq c_{0}, c_{1} \leq 1$ and $c_{0}^2 + c_{1}^2 = 1$. The state (\ref{eq21}) is precisely the GHZ-state for $c_{0}= c_{1} = 1/\sqrt{2}$ when it is maximally entangled.

We consider two of the most commonly discussed EMs in tripartite system. One is the Tangle introduced by Coffman \textit{et al.}  [\ref{3.25}] which can be regarded as a hyperdeterminant of second order [\ref{3.26}]. Tangle for the state (\ref{eq21}) is given by
\begin{align}
\tau = 4c_{0}^2c_{1}^2 .
\end{align}

Another relevant EM is global measure of entanglement($G$) introduced by Meyer \textit{et al.} [\ref{3.27}] in Brennon form [\ref{3.28}-\ref{3.30}]. The value of this measure for the state (\ref{eq21}) is given by
\begin{align}
G = 4c_{0}^2c_{1}^2.
\end{align} 

It is found that the expressions for EMs $G$ and $\tau$ are same for the class of states given by Eq. (\ref{eq21}). Hence, estimation of the deviation of any given entangled state of form (\ref{eq21}) from the maximally entangled state using these two EMs will be the same, while both these measures give the same value. Both these measures give the same value one [\ref{3.27}] for the GHZ state.

Next, we consider another commonly discussed tripartite pure qubit state of the form 
\begin{align}
\ket{\xi_2} = c_{0}\ket{001} + c_{1}\ket{010} + c_{2}\ket{100},  
\label{eq24}
\end{align}
where $0 \leq c_{0}, c_{1}, c_{2} \leq 1$ and $c_{0}^2$ + $c_{1}^2$ + $c_{2}^2$ = 1. The above state is a W-state for $c_{0}$ = $c_{1}$ = $c_{2}$ = 1/$\sqrt{3}$ when it is maximally entangled. 

On calculation, it is found that for the state (\ref{eq24}) Tangle vanishes for all the values of the state parameters, whereas the other measure $G$ is non-zero and given by 
\begin{align}
 G = 8/3(c_{0}^2c_{1}^2+c_{1}^2c_{2}^2+c_{2}^2c_{0}^2).
\end{align}
 
It can be seen from above that for W-state, $G = 8/9$ whereas $\tau$ remains zero [\ref{3.25}-\ref{3.27}]. Hence, the deviation of any non-maximally entangled state of the form (\ref{eq24}) from the maximally entangled W-state can only be estimated in terms of $G$, not using tangle. A curious observation is that although the two entanglement measures $\tau$ and $G$ are defined differently, but for the particular class of state in (\ref{eq21}), they capture same amount of entanglement as well as give same amount of deviation from the maximally entangled state. This example is in stark contrast to the two-qubit case where different EMs give different deviations. On the other hand, for state (\ref{eq24}), $\tau$ vanishes whereas $G$ remains non-zero. This highlights the importance of this enterprise of comparing and contrasting different EMs. It would be worth further developing this line of study using other EMs as well for the tripartite qubit case.

\section{Outlook} \label{sec3.5}
The appreciable quantitative disagreement between estimates of the percentage deviation of a given state from the maximally entangled state using different EMs, thoroughly shown in this chapter using both theoretical and experiment-based (for bipartite qubit states) analyses, underscores the need for an appropriate quantifier for addressing such an empirically relevant issue which is of significance for evaluating the efficacy of a given entangled state for its various applications. In this context, it is relevant to note that the question of quantifying `distance' or deviation of a given state from the separable state has been raised earlier and for this purpose the concept of `distance measures'  has been suggested in two different ways; one of which is in terms of relative entropy of entanglement [\ref{3.04}] and the other using the notion of `robustness' [\ref{3.31}] in terms of the noise that is required to be added to a given state to make it a separable state. Taking clue from such studies, one line of study may be to formulate a suitable `distance measure' for capturing the departure of a given non-maximally entangled state from the maximally entangled state and compare the results obtained with the relevant estimates using different EMs.  On the other hand, one can take cue from the concept of `teleportation distance'  [\ref{3.10}] that has been used to quantify the degree of performance of the resource channel used for teleportation, where distance of the teleported state from the target state has been quantified in terms of the trace norm (T = $\mathrm{Tr}[\sqrt{A^{\dagger}A}]$). Adapting this measure, by using the Frobenius norm  [\ref{3.32}], one may invoke the following measure to signify how close a given state ($\psi$) is to the intended maximally entangled state ($\psi_{\text{max}}$) (the correct choice of the maximally entangled state $\psi_{\text{max}}$ is the one which has minimum distance from the given state)
\begin{align}
\text{D} = || \ket{\psi}\bra{\psi}-\ket{\psi_{\text{max}}}\bra{\psi_{\text{max}}}||~,
\end{align}

where the Frobenius norm for a density matrix $A$, $||A||$ used above is given by $\sqrt{\mathrm{Tr}[ A^{\dagger}A]}$. Here, the Frobenius norm is preferred over trace norm in order to ensure that the parameter $D$ ranges from 0 to 1 for two-qubit pure states, with $D=0$ for the maximally entangled state and $D=1$ for the separable state.

Another possible approach is in terms of the notion of fidelity  [\ref{3.33}] as a measure of `closeness' between two pure states. The metric defined using this notion of fidelity like Bures distance [\ref{3.34}, \ref{3.35}], or the Fubini-Study metric used by Anandan and Aharonov [\ref{3.36}] does not range from 0 to 1 for two-qubit pure states and hence is not useful for comparison with the results obtained using the deviation parameters (ranging from 0 to 1) defined in Section \ref{sec3.2}. On the other hand, by appropriately scaling the metric defined in terms of fidelity by Gilchrist \textit{et al.}, [\ref{3.37}], the following measure `$C$' which ranges from 0 to 1 for two-qubit pure states can be used for the purpose of comparison with the results obtained from the deviation parameters defined in terms of different EMs.
\begin{align}
C = \sqrt{2-2(\langle\psi|\psi_\text{max}\rangle)^2}~,
\end{align}

where $\psi_{\text{max}}$ is the intended maximally entangled two-qubit pure state. Note that $C=0$ for the maximally entangled state and $C=1$ for the separable state. Now, we note that for any given $\ket{\psi}$, the two measures $D$ and $C$ defined above, can be shown analytically to be equivalent. The numerical study in this case by varying values of the state parameter $c_{0}$ in Eq~(\ref{eq3.1}), interestingly, shows that the above mentioned distance measure $D(C)$  provides  an upper bound to the fractional deviations of various states from the maximally entangled state, calculated using N and LN; i.e.,
\begin{align}
D(C)\geq Q_{\text{N}}, Q_{\text{L}}.
\end{align}

However, there are certain states for which the respective fractional deviation from the maximally entangled state in terms of EOF, i.e., the quantity $Q_{\text{E}}$ is greater than $D(C)$, thus, restricting the use of $D(C)$ as an upper bound of such fractional deviations using EOF. These features pertaining to the distance measure $D(C)$ from the maximally entangled state are illustrated in Fig.~\ref{fig3.10}, whose implications may further be studied.

\begin{figure}[H]
\centering
\includegraphics[width = 0.8\linewidth]{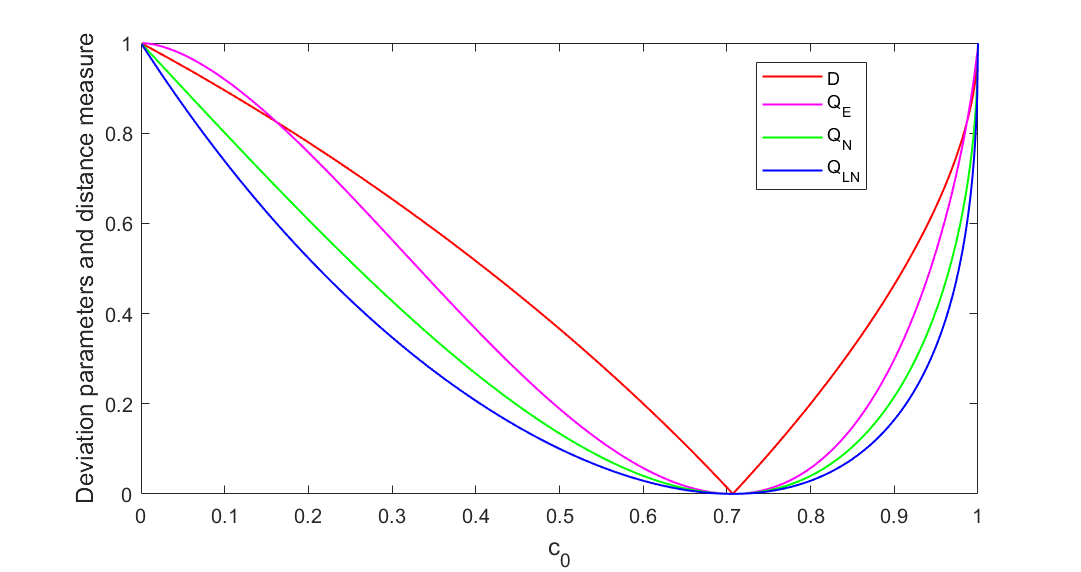}
\caption {\textit{ This figure illustrates that the curve for the distance measure $D$ provides an upper bound to the fractional deviation curves for N and LN, but not for EOF, since there are certain states for which the fractional deviation curve for EOF is not upper bounded by the curve related to $D$.}\label{fig3.10}}
\end{figure}

A line of study different from the above mentioned approaches in terms of `distance measures' could be from the operational perspective of the use of quantum entanglement as resource; in other words, one may try to assess how close a non-maximally entangled state is to the maximally entangled state in terms of how useful it is as a resource. For example, in the context of demonstrating non-locality, in order to quantify the fractional deviation of a given non-maximally entangled state from the maximally entangled state, one may compute the maximum possible quantum mechanical violation of the Bell-CHSH inequality ($B_{V}$), i.e., the difference between the maximum quantum mechanical value of the Bell-CHSH expression for the given state and the lower bound of the Bell-CHSH expression. One can then estimate the fractional deviation of the parameter $B_{V}$ from its maximum value $(B_{V})_{\text{max}}$  = $2\sqrt{2}-2$ which is achieved for the maximally entangled state. Note that for any the  two-qubit pure state characterized by the Schmidt coefficients $c_{0}$ and $c_{1}$, $B_{V}$ for the two outcomes-two settings scenario is given by [\ref{3.38}-\ref{3.40}]
\begin{align}
B_{V} = 2\sqrt{1+(2c_{0}c_{1})^2}-2. 
\end{align} 

Now, similar to the other fractional deviation parameters defined in Section \ref{sec3.2}, here we define the following parameter as a measure of the fractional deviation of $B_{V}$ from its value $(B_{V})_\text{max}$ for the maximally entangled state.
\begin{align}
Q_{B_{V}} = \frac{(B_{V})_\text{max}-B_{V}}{(B_{V})_\text{max}}.
\end{align}

Note that the values of $Q_{{B}_{V}}$ range from 0 to 1, with 0 for the maximally entangled state and 1 for the separable state. A numerical study of $Q_{B_{V}}$ by varying  values of the state parameter $c_{0}$ shows an interesting result that for any non-maximally entangled state, this fractional deviation parameter is greater than all other such fractional deviation parameters evaluated using different EMs like N, LN and EOF; i.e.,
\begin{align}
Q_{B_{V}} \geq Q_\text{N}, Q_\text{L},Q_\text{E},
\end{align}

where the equality holds good for the maximally entangled state and the separable state. This means that the measure of deviation of a given non-maximally entangled state from the maximally entangled state as quantified by the parameter $Q_{{B}_V}$ is always greater than that obtained from different EMs. For example, for a state with $c_{0}=0.4$, the fractional deviation of $B_{V}$ from its value corresponding to the maximally entangled state in percentage is 42.06\%; while the fractional deviations of N, LN and EOF from their values corresponding to the maximally entangled state in percentages are 26.68\%, 20.66\% and 36.57\%, respectively.  Illustration of this feature provided in Fig.~\ref{fig3.11}, thus, suggests nuances in the quantitative relationship between EMs and the amount of non-locality shown by the Bell-CHSH violation present even in the simplest two outcomes - two settings scenario involving two-qubit pure states; on the other hand, aspects of quantitative non-equivalence between entanglement and non-locality have so far been discussed essentially for high dimensional systems or scenarios involving larger number of settings [\ref{3.41}-\ref{3.44}].

\begin{figure}[H]
\centering
\includegraphics[width =0.8\linewidth]{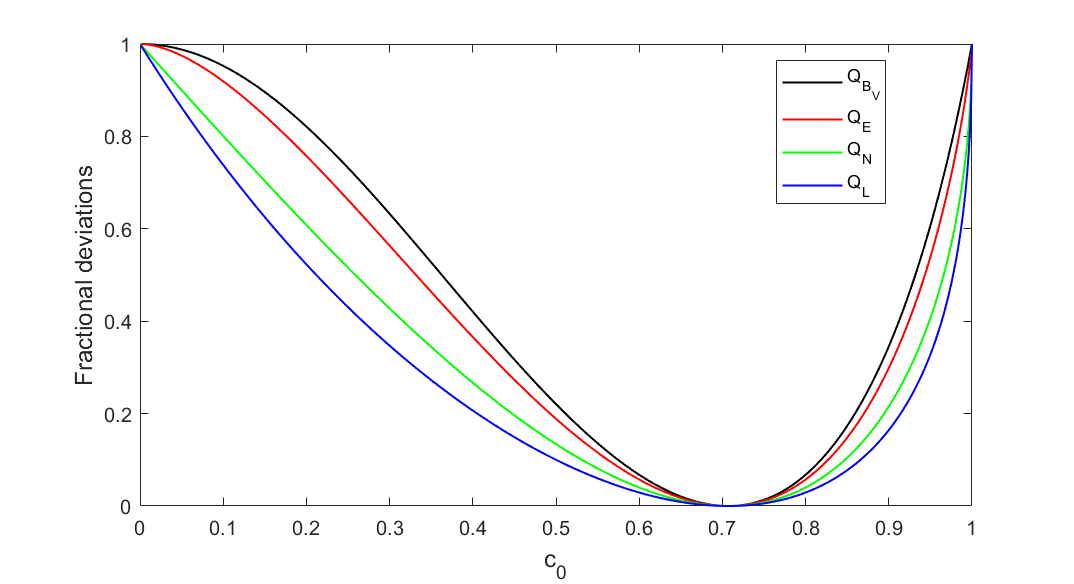}
\caption{{\textit{This figure shows that the fractional deviation curve for the quantity $B_{V}$ provides an upper bound to the other fractional deviation curves corresponding to the quantities N, LN and EOF, signifying that in terms of the amount of  Bell-CHSH violation, any given state is further away from the maximally entangled state than that estimated using different entanglement measures.}}\label{fig3.11}} 
\end{figure}

\section{Concluding Remarks} \label{sec3.6}
In sum, the results of studies of the present chapter bring out the need for exploring different ideas for quantifying how close (far) a given entangled state is to (from) the maximally entangled state and comparing the results obtained by using such quantifiers with that based on different entanglement measures. This leads to the following question: Is there any fundamental criterion for assessing which quantifier is the appropriate one to be used for addressing questions such as the one posed in this chapter, or whether such a criterion would have to be operationally defined essentially dependent on the specific context in which the entangled state is used as a resource? A comprehensive study is required for shedding further light on this issue as well as for gaining a deeper understanding of the comparison between different entanglement measures taking into account the studies probing their respective physical significance [\ref{3.02},\ref{3.03},\ref{3.45}]. 

The published version of the research work reported in this chapter can be found in Ref. [\ref{3.46}].

\section{References} \label{3.7}
\begin{enumerate}
\item \textit{C. H. Bennett, H. J. Bernstein, S. Popescu, and B. Schumacher, ``Concentrating partial entanglement by local operations," \href{https://doi.org/10.1103/PhysRevA.53.2046}{Phys. Rev. A \textbf{53}, 2046 (1996)}}. \label{3.01}
\item \textit{C. H. Bennett, D. P. DiVincenzo, J. A. Smolin, and W. K. Wootters, ``Mixed-state entanglement and quantum error correction," \href{https://doi.org/10.1103/PhysRevA.54.3824}{Phys. Rev. A \textbf{54}, 3824 (1996)}}.  \label{3.02}
\item \textit{S. Popescu and D. Rohrlich, ``Thermodynamics and the measure of entanglement," \href{https://doi.org/10.1103/PhysRevA.56.R3319}{Phys. Rev. A \textbf{56}, R3319 (1997)}}.\label{3.03}
\item \textit{V. Vedral, M. B. Plenio, M. A. Rippin, and P. L. Knight, ``Quantifying entanglement," \href{https://doi.org/10.1103/PhysRevLett.78.2275}{Phys. Rev. Lett. \textbf{78}, 2275 (1997)}}.\label{3.04}
\item \textit{V. Vedral and M. B. Plenio, ``Entanglement measures and purification procedures," \href{https://doi.org/10.1103/PhysRevA.57.1619}{Phys. Rev. A \textbf{57}, 1619 (1998)}}.\label{3.05}
\item \textit{G. Vidal, ``Entanglement monotones," \href{https://doi.org/10.1080/09500340008244048}{J. Mod. Opt. \textbf{47}, 355–376 (2000)}}.\label{3.06}
\item \textit{K. Zyczkowski, P. Horodecki, A. Sanpera, and M. Lewenstein,
``Volume of the set of separable states," \href{https://doi.org/10.1103/PhysRevA.58.883}{Phys. Rev. A \textbf{58}, 883 (1998)}}. \label{3.07}
\item \textit{K. Zyczkowski, ``Volume of the set of separable states. II," \href{https://doi.org/10.1103/PhysRevA.60.3496}{Phys. Rev. A \textbf{60}, 3496 (1999)}}. \label{3.08}
\item \textit{J. Lee, M. Kim, Y. Park, and S. Lee, ``Partial teleportation of entanglement in a noisy environment," \href{https://doi.org/10.1080/09500340008235138}{J. Mod. Opt. \textbf{47}, 2151–2164 (2000)}}.\label{3.09}
\item \textit{G. Vidal and R. F. Werner, ``Computable measure of entanglement," \href{https://doi.org/10.1103/PhysRevA.65.032314}{Phys. Rev. A \textbf{65}, 032314 (2002)}}. \label{3.10}
\item \textit{J. Eisert, ``Entanglement in quantum information theory," \href{https://arxiv.org/abs/quant-ph/0610253}{quant-ph/0610253 (2006)}}.   \label{3.11}
\item \textit{W. K. Wootters, ``Entanglement of formation of an arbitrary state of two qubits," \href{https://doi.org/10.1103/PhysRevLett.80.2245}{Phys. Rev. Lett. \textbf{80}, 2245 (1998)}}.  \label{3.12}
\item \textit{J. Eisert and M. B. Plenio, ``A comparison of entanglement measures," \href{https://doi.org/10.1080/09500349908231260
}{J. Mod. Opt. \textbf{46}, 145–154 (1999)}}. \label{3.13}
\item \textit{P. G. Kwiat, E. Waks, A. G. White, I. Appelbaum, and P. H. Eberhard, ``Ultrabright source of polarization-entangled photons," \href{https://doi.org/10.1103/PhysRevA.60.R773}{Phys. Rev. A \textbf{60}, R773 (1999)}}. \label{3.14}
\item \textit{R. Rangarajan, M. Goggin, and P. Kwiat, ``Optimizing type-I
polarization-entangled photons," \href{https://doi.org/10.1364/OE.17.018920}{Opt. Express \textbf{17}, 18920–18933 (2009)}}.  \label{3.15}
\item \textit{D. F. V. James, P. G. Kwiat, W. J. Munro, and A. G. White, ``Measurement of qubits," \href{https://doi.org/10.1103/PhysRevA.64.052312}{Phys. Rev. A \textbf{64}, 052312 (2001)}}.\label{3.16}
\item \textit{C. Eltschka, G. Tóth, and J. Siewert, ``Partial transposition as a direct link between concurrence and negativity," \href{https://doi.org/10.1103/PhysRevA.91.032327}{Phys. Rev. A \textbf{91}, 032327 (2015)}}. \label{3.17}
\item\textit{W. K. Wootters, ``Entanglement of formation and concurrence," \href{}{Quantum Information \& Computation \textbf{1}, 27–44 (2001)}}.\label{3.18}
\item \textit{Y. Maleki and B. Ahansaz, ``Quantum correlations in qutrit-like superposition of spin coherent states," \href{https://doi.org/10.1088/1612-202X/ab12e5}{Laser Phys. Lett. \textbf{16}, 075205 (2019)}}.\label{3.19}
\item \textit{Y. Maleki and A. M. Zheltikov, ``Linear entropy of multiqutrit nonorthogonal states," \href{ttps://doi.org/10.1364/OE.27.008291}{Opt. express \textbf{27}, 8291–8307 (2019)}}.\label{3.20}
\item \textit{D. Ghosh, T. Jennewein, and U. Sinha, ``Entanglement certification and quantification in spatial-bin photonic qutrits," \href{https://arxiv.org/abs/1909.01367}{arXiv:1909.01367 (2019)}}. \label{3.21}
\item \textit{D. M. Greenberger, M. A. Horne, A. Shimony, and A. Zeilinger, ``Bell’s theorem without inequalities," \href{https://doi.org/10.1119/1.16243}{Am. J. Phys. \textbf{58}, 1131–1143 (1990)}}. \label{3.22}
\item \textit{W. Dür, G. Vidal, and J. I. Cirac, ``Three qubits can be entangled in two inequivalent ways," \href{https://doi.org/10.1103/PhysRevA.62.062314}{Phys. Rev. A \textbf{62}, 062314 (2000)}}. \label{3.23}
\item \textit{H. J. Briegel and R. Raussendorf, ``Persistent entanglement in arrays of interacting particles," \href{https://doi.org/10.1103/PhysRevLett.86.910}{Phys. Rev. Lett. \textbf{86}, 910 (2001)}}. \label{3.24}
\item \textit{V. Coffman, J. Kundu, and W. K. Wootters, ``Distributed entanglement," \href{https://doi.org/10.1103/PhysRevA.61.052306}{Phys. Rev. A \textbf{61}, 052306 (2000)}}. \label{3.25}
\item \textit{A. Miyake, ``Classification of multipartite entangled states by multidimensional determinants," \href{https://doi.org/10.1103/PhysRevA.67.012108}{Phys. Rev. A \textbf{67}, 012108 (2003)}}. \label{3.26}
\item \textit{D. A. Meyer and N. R. Wallach, ``Global entanglement in multiparticle systems," \href{https://doi.org/10.1063/1.1497700}{J. Math. Phys. \textbf{43}, 4273–4278 (2002)}}. \label{3.27}
\item \textit{G. K. Brennen, ``An observable measure of entanglement for pure states of multi-qubit systems," \href{}{Quantum Information \& Computation \textbf{3}, 619–626 (2003)}}. \label{3.28}
\item \textit{Y. Maleki, F. Khashami, and Y. Mousavi, ``Entanglement of three spin states in the context of su (2) coherent states," \href{https://doi.org/I 10.1007/s10773-014-2215-5}{Int. J. Theor. Phys. \textbf{54}, 210–218 (2015)}}. \label{3.29}
\item \textit{Y. Maleki and A. Maleki, ``Entangled multimode spin coherent states of trapped ions," \href{https://doi.org/10.1364/JOSAB.35.001211}{J. Opt. Soc. Am. B \textbf{35}, 1211–1217 (2018)}}. \label{3.30}
\item \textit{G. Vidal and R. Tarrach, ``Robustness of entanglement," \href{https://doi.org/10.1103/PhysRevA.59.141}{Phys. Rev. A \textbf{59}, 141 (1999)}}.  \label{3.31}
\item \textit{R. A. Horn and C. R. Johnson, Matrix Analysis (Cambridge University Press, 2012), \href{http://www.cambridge.org/9780521548236}{chap. 5, pp. 321}}. \label{3.32}
\item \textit{R. Jozsa, ``Fidelity for mixed quantum states," \href{https://doi.org/10.1080/09500349414552171}{J. Mod. Opt. \textbf{41}, 2315–2323 (1994)}}.  \label{3.33}
\item \textit{D. Bures, ``An extension of kakutani’s theorem on infinite product measures to the tensor product of semifinite w*-algebras," \href{https://www.jstor.org/stable/1995012}{Trans. Am. Math. Soc. \textbf{135}, 199–212 (1969)}}.  \label{3.34}
\item \textit{K. Zyczkowski and I. Bengtsson, ``Relativity of pure states entanglement," \href{https://doi.org/10.1006/aphy.2001.6201}{Ann. Phys. \textbf{295}, 115–135 (2002)}}. \label{3.35}
\item \textit{J. Anandan and Y. Aharonov, ``Geometry of quantum evolution," \href{https://doi.org/10.1103/PhysRevLett.65.1697}{Phys. Rev. Lett. \textbf{65}, 1697 (1990)}}. \label{3.36}
\item \textit{A. Gilchrist, N. K. Langford, and M. A. Nielsen, ``Distance measures to compare real and ideal quantum processes," \href{https://doi.org/10.1103/PhysRevA.71.062310}{Phys. Rev. A \textbf{71}, 062310 (2005)}}. \label{3.37}
\item \textit{N. Gisin, ``Hidden quantum nonlocality revealed by local filters," \href{https://doi.org/10.1016/S0375-9601(96)80001-6}{Phys. Lett. A \textbf{210}, 151–156 (1996)}}.  \label{3.38}
\item \textit{R. Horodecki, P. Horodecki, and M. Horodecki, ``Violating bell inequality by mixed spin-1/2 states: necessary and sufficient condition," \href{https://doi.org/10.1016/0375-9601(95)00214-N}{Phys. Lett. A \textbf{200}, 340–344 (1995)}}. \label{3.39}
\item \textit{B. Horst, K. Bartkiewicz, and A. Miranowicz, ``Two-qubit mixed states more entangled than pure states: Comparison of the relative
entropy of entanglement for a given nonlocality," \href{https://doi.org/10.1103/PhysRevA.87.042108}{Phys. Rev. A \textbf{87}, 042108 (2013)}}. \label{3.40}
\item \textit{S. Zohren and R. D. Gill, ``Maximal violation of the collins-gisinlinden-massar-popescu inequality for infinite dimensional states," \href{https://doi.org/10.1103/PhysRevLett.100.120406}{Phys. Rev. Lett. \textbf{100}, 120406 (2008)}}. \label{3.41}
\item \textit{C. Bernhard, B. Bessire, A. Montina, M. Pfaffhauser, A. Stefanov, and S. Wolf, ``Non-locality of experimental qutrit pairs," \href{https:/doi.org/10.1088/1751-8113/47/42/424013}{J. Phys. A \textbf{47}, 424013 (2014)}}.  \label{3.42}
\item \textit{A. Acín, R. Gill, and N. Gisin, ``Optimal bell tests do not require maximally entangled states," \href{https://doi.org/10.1103/PhysRevLett.95.210402}{Phys. Rev. Lett. \textbf{95}, 210402 (2005)}}.  \label{3.43}
\item \textit{N. Brunner, N. Gisin, and V. Scarani, ``Entanglement and nonlocality are different resources," \href{https://doi.org/10.1088/1367-2630/7/1/088}{New J. Phys. \textbf{7}, 88 (2005)}}.  \label{3.44}
\item \textit{C. Eltschka and J. Siewert, ``Negativity as an estimator of entanglement dimension," \href{https://doi.org/10.1103/PhysRevLett.111.100503}{Phys. Rev. Lett. \textbf{111}, 100503 (2013)}}.  \label{3.45}
\item \textit{Ashutosh Singh, Ijaz Ahamed, Dipankar Home, and Urbasi Sinha, ``Revisiting comparison between entanglement measures for two-qubit pure states," \href{https://doi.org/10.1364/JOSAB.37.000157}{J. Opt. Soc. Am. B \textbf{37}, 157-166 (2020)}}.\label{3.46}
\end{enumerate}

%% file: chapter4.tex
\setcounter{equation}{0}
\chapter{Manipulation of entanglement sudden death in an all-optical setup}

The unavoidable and irreversible interaction between an entangled quantum system and its environment causes decoherence of the individual qubits as well as degradation of the entanglement between them. Entanglement sudden death (ESD) is the phenomenon wherein disentanglement happens in finite time even when individual qubits decohere only asymptotically in time due to noise. Prolonging the entanglement is essential for the practical realization of entanglement-based quantum information and computation protocols. For this purpose, the local NOT operation in the computational basis on one or both qubits has been proposed. Here, we formulate an all-optical experimental set-up involving such NOT operations that can hasten, delay, or completely avert ESD, all depending on when it is applied during the process of decoherence. Analytical expressions for these are derived in terms of parameters of the initial state's density matrix, whether for pure or mixed entangled states. After a discussion of the schematics of the experiment, the problem is theoretically analyzed, and simulation results of such manipulations of ESD are presented.

\section{Introduction} \label{S4.1}

Quantum entanglement [\ref{4.01},\ref{4.02}] is a non-classical correlation shared among quantum systems which could be non-local [\ref{4.03},\ref{4.04}] in some cases. It is a fundamental trait of quantum mechanics. Like classical correlations, entanglement also decays with time in the presence of noise in the ambient environment. The decay of entanglement depends on the initial state and the type and amount of noise (Amplitude damping, Phase damping, etc.) acting on the system [\ref{4.05}-\ref{4.07}]. The entangled states: $|\psi^\pm\rangle=|\alpha| |ge\rangle\pm|\beta| \exp(\iota \delta) |eg\rangle)$ and $|\phi^\pm\rangle=|\alpha| |gg\rangle\pm|\beta|  \exp(\iota \delta) |ee\rangle$ (maximally entangled ``Bell states" for $|\alpha|=|\beta|=1/\sqrt{2},~\delta=0$) being the simplest and most useful entangled states in quantum information processing receive special attention. The maximally entangled states $|\phi^\pm\rangle$ and $|\psi^\pm\rangle$ undergo asymptotic decay of entanglement in the presence of an amplitude damping channel (ADC). The non maximally entangled states $|\psi^\pm\rangle$ always undergo asymptotic decay of entanglement, whereas $|\phi^\pm\rangle$ undergo asymptotic decay for $|\alpha|>|\beta|$ and a finite time end called entanglement sudden death (ESD) for $|\alpha|<|\beta|$ in the presence of ADC. On the other hand, a pure phase damping channel (PDC) causes entanglement to always decay asymptotically. Two different initial states ($|\psi\rangle=|\alpha|~|gg\rangle+|\beta| \exp(\iota \delta)~ |ee\rangle, ~|\alpha|^2+|\beta|^2=1~$, where $ ~ (i)~|\beta|=k |\alpha|~~\text{and}~~(ii) ~|\alpha|=k|\beta|,~k>1~$), which share the same amount of initial entanglement (measured through Negativity) being affected by the same type of noise may follow very different trajectories of entanglement decay. In the presence of multiple stochastic noises, although the decoherence of individual qubits follows the additive law of relaxation rates, the decay of entanglement, does not. In fact, entanglement may not decay asymptotically at all, and disentanglement can happen in finite time (ESD). ESD has been experimentally demonstrated in atomic [\ref{4.08}] and photonic systems [\ref{4.09},\ref{4.10}].

Since entanglement is a resource in  quantum information processing [\ref{4.11}-\ref{4.13}], manipulation that prolongs entanglement will help realize protocols that would otherwise suffer due to short entanglement times. Also, entanglement purification [\ref{4.14}] and distillation [\ref{4.15}] schemes could possibly recover the initial correlation from the ensemble of noise-degraded correlation so long as the system has not completely disentangled. Therefore, the delay or avoidance of ESD is important. Several proposals exist to suppress the decoherence; for example, decoherence-free subspaces [\ref{4.16}-\ref{4.19}], quantum error correction [\ref{4.20},\ref{4.21}], dynamical decoupling [\ref{4.22}-\ref{4.24}], quantum Zeno effect [\ref{4.25}-\ref{4.27}], quantum measurement reversal [\ref{4.28}-\ref{4.33}], and delayed-choice decoherence suppression [\ref{4.34}]. Protecting entanglement using weak measurement and quantum measurement reversal [\ref{4.32},\ref{4.33}], and delayed choice decoherence suppression [\ref{4.34}] have been experimentally demonstrated. Both of these schemes, however, have the limitation that the success probability of decoherence suppression decreases as the strength of the weak interaction increases.

The practical question we want to address here is; whether, given a two-qubit entangled state in the presence of amplitude damping channel which causes disentanglement in finite time, can we alter the time of disentanglement by a suitable operation during the process of decoherence?  A theoretical proposal exists in the literature for such manipulation of ESD [\ref{4.35}] through a local unitary operation (NOT operation in computational basis: $\sigma_x$) performed on the individual qubits which swaps their population of ground and excited states. Depending on the time of application of this NOT operation, it can avoid, delay, or hasten the ESD. Based on this proposal [\ref{4.35}], we have extended the experimental set up [\ref{4.09}] for ESD and propose here an all-optical experimental set up for manipulating ESD involving the NOT operation on one or both the qubits of a bipartite entangled state in a photonic system.

The system consists of polarization-entangled photon pairs: $|\psi\rangle=|\alpha||HH\rangle + |\beta| \exp(i\delta) |VV\rangle$ produced in the sandwich configuration Type-I spontaneous parametric down conversion (SPDC)[\ref{4.38}]. These photons are sent to two displaced identical Sagnac interferometers, where ADC is simulated using rotating HWPs (half wave plates) placed in the path of incoming photons (See Fig.~\ref{fig4.2}). The HWP selectively causes a $|V\rangle$ polarized photon to ``decay" to $|H\rangle$ ($|H\rangle$ and $|V\rangle$ serve as ground and excited states of the system, the two states of a qubit) [\ref{4.09}]. The NOT operation is implemented by a HWP with fast axis at $45^o$ relative to $|V\rangle$, placed right after the ADC. This HWP  is followed by PBS (polarizing beam splitter) to segregate the $|H\rangle$ and $|V\rangle$ polarizations and, with subsequent ADC after the NOT operation implemented by a set of secondary  HWPs acting on $|V\rangle$ only. Such a set of secondary HWPs simulating the ADC (or evolution of qubits in noisy environment) is essential to our study as the ADC (for example, spontaneous emission in case of a two-level atomic system) continues to act even after the NOT operation is applied [\ref{4.35}] and these secondary HWPs simulate it in our proposed experiment. The orientation of the HWP ($\theta$) plays the role of time ($t$) with $\theta\rightarrow 45^o$ ($p=\sin^2(2\theta)\rightarrow 1$) analogous to $t\rightarrow \infty$ ($p=1-\exp(-\Gamma t)\rightarrow 1$) for a two-level atomic system decaying to the ground state due to spontaneous emission.

We use Negativity as a measure of entanglement. It is defined as the sum of absolute values of negative eigen values of the partially transposed density matrix [\ref{4.36},\ref{4.37}]. We find that our simulation results for the manipulation of ESD involving NOT operations on one or both the qubits of a polarization entangled photonic system in presence of ADC are completely consistent with the theoretical predictions of the reference [\ref{4.35}] which has analyzed an atomic system. The merit of our scheme is that it can delay or avoid ESD (provided the NOT operation is performed sufficiently early) unlike previous experiments [\ref{4.32}-\ref{4.34}] where success probabilities scaled with the strength of the weak interaction. Since the photonic system is time independent and noise is simulated using HWPs, it gives experimentalists complete  freedom to  study and manipulate the disentanglement dynamics in a controlled manner. In this, our photonic system through a controllable HWP offers an advantage over others such as atomic states where the decay occurs through noise sources lying outside experimental control. The NOT operation that we apply through a HWP is the analogue of flipping spin in a nuclear magnetic system, achieved through what is referred to as a $\pi$-pulse.

The chapter is organized as follows: In Section~(\ref{S4.2}), we discuss the all-optical implementation of the proposed ESD experiment and analyze it theoretically using the Kraus operator formalism. In Section~(\ref{S4.3}) and (\ref{S4.4}), we discuss and theoretically analyze the proposed ESD-manipulation experiment involving NOT operation on both or on only one of the qubits, respectively. In Section~(\ref{S4.5}), we give analytical expressions for probabilities $p_0,p_A,~\text{and}~p_B$ and also for ESD and its manipulation curves in terms of the parameters of the initial state (density matrix). The first of these is the setting (``time") for ESD, the next setting for the NOT marking the  border between hastening and delay; that is, if the NOT is applied after $p_A$ (and of course before $p_0$), it actually hastens, ESD happening before $p_0$, whereas application before delays ESD to stretch past $p_0$ to larger but still finite value less than one. The third, $p_B$, marks the border between delaying or completely avoiding ESD. Applying the NOT after $p_B$ delays to a larger $p_0$ value whereas applying before avoids ESD altogether. In Section~(\ref{S4.6}), we summarize the results of manipulation of ESD for different pure and mixed initial entangled states, giving  numerical values of $p_0,p_A$, and $p_B$. Section~(\ref{S4.7}) concludes with pros and cons of the proposed scheme for the manipulation of ESD and the future scope of this work.

\section{Experimental set up for ESD and its analysis} \label{S4.2}

\begin{figure} [!htb]
\begin{center}
\includegraphics[clip, trim=3.5cm 9cm 0.5cm 2cm, width=1\textwidth]{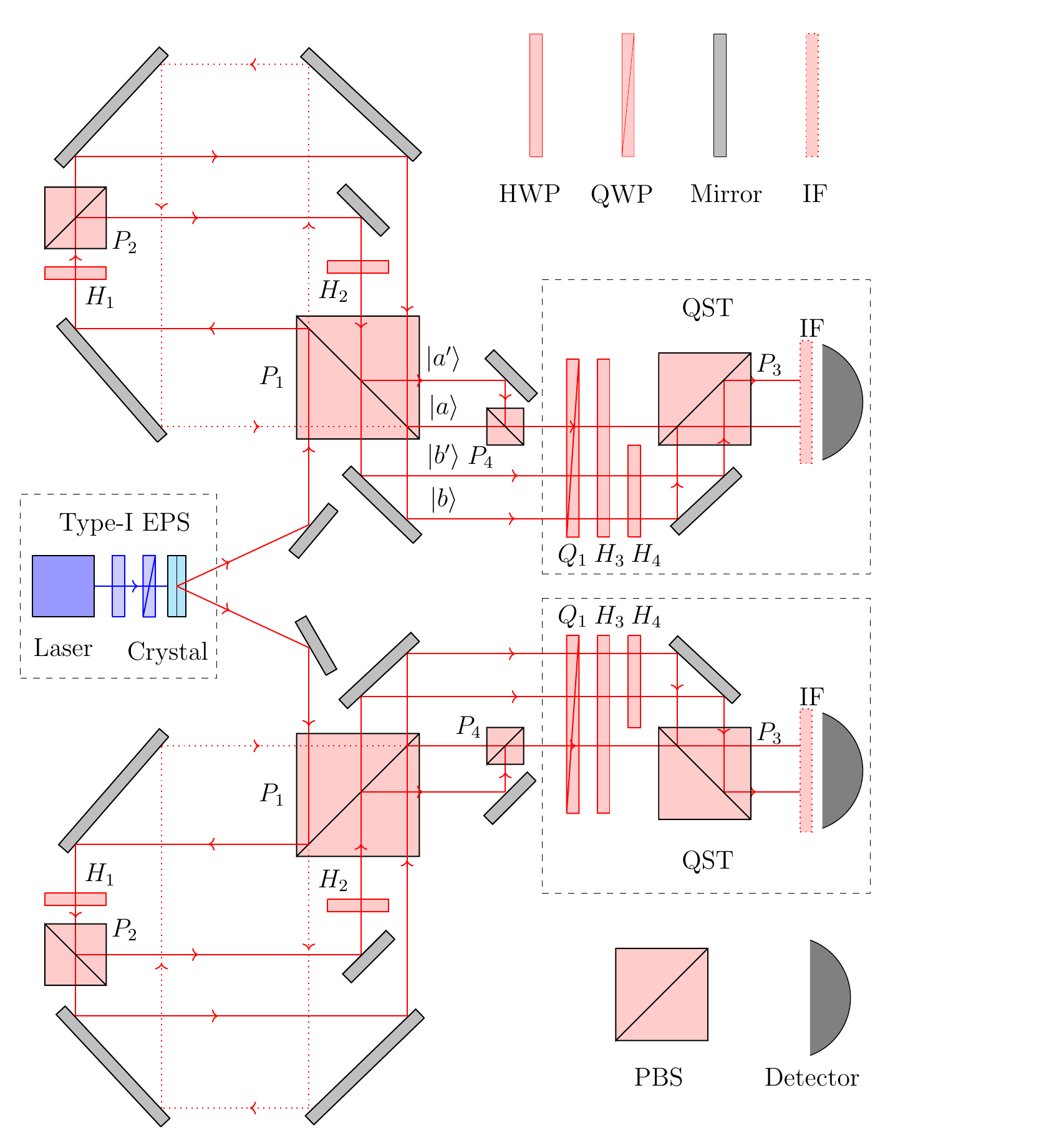}
\caption{\textit{The proposed experimental set up for ESD in the presence of ADC in a photonic system.}}
\label{fig4.1}
\end{center}
\end{figure}

The proposed experimental setup for ESD is shown in Fig.~(\ref{fig4.1}), which is a generalization of the scheme used in Ref.~[\ref{4.09}]. The type-I polarization entangled photons ($|\psi\rangle=|\alpha||HH\rangle+|\beta|~  \exp(\iota\delta)|VV\rangle$) can be prepared by standard methods [\ref{4.38}]; the amplitudes $|\alpha|$ and $|\beta|$ and relative phase $\delta$  are controlled by the HWP and QWP (quarter wave plate). These entangled photons are sent to two displaced Sagnac interferometers with HWPs simulating the ADC, where decoherence takes place, and finally these photons are sent for tomographic reconstruction of the quantum state [\ref{4.39}]. The $|H\rangle$ and $|V\rangle$ polarizations of the photon serve as the ground and excited states of the analogous atomic system, while output spatial modes of the reservoir $|a\rangle,~|a'\rangle~\text{and}~|b\rangle,~|b'\rangle$ serve as the ground and excited states of the reservoir. Asymmetry between degenerate polarization states of a photon ( $|H\rangle~\text{and}~|V\rangle$) is introduced by the HWP rotation such that it selectively causes an incident $|V\rangle$ polarization to ``decay" to $|H\rangle$, while leaving $|H\rangle$ polarization intact. Thus, the $|H\rangle$ and $|V\rangle$ polarization states are analogous to the ground and excited states, respectively, of a two-level atomic system.

The polarization entangled photons from the mid-left of the figure are sent to two displaced identical Sagnac interferometers. The Fig.~(\ref{fig4.1}) shows the Sagnac interferometers and its output spatial modes: $|a\rangle,|a'\rangle, |b\rangle ~\text{and}~ |b'\rangle$, which serve as the modes of the reservoir. Following an initial photon in the upper arm, an incident $|V\rangle$ polarized photon is reflected by PBS $P_1$ and traverses the interferometer in clockwise sense where HWP $H_1$ acts as ADC causing $|V\rangle$ polarized photon to decay to $\cos(2\theta)|V\rangle+\sin(2\theta)|H\rangle$. The PBS $P_2$ transmits $|H\rangle$ component and reflects $|V\rangle$ component. The transmitted $|H\rangle$ component comes back to PBS $P_1$ and is transmitted to spatial mode $|b\rangle$. The reflected $|V\rangle$ component from $P_2$ sees another ADC HWP $H_2$ which causes it to decay to $\cos(2\theta')|V\rangle+\sin(2\theta')|H\rangle$ and when it comes to PBS $P_1$, the $|V\rangle$ component gets reflected to mode $|a'\rangle$ and $|H\rangle$ component gets transmitted to mode $|b'\rangle$. On the other hand, an incident $|H\rangle$ polarized photon is transmitted by the PBS $P_1$ and traverses the interferometer in counter-clockwise sense, comes back to PBS $P_1$ and gets transmitted to mode $|a\rangle$. The path lengths of the photons reaching in modes $|a'\rangle$ and $|a\rangle$ are compensated for coherent recombination of polarization amplitudes at PBS $P_4$. This ensures that when all the ADC HWPs are set to zero, and an initial entangled state $|\alpha||HH\rangle+|\beta|\exp(\iota\delta)|VV\rangle$ is incident at the input ports of the interferometers, the initial state is reconstructed at the output ports. These photons are finally sent for quantum state tomography (QST). The HWP $H_4$ is used to flip the polarization of photons passing through it such that the QST settings remain the same for the photons in all spatial modes. The Q, H and P stand for Quarter wave plate, Half wave plate and Polarizing beam splitter. EPS and IF stand for Entangled Photon Source and Interference Filter respectively. For tracing photon paths with polarizations and spatial modes, see also Appendix [\ref{S4.8a}].

The ADC is implemented using two HWPs: $H_1~\text{and}~H_2$ oriented at $\theta~\text{and}~\theta'$ respectively, such that incident $|V\rangle$ polarization amplitude ``decays`` to $|H\rangle$. For different fixed orientations of $H_1$, evolution in ADC is completed by rotating $H_2$. The PBS $P_2$ is used to segregate the $|H\rangle$ and  $|V\rangle$ polarization amplitudes such that the HWP $H_2$ is applied only to $|V\rangle$ polarization for it to serve as excited state of the system and leaving $|H\rangle$ polarization (ground state of the system) undisturbed. 

The single qubit Kraus operators for ADC are given by,
\begin{equation} \label{E4.01}
M_1= \left( \begin{array}{cccc} 1 & 0\\
0 & \sqrt{1-p} \end{array} \right)~~,~~
M_2=\left( \begin{array}{cccc} 0 &\sqrt{p}\\
0 & 0  \end{array} \right),
\end{equation}
where $p=\sin^2(2\theta)$ for ADC mimicked by HWP in a photonic system [\ref{4.09}].

These operators satisfy the completeness condition,
\begin{equation} \label{E4.02}
M_1^\dag M_1+M_2^\dag M_2=\mathbb{I} ,
\end{equation}
where $\mathbb{I}$ is the identity matrix.

The Kraus operators for the two-qubits are obtained by taking appropriate tensor products of single qubit Kraus operators as follows, 
\begin{equation} \label{E4.03}
M_{ij}=M_i \otimes M_j~~;~~i,j=1,2 .
\end{equation}

Label another set of Kraus operators by $M'_{ij}~;~i,j=1,2$, with variable $p$ replaced by $p'$ ($p'=\sin^2(2\theta')$) to distinguish it from the former, with the form of Kraus operators remaining similar to that in Eq.~(\ref{E4.01}). Such a splitting into two angles or two values of probability will prove convenient for later applications in the section of manipulation using optical elements in between. 

Let the initial state of the system be
\begin{equation} \label{E4.04}
|\psi\rangle=\alpha |HH\rangle + \beta \exp(\iota\delta)|VV\rangle ,
\end{equation}
with a corresponding density matrix given by,
\begin{equation} \label{E4.05}
\rho(0,0)= \left( \begin{array}{cccc} 
u & 0 & 0 & v\\
0 & 0 & 0 & 0 \\
0 & 0 & 0 & 0 \\
v^* & 0 & 0  & x 
\end{array} \right),
\end{equation}
where $u=|\alpha|^2~,~v=\alpha\beta^*~,~v^*=\beta\alpha^*~,~x=|\beta|^2$ and $u+x=1$. In general, if $|v|^2=ux$, this represents a pure entangled state, otherwise a mixed entangled state. A more general mixed state with non-zero entries in the other two diagonal positions is considered in appendix [\ref{S4.8c}].   

The initial state of the system (\ref{E4.05}) in the presence of ADC (due to $H_1$ at $\theta$)  evolves as follows,
\begin{equation} \label{E4.06}
\rho^{(1)}(p,0)=\sum _{i,j} M_{ij}~\rho(0,0)~M_{ij}^\dag~~;~~i,j=1,2.
\end{equation} 

Apply the Kraus operators $M'_{ij}$ to complete the evolution in the presence of ADC (due to $H_2$ at $\theta'$) as follows,
\begin{equation} \label{E4.07}
\begin{aligned}
&\rho^{(1)}(p,p')=\sum _{i,j} M'_{ij}~\rho(p,0)~M_{ij}^{\prime\dag}~~;~~i,j=1,2~, \\
&=\left(\begin{array}{cccc}
 \rho^{(1)}_{11}(p,p') & 0 & 0 &  \rho^{(1)}_{14}(p,p') \\
 0 &  \rho^{(1)}_{22}(p,p') & 0 & 0 \\
 0 & 0 &  \rho^{(1)}_{33}(p,p') & 0 \\
  \rho^{(1)}_{41}(p,p')& 0 & 0 &  \rho^{(1)}_{44}(p,p') \\
\end{array}\right),
\end{aligned}
\end{equation} 

where,
\begin{equation} \label{E4.08}
\begin{aligned}
\rho^{(1)}_{11}(p,p')&=u + p^2 x+ p'^2 (1 - p)^2 x + 2 p' (1 - p) p x ,\\
\rho^{(1)}_{22}(p,p')&=(1 - p') p' (1 - p)^2 x + (1 - p') (1 - p) p x,\\
\rho^{(1)}_{33}(p,p')&=(1 - p') p' (1 - p)^2 x + (1 - p') (1 - p) p x,\\
\rho^{(1)}_{44}(p,p')&=(1 - p')^2 (1 - p)^2 x,\\
\rho^{(1)}_{14}(p,p')&=(1 - p') (1 - p) v,\\
\rho^{(1)}_{41}(p,p')&=(1 - p') (1 - p) v^*.\\    
\end{aligned}
\end{equation}
 
 \section{Manipulation of ESD using the NOT operation on both qubits}  \label{S4.3}

\begin{figure}[!htb]
\begin{center}
\includegraphics[clip, trim=3.5cm 9cm 2.5cm 2cm, width=1 \textwidth]{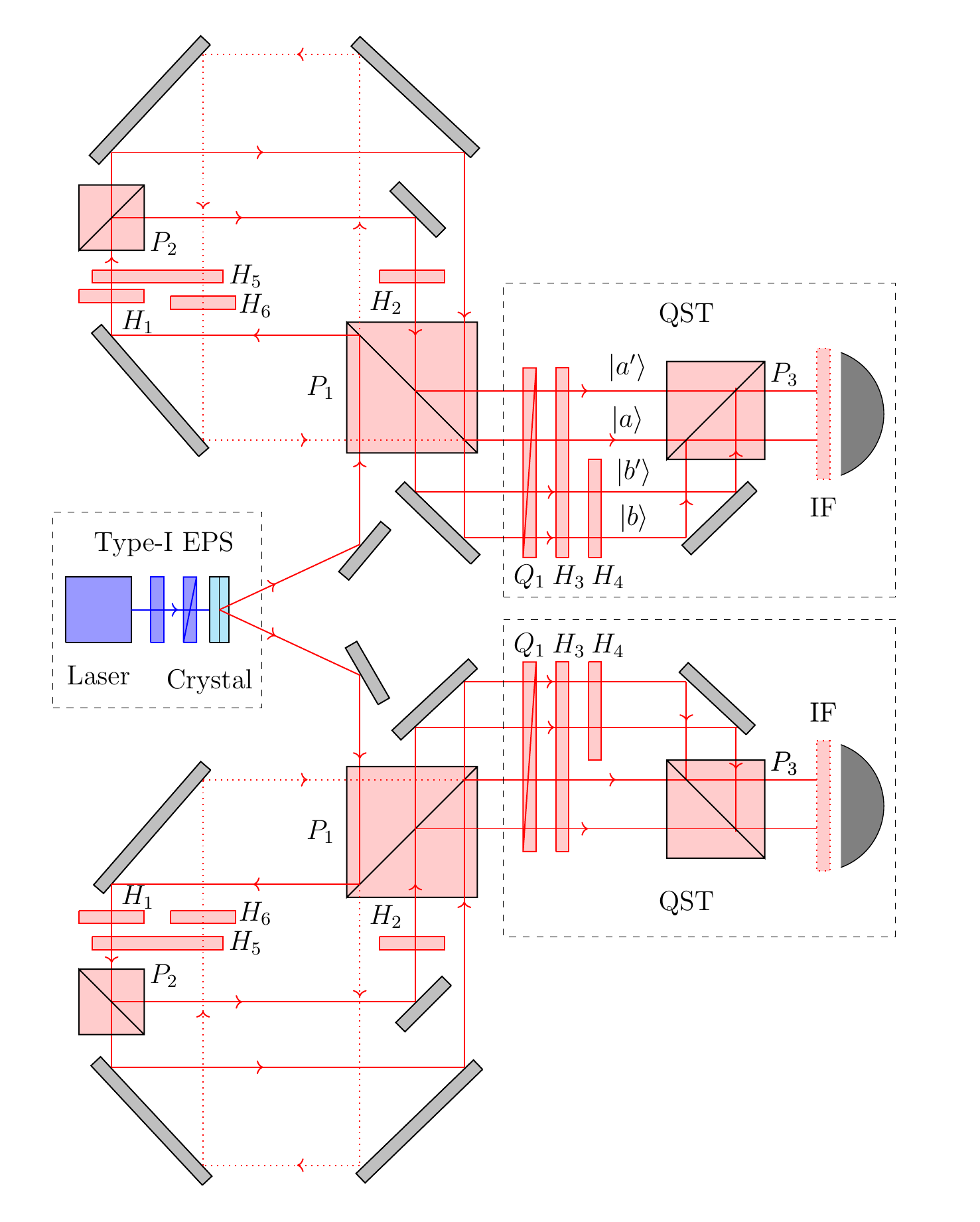}
\caption{\textit{The proposed experimental set up for manipulation of ESD involving NOT operation on both the qubits of a bipartite entangled state in the presence of ADC in a photonic system.}}
\label{fig4.2}
\end{center}
\end{figure}

The experimental set up for manipulation of ESD based on the local NOT operation performed on both the qubits of a bipartite entangled state (\ref{E4.05}) is shown in Fig.~(\ref{fig4.2}). The HWP $H_1$ acts as ADC for incident $|V\rangle$ polarized photon and then NOT operation is performed by $H_5$ at $45^o$, which swaps the $|H\rangle$ and $|V\rangle$ amplitudes, which are then segregated by PBS $P_2$. The ADC is continued by synchronous rotation of $H_2~\text{and}~H_6$ oriented at $\theta'$, which causes the swapped $|V\rangle$ amplitude to ``decay" to $|H\rangle$. The photons from the output spatial modes of the interferometer are sent for tomographic reconstruction of the quantum state [\ref{4.39}].

The polarization entangled photons are sent to two displaced Sagnac interferometers where ADC is simulated by rotating HWP $H_1$ and the NOT operation is performed by HWP $H_5$, and then ADC is continued by a set of secondary HWPs $H_2~\text{and}~H_6$. The path lengths of the photons reaching in mode $|b\rangle$ are compensated for coherent recombination of polarization amplitudes at PBS $P_1$. This ensures that when all the ADC HWPs are set to zero, except for NOT operation, and an initial entangled state $|\alpha||HH\rangle+|\beta|\exp(\iota\delta)|VV\rangle$ is incident at the input ports of the interferometers, we reconstruct the state $|\alpha||VV\rangle+|\beta|\exp(\iota\delta)|HH\rangle$ at the output ports. These photons are finally sent for quantum state tomography (QST). The HWP $\text{H}_4$ is used to flip the polarization of photons passing through it such that the QST settings remain the same for the photons in all spatial modes.

The initial state of the system (\ref{E4.05}) in the presence of ADC (due to $H_1$ at $\theta$)  evolves as follows,
\begin{equation}
\rho^{(2)}(p,0)=\sum _{i,j} M_{ij}~\rho(0,0)~M_{ij}^\dag~~;~~i,j=1,2 .
\end{equation} 

Apply the NOT operation on both the qubits at $p=p_n$ as follows,
\begin{equation}
\rho^{(2)}(p_n,0)=(\hat{\sigma_x} \otimes \hat{\sigma_x}) \rho^{(2)}(p,0)~(\hat{\sigma_x} \otimes \hat{\sigma_x})^\dag ,
\end{equation}
where $\hat{\sigma_x}$ is the Pauli matrix. This amounts to switching the elements $\rho_{11}$ and  $\rho_{44}$ and  $\rho_{22}$ and  $\rho_{33}$ and interchanging (complex conjugation) the off-diagonal elements.

Apply the Kraus operators $M'_{ij}$ to complete the evolution of the system in the presence of ADC (due to $H_2$ and $H_6$ at $\theta'$) after the NOT operation as follows,
\begin{equation} \label{E4.11}
\rho^{(2)}(p_n,p')=\sum _{i,j} M'_{ij}~\rho^{(2)}(p_n,0)~M_{ij}^{\prime\dag}~~;~~i,j=1,2, 
\end{equation}

with entries now in the form of (7) given by,
\begin{equation}
\begin{aligned}
\rho^{(2)}_{11}(p_n,p')&=(1 - p_n)^2 x + 2 p' (1 - p_n) p_n x + p'^2 (u + p_n^2 x),\\
\rho^{(2)}_{22}(p_n,p')&=(1 - p') (1 - p_n) p_n x + (1 - p') p' (u + p_n^2 x),\\
\rho^{(2)}_{33}(p_n,p')&=(1 - p') (1 - p_n) p_n x + (1 - p') p' (u + p_n^2 x),\\
\rho^{(2)}_{44}(p_n,p')&=(1 - p')^2 (u + p_n^2 x),\\
\rho^{(2)}_{14}(p_n,p')&=(1 - p') (1 - p_n) v^*,\\
\rho^{(2)}_{41}(p_n,p')&=(1 - p') (1 - p_n) v.\\    
\end{aligned}
\end{equation}

\section{Effect of the NOT operation applied on only one of the qubits}  \label{S4.4}

The experimental set up for studying the effect of a NOT operation applied on only one of the qubits of a bipartite entangled state in the presence of ADC on the dynamics of entanglement is to retain one half, say the lower, as in Fig.~(\ref{fig4.1}) and have only the upper half as in Fig.~(\ref{fig4.2}), the optical elements $H_5$ and $H_6$ occurring only in the upper arm.

The initial state of the system (\ref{E4.05}) in the presence of ADC (due to $H_1$ at $\theta$)  evolves as follows
\begin{equation}
\rho^{(3)}(p,0)=\sum _{i,j} M_{ij}~\rho(0,0)~M_{ij}^\dag~~;~~i,j=1,2.
\end{equation} 

Apply the NOT operation on only one of the qubits by $H_5$ at $45^\circ$, let us say first qubit, at $p=p_n$ as follows
\begin{equation} \label{E4.14}
\rho^{(3)}(p_n,0)=(\hat{\sigma_x} \otimes \hat{\mathbb{I}})~ \rho^{(3)}(p,0)~(\hat{\sigma_x} \otimes \hat{\mathbb{I}})^\dag .
\end{equation}

Apply next the Kraus operators $M'_{ij}$ to complete the evolution of the system in the presence of ADC (due to $H_2~\text{and}~H_6$ at $\theta'$) after the NOT operation to give
\begin{equation} \label{E4.15}
\rho^{(3)}(p_n,p')=\sum _{i,j} M'_{ij}~\rho^{(3)}(p_n,0)~M_{ij}^{\prime\dag}~~;~~i,j=1,2,
\end{equation}

with entries now in the form of (\ref{E4.07}) given by,
\begin{equation} \label{E4.16}
\begin{aligned}
\rho^{(3)}_{11}(p_n,p')&= p' (1 - p_n)^2 x + (1 - p_n) p_n x + p'^2 (1 - p_n) p_n x + p' (u + p_n^2 x),\\
\rho^{(3)}_{22}(p_n,p')&=(1 - p') (1 - p_n)^2 x + (1 - p') p' (1 - p_n) p_n x,\\
\rho^{(3)}_{33}(p_n,p')&=(1 - p') p' (1 - p_n) p_n x + (1 - p') (u + p_n^2 x),\\
\rho^{(3)}_{44}(p_n,p')&=(1 - p')^2 (1 - p_n) p_n x,\\
\rho^{(3)}_{23}(p_n,p')&=(1 - p') (1 - p_n) v^*,\\
\rho^{(3)}_{32}(p_n,p')&=(1 - p') (1 - p_n) v.\\    
\end{aligned}
\end{equation}

\section{Some analytical expressions}  \label{S4.5}
 
Let the two polarization entangled qubits constitute the system, as given by Eq.~(\ref{E4.05}), and the action of the rotating HWPs simulate the ADC. This causes a $|V\rangle$ polarized photon to probabilistically ``decay" to $|H\rangle$ with probability  $p=\sin^2(2\theta)$ ($p'=\sin^2(2\theta')$), where $\theta$ ($\theta'$) is the angle between the fast axis of the HWP and $|V\rangle$. The ADC probability $p'_0$ at which ESD happens, depends on the initial state parameters of the entangled system and the ADC setting of first HWP $p$. The criterion for ESD as indicated by a switch in sign of the eigenvalues of the partial transpose of Eq.~(\ref{E4.07}) is given by $\rho_{22}\rho_{33}=|\rho_{14}|^2$. For the initial state (\ref{E4.05}), the condition for ESD is obtained by computing the Negativity of the state (\ref{E4.07}) and equating it to zero. The condition for ESD is given by, 
\begin{equation} \label{E4.17}
p'_0=\frac{|v|-xp}{ x(1-p)}.
\end{equation}

Let us denote the effective end of entanglement due to combined evolution through two HWPs by $p_{end}$. The $p_{end}$ involves a multiplication of survival probabilities to give,
\begin{equation} \label{E4.18}
1-p_{end}=(1-p)(1-p'_0)~~ \text{with}~~ p_{end} =|v|/x.
\end{equation}
depending only on the initial state parameters in (\ref{E4.05}).

For the manipulation of ESD using NOT operation on both the qubits, this operation switches $\rho_{11}$ and  $\rho_{44}$, $\rho_{22}$ and  $\rho_{33}$  , and interchanges the off-diagonal elements $\rho_{14}$ and $\rho_{41}$ in Eq.~(\ref{E4.07}). With subsequent evolution, the criterion for when ESD now happens, can be used to determine the value of $p_A$ that marks the boundary between hastening or not relative to $p_{end}$, and similarly the value $p_B$ that is the boundary between delaying $p_{end}$ past $p_0$ or averting ESD completely. We get,
\begin{equation} \label{E4.19}
p_A=\frac{1-2 u}{2 (1-u)}~~,~~p_B=\frac{|v|-u}{1+|v|-u}.
\end{equation}

For the manipulation of ESD using NOT operation on only one of the qubits, this operation now switches $\rho_{11}$ and $\rho_{33}$, $\rho_{22}$  and $\rho_{44}$, and moves $\rho_{14}$  into the $\rho_{23}$  position. Following subsequent evolution, the ESD criterion through the partial transpose matrix now becomes $\rho_{11} \rho_{44}=|\rho_{23}|^2$. We now get,
\begin{equation} \label{E4.20}
p_A=\frac{|v|}{u+ 2|v|}~~,~~p_B=\frac{|v|^2}{|v|^2-u+1}.
\end{equation}

These simple expressions defining the time $p_0$ for ESD, and the times for NOT operation that define the delay/hasten and avert/delay boundaries may also be given for a more general mixed state density matrix with also non-zero entries in the two other diagonal position in (\ref{E4.05}) and are recorded in Appendix [\ref{S4.8c}]. The Appendix [\ref{S4.8d}] records similar expressions for a density matrix with non-zero values in the other off-diagonal position as in [\ref{4.35}].

The NOT operation applied on both the qubits at $p=p_n$ of a bipartite entangled state leads to the end of entanglement given by,
\begin{equation} \label{E4.21}
p_{end}=\frac{p_n^2 (2x+|v|)+p_n (1-2x-2|v|)+|v|}{x \left(p_n^2-1\right)+1}.
\end{equation}

When NOT operation is applied on only one of  the qubits at $p=p_n$, the end of entanglement is given by,
\begin{equation}  \label{E4.22}
p_{end}=\frac{[4 x (p_n-1) p_n + 4 (p_n-1)^2 |v|^2+1]^{1/2}+2 x p_n-1}{2x p_n}.
\end{equation}

\section{Results and Discussion}  \label{S4.6}

As an example, we choose the initial state: $|\psi\rangle=|\alpha||HH\rangle+|\beta|\exp(\iota\delta)|VV\rangle$ with $|\alpha|=1/\sqrt{5},|\beta|=2/\sqrt{5}$ and $\delta=0$ and report the results for ESD, and ESD-manipulation using the NOT operation on one or both the qubits of the bipartite entangled state.

\begin{figure}[!htb]
\begin{center}
\includegraphics[width=0.7\textwidth]{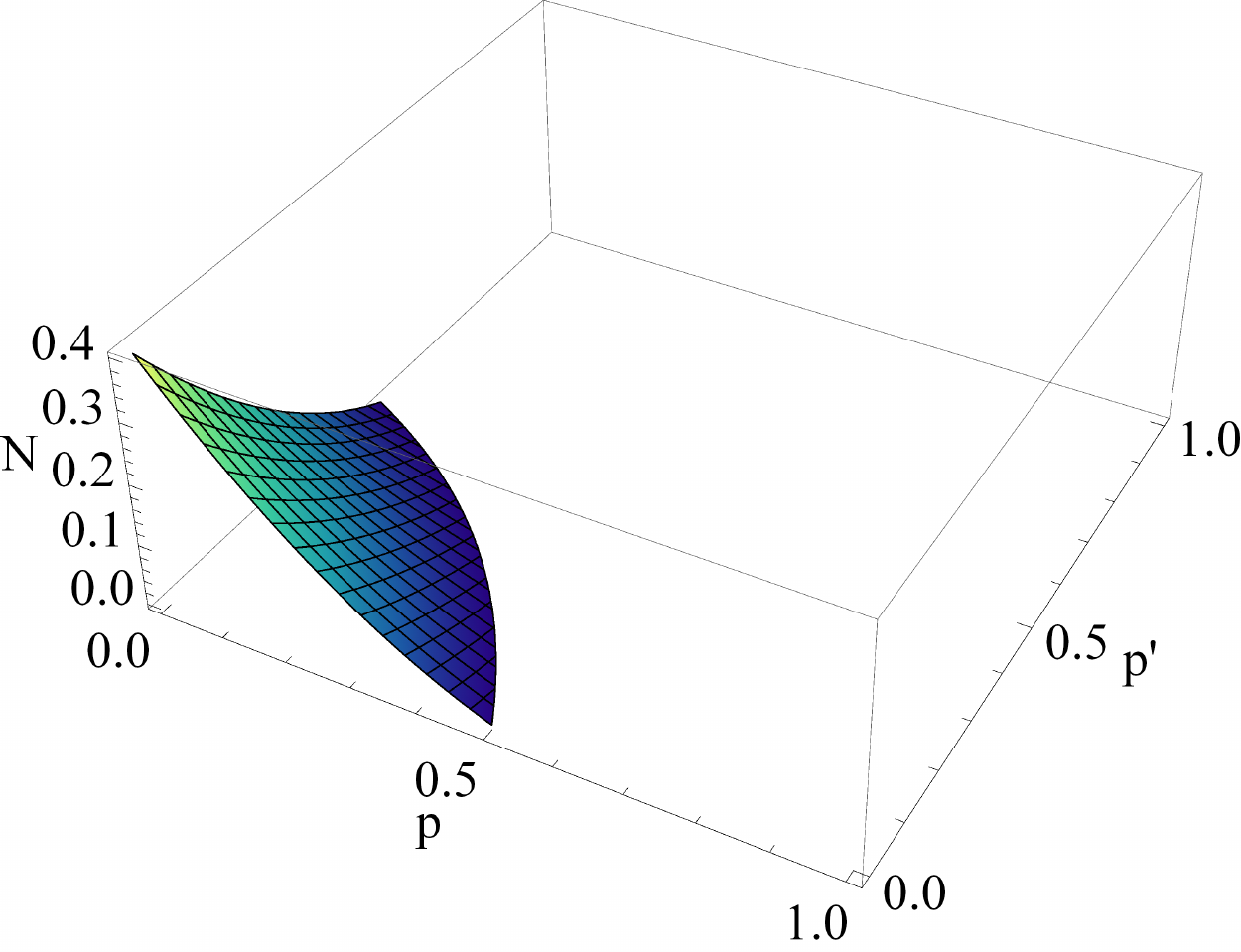}  
\caption{\textit{Plot of Negativity N vs. ADC probability ($p,p'$) for ESD using two HWPs for a bipartite entangled state. The plot implies ESD for $p=0,~p'=0.5$ and $p'=0,~ p=0.5$. For intermediate values of $p~(p')$, the curvature reflects the non-linear relation between them as per (17), the action of two HWPs oriented at $\theta$ and $\theta'$ applied one after another is not equivalent to that of one HWP oriented at $\theta+\theta'$.}}
\label{fig4.3}
\end{center}
\end{figure}

\begin{figure}[!htb]
\begin{center}
\includegraphics[width=0.7\textwidth]{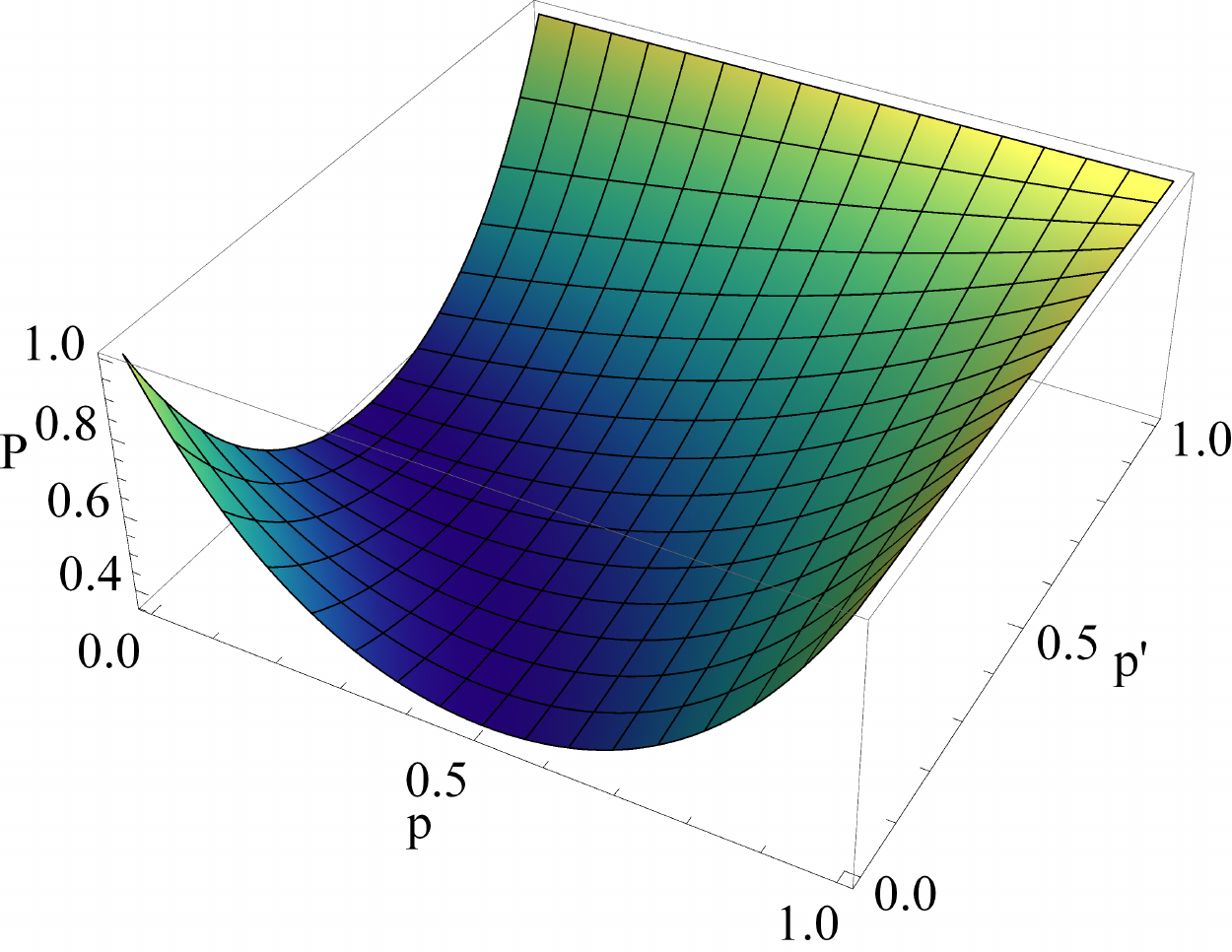} 
\caption{\textit{Plot of Purity P vs. ADC probability ($p,p'$) for ESD using two HWPs for a bipartite entangled state. The state (\ref{E4.07}) is initially pure, gets mixed at intermediate ADC probabilities and becomes pure again at $p=1$ or $p'=1$.}}
\label{fig4.4}
\end{center}
\end{figure}

For ESD using two HWPs, the  disentanglement happens for $p=0$ at $p'_0=0.5$  and for any other combination of $p$ and $p'$, $p'_0$ follows the non-linear Eq.~(\ref{E4.17}) in $p$, and the effective end due to two HWPs is given by non-linear Eq.~(\ref{E4.18}) with $p_{end}=0.5$. The plot of Negativity N vs. probability of decay of qubits ($p, p'$) for the state (\ref{E4.07}) is shown in Fig.~(\ref{fig4.3}). The plot of Purity (defined as $\mathrm{Tr}[\rho^2]$) vs. probability of decay of qubits ($p,p'$) for ESD is shown in Fig.~(\ref{fig4.4}). The two-qubit entangled state~(\ref{E4.05}), initially in a pure state, gets mixed at intermediate stages of amplitude damping and finally becomes pure again when both the qubits have decohered  down to the ground state ($|HH\rangle$) at $p=1$ or $p'=1$. However, at ESD for $p_0=0.5$, it ends as a mixed disentangled state.

\begin{figure}[!htb]
\begin{center}
\includegraphics[width=0.7\textwidth]{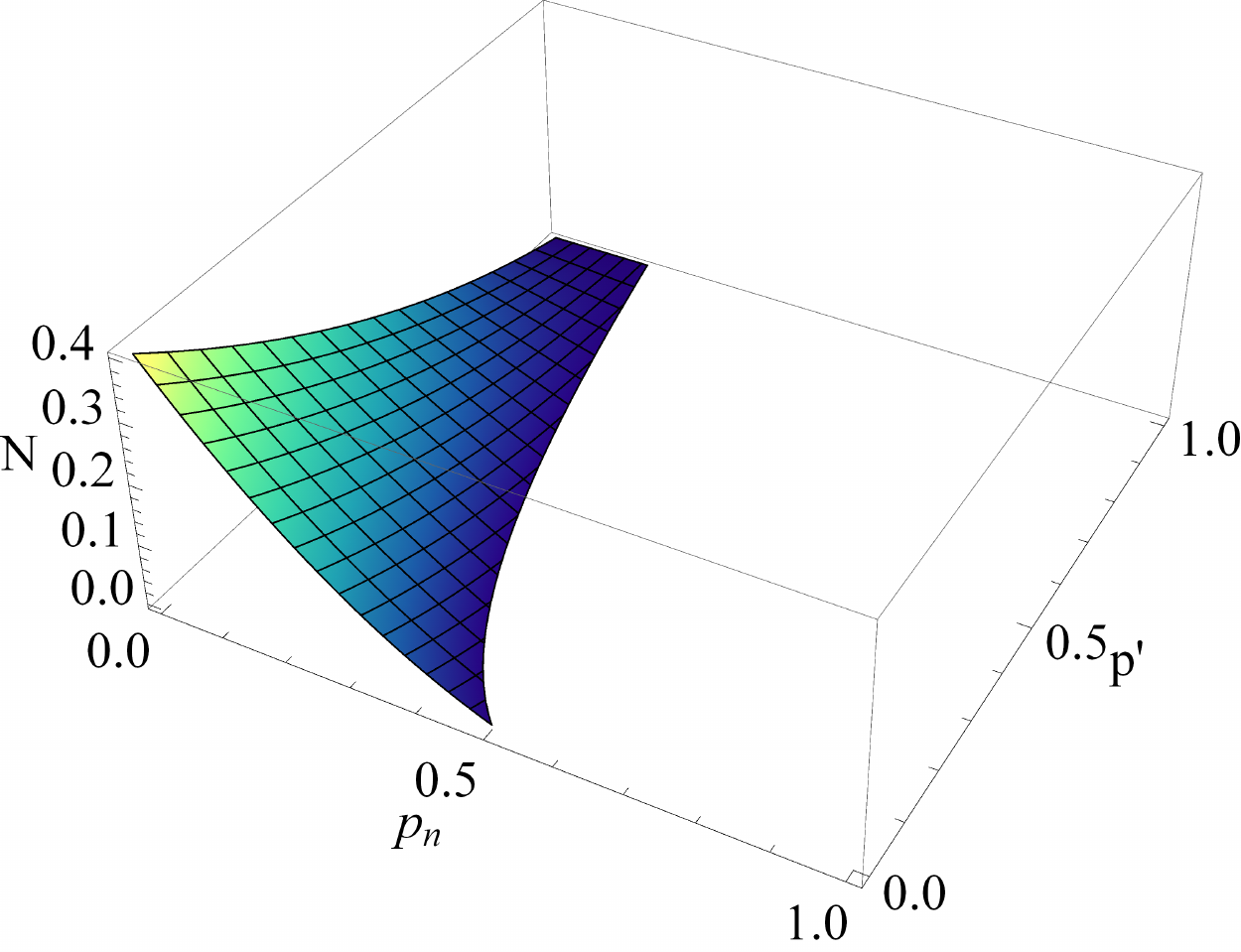}
\caption{\textit{Plot of Negativity N vs. ADC probability ($p_n,p'$) such that NOT operation is applied on both the qubits at $p=p_n$ for manipulation of ESD of a bipartite entangled state. The NOT operation leads to hastening for $0.375 < p_n < 0.5$, delay for $0.1667 < p_n < 0.375$, and  avoidance of ESD for $0\leq p_n \leq 0.1667$.}}
\label{fig4.5}
\end{center}
\end{figure}

\begin{figure}[!htb]
\begin{center}
\includegraphics[width=0.7\textwidth]{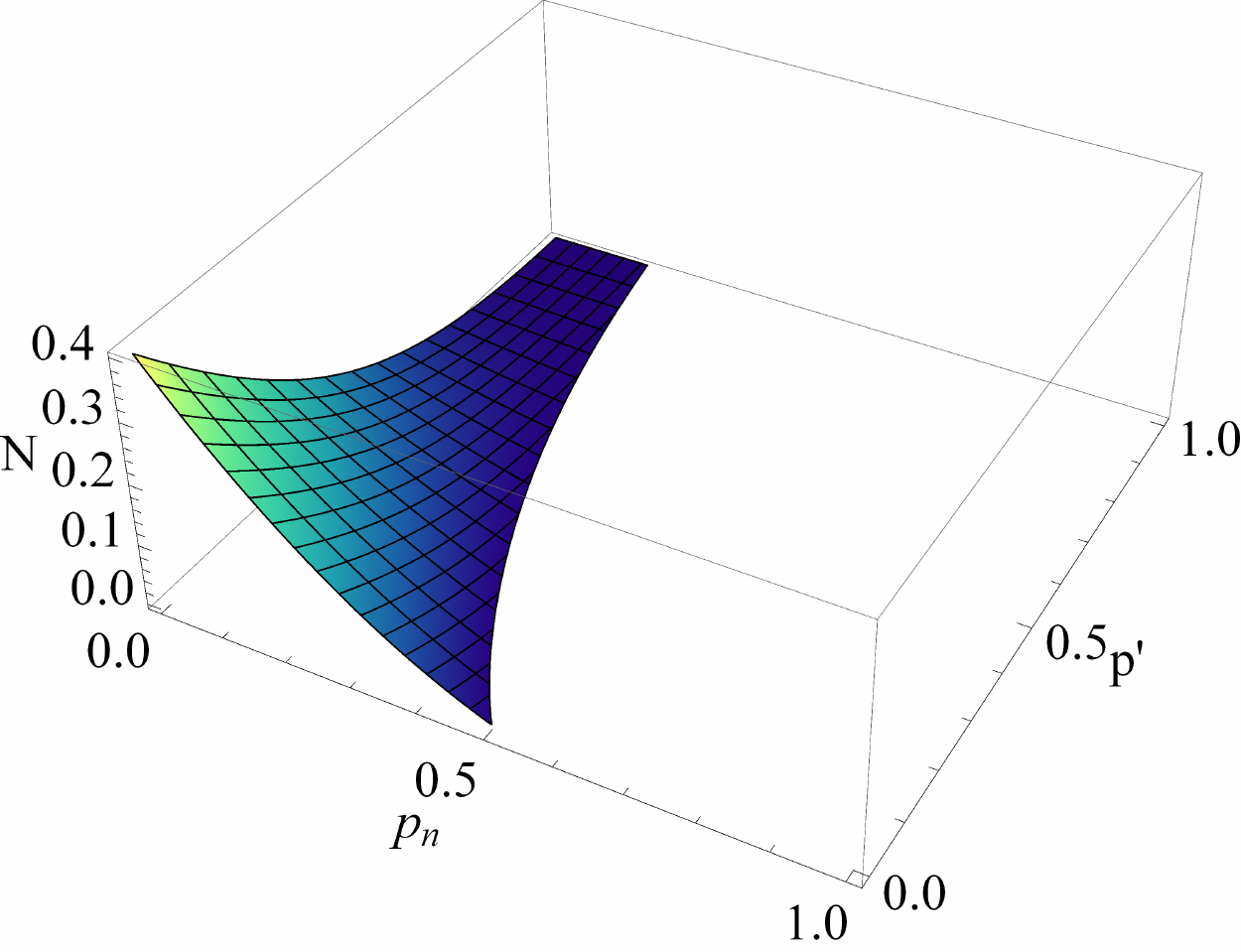} 
\caption{\textit{Plot of Negativity N vs. ADC probability ($p_n,p'$) such that NOT operation is applied on only one of the qubits at $p=p_n$ for manipulation of ESD of a bipartite entangled state. The NOT operation leads to hastening for $0.4 < p_n <0.5$, delay for $0.1667 < p_n < 0.4$, and avoidance of ESD for $0\leq p_n \leq 0.1667$.}}
\label{fig4.6}
\end{center}
\end{figure}

For the manipulation of ESD using NOT operation on both the qubits: we get $p_A=0.375$, and $p_B=0.1667$. The corresponding plot of Negativity $N$ vs. ADC probability ($p_n , p'$) for the state (\ref{E4.11}) is  shown in Fig.~(\ref{fig4.5}). For the  manipulation of ESD using NOT operation on only one of the qubits: we get $p_A=0.4,p_B=0.1667$. The corresponding plot of Negativity $N$ vs. ADC probability ($p_n , p'$) for the state (\ref{E4.15}) is  shown in Fig.~(\ref{fig4.6}).

\begin{figure}[!htb]
\begin{center}
\includegraphics[width=0.7\textwidth]{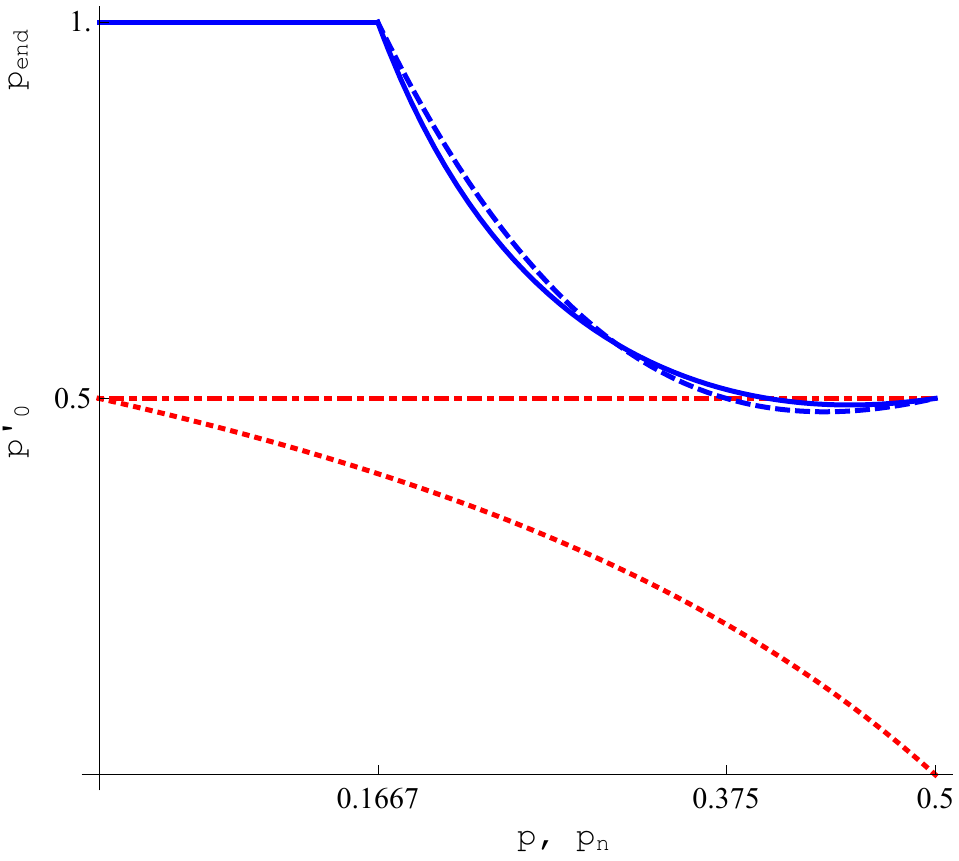} 
\caption {\textit{ Plot shows the effect of NOT operation on manipulation of ESD at various $p_n$ values for density matrix parameters $u=0.2, |v|=0.4$. The plot for ESD, Eq.~(\ref{E4.17}) and a rendering of (\ref{E4.18}), for every $p$, giving the value of $p'_0$, the compounding of them giving the flat line $p_{end}=0.5$  are shown by dotted red curve and dot-dashed red line. The plot for manipulation of ESD using NOT on both or only one qubit are shown by dashed and solid blue curves, respectively. For $0\leq p_n \leq 0.1667$, $p_{end}=1$ means avoidance of ESD, the $p_B$ value for single and double NOT coinciding for these particular parameters but $p_B$ is in general different. For $0.1667 < p_n < 0.4$ ( $0.1667 < p_n < 0.375$) the dashed (solid) blue curve lies above 0.5, implies delays of ESD for double (single) NOT manipulation of ESD. The dashed (solid) blue curve dipping below 0.5 for $0.375 < p_n < 0.5$ ($0.4 < p_n < 0.5$) for manipulation using double (single) NOT operation implies hastening of ESD in this range.}}
\label{fig4.7}
\end{center}
\end{figure}

 The plot of $p_{end}$ vs. $p_n$ for Eqs.~(\ref{E4.21}) and (\ref{E4.22}) such that the NOT operations applied on both (only one of) the qubits at $p_n$ leads to disentanglement at $p_{end}$ is shown by dashed (solid) blue curve in Fig.~(\ref{fig4.7}). In the avoidance range $0 \leq p_n \leq 0.1667$, the $p_{end}$ vs. $p_n$ curves are cut off at $p_{end}=1$ to signify the asymptotic decay with probabilities remaining in the physical domain. For comparison, we have also included the results of ESD; Eq.~(\ref{E4.17}) and a rendering of (\ref{E4.18}), for every $p$, giving the value of $p'_0$, the compounding of them giving the flat line at  $p_{end}=0.5$ as shown by dotted red curve and dot-dashed red line in Fig.~(\ref{fig4.7}). The role of NOT operation on manipulation of ESD is evident as for (i) $0 \leq  p_n \leq 0.1667$, we get avoidance of ESD with $p_{end}=1$ in this range for NOT operation on only one or both the qubits (ii) $0.1667 < p_n < 0.375$ ( $0.1667 < p_n < 0.4$), we get delay of ESD as the dashed (solid) blue curve lies above 0.5 but less than 1, and (iii) $0.375 < p_n < 0.5$ ($0.4 < p_n < 0.5$), we get hastening of ESD as the dashed (solid) blue curve dips below 0.5.

The discussion so far, and Figs.~(\ref{fig4.3} - \ref{fig4.7}), pertain to the choice $u=0.2, |v|=0.4$, and result in $p_0=0.5, p_B=0.1667, p_A=0.375$ for NOT applied to both whereas $p_A=0.4$ when applied to just one qubit. This is an example when $p_A > p_B$ and both lie in the physically relevant interval ($0, p_0$). All three phenomena, of hastening ($p_A < p_\text{NOT} < p_0$), delaying ($p_B < p_\text{NOT} < p_A$), and averting ($0 \leq p_\text{NOT}\leq p_B$) ESD then occur. The appearance of the various manipulation regimes (hastening, delay, and avoidance of ESD) critically depends on the choice of the parameters of the initial state (density matrix) of the system as expressed in Eqs.~(\ref{E4.17} - \ref{E4.20}).

Consider a general initial state (\ref{E4.05}), with $|v|\leq \sqrt{u(1-u)}$, which captures pure as well as mixed entangled states. The condition for the existence of hastening regime is $u+|v| > 0.5$ for manipulation of ESD using single or double NOT operation. The condition for existence of avoidance regime is $|v|>0$ ($|v|>u$) for manipulation of ESD using single (double) NOT operation. For the pure entangled state (\ref{E4.04}), the condition for a physically relevant $p_A$ is $u\geq 0.1464$. Thus, pure entangled states (\ref{E4.04}) with $0.1464\leq u <0.5$ give rise to hastening, delay as well as avoidance of ESD, whereas states with $0<u < 0.1464$ give rise to delay and avoidance of ESD only. For all values of initial parameters, the analytical expressions in (\ref{E4.17} - \ref{E4.20}) provide $p_0$ for ESD, and $p_A$ and $p_B$. When these lie within the domain $(0,p_0)$, all three regimes are realized. Otherwise, one may have only two or one of the three regimes of avoidance, delay or hastening of ESD. More general expressions for a wider class of density matrices than (\ref{E4.05}) are  given in Appendix [\ref{S4.8c}, \ref{S4.8d}]. 

\begin{figure}[!htb]
\begin{center}
\includegraphics[width=0.7\textwidth]{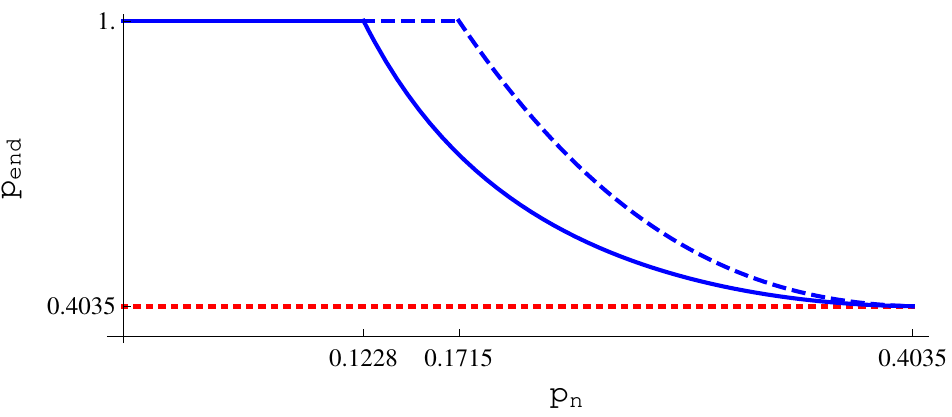} 
\caption {\textit{Plot of $p_{end}$ vs. $p_n$ for manipulation of ESD using NOT operation on one (solid blue curve) or both (dashed blue curve) the qubits of a bipartite entangled state in the presence of ADC for density matrix parameters $u=0.14, |v|=0.347$. The NOT operation leads to avoidance for $0 \leq p_n \leq 0.1228$  ($0 \leq p_n \leq 0.1715$) with disentanglement happening at $p_{end}=1$ in this range, and  delay of ESD for $0.1228 <  p_n < 0.4035$ ($0.1715 < p_n < 0.4035$) as $p_{end}$ curve lies above the $p_{end}=0.4035$ dotted red ESD line for single (double) NOT operation. There is no hastening of ESD for this particular choice of parameters for single or double NOT operation.}}
\label{fig4.8}
\end{center}
\end{figure}

\begin{figure}[!htb]
\begin{center}
\includegraphics[width=0.7\textwidth]{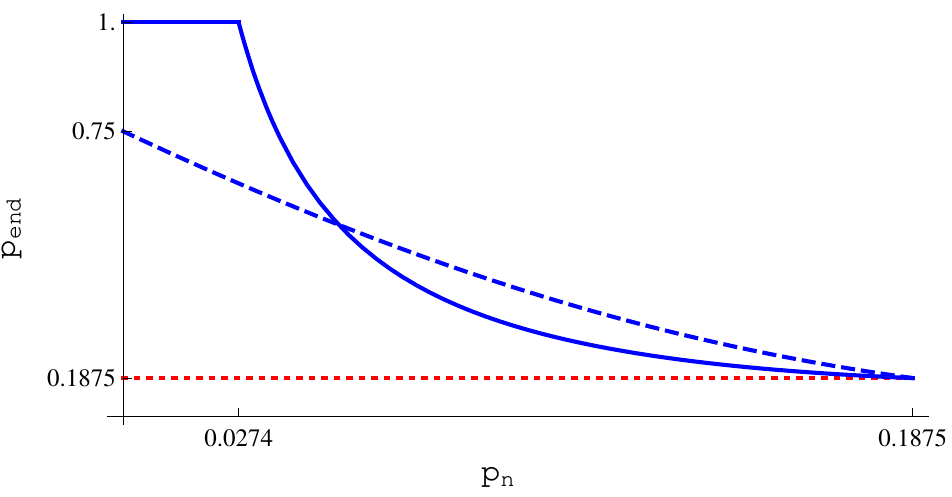} 
\caption {\textit{Plot of $p_{end}$ vs. $p_n$ for manipulation of ESD using NOT operation on one (solid blue curve) or both (dashed blue curve) the qubits of a bipartite entangled state in the presence of ADC for density matrix parameters $u=0.2, |v|=0.15$. The NOT operation when applied on only one of the qubits leads to avoidance of ESD for $ 0 < p_n < 0.0274$  with disentanglement happening at $p_{end}=1$ in this range, and delay for $0.0274 < p_n < 0.1875$ as the  solid blue curve lies above the $p_{end}=0.1875$ dotted red ESD line. The NOT operation when applied on both the qubits leads to delay for $0 < p_n < 0.1875$ as the dashed blue curve lies above the $p_{end}=0.1875$ dotted red ESD line. There is no hastening of ESD for single or double NOT operation and no avoidance of ESD for double NOT operation for this particular choice of parameters.}}
\label{fig4.9}
\end{center}
\end{figure}

Consider another example of pure entangled state of the form (\ref{E4.05}) with $u=0.14$ and $|v|=0.347$. For this state, $p_0=0.4035$,  $p_B=0.1228$ ($p_B=0.1715$) and $p_A$ does not exist in the physical domain for single (double) NOT operation. Therefore, we get only delay and avoidance of ESD. The corresponding plot of $p_{end}$ vs. $p_n$ is shown by solid (dashed) blue curve in Fig.~(\ref{fig4.8}). Next, consider an example of mixed entangled state of the form (\ref{4.05}) with $u=0.2$ and $|v|=0.15$. For this state, $p_0=0.1875$, $p_B=0.0274$ ($p_B$ does not exist), and  $p_A$ does not exist in the physical domain for single (double) NOT operation. Therefore, NOT operation applied on only one (both) of the qubits delays as well avoids (only delays) the ESD. The corresponding plot of $p_{end}$ vs. $p_n$ is shown by solid (dashed) blue curve in Fig.~(\ref{fig4.9}).

\newpage
\section{Conclusions and future work} \label{S4.7}

We have proposed an all-optical experimental setup for the demonstration of hastening, delay, and avoidance of ESD in the presence of ADC in a photonic system. The simulation results of the manipulation of ESD considering a photonic system, when NOT operations are applied on one or both the qubits, are completely consistent with the theoretical predictions of reference [\ref{4.35}] for the two-level atomic system where spontaneous emission is the ADC. We give analytical expressions for $p_0,~p_A~\text{and}~p_B$ which depend on the parameters of the density matrix of the system for both the forms considered here in (5) and that in [\ref{4.35}].

Our proposal also has an advantage over decoherence suppression using weak measurement and quantum measurement reversal, and delayed choice decoherence suppression. There, as the strength of weak interaction increases, the success probability of decoherence suppression decreases. In our scheme, however, we can manipulate the ESD, in principle, with unit success probability as long as we perform the NOT operation at the appropriate wave plate angle which is analogous to time in the atomic system. Delay and avoidance of ESD, in particular, will find application in the practical realization of quantum information and computation protocols which might otherwise suffer a short lifetime of entanglement. Also, it will have implications towards such control over other physical systems. The advantage of the manipulation of ESD in a photonic system is that one has complete control over the damping parameters, unlike in most atomic systems. An experimental realization of our proposal will be important for practical noise engineering in quantum information processing, and is under way. Further work in the future could study the dynamics of entanglement in the presence of the generalized ADC [\ref{4.40}-\ref{4.43}] and the squeezed generalized ADC [\ref{4.44}] and the possible schemes for manipulation of entanglement sudden death in the presence of such damping channels. 

\section{An alternative approach to analyze the ESD and manipulation experiments}  \label{S4.8}
We provide here an alternative and intuitive approach to analyze the ESD and its manipulation set up by tagging the photon polarization states with the spatial modes of the interferometer upon the action of each of optical component encountered in the photon's path. The evolution of system plus reservoir is represented  by a unitary operator $U_{SR}$. The degrees of freedom of the reservoir can be traced out from $U_{SR}$ to get the Kraus operators which govern the evolution of the system by itself.

\subsubsection{\Large{ESD experiment}}  \label{S4.8a}
Consider the experimental set up for ESD as shown in Fig.~(\ref{fig4.1}). An incident $|H\rangle$ polarized photon is transmitted though the PBS $P_1$ and traverses the interferometer in a counter-clockwise direction, returns to $P_1$ and is transmitted into spatial mode $|a\rangle$ of the reservoir. The corresponding quantum map is given by, 
\begin{equation}
\begin{aligned}
U_{SR}|H\rangle_S|a\rangle_R \rightarrow |H\rangle_S|a\rangle_R.\\
\end{aligned}
\end{equation}

An incident  $|V\rangle$ polarized  photon is reflected by PBS $P_1$ and traverses the interferometer in a clockwise direction. The action of HWPs $H_1$ and $H_2$ and  PBS $P_1$ and $P_2$ are represented by the quantum map, 
\begin{equation}
\begin{aligned}
U_{SR}|V\rangle_S|a\rangle_R \xrightarrow [P_2]{H_1@\theta} &\sqrt{1-p} |V\rangle_S|a'\rangle_R+\sqrt{p} |H\rangle_S|b\rangle_R,\\
\xrightarrow[in~|V\rangle~arm]{H_2@\theta'}& \sqrt{1-p}[\sqrt{1-p'}~|V\rangle_S|a'\rangle_R\\& + \sqrt{p'}~ |H\rangle_S|b'\rangle_R] +\sqrt{p}|H\rangle_S|b\rangle_R],\\
\end{aligned}
\end{equation}
where $p=\sin^2(2\theta)~,~p'=\sin^2(2\theta')$.

Combining above two Eqs., the unitary operator $U_{SR}$ can be written as:
\begin{equation}
\begin{split}
U_{SR} = & {|H\rangle_S |a\rangle_R}~{_S \langle H| _R\langle a|} + [\sqrt{p}|H\rangle_S|b\rangle_R+  \sqrt{(1-p)(1-p')}|V\rangle_S|a'\rangle_R  \\ &
+ \sqrt{(1-p)p'|H\rangle_S|b'\rangle_R}] _S\langle V| _R\langle a|.
\end{split}
\end{equation}

Since we coherently recombine the spatial modes $|a\rangle$ and $|a'\rangle$ due to physically motivated reasons, i.e., $|a'\rangle\equiv|a\rangle$, the above unitary operator reduces to
\begin{equation}
\begin{split}
U_{SR} = & {|H\rangle_S |a\rangle_R}~{_S \langle H| _R\langle a|} + [\sqrt{p}|H\rangle_S|b\rangle_R+  \sqrt{(1-p)(1-p')}|V\rangle_S|a\rangle_R   \\ &
+ \sqrt{(1-p)p'|H\rangle_S|b'\rangle_R}] _S\langle V| _R\langle a|.
\end{split}
\end{equation}

The Kraus operator for the ESD are obtained by tracing out the degrees of freedom of the reservoir as follows:
\begin{equation}
K_\mu=~_R\langle\mu|U_{SR}|a\rangle_R~~;~\mu=a,b,b'.
\end{equation}

The Kraus operators, thus, obtained are
\begin{subequations}
\begin{align}
\begin{split}
K_1=&~_R\langle a|U_{SR}|a\rangle_R=\left( \begin{array}{cc}
1 & 0\\
0 & \sqrt{(1-p)(1-p')}
\end{array}\right),
\end{split}\\
\begin{split}
K_2=&~_R\langle b|U_{SR}|a\rangle_R=\left( \begin{array}{cc}
0 & \sqrt{p}\\
0 & 0
\end{array}\right),
\end{split}\\
\begin{split}
K_3=&~_R\langle b'|U_{SR}|a\rangle_R=\left( \begin{array}{cc}
0 & \sqrt{(1-p')p}\\
0 & 0
\end{array}\right).
\end{split}
\end{align}
\end{subequations}

The Kraus operators for two-qubit system is obtained by
\begin{equation}
K_{ij}=K_i\otimes K_j~~;i,j=1,2,3.
\end{equation}

These Kraus operators govern the evolution of the system from the initial state $\rho(0)$ to the final state $\rho(p,p')$ as follows. 
\begin{equation}
\rho(p,p')=\sum_{i,j=1}^3 K_{ij}\rho(0,0)K_{ij}^\dag.
\end{equation}

\subsubsection{\Large{ESD manipulation experiment}}  \label{S4.8b}
Consider the experimental setup for ESD manipulation using double NOT operation   as shown in Fig.~(\ref{fig4.2}). An incident $|H\rangle$ polarized photon is transmitted through PBS $P_1$ and traverses the interferometer in a counter-clockwise direction where the NOT operation is applied by HWP $H_5$ and ADC afterwards is simulated by $H_6$. The corresponding quantum map is given by
\begin{equation}
\begin{aligned}
U_{SR}|H\rangle_S|a\rangle_R&\xrightarrow{H_5~\text{at}~45}|V\rangle_S|b\rangle_R~,\\
&\xrightarrow{H_6~\text{at}~\theta'}\sqrt{1-p'}|V\rangle_S|b\rangle_R+\sqrt{p'}|H\rangle_S|a\rangle_R~.
\end{aligned}
\end{equation}

An incident $|V\rangle$ polarized photon is reflected by PBS $P_1$ and traverses the interferometer in a clockwise direction where ADC is introduced by HWP $H_1$ followed by NOT operation by HWP $H_5$ and then ADC is continued by HWP $H_2$. The corresponding quantum map is given by
\begin{equation}
\begin{aligned}
U_{SR}|V\rangle_S|a\rangle_R \xrightarrow{H_1~\text{at}~\theta}&\sqrt{1-p}|V\rangle_S|a\rangle_R+\sqrt{p}|H\rangle_S|b\rangle_R~,\\
\xrightarrow[H_5~\text{at}~45]{P_2}&\sqrt{1-p}|H\rangle_S|b\rangle_R+\sqrt{p}|V\rangle_S|a'\rangle_R~,\\
\xrightarrow[\text{in}~|V\rangle~\text{arm}]{H_2~\text{at}~\theta'} &\sqrt{1-p}|H\rangle_S|b\rangle_R
+\sqrt{p}[\sqrt{1-p'}|V\rangle_S|a'\rangle_R +\sqrt{p'}|H\rangle_S|b'\rangle_R]~.\\
\end{aligned}
\end{equation}

Combining above two Eqs., the unitary operator $U_{SR}$ can be written as:
\begin{equation}
\begin{aligned}
\begin{split}
U_{SR} =& \left[{\sqrt{1-p'}|V\rangle_S|b\rangle_R+\sqrt{p'}|H\rangle_S |a\rangle_R}\right]~{_S \langle H| _R\langle a|}~~ + \\
&  \left[ \sqrt{1-p}|H\rangle_S|b\rangle_R+ 
  \sqrt{p}\left(\sqrt{1-p'}|V\rangle_S|a'\rangle_R +  
\sqrt{p'}|H\rangle_S|b'\rangle_R~\right)\right]~ _S\langle V| _R\langle a|.
\end{split}
\end{aligned}
\end{equation}

The Kraus operators which govern the evolution of the system are obtained by tracing out the degrees of freedom of the reservoir as follows:
\begin{equation}
K_\mu=_R\langle\mu|U_{SR}|a\rangle_R~~;~\mu=a,a',b,b'.
\end{equation}
The Kraus operators, thus, obtained are
\begin{subequations}
\begin{align}
\begin{split}
K_1=&~_R\langle a|U_{SR}|a\rangle_R=\left(\begin{array}{cc}
\sqrt{p'} & 0\\
0 & 0
\end{array}\right),
\end{split}\\
\begin{split}
K_2=&~_R\langle b|U_{SR}|a\rangle_R=\left(\begin{array}{cc}
0 & \sqrt{1-p}\\
\sqrt{1-p} & 0
\end{array}\right),
\end{split}\\
\begin{split}
K_3=&~_R\langle a'|U_{SR}|a\rangle_R=\left(\begin{array}{cc}
0 & 0\\
0 & \sqrt{p(1-p')}
\end{array}\right),
\end{split}\\
\begin{split}
K_4=&~_R\langle b'|U_{SR}|a\rangle_R=\left(\begin{array}{cc}
0 & \sqrt{p p'}\\
0 & 0
\end{array}\right).
\end{split}
\end{align}
\end{subequations}

The Kraus operators for two-qubit system is obtained by
\begin{equation}
K_{ij}=K_i\otimes K_j~;~i,j=1,2,3,4.
\end{equation}
These Kraus operators govern the evolution of the system from the initial state $\rho(0)$ and give the final state $\rho(p,p')$. 
\begin{equation}
\rho(p,p')=\sum_{i,j=1}^4 K_{ij}\rho(0,0)K_{ij}^\dag~.
\end{equation}

\subsection{Analytical expressions for X-state with non-zero entries in the other diagonal terms}   \label{S4.8c}

As discussed in Section~(\ref{S4.5}), we again use the negativity criterion for the partial transposed density matrix to determine the occurrence of ESD. By following, as in [\ref{4.35}], the evolution of the parameters in Eq.~(\ref{E4.02}), and double NOT switching $a$ and $d$, $b$ and $c$, and swapping the off-diagonal terms $z$ and $z^*$, whereas a single NOT at one end switches $a$ and $c$, $b$ and $d$, and moves $z$ and $z^*$ inward along the anti-diagonal, we again obtain analytical expressions for the various $p$ (or equivalently, $\gamma$ and $t$) of interest. Consider the initial mixed entangled state with the density matrix given in a form more general than (\ref{E4.05}) given below.
\begin{equation}
\rho_2= \left( \begin{array}{cccc} 
a & 0 & 0 & z\\
0 & b & 0 & 0 \\
0 & 0 & c & 0 \\
z^* & 0 & 0  & d \end{array} \right),
\end{equation}

where $a+b+c+d=1$. As per the convention for ground and excited states in the reference [\ref{4.35}], we choose $`a'$ as the population of both qubits being in excited states and $`d'$ being the population of both qubits in the ground states unlike the (reversed) convention in the main body of this paper.

In the presence of ADC, the condition for ESD is given by
\begin{equation}
p_0=\frac{-b-c+[(b-c)^2+4|z|^2]^{1/2}}{2a}.
\end{equation}

For manipulation of ESD using NOT operation on both the qubits, the condition for hastening ESD is given by
\begin{equation}
p_A=\frac{a-d}{1+a-d}.
\end{equation}

The condition for avoidance of ESD is given by
\begin{equation}
p_B =1-\frac{2a+b+c-[(b-c)^2+4|z|^2)]^{1/2}}{2[(a+b)(a+c)-|z|^2]}.
\end{equation}

For manipulation of ESD using NOT operation on only one of the qubits, the condition for hastening ESD is given by
\begin{equation}
p_A=1-\frac{(c+a)[(c+a)(1-p_0)-1]}{(a+b)\{(a+b)-[(b-c)^2+4|z|^2]^{1/2} \}-a}.
\end{equation}
 
 and the condition for avoidance of ESD is given by
 \begin{equation}
p_B=\frac{|z|^2-c}{|z|^2+a}.
\end{equation}

\subsection{Analytical expressions for the X-state in Reference [35]}  \label{S4.8d}

Consider the initial mixed entangled state with the form of density matrix as in reference [\ref{4.35}],
\begin{equation}
\rho_1= \left( \begin{array}{cccc} 
a & 0 & 0 & 0 \\
0 & b & z & 0 \\
0 & z^* & c & 0 \\
0 & 0 & 0  & d \end{array} \right),
\end{equation}

for simplicity, the 1/3 factor in [\ref{4.35}] has been absorbed into the density matrix elements.

In the presence of ADC, the condition for ESD is given by
\begin{equation}
p_0=\frac{-b-c+[(b+c+2a)^2-4(a-|z|^2)]^{1/2}}{2a},
\end{equation}
where $p=1-\gamma^2=1-\exp(-\Gamma t)$, $\Gamma$ is the spontaneous decay rate of the two level atomic qubit as introduced in [\ref{4.35}].

For manipulation of ESD using NOT operation on both the qubits, the condition for hastening ESD is given by
\begin{equation}
p_A=\frac{a-d}{1+a-d}.
\end{equation}

The condition for avoidance of ESD is given by
\begin{equation}
p_B =\frac{2(a-|z|^2)-(2a+b+c)+[(b+c+2a)^2-4(a-|z|^2)]^{1/2}}{2(a-|z|^2)}.
\end{equation}

For manipulation of ESD using NOT operation on only one of the qubits, the condition for hastening ESD is given by
\begin{equation}
p_A=\frac{(c+a)[2a(1-p_0)-(c+a)(1-p_0)+c+d]-a}{(c+a)[2a(1-p_0)-(b+a)]-a}.
\end{equation}
 
The condition for avoidance of ESD is given by
\begin{equation}
p_B =1-\frac{a+c}{(a+b)(a+c)+|z|^2}.
\end{equation}
 
For hastening, delay and avoidance to exist in a physical region, the corresponding parameters must satisfy the condition $0<p_B,~p_A<p_0$. As an example, the choice $a=0.4, b=c=0.2, z=0.25$ gives $ p_0=0.125$, $p_A=0.1667$, and an unphysical negative value of $p_B$. This means that neither hastening nor averting ESD is possible, only delaying it by applying NOT between $0$ and $p_0$.

The published version of the research work reported in this chapter can be found in Ref. [\ref{4.45}].

\section{References}
\begin{enumerate}
\item \textit{E. Schr$\ddot{o}$dinger, ``Naturwissenschaften", \href{https://doi.org/10.1007/BF01491891}{\textbf{23}, 807 (1935)}}. \label{4.01}
\item \textit{A. Einstein, B. Podolsky, N. Rosen, ``Can the Quantum-mechanical description of a physical reality be considered complete?" \href{https://doi.org/10.1103/PhysRev.47.777}{Phys. Rev. \textbf{47}, 777-780 (1935).}} \label{4.02}
\item \textit{ J. S. Bell, ``On the Einstein Podolsky Rosen paradox," Physics (Long Island City, N.Y.) \href{https://doi.org/10.1103/PhysicsPhysiqueFizika.1.195}{\textbf{1}, 195-200 (1964).}} \label{4.03}
\item \textit{S. J. Freedman and J. F. Clauser, ``Experimental Test of Local Hidden-Variable Theories," \href{https://doi.org/10.1103/PhysRevLett.28.938}{Phys. Rev. Lett. \textbf{28}, 938 (1972).}}\label{4.04}
\item \textit{T. Yu  and J. H. Eberly , ``Finite-Time Disentanglement Via Spontaneous Emission," \href{https://doi.org/10.1103/PhysRevLett.93.140404}{Phys. Rev. Lett. \textbf{93}, 140404 (2004).}} \label{4.05}
\item \textit{T. Yu  and J. H. Eberly , ``Quantum Open System Theory: Bipartite Aspects," \href{https://doi.org/10.1103/PhysRevLett.97.140403}{Phys.Rev. Lett. \textbf{97}, 140403 (2006).}} \label{4.06}
\item \textit{T. Yu  and J. H. Eberly , ``Sudden Death of Entanglement," \href{https://doi.org/10.1126/science.1167343}{Science \textbf{ 323} , 598 (2009).}} \label{4.07}
\item \textit{J. Laurat, K. S. Choi, H. Deng, C. W. Chou, H. J. Kimble, ``Heralded Entanglement between Atomic Ensembles: Preparation, Decoherence, and Scaling," \href{https://doi.org/10.1103/PhysRevLett.99.180504}{Phys. Rev. Lett. \textbf{99}, 180504 (2007).}} \label{4.08}
\item \textit{M. P. Almeida, F. de Melo, M. Hor-Meyll, A. Salles, S. P. Walborn, P. H. Souto Ribeiro, L. Davidovich,``Environment-Induced Sudden Death of Entanglement," \href{https://doi.org/10.1126/science.1139892}{Science \textbf{316}, 555(2007).}} \label{4.09}
\item \textit{Jin-Shi Xu, Chuan-Feng Li, Ming Gong, Xu-Bo Zou, Cheng-Hao Shi, Geng Chen, and Guang-Can Guo, ``Experimental demonstration of photonic entanglement collapse and revival", \href{https://doi.org/10.1103/PhysRevLett.104.100502}{Phys. Rev. Lett.\textbf{ 104}, 100502 (2010).}} \label{4.10}
\item \textit{R. Horodecki, P. Horodecki, M. Horodecki, K. Horodecki, ``Quantum entanglement",  \href{https://doi.org/10.1103/RevModPhys.81.865}{Rev. Mod. Phys.\textbf{ 81}, 865 (2009).}} \label{4.11}
\item \textit{C. H. Bennett and D. P. DiVincenzo,\href{https://doi.org/10.1038/35005001}{``Quantum information and computation", Nature \textbf{404}, 247 (2000).}} \label{4.12}
\item \textit{M. A. Nielsen  and I. L. Chuang, ``Quantum Computation and Quantum Information" (Cambridge University Press, Cambridge, 2000).} \label{4.13}
\item \textit{J.W. Pan, S. Gasparoni, R. Ursin, G. Weihs, \& A. Zeilinger,  ``Experimental entanglement purification of arbitrary unknown states", \href{https://doi.org/10.1038/nature01623}{Nature \textbf{423}, 417 (2003).}} \label{4.14}
\item\textit{P. G. Kwiat, S. Barraza-Lopez, A. Stefanov, and N. Gisin, ``Experimental entanglement distillation and ‘hidden’ non-locality", \href{https://doi.org/10.1038/35059017}{Nature \textbf{ 409}, 1014 (2001).}} \label{4.15}
\item \textit{D. A. Lidar, I. L. Chuang, and K. B. Whaley, ``Decoherence-free subspaces for quantum computation", \href{https://doi.org/10.1103/PhysRevLett.81.2594}{Phys. Rev. Lett.\textbf{ 81}, 2594 (1998)}}. \label{4.16}
\item \textit{P. G. Kwiat, A. J. Berglund, J. B. Altepeter, and A. G. White, ``Experimental verification of decoherence-free subspaces",  \href{https://doi.org/10.1126/science.290.5491.498}{Science\textbf{ 290}, 498 (2000)}}. \label{4.17}
\item \textit{D. Kielpinski, V. Meyer, M. A. Rowe, C. A. Sackett, W. M. Itano, C. Monroe, and D. J. Wineland, ``A decoherence-free quantum memory using trapped ions", \href{https://doi.org/10.1126/science.1057357}{Science\textbf{ 291}, 1013 (2001)}}. \label{4.18}
\item \textit{L. Viola, E. M. Fortunato, M. A. Pravia, E. Knill, R. Laflamme, and D. G. Cory, ``Experimental realization of noiseless subsystems for quantum information processing",  \href{https://doi.org/10.1126/science.1064460}{Science \textbf{293}, 2059 (2001)}}. \label{4.19}
\item \textit{P.W. Shor, ``Scheme for reducing decoherence in quantum computer memory",  \href{https://doi.org/10.1103/PhysRevA.52.R2493}{Phys. Rev. A \textbf{52}, R2493  (1995)}}. \label{4.20}
\item \textit{A.M. Steane, ``Error Correcting Codes in Quantum Theory",  \href{}{Phys. Rev. Lett. \textbf{77} , 793 (1996)}}. \label{4.21}  
\item \textit{Lorenza Viola, Emanuel Knill and Seth Lloyd, ``Dynamical Decoupling of Open Quantum Systems",  \href{https://doi.org/10.1103/PhysRevLett.82.2417}{Phys. Rev. Lett. \textbf{82}, 2417 (1999)}}. \label{4.22} 
\item \textit{Michael J. Biercuk, Hermann Uys, Aaron P. VanDevender, Nobuyasu Shiga, Wayne M. Itano \& John J. Bollinger, ``Optimized dynamical decoupling in a model quantum memory", \href{https://doi.org/10.1038/nature07951}{Nature \textbf{458}, 996 (2009)}}. \label{4.23} 
\item\textit{Jiangfeng Du, Xing Rong, Nan Zhao, Ya Wang, Jiahui Yang \& R. B. Liu, ``Preserving electron spin coherence in solids by optimal dynamical decoupling", \href{https://doi.org/:10.1038/nature08470}{Nature \textbf{461}, 1265 (2009)}}. \label{4.24} 
\item \textit{P. Facchi, D. A. Lidar, and S. Pascazio, ``Unification of dynamical decoupling and the quantum Zeno effect", \href{https://doi.org/10.1103/PhysRevA.69.032314}{Phys. Rev. A \textbf{69}, 032314 (2004)}}. \label{4.25} 
\item \textit{ S. Maniscalco, F. Francica, R. L. Zaffino, N. L. Gullo, and F. Plastina, ``Protecting entanglement via the quantum Zeno effect", \href{https://doi.org/10.1103/PhysRevLett.100.090503}{Phys. Rev. Lett. \textbf{100}, 090503 (2008)}}. \label{4.26} 
\item \textit{J. G. Oliveira, Jr., R. Rossi, Jr., and M. C. Nemes, ``Protecting, enhancing, and reviving entanglement", \href{https://doi.org/10.1103/PhysRevA.78.044301}{Phys. Rev. A \textbf{78}, 044301 (2008)}}. \label{4.27} 
\item \textit{Y.S. Kim, Y.W. Cho, Y.-S. Ra, and Y.-H. Kim, ``Reversing the weak quantum measurement for a photonic qubit", \href{https://doi.org/10.1364/OE.17.011978}{Opt. Express\textbf{ 17}, 11978 (2009)}}. \label{4.28} 
\item \textit{A. N. Korotkov and K. Keane, ``Decoherence suppression by quantum measurement reversal", \href{https://doi.org/10.1103/PhysRevA.81.040103}{Phys. Rev. A \textbf{81},040103(R) (2010)}}. \label{4.29} 
\item \textit{ J.C. Lee, Y.C. Jeong, Y.S. Kim, and Y.H. Kim, ``Experimental demonstration of decoherence suppression via quantum measurement reversal", \href{https://doi.org/10.1364/OE.19.016309}{Opt. Express \textbf{19}, 16309 (2011)}}. \label{4.30} 
\item \textit{Q. Sun, M. Al-Amri, L. Davidovich, and M. S. Zubairy, ``Reversing entanglement change by a weak measurement", \href{https://doi.org/10.1103/PhysRevA.82.052323}{Phys. Rev. A \textbf{ 82}, 052323 (2010)}}. \label{4.31} 
\item \textit{Y.S. Kim, J.C. Lee, O. Kwon, and Y.-H. Kim, ``Protecting entanglement from decoherence using weak measurement and quantum measurement reversal", \href{https://doi.org/10.1038/NPHYS2178}{Nature Phys. \textbf{8}, 117 (2012)}}. \label{4.32} 
\item \textit{H.T. Lim, J.C. Lee, K.H. Hong, and Y.H. Kim, ``Avoiding entanglement sudden death using single-qubit quantum measurement reversal", \href{https://doi.org/10.1364/OE.22.019055}{Opt. Express \textbf{22}, 19055 (2014)}}. \label{4.33} 
\item \textit{Jong-Chan Lee, Hyang-Tag Lim, Kang-Hee Hong, Youn-Chang Jeong, M.S. Kim \& Yoon-Ho Kim, ``Experimental demonstration of delayed-choice decoherence suppression", \href{https://doi.org/10.1038/ncomms5522}{Nature Communications \textbf{5}, 4522 (2014)}}. \label{4.34} 
\item \textit{A.R.P. Rau, M. Ali, and G. Alber, ``Hastening, delaying or averting sudden death of quantum entanglement", \href{https://doi.org/10.1209/0295-5075/82/40002}{EPL \textbf{82}, 40002 (2008)}}. \label{4.35} 
\item \textit{A. Peres , ``Separability Criterion for Density Matrices",  \href{https://doi.org/10.1103/PhysRevLett.77.1413}{Phys. Rev. Lett. \textbf{77}, 1413 (1996)}}. \label{4.36} 
\item \textit{M.Horodecki, P. Horodecki  and R. Horodecki , ``Separability of mixed states: necessary and sufficient conditions", \href{https://doi.org/10.1016/S0375-9601(96)00706-2}{Phys. Lett. A \textbf{223}, 1 (1996)}}. \label{4.37} 
\item \textit{P.G. Kwiat, Edo Waks, Andrew G. White, Ian Appelbaum, and Philippe H. Eberhard, ``Ultra bright source of polarization entangled photons", \href{https://doi.org/10.1103/PhysRevA.60.R773}{Phys. Rev. A \textbf{60}, 773-776(R) (1999)}}. \label{4.38} 
\item \textit{D. F. V. James, P. G. Kwiat, W. J. Munro, A. G. White, ``Measurement of qubits", \href{https://doi.org/10.1103/PhysRevA.64.052312}{Phys. Rev. A \textbf{64}, 052312 (2001)}}. \label{4.39} 
\item \textit{Akio Fujiwara, ``Estimation of a generalized amplitude-damping channel", \href{https://doi.org/10.1103/PhysRevA.70.012317}{Phys. Rev. A \textbf{70}, 012317 (2004)}}. \label{4.40} 
\item \textit{Asma Al-Qasimi and Daniel F. V. James, ``Sudden death of entanglement at finite temperature", \href{https://doi.org/10.1103/PhysRevA.77.012117}{Phys. Rev. A \textbf{77},012117 (2008)}}.  \label{4.41} 
\item \textit{M. Ali, A. R. P. Rau, and G. Alber, ``Manipulating entanglement sudden death of two-qubit X-states in zero- and finite-temperature reservoirs",  \href{https://doi.org/10.1088/0953-4075/42/2/025501}{J. Phys. B: At. Mol. Opt. Phys.\textbf{ 42}, 025501(8)(2009)}}.  \label{4.42} 
\item\textit{Mahmood Irtiza Hussain, Rabia Tahira and Manzoor Ikram, ``Manipulating the Sudden Death of Entanglement in Two-qubit Atomic Systems", \href{10.3938/jkps.59.2901}{Journal of the Korean Physical Society, \textbf{59}, 2901-2904(2011)}}. \label{4.43} 
\item \textit{R. Srikanth and Subhashish Banerjee, ``Squeezed generalized amplitude damping channel", \href{https://doi.org/10.1103/PhysRevA.77.012318}{Phys. Rev. A \textbf{77}, 012318 (2008)}}. \label{4.44}  
\item \textit{Ashutosh Singh, Siva Pradyumna, A. R. P. Rau, and Urbasi Sinha, "Manipulation of entanglement sudden death in an all-optical setup,"  \href{https://doi.org/10.1364/JOSAB.34.000681}{ J. Opt. Soc. Am. B \textbf{34}, 681-690 (2017)}}.\label{4.45}  
\end{enumerate}

%% file: chapter5.tex
\setcounter{equation}{0}
\chapter{Experimental demonstration of ESD and its manipulation}

\section{Introduction}

In this chapter, we will present a detailed discussion on the experimental methods involved in setting up the Entanglement Sudden Death (ESD) experiment from scratch, issues encountered in the process, and their resolution. Then, we will present experimental results on the demonstration of ESD in the presence of a simulated Amplitude Damping Channel (ADC) in an all-optical experimental setup. For completeness, we will also demonstrate the Asymptotic Decay of Entanglement (ADE) in the presence of an ADC for a different initial state. This will be followed by a discussion on our attempts in setting up the ESD-manipulation (ESDM) experiment using local unitary operations (to be specific, NOT operation: Pauli $\sigma_x$ operator) on both the qubits during the process of decoherence as proposed in the previous chapter. It is worth noting that the action of local unitary operations on individual subsystems cannot change the amount of entanglement in the system but subsequent dynamics can be altered. We will outline various constraints in the ESDM setup that need to be simultaneously satisfied for the ESDM experiment to work. Then we will discuss our efforts towards setting up the manipulation experiment, current status of the experiment, and the way forward. We will conclude this chapter with a brief discussion on the possible impact of ESD-manipulation on the other physical systems and Quantum Information Processing (QIP) tasks. 

One important element of this experiment is the Sagnac Interferometer in the displaced configuration, known as Displaced Sagnac Interferometer (DSI), which consists of one PBS and three mirrors. For any input polarization state, DSI splits the H- and V-polarization amplitudes in orthogonal directions and spatially segregates their path inside the interfereometer and thus gives us separate access to these polarization components for controlled manipulation. Since, both the clockwise as well as counter-clockwise propagating beams pass through the same set of optics (PBS and mirrors), it is fairly stable against vibrations as compared to the other interferometers such as  Michelson or Mach-Zhender interferometers. Inherent stability of the SI is one of the reasons why it is a preferred choice in many interferometric applications such as in preparation of SPDC based polarization entangled photon sources, etc.

The ESD experiment involves setting up of two DSIs using an alignment laser beam and then fine alignment when the entangled photon source is turned on to optimize the coincidence in different bases. Such an optimization is required to ensure that the input state to DSI is reconstructed at the output port with a high-fidelity for zero-degree setting of the damping channel parameters. During the experiment, it was found that one of the DSIs had some instability as a result of which we observed periodic fluctuations and drift in the single counts of one side and thereby coincidence counts in a given projection bases. Such instabilities were quantified by correlation-noise based analyses using Pearson correlation coefficient (PCC). It was conjectured that such instability could arise due to two reasons: (i) air currents flowing in the lab due to air-conditioning and fan-filter units, and/or (ii) expansion and contraction of the Holmarc platform mount springs/knobs used for mounting PBS in the corresponding DSI due to temperature change of $1-2^\circ \text{c}$ from morning to evening. To isolate the interferometers from air currents, we covered the two DSIs using cardboard and the poor-quality beam splitter platform mount from Holmarc was replaced by a 2-inch beam splitter mount from Thor Labs. These solutions resolved the coincidence drift issue to a great extent and then we performed the ESD and ADE experiments which was followed by ESDM experiment.

\section{Setting up the ESD experiment}

\subsection{ESD with one HWP vs. two HWPs}

The ESD experiment using two HWPs as proposed in the previous chapter is very difficult to set up as it is very challenging and cumbersome to ensure the coherent recombination of modes $|a\rangle$ and $|a'\rangle$ as in Fig.~\textcolor{darkblue}{4.01} due to macroscopic path differences between these spatial modes. Here, a mathematical trick comes to our rescue. From Eq.~(\textcolor{darkblue}{4.18}), it is evident that the effective end of entanglement due to combined evolution through two HWPs is given by $p_\text{end}$. The $p_\text{end}$ involves a multiplication of survival probabilities due to individual HWPs to give
\begin{equation}
1-p_\text{end}=(1-p)(1-p'_0)~~ \text{with}~~ p_\text{end} =|v|/x.
\label{eq5.01}
\end{equation}
depending only on the initial state parameters of the density matrix.

The end of entanglement due to two HWPs mimicking the ADC oriented at $\theta=\arcsin(\sqrt{p})/2$  and $\theta_0'=\arcsin(\sqrt{p_0'})/2$, respectively, as given by Eq.~(\ref{eq5.01}) is same as that due to a single HWP oriented at $\theta_\text{end}=\arcsin(\sqrt{p_\text{end}})/2$ in the ESD setup in Ref.~[\ref{5.02}]. We will exploit the equivalence due to Eq.~(\ref{eq5.01}) and  set up the ESD experiment for polarization entangled photonic qubits using DSI as in Ref.~[\ref{5.01}]. These results will set the reference point for the ESD manipulation experiment. These results will be then compared with the ESD manipulation experiment and then phenomenon of hastening, delay, and avoidance of ESD will be demonstrated for an initially entangled state. The difference between what we call $p'$ and $p$ is that factor $(1-p)$, that is, any $p'$ such as $p_{end}$ can be  converted into the unprimed $p_{end}$ through an expression such as Eq.~(\textcolor{darkblue}{4.18}).  What is multiplicative is the survival $(1-p)$, not $p$, itself, so that to  make the passage you take $(1-p')$ and multiply by $(1-p)$ or $(1-p_n)$ and  that gives the corresponding $(1-p)$ without the prime.

\subsection{Alignment of DSIs using He-Ne Laser and coincidence optimization} 

\begin{figure} [H] 
\centering
\includegraphics[clip, trim=3.4cm 9cm 2.8cm 2.5cm, width=0.8\textwidth]{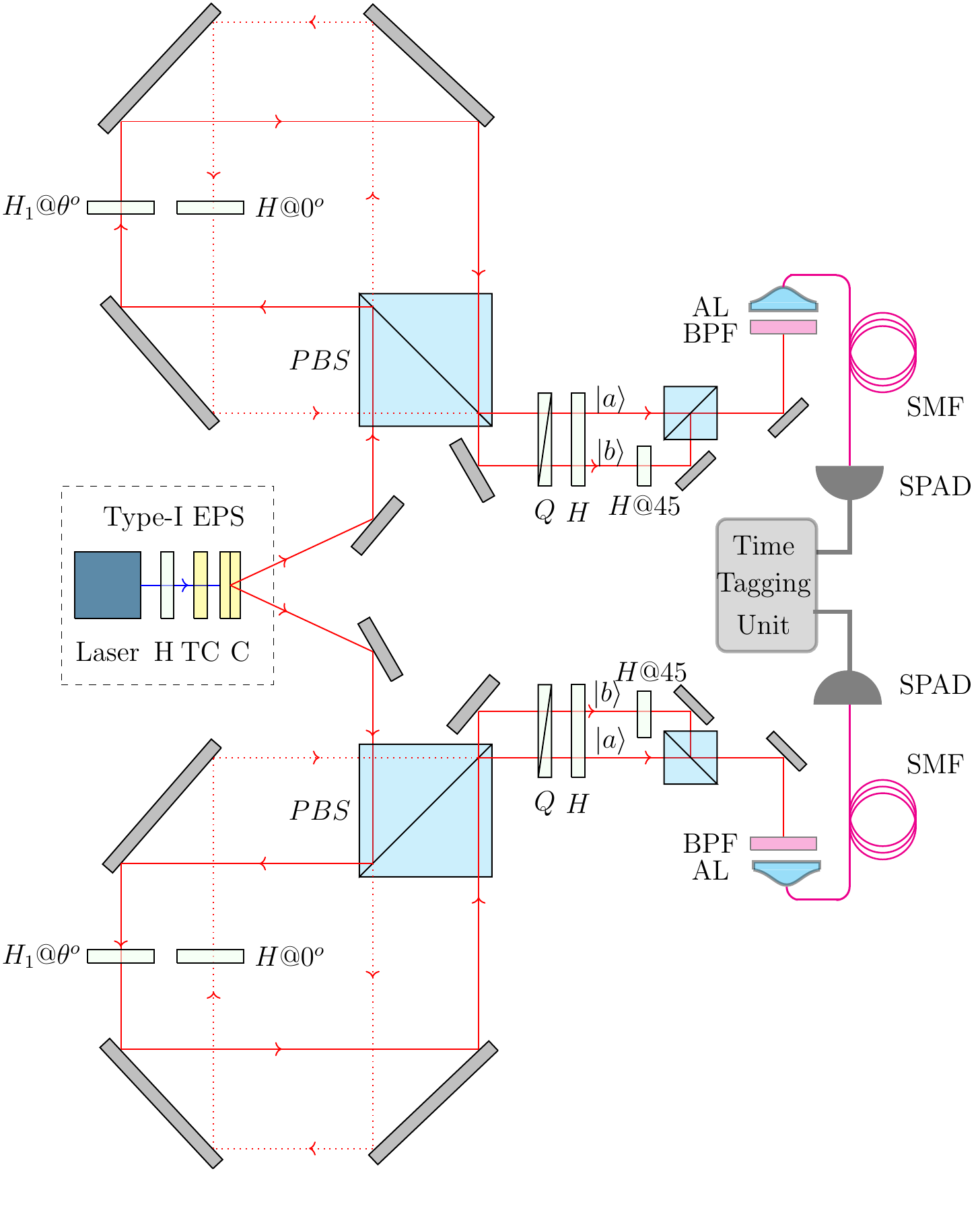}
\caption{\textit{Experimental setup for demonstrating ESD in the photonic system [\ref{5.01}]. Polarization entangled photons are sent to two DSI where a half-waveplate ($HWP1~\text{at}~\theta$) acts on $|V\rangle$ polarization only. Upon tracing out the spatial modes; mode $|a\rangle$ (transmitted, top) and $|b\rangle$(reflected, bottom), of the interferometer at PBS, the HWP action mimics ADC.}\label{fig5.01}} 
\end{figure}

Schematic of the experimental setup that we have used for demonstrating ESD in an all-optical setup is shown in Fig.~\ref{fig5.01}.

To give a feel for the experiment as well as a visual aid to the interested readers, a picture of the experimental setup on the optical table is shown in Fig.~\ref{fig5.02} below. The entire experimental setup including entangled photon source and displaced Sagnac interferometer for ESD experiment is built on a $6 \times 4$ square-feet optical table. Sagnac interferometer paths are approximately 10 inch long.

\begin{figure} [H] 
\centering
\includegraphics[width=\textwidth]{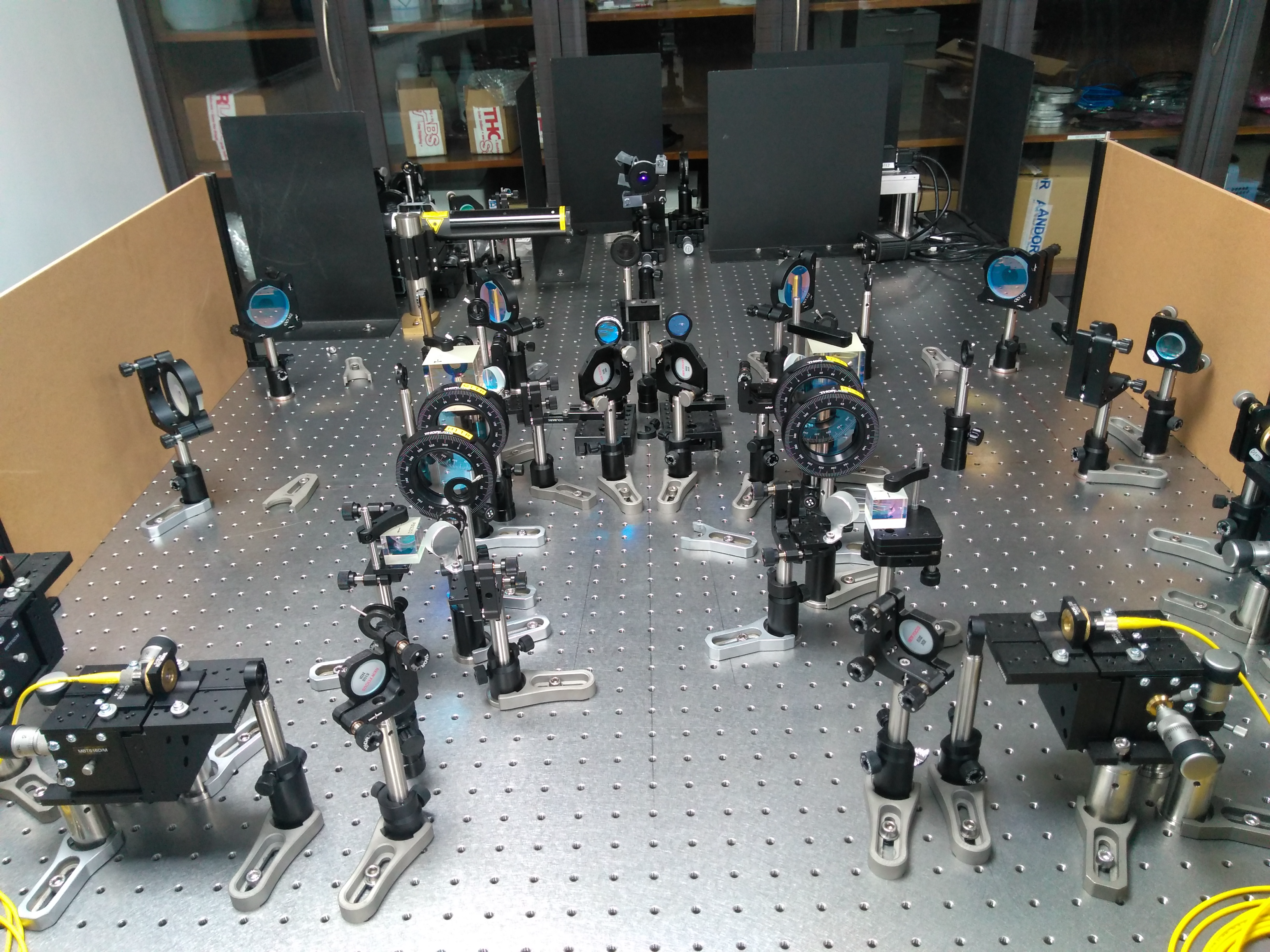}
\caption{\textit{A piture of the experimental setup for demonstrating ESD in the photonic system captured from our laboratory.}\label{fig5.02}} 
\end{figure}

Starting from the entangled photon source setup, one DSI is set up at a time using the alignment laser. The other aligned side serves as a reference for the DSI alignment to get the single photon coupling and coincidence optimization when source is turned on. When a diagonally polarized laser beam is incident on the PBS of DSI, half of the incident light gets transmitted and remaining half gets reflected. These two light beams travel the interferometer in counter-clockwise and clockwise directions, respectively, get reflected from the three mirrors and meet at a different point on the PBS. This is where coherent recombination of the two beams (polarization amplitudes in the cases of single/entangled photons) takes place and ideally entire light beam should get transmitted to spatial mode $|a\rangle$. Experimentally, however, one would observe a very small leakage to spatial mode $|b\rangle$ due to imperfections in optical components (such as mirror reflectivity or PBS extinction ratio) or due to error in aligning the wave-plates at a given angle. If DSI is balanced and perfectly phase stable, then the polarization state of light at the output port remains same as that at the input port. 

The relative path delay between clockwise and counter-clockwise paths, which are supposed to be coherently recombined at the PBS of displaced Sagnac interferometer, should be ideally zero. However, due to manufacturing limitations on the surface flatness of 2" optics (mirrors, polarizing beam splitters, and waveplates), there will be a residual path difference between the two arms. Since upon performing tomography (see Sec. 5.2.3 and 5.3.1, for instance), we got better than 95\% fidelity of the entanglement for zero settings of ADC HWPs (for ESD experiment) with the interferometer. Thus, we infer that the path difference between the two arms would be significantly lower than the coherence length of the down-converted photons: $l_c=\lambda^2/\Delta \lambda=67$ microns (for a 10 nm band pass filter used for spectral filtering at 810 nm). 

We used an alignment laser (He-Ne laser, 632.8 nm), apertures and beam profiler for setting up the DSI for the ESD experiment. Overlap of the He-Ne beam traversing the DSI clockwise and counter-clockwise directions was checked at the output port (mode $|a\rangle$) at different positions and it was ensured that two beams remain on-top of each other. This ensures the overlap of the  $\textbf{K}$ vectors of the two beams. In collinear configuration of the DSI, there will be a single Gaussian spot at the output port as shown in Fig.~\ref{fig5.03} below.

\begin{figure}[!htb]
\begin{center}
\includegraphics[width=0.9\textwidth]{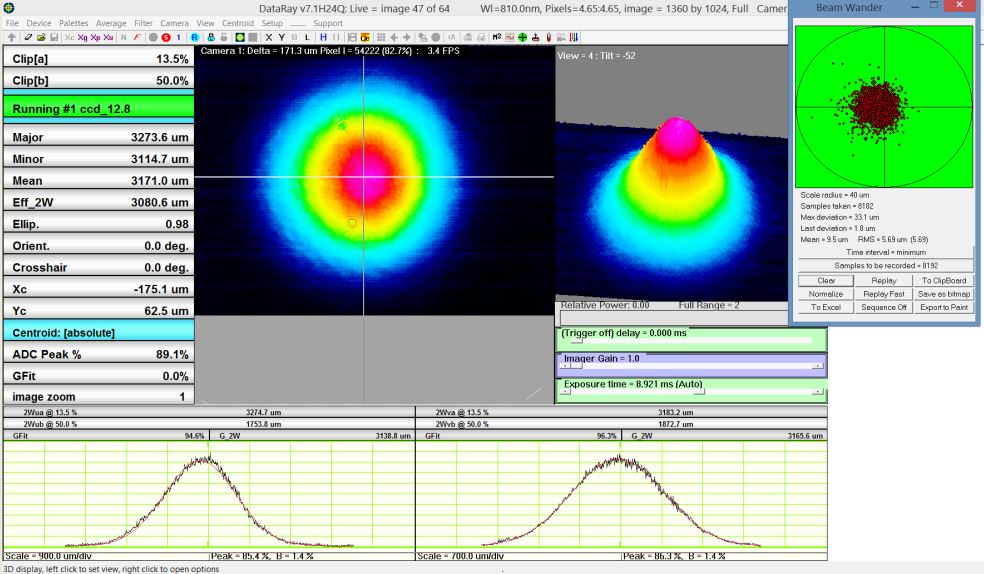}
\caption{\textit{Overlap of the clockwise and counter-clockwise propagating Gaussian beams at the output port (spatial mode $|a\rangle$) of the DSI in collinear configuration captured by DataRay beam profiler.}\label{fig5.03}}
\end{center}
\end{figure}

When a polarizer with transmission axis at $45^\circ$ was placed at the output port in spatial mode $|a\rangle$ of the DSI to see the interference due to diagonal components of the two beams in the non-collinear configuration (by slightly misaligning one of the mirrors so that $k$-vectors of transmitted and reflected components exit at a small angle), we observe interference fringes as shown in Fig.~\ref{fig5.04} below. In the end, DSI is brought back to the collinear configuration.

\begin{figure}[H]
\begin{center}
\includegraphics[width=0.9\textwidth]{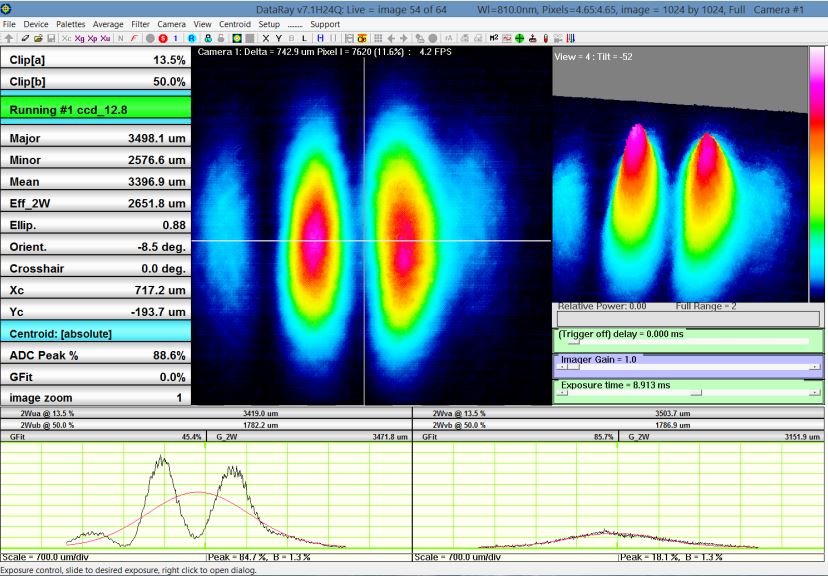}
\caption{\textit{Interference due to clockwise and counter-clockwise propagating Gaussian beams on a polarizer at $45^\circ$ at the output port $|a\rangle$ of the DSI in non-collinear configuration captured by DataRay beam profiler.}\label{fig5.04}}
\end{center}
\end{figure}

To introduce the spatial mode $|b\rangle$, HWP in the reflected arm of DSI is set to $45^\circ$ so that the polarization state of light after passing through this HWP becomes H-polarized and thereby gets transmitted to spatial mode $|b\rangle$. This helps us in aligning the mirrors in mode $|b\rangle$.  The light beam in mode $|b\rangle$ is incoherently recombined with that in mode $|a\rangle$ (as path difference between mode $|a\rangle$ and $|b\rangle$ is much larger than the coherence length of the signal) using a HWP at $45^\circ$ onto PBS and gets reflected towards the detector. The HWP at $45^\circ$ in mode $|b\rangle$ ensures that the QST settings for both modes remain the same.

\subsection{Coincidence optimization and QST}

After aligning one of the DSIs using He-Ne laser, ADC HWP ($\text{H}_1~\text{at}~\theta^\circ$) as well as optical path compensating HWP ($\text{H}~\text{at}~0^\circ$) in the DSI were placed with fast axes making zero-degree with vertical. At the output port of the DSI, QST setup was already placed. On the other side, only QST setup was placed. In such case, when entangled photon source is turned on, one of the photons (say, signal) passes through QST setup directly whereas other one (say, idler) passes through the DSI. In the DSI, $|H\rangle$ polarization amplitude is transmitted and traverses the interferometer in counter-clockwise direction, and $|V\rangle$ polarization amplitude is reflected and traverses the interferometer in clockwise direction. Both of these polarization amplitudes then coherently recombine at a different point on the PBS and get transmitted to mode $|a\rangle$. Thus, only mode $|a\rangle$ is populated when ADC HWPs are set to zero-degree.

Coincidence was optimized for $|HH\rangle\langle HH|$ and $|VV\rangle\langle VV|$ projections to maximize and make them equal, and at the same time for  $|HV\rangle\langle HV|$ and $|VH\rangle\langle VH|$ projections coincidence should be close to zero for a maximally entangled state as the input. Next, ADC HWP in the reflected arm of the DSI was set to $45^\circ$ and coincidence was optimized from mode $|b\rangle$ for $|HH\rangle\langle HH|$ projection, preferably by using mirrors in the mode $|b\rangle$ alone. Ideally, sum of the coincidences from mode $|a\rangle$ for $|HH\rangle\langle HH|$ and $|VV\rangle\langle VV|$ projections for ADC HWPs at zero-degree should be equal to the coincidence from mode $|a\rangle$ and $|b\rangle$ for $|HH\rangle\langle HH|$ projection with ADC HWP at $45^\circ$. Also, the coincidence contribution from mode $|b\rangle$ should be equal to that from mode $|a\rangle$ for $|HH\rangle\langle HH|$ projection when ADC HWPs are set to $45^\circ$. 

 In the next step, DSI was aligned on the other side as well using alignment laser followed by coincidence optimization with entangled photon source on, following exactly the same procedure as discussed above. 

Then QST was performed for pump HWP at $22.5^\circ$ and physical state (density matrix) was reconstructed using Maximum Likelihood Estimation (MLE). In the ideal case, we expect that the initial state incident at the input ports of the DSI should be reconstructed at the output ports. The real and imaginary parts of the density matrix obtained upon reconstructing the state, is shown in the Fig.~\ref{fig5.05} below.

\begin{figure}[H]
\begin{center}
\includegraphics[scale=0.33]{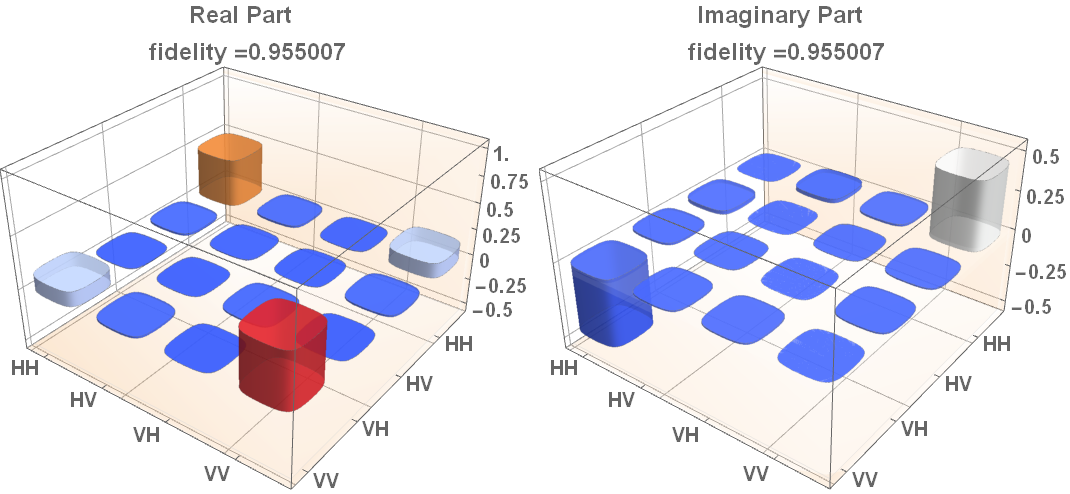}
\end{center}
\caption{\textit{The two-qubit entangled state reconstructed through QST with perfect temporal compensation and narrow spectral filtering using 810-10 nm band pass filters for pump HWP at $22.5^\circ$. Left (right) figure shows the real (imaginary) part of the density matrix. The reconstructed state has a fidelity of $95.50\%$ with the ideal state.}\label{fig5.05}}
\end{figure}

The reconstructed state had the following properties: 
\begin{itemize}
\item Concurrence of the ideal state = 1.
\item Concurrence of the reconstructed state = 0.9170.
\item Fidelity with the ideal state = 0.9550.
\item Purity of the state: 0.9179.
\end{itemize}

At this stage, many QSTs were performed for pump HWP at 22.5$^\circ$ with ADC HWPs at zero deg to see its reproducibility but state properties were found to be significantly different for different runs of the experiment. For a sanity check, we changed the pump HWP to $\approx 45.1^\circ$ so that down-converted state becomes nearly $|HH\rangle$ and reconstructed the state using QST. We found that the reconstructed state had following properties:
\begin{itemize}
\item  Fidelity with the ideal state: 99.68\%.
\item Concurrence of the ideal state: 0.00698 .
\item Concurrence of the reconstructed state: 0.00724 .
\item Purity of the state: 0.99999 .
\end{itemize}

The real and imaginary parts of the density matrix are shown in the Fig.~\ref{fig5.06} below.

\begin{figure}[H]
\begin{center}
\includegraphics[scale=0.33]{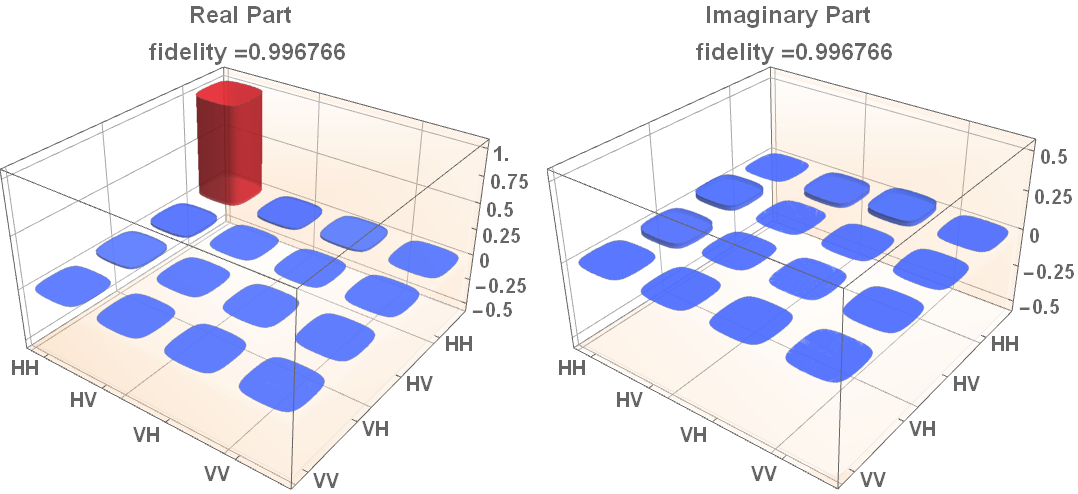}
\end{center}
\caption{\textit{The two-qubit product state $|HH\rangle$ reconstructed through QST with narrow spectral filtering using 810-10 nm band pass filters for pump HWP at $45^\circ$. Left (right) figure shows the real (imaginary) part of the density matrix. The reconstructed state has a fidelity of $99.67\%$ with the ideal state.}}\label{fig5.06}
\end{figure}

Thus, we were able to reconstruct the product state $|HH\rangle$ with a very high purity ($>99.99\%$) and fidelity ($99.67\%$) but for entangled states, purity and fidelity were significantly lower. It should be noted here that the product state $|HH\rangle$ (or $|VV\rangle$) does not involve any recombination of the counter propagating polarization amplitudes at PBS in the DSI. Any entangled state, however, would have both clockwise and counter-clockwise propagating polarization amplitudes in the DSI which should coherently recombine at the PBS in order to reconstruct the input state at the output ports. The relative phase $\phi_o$ of the reconstructed state would depend on the input state phase $\phi_i$ (where $|\psi_{i}\rangle=\alpha|HH\rangle+\beta \exp(i\phi_i)|VV\rangle$) as well as phase accrued due to propagation in the DSI due to the infinitesimal optical path difference in the two arms. We need to take a total 36-measurements to fully reconstruct the two-qubit state. If DSI had some phase instability, say optical path difference changes by a few nm in different QST runs, then net phase $\phi$ would be different for different projections. When density matrix is reconstructed from such projections, this would lead to a phase averaging, and thus net dephasing effect on the reconstructed state, leading to drop in coherence term and state purity. Furthermore, if the spatial overlap of clockwise and counter-clockwise propagating components is not perfect at PBS, then coherent recombination condition would not be satisfied which would again lead to drop in the state purity. All these conjectures lead us to investigate the (in)stability of the DSI in detail as discussed in the next section.
 
\subsection{Instability and drift in the DSI}
 
When the initial entangled state $|\psi_i\rangle=\left[ |HH\rangle+\exp(i\phi_i)|VV\rangle\right]/\sqrt{2}$ is inputted to the DSI, it showed short term oscillations and long term drift in the coincidence for a given projection setting. Projections which do not depend on the relative phase $\phi_i$ (such as $|HH\rangle\langle HH|,~|HV\rangle\langle HV|$, etc.) show short oscillation and long terms drift in the coincidence and singles of one side where singles/coincidence increased by nearly $30\%$ from morning to evening.  Whereas  projections dependent on the relative phase $\phi_i$ (such as $|RR\rangle\langle RR|,~|RL\rangle\langle RL|$, etc.) show a nearly periodic short term oscillations and a long term oscillation as shown in Figs.~\ref{fig5.07} and ~\ref{fig5.08} below.
 
\begin{figure}[H]
\begin{center}
\includegraphics[width=0.95\textwidth]{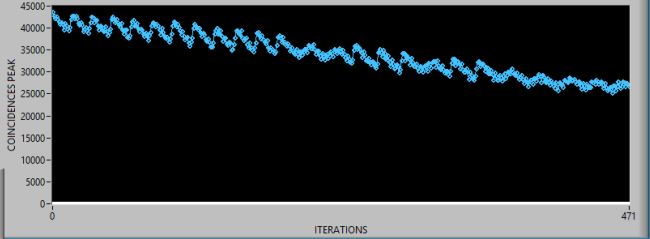}
\caption{\textit{Coincidence measured in $|VV\rangle\langle VV|$ bases vs. time (100s iterations) for a maximally entangled state input to DSIs. It clearly shows short time scale oscillation and long term drift.}\label{fig5.07}}
\end{center}
\end{figure}

\begin{figure}[!htb]
\begin{center}
\includegraphics[width=0.95\textwidth]{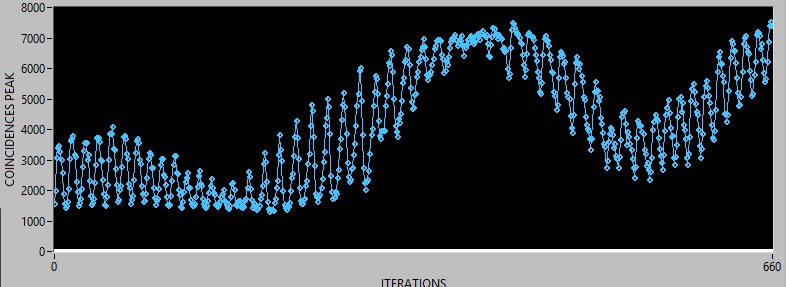}
\caption{\textit{Coincidence measured in $|LL\rangle\langle LL|$ bases vs. time (100s iteration) for maximally entangled state input to the DSIs. It shows both both short time scale as well as long time scale oscillations.}\label{fig5.08}}
\end{center}
\end{figure}

To understand this further, Pearson correlation coefficient (PCC) based analysis was carried out where singles, coincidence, temperature and laser power were recorded simultaneously. PCC between two variables X and Y is a measure of linear correlation between them and defined as follows:

\begin{equation}
P_{XY}=\frac{\text{cov}(X,Y)}{\sigma_X \sigma_Y}
\end{equation}
where $\text{cov}(X,Y)$ is the covariance, and $\sigma_X$ and $\sigma_Y$ are standard deviation of X and Y, respectively. 

Expressing covariance between X and Y as 
\begin{equation}
\text{cov}(X,Y)=E[(X-\bar{X})(Y-\bar{Y})],
\end{equation}

where E denotes expectation value, and $\bar{X},\bar{Y}$ denote mean of X and Y respectively. Expression for PCC can be rewritten as
 
\begin{equation}
P_{XY}=\frac{E[(X-\bar{X})(Y-\bar{Y})]}{\sigma_X \sigma_Y}
\end{equation}

\begin{figure}[H]
\begin{center}
\includegraphics[scale=0.8]{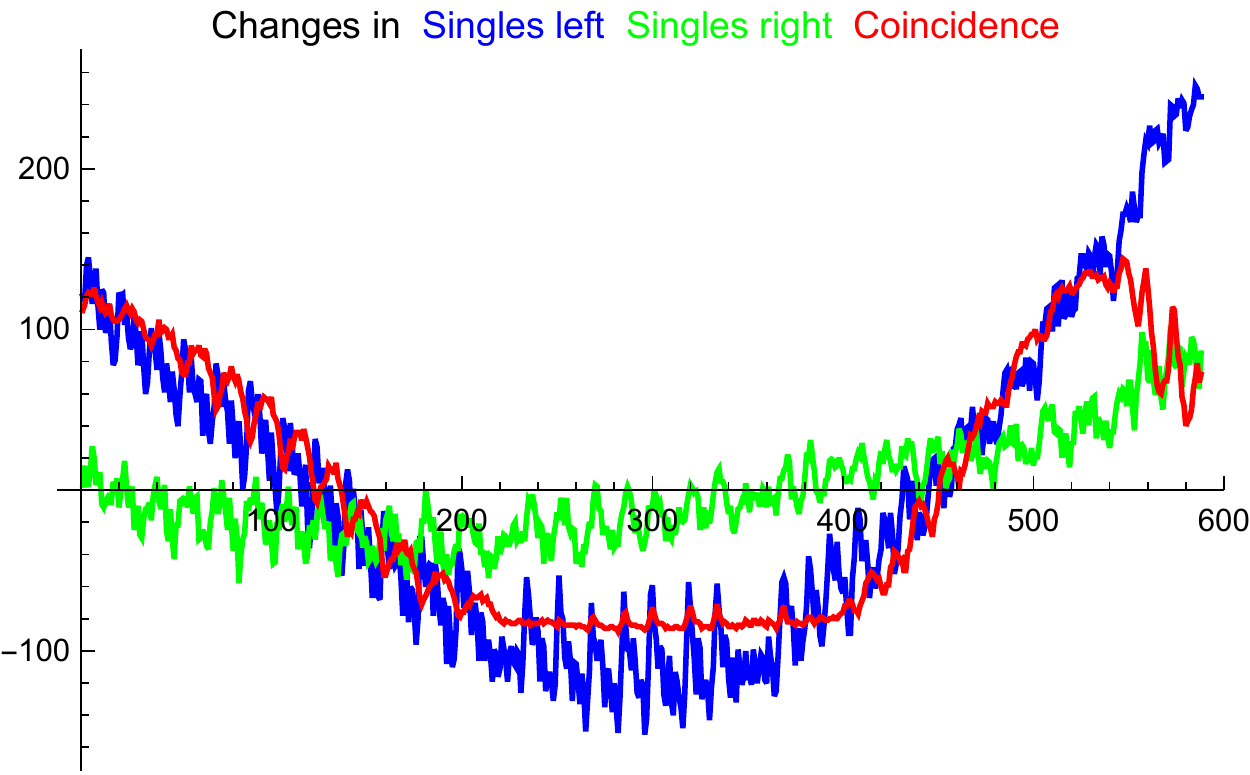}
\caption{\textit{The horizontal axis denote the number of 100s iterations for which data was recorded and vertical axis denotes corresponding changes in singles counts on the left side, right side, and the coincidence counts per iteration.}\label{fig5.09}}
\end{center}
\end{figure}

The coincidence data was collected for 600 iterations of 100 s acquisition time each, along with lab temperature and laser power. It can be seen from the plot in Fig.~\ref{fig5.09} that change in singles on left side correlates well with the change in coincidence.

 The coincidence data was collected for 100 s accumulation time and temperature data was also recorded at these intervals, and PCC is calculated for every 100 s data. Corresponding PCC plot between coincidence and temperature is shown in the Fig.~\ref{fig5.10} below. Since both the data (coincidence as well as temperature) were recorded at fixed time intervals, time was an appropriate parameter to plot on the x-axis.

\begin{figure}[H]
\begin{center}
\includegraphics[scale=0.8]{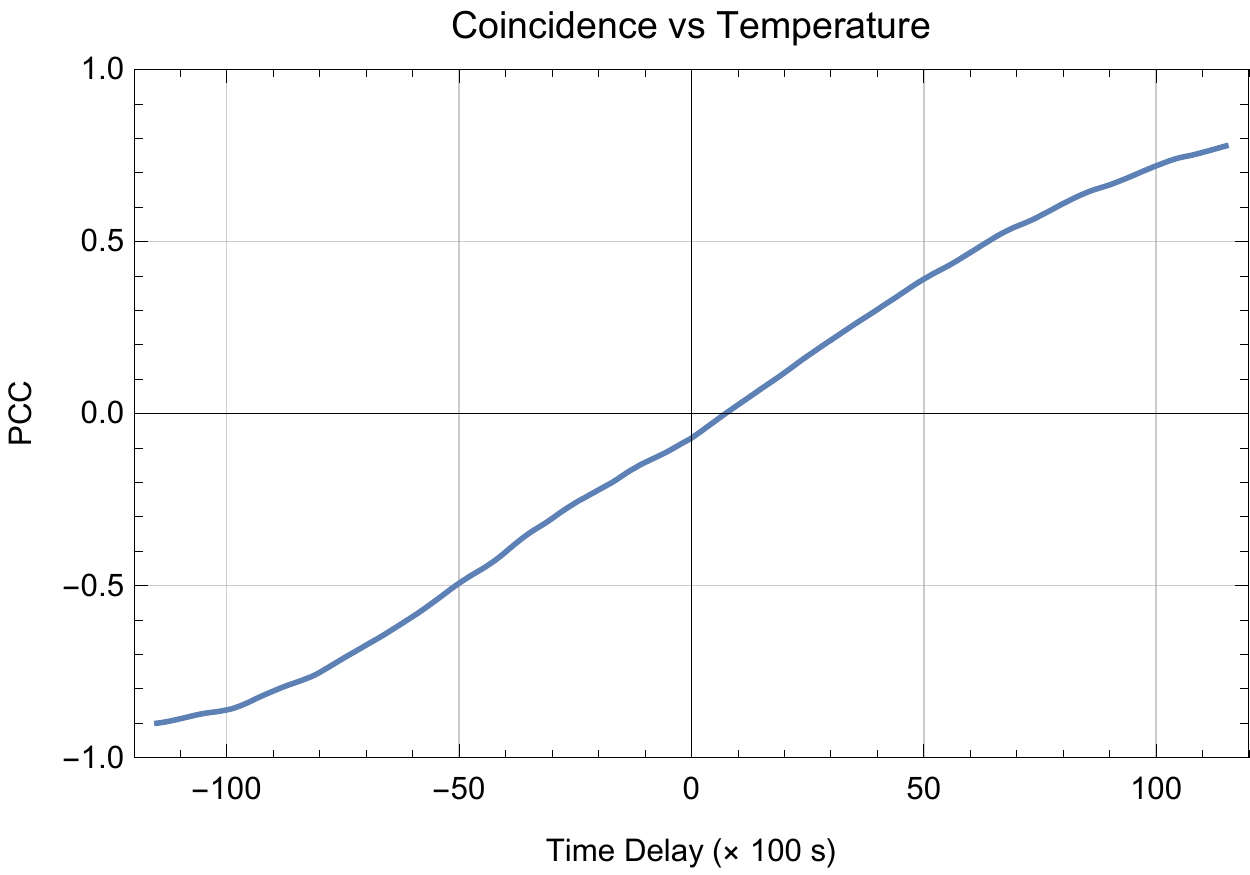}
\caption{\textit{The plot of PCC vs. time delay. Since each coincidence count measurement as well as temperature data was recorded in the steps of 100 s , hence PCC is calculated at the interval of 100 s.}\label{fig5.10}}
\end{center}
\end{figure}

It is evident from the plots in Fig.~\ref{fig5.09} that the changes in the coincidence counts nearly mimic the changes in single counts on left side.  The PCC plot between temperature and coincidence shown in Fig.~\ref{fig5.10} further reinforces the conjecture that the changes in single counts on left and coincidence counts are strongly correlated with the temperature. Such instability could arise because of the expansion/contraction of the platform mount used for PBS in the DSI. Interestingly, on the left side, PBS mount was from a local company (Holmarc Inc., India), hence we replaced it with the Thor Labs 2" kinematic platform mount which led to the marked improvement. After achieving the stability of SI, it was found that  the QST data were pretty much reproducible. Then we move onto performing the ESD experiment.

\section{ESD experiment and results}

The experimental demonstration of ESD requires preparation of a non-maximally entangled state ($|\psi\rangle=\alpha|HH\rangle+\beta|VV\rangle$, where $|\beta|>|\alpha|$) with high purity as the initial state. The individual subsystems (photons) of the initial entangled state are then sent through two separate DSIs where local, identical but independent amplitude damping channels (ADC) act on the subsystems and evolution of entanglement is studied as a function of the ADC parameter. The ADC is mimicked by a half-wave plate (HWP) which acts only on the $|V\rangle$ polarization state which is treated as excited state ($|H\rangle$ polarization state remains unaffected as it is treated as ground state) and thus transforms the polarization state of the input photon as: 
\begin{equation}
\begin{aligned}
\begin{split}
|H\rangle|a\rangle & \rightarrow |H\rangle|a\rangle~,\\
|V\rangle|a\rangle & \rightarrow \cos(2\theta) |V\rangle|a\rangle +\sin(2\theta) |H\rangle|b\rangle~,
\end{split}
\end{aligned}
\label{eq5.05}
\end{equation}
where $|a\rangle$ and $|b\rangle$ are the output spatial modes of DSI and $\theta$ is the orientation  of the fast axis of the HWP with respect to vertical. The decay probability of a photon in $|V\rangle$  state to $|H\rangle$ state due to ADC is given by $p=\sin^2(2\theta)$.

Eq.~(\ref{eq5.05}) is a quantum map that represents the unitary evolution of the system and environment (output spatial mode of the DSI). When the spatial modes ($|a\rangle$ and $|b\rangle$) of the interferometer are traced out by incoherently recombining them on the PBS, it acts as the ADC on the polarization state of the photon pairs. For detailed discussion on this, please refer to the theory chapter [\textcolor{darkblue}{4}]. In the end, quantum state tomography is done by projecting the polarization state of photons onto different bases then these photons are coupled into single-mode fiber (SMF) and coincidence measurement is performed (see Fig.~\ref{fig5.01}).

\subsection{Results}

\begin{figure}[htb]
\begin{center}
\includegraphics[scale=0.33]{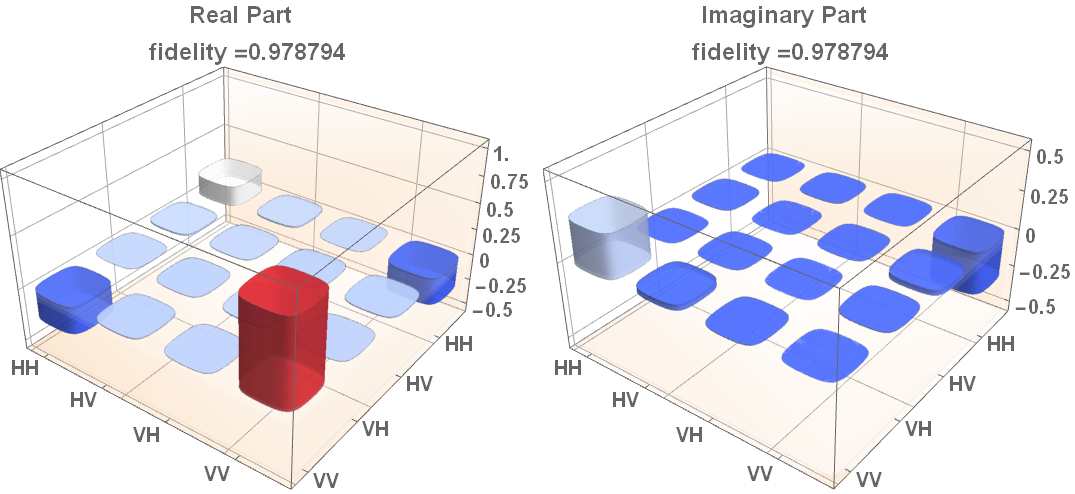}
\end{center}
\caption{\textit{The two-qubit entangled state reconstructed through QST with nearly perfect temporal compensation and spectral filtering using 810-10 nm band pass filters for pump HWP at $13.1^\circ$. The left (right) figure shows the real (imaginary) part of the density matrix. The reconstructed state had a fidelity of $ 97.88\%$ with the ideal state.}\label{fig5.11}}
\end{figure}

We set pump HWP at $13.1^\circ$ from vertical for the initial state preparation for ESD experiment. Theoretically, concurrence is expected to be 0.792 but the best reconstructed state through QST and MLE (after the initial state passed through DSIs, with ADC HWPs and optical path compensating HWPs at zero-deg) was found to be $96.17\%$ pure and had a concurrence value of 0.766. Five such initial states were prepared and collective evolution of all the states were used to obtain the theoretical plot for the ESD experiment. The initial state density matrix  for one of the experimentally reconstructed states is given in Eq.~(\ref{eq5.06}) and its real and imaginary parts are shown in Fig.~\ref{fig5.11} in a 3D plot.

\begin{equation}\resizebox{1\hsize}{!}{$
\renewcommand{\arraystretch}{1.5}
\rho_\text{exp}=\left(\begin{array}{cccc}
 0.195 & 0.0148 -0.0044 i & 0.003 +0.0004 i & -0.2635-0.2835 i \\
 0.0148 +0.0044i & 0.0067 & -0.0054-0.0024 i & -0.0092-0.0366 i \\
 0.003-0.0004 i & -0.0054+0.0024 i & 0.0105 & -0.0045+0.0054 i \\
 -0.2635+0.2835i & -0.0092+0.0366i & -0.0045-0.0054i & 0.788 \\
\end{array}\right).$}
\label{eq5.06}
\end{equation}

The reconstructed state has the following properties: 
\begin{itemize}
\item Concurrence of the ideal state = 0.7923.
\item Concurrence of the reconstructed state = 0.7659.
\item Fidelity with the ideal state = 0.9788.
\item Purity of the state = 0.9617.
\end{itemize}

Results of the ESD experiment obtained from five runs of the experiment are shown in Figs.~\ref{fig5.12} and \ref{fig5.13} below. The Fig.~\ref{fig5.12} demonstrates the phenomenon of ESD as the initial state having average Concurrence = 0.755(7) gets disentangled at $p=0.48$ and remains separable afterwards. The Fig.~\ref{fig5.13} demonstrates the change in purity of the state as a function of ADC parameter. Initial state has a purity = 0.959(3), which decreases as the strength of ADC parameter ($p$) increases until ESD occurs at $p=0.48$, where purity of the states reach a minimum value of 0.354(4). After this point, purity would increase as more and more population get pumped into $|HH\rangle$ bases state due to the action of ADC and finally joint state would  end up as a pure state when both the qubits of two-qubit system decay down to ground state $|HH\rangle$ at $p=1$. In both the plots, shaded blue curve indicates the theoretically expected evolution in the ADC obtained for the five initial states reconstructed via QST and MLE. The red dots indicate average value of the experimental data for the corresponding ADC value and vertical bars represent error in the state reconstruction due to statistical fluctuation in the counts.

\begin{figure}[!htb] 
    \centering
\includegraphics[width=0.65\textwidth]{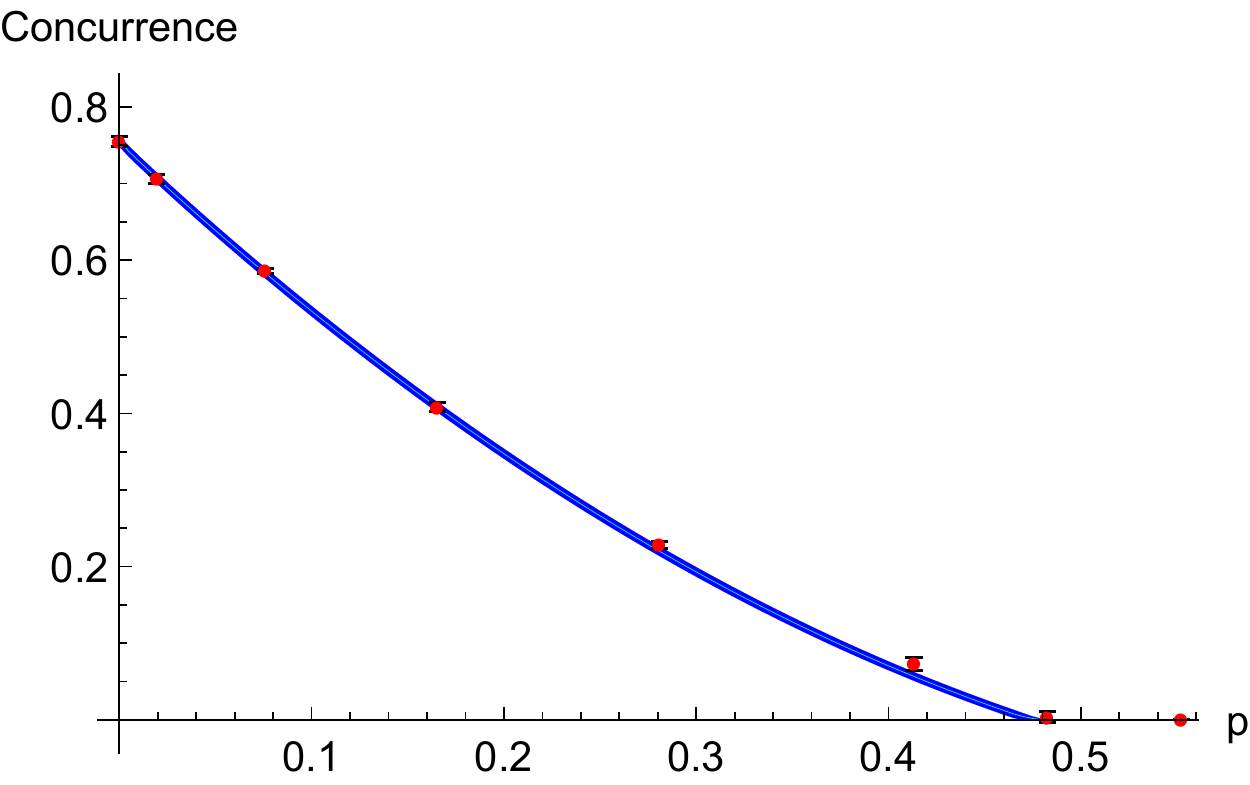}
    \caption{\textit{Plot of Concurrence vs. ADC probability $p$ shows the pheonomenon of ESD. The initial state $|\psi_i\rangle = \alpha|HH\rangle + \beta|VV\rangle$ with $|\beta| = 4 |\alpha|$ undergoes ESD (solid blue shaded line) as concurrence drops to zero at $p=0.48$. The red dots indicate experimental data points and vertical bars represent error in state reconstruction due to statistical fluctuation in the counts.}\label{fig5.12}} 
\end{figure}

\begin{figure}[!htb]
\centering
\includegraphics[width=0.65\textwidth]{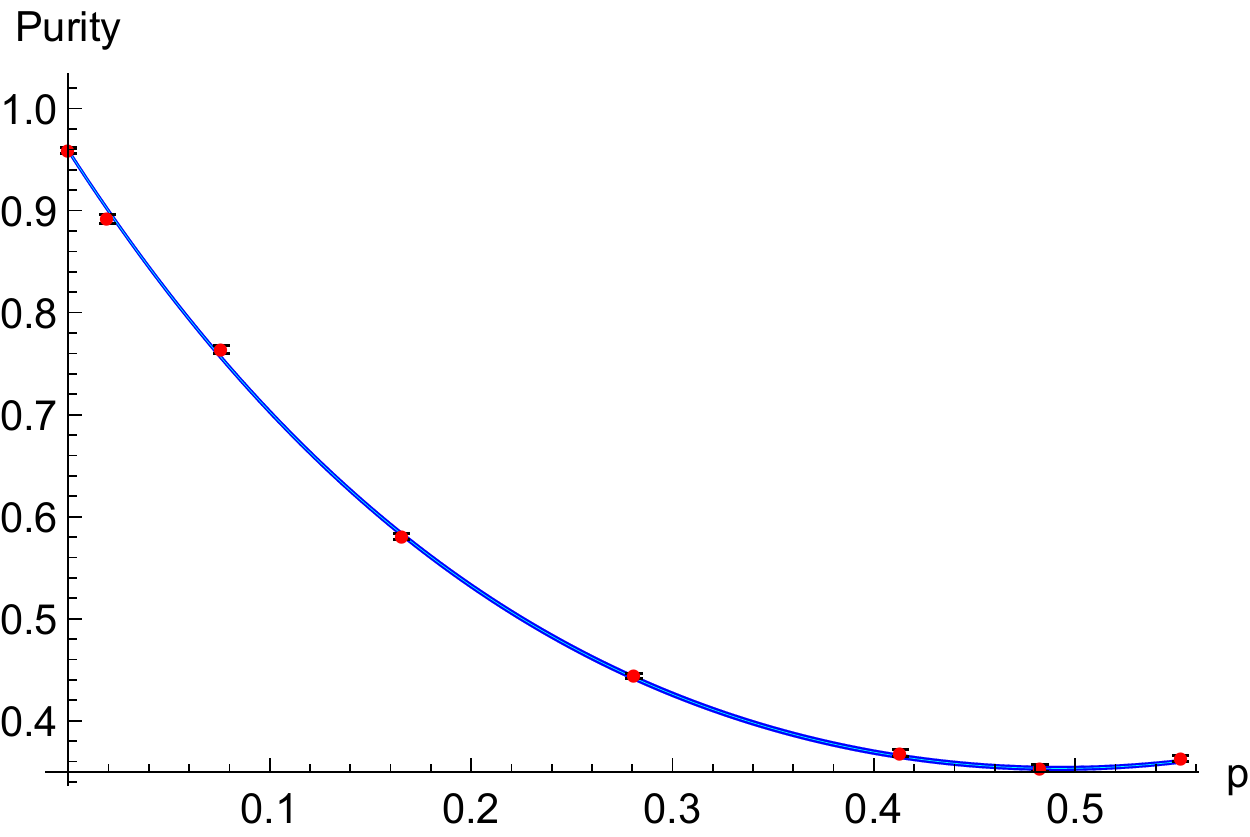}
\caption{\textit{Plot of purity vs. ADC probability p for the ESD experiment. The average purity of the initial state $|\psi_i\rangle = \alpha|HH\rangle + \beta|VV\rangle$ with $|\beta| = 4.12 |\alpha|$ drops from 0.959(3) to a minimum value of 0.353(4) at the time of ESD and then starts increasing again after this point.  The red dots indicate experimental data points and vertical bars represent error in state reconstruction due to statistical fluctuation in the counts.}\label{fig5.13}} 
\end{figure}
\newpage
\section{ADE experiment and results}
\begin{figure}[H]
\begin{center}
\includegraphics[scale=0.33]{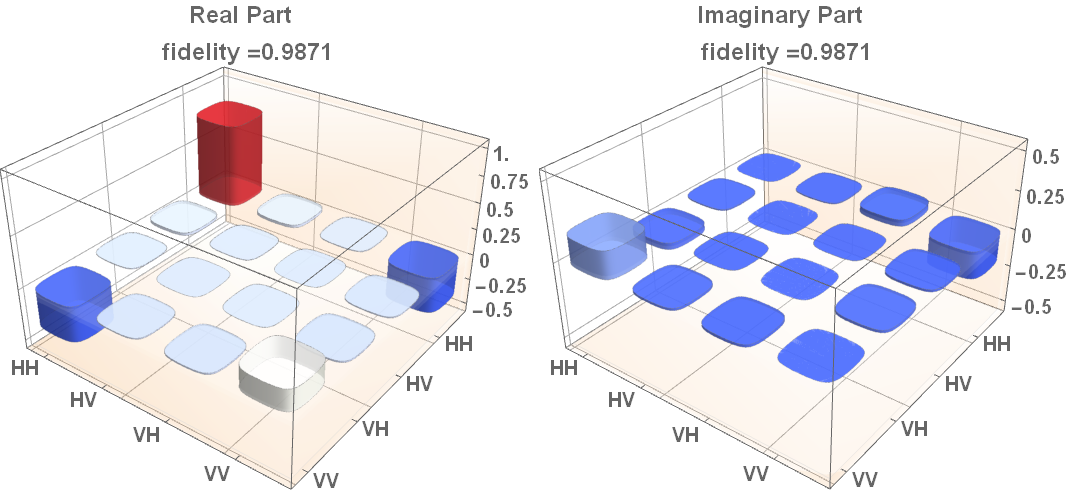}
\end{center}
\caption{\textit{The two-qubit entangled state reconstructed through QST with perfect temporal compensation and narrow spectral filtering using 810-10 nm band pass filters for pump HWP at $13.1^\circ$. Left (right) figure shows the real (imaginary) part of the density matrix. The reconstructed state had a fidelity of $ 98.71\%$ with the ideal state.}\label{fig5.14}}
\end{figure}

The experimental demonstration of ADE requires preparation of a non-maximally entangled state ($|\psi\rangle=\alpha|HH\rangle+\beta|VV\rangle$, where $|\alpha|>|\beta|$) with high purity as the initial state. We set the pump HWP at $31.9^\circ$ from vertical for the initial state preparation with $|\alpha|=4.12~|\beta|$. Theoretically, concurrence is expected to be $0.792$ but the best reconstructed state through QST and MLE (after the initial state was passed through DSIs, with ADC HWPs and optical path compensating HWPs at zero-deg) was $98.12\%$ pure and had a concurrence value of $0.778$. Five such initial states were prepared and collective evolution of all the states were used to obtain the theoretical plot for the ESD experiment. The initial state density matrix  for one of the experimentally reconstructed states is given in Eq.~(\ref{eq5.07}) and its real and imaginary parts are shown in Fig.~\ref{fig5.14} in a 3D plot.

\begin{equation}\resizebox{1\hsize}{!}{$
\rho_\text{exp}= \left(\begin{array}{cccc}
 0.8073 & 0.0228-0.0189 i & 0.007 +0.037 i & -0.3737-0.0048 i \\
 0.0228 +0.0189 i & 0.0058 & 0.0021 +0.0004 i & -0.0151-0.0129 i \\
 0.007 -0.037i & 0.0021 -0.0004 i & 0.0045 & -0.0063+0.0131 i \\
 -0.3737+0.0048 i & -0.0151+0.0129 i & -0.0063-0.0131 i & 0.1825 \\
\end{array}\right)$}
\label{eq5.07}
\end{equation}

The experimentally reconstructed state had following properties:
\begin{itemize}
\item Fidelity with the ideal state = 0.9871
\item Concurrence of the ideal state = 0.7923
\item Concurrence of the reconstructed state = 0.7781
\item Purity of the state = 0.9812
\item Relative phase = $153^\circ$
\end{itemize}

\subsection{Results}

Results of the ADE experiment obtained from five runs of the experiment are shown in Figs.~\ref{fig5.15} and \ref{fig5.16} below. The Fig.~\ref{fig5.15} demonstrates the phenomenon of ADE as the initial state having Concurrence=0.76(1) gets disentangled only at $p=1$, i.e., when both the polarization qubits decay down to the ground state $|HH\rangle$. The Fig.~\ref{fig5.16} demonstrates the change in purity of the state function of ADC parameter. Initial state has a purity = 0.97(1), which initially decreases as the strength of ADC parameter ($p$) increases until $p=0.5$. After this point, purity starts increasing as more and more population gets pump into $|HH\rangle$ state due to the action of ADC and finally joint state ends up as a pure state when both the qubits of two-qubit system decay down to ground state $|HH\rangle$ at $p=1$. In both the plots, shaded blue curve indicates the theoretically expected evolution in the ADC obtained for the five initial states reconstructed via QST and MLE. The red dots indicate average value of the experimental data for the corresponding ADC value and vertical bars represent error in state reconstruction due to statistical fluctuation in the counts.

\begin{figure} [!htb]
    \centering
\includegraphics[width=0.65\textwidth]{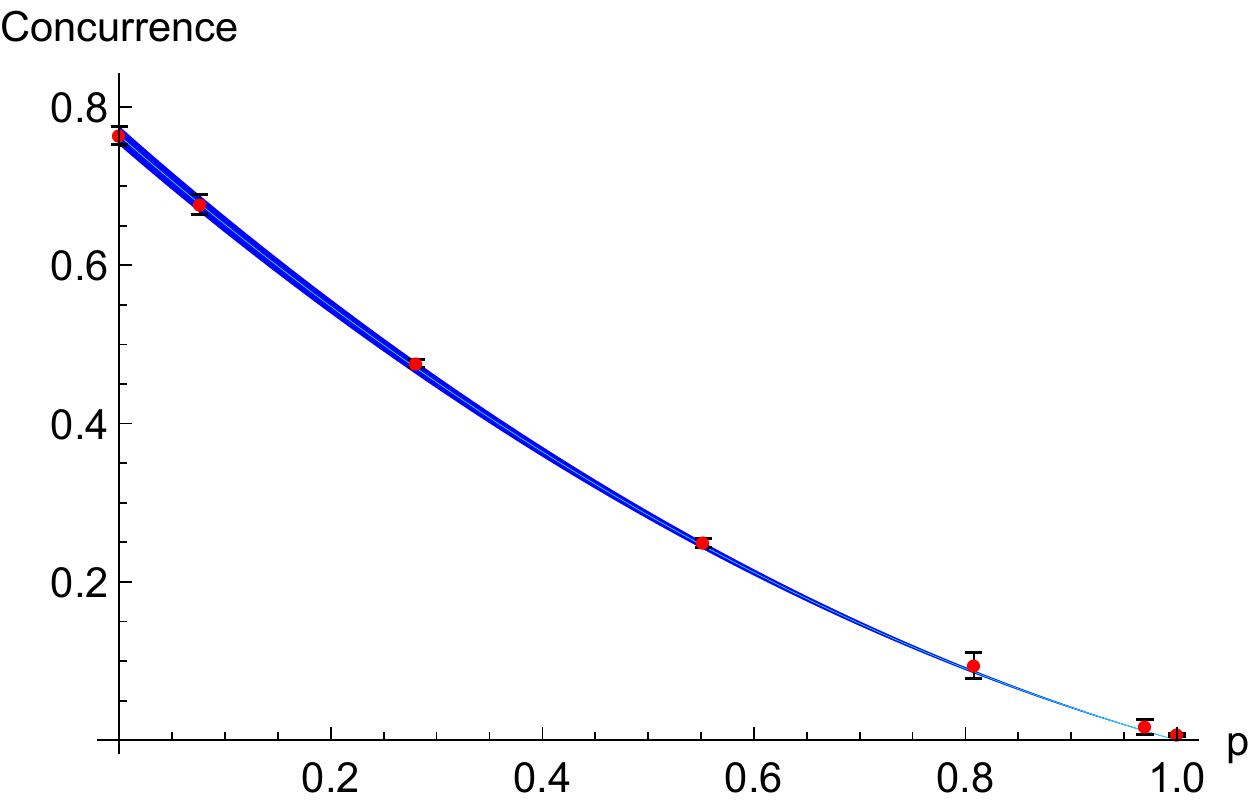}
    \caption{\textit{Plot of concurrence vs. ADC probability p for the ADE experiment. The state $|\psi_i\rangle = \alpha|HH\rangle + \beta|VV\rangle$ with $| \alpha|=4.12|\beta|$ undergoes ADE (solid blue shaded curve) as concurrence drops to zero only at $p=1$. The red dots indicate experimental data points and vertical bars represent error in state reconstruction due to statistical fluctuation in the counts.}\label{fig5.15}} 
\end{figure}

\begin{figure} [!htb] 
\centering
\includegraphics[width=0.65\textwidth]{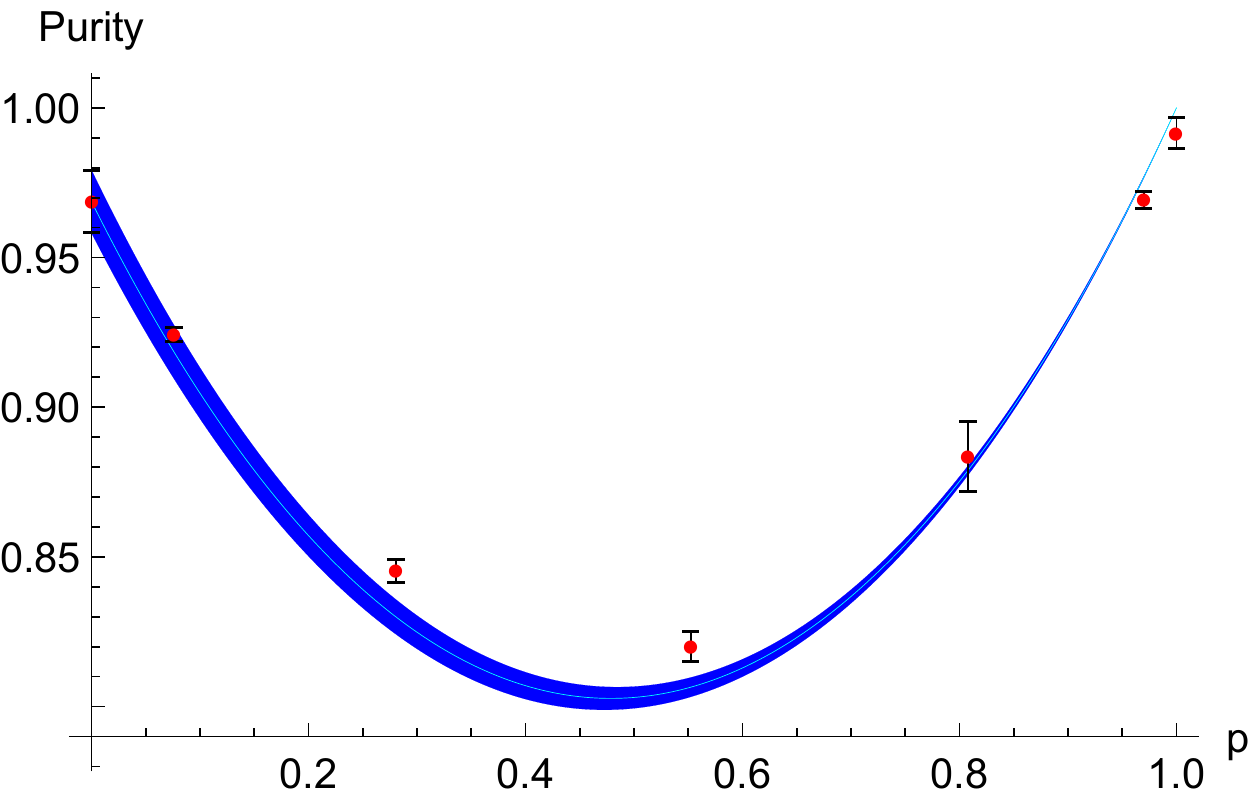}
\caption{\textit{Plot of purity vs. ADC probability p for ADE experiment. The average purity of the initial state: $|\psi_i\rangle = \alpha|HH\rangle + \beta|VV\rangle$ with $|\alpha|=4.12|\beta|$, drops from 0.97(1) to a minimum value of 0.80(1) and then starts increasing again and finally end up as a pure state when the joint state of the two-qubits becomes $|HH\rangle$. The red dots indicate experimental data points and vertical bars represent error in state reconstruction due to statistical fluctuation in the counts.}\label{fig5.16}} 
\end{figure}

An interesting observation is that the blue shaded curves, in both Fig.~\ref{fig5.15} as well as Fig.~\ref{fig5.16}, have larger thickness in the beginning (for lower ADC values) and then it narrows down as $p$ approaches 1. When initial state is reconstructed for lower ADC values, we get slightly different states (density matrices) in different runs of the QST. This can be attributed to the statistical fluctuations in the coincidences for different projections as well as imperfect projections due to the error in half- and quarter- waveplate orientations as compared to the ideal value which leads to different results in different runs of the QST. Since all the initial states, starting from different points in the Hilbert space approach towards a common final state $|HH\rangle$ in the presence of ADC, hence curve narrows down as $p$ increases. Also, as pointed out earlier, entangled states get more affected with slight interferometric instabilities as compared to the separable state such as $|HH\rangle$ and $|VV\rangle$ as they do not involve recombination of counter-propagating polarization amplitudes at PBS in DSI.

\section{Attempts towards manipulation of ESD using NOT operation on both the qubits}

\subsection{Setting up the ESD manipulation experiment}

The ESD manipulation experiment [\ref{5.02}] requires setting up the  two DSIs with added optical components for the manipulation as shown in the Fig.~\ref{fig5.17} below.

\begin{figure} [H]
\begin{center}
\includegraphics[clip, trim=3cm 8.5cm 0.5cm 2cm, width=0.8\textwidth]{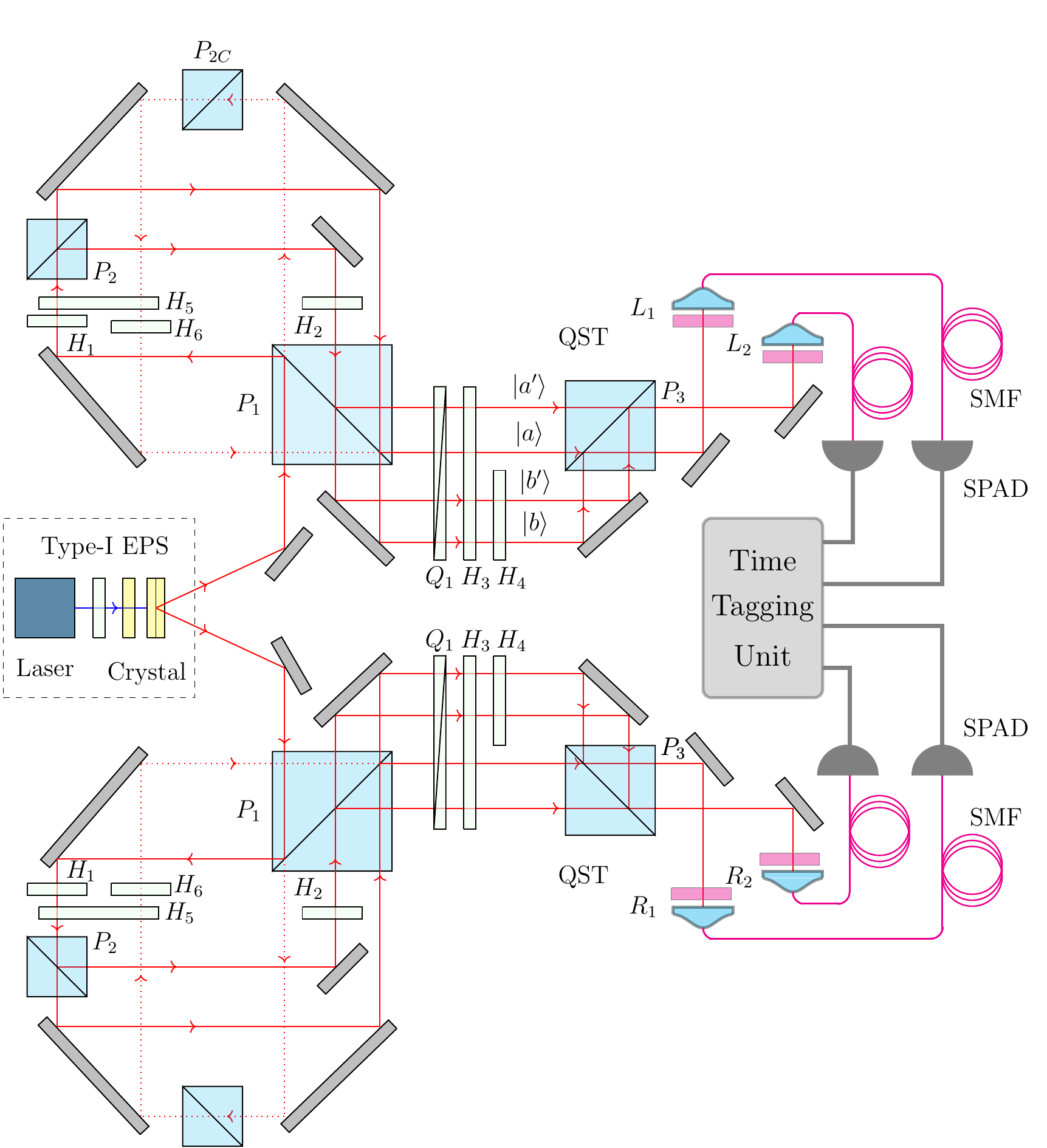}
\caption{\textit{Experimental setup for demonstrating manipulation of ESD using NOT operation on both the qubits of a bipartite entangled state in the presence of an ADC in a photonic system [\ref{5.02}]. Entangled photons are prepared on the left side which are then fed into two DSIs (top and bottom) where ADC is mimicked using HWPs acting on V-polarization component and NOT operations are applied by a HWP at $45^\circ$ on both H- as well as V-polarization components for the manipulation of ESD and system is evolved in the ADC afterwards.  In the end, output spatial modes are traced out and QST is performed (right side).}\label{fig5.17}}
\end{center}
\end{figure}

\subsection{Aligning ESD manipulation setup using He-Ne laser}\label{S5.5.2}
It is evident from the Fig.~\ref{fig5.17} that many additional optical components such as ADC HWPs, NOT operation HWP ($\text{HWP}~\text{at}~45$), PBS, mirrors, and couplers need to be suitably introduced in the ESD setup (Fig.~\ref{fig5.01}) for the manipulation experiment. Initially, these components were aligned using a forward propagating alignment laser (He-Ne, 632.8 nm) such that the adjacent spatial modes ($|a'\rangle~|a\rangle,~|b'\rangle,~\text{and}~|b\rangle$) are equidistant (approx. 0.5 inch). He-Ne laser beam was then coupled into SMF to ensure proper positioning of the new coupler $L_2$ ($R_2$). The first step to ensure the single photons coupling into the new coupler which had a SMF, a backward propagating laser beam (Ti-Saph laser, 810 nm) was passed through the couplers $L_1$ and $L_2$ (and likewise $R_1$ and $R_2$). The spatial mode overlap of the forward propagating He-Ne beam with the corresponding backward propagating Ti-Saph beam from $L_1$ and $L_2$ was ensured throughout the interferometer and near the crystal for appropriate settings of the ADC HWPs. Since $L_1$ and $R_1$ were already coupled for ESD experiment, they served as a reference for aligning $L_2$ and $R_2$.

Next, overlap of the $\mathbf{K}$-vectors of the backward propagating beams from $L_1$ and $L_2$ was ensured by placing a beam profiler near the crystal and looking at the interference pattern of the co-polarized beams. This step is necessary to guarantee the single photon coupling into SMF. The power levels of the two backward propagating beams were adjusted to make them equal by choosing appropriate setting of the waveplates (or neutral density filters) and beam collimation was ensured by focusing screw of the coupler. Initially, non-collinear interference fringes were observed which was iteratively made collinear by tweaking the new coupler $L_2$ and the mirror next to it until fringes disappeared and only two overlapping Gaussian spots were seen. For tweaking, we used H- and V-screws of the mirror and coupler $L_2$ iteratively and looked at the centroid of the individual beam (from $L_1$ and $L_2$) and minimization of the interference pattern when both the beams were present together. When horizontal and vertical positions of the centroid of individual beam matched (within 10 $\mu m$) with that when both beams were present, the two beams were assumed to have matching $\mathbf{K}$-vectors. Furthermore, no interference fringes were observed in this configuration. One can also check the beam wander profile of the individual beams as well as both together, which should match for the two beams with overlapping $\mathbf{K}$ vectors.

\subsection{Spatial mode tracing and constraints}

In order to understand the output mode population for different ADC settings, we represent the action of different optical elements on the input state $|\psi\rangle=\alpha|HH\rangle+\beta|VV\rangle$ as given below.

\begin{itemize}
\item An input $|HH\rangle$ polarization state gets transformed as follows:
\begin{equation}
\begin{aligned}
\alpha |HH\rangle \xrightarrow{H~\text{at}~45} & \alpha |VV\rangle~,\\
\xrightarrow {~~p'_H~~} & \alpha [(1-p'_H)|VV\rangle +p'_H|HH\rangle],\\
\xrightarrow[\text{path modes}]{\text{Tagging}} & \alpha [(1-p'_H)|VV\rangle |b\rangle+p'_H|HH\rangle |a\rangle] .
\end{aligned}
\label{eq5.08}
\end{equation}
where $p'_H=\sin^2(2\theta'_H)$. This is due to the action of an ADC HWP after the NOT operation in the path of counter-clockwise propagating (initially) H-polarized beam.

\item An input $|VV\rangle$ polarization state gets transformed as follows:
\begin{equation}
\begin{aligned}
\beta |VV\rangle \xrightarrow{p_V~} & \beta[(1-p_V)|VV\rangle+p_V|HH\rangle],\\
 \xrightarrow{H~\text{at}~45} & \beta[(1-p_V)|HH\rangle+p_V|VV\rangle],\\
 \xrightarrow{~~p'_V~~} & \beta[(1-p_V)|HH\rangle+p_V \left((1-p'_V)|VV\rangle+p'_V|HH\rangle\right)], \\
 \xrightarrow[\text{path modes}]{\text{Tagging}} &  \beta[(1-p_V)|HH\rangle|b\rangle+  p_V(1-p'_V)|VV\rangle|a'\rangle + p_V p'_V|HH\rangle|b'\rangle].
\end{aligned}
\label{eq5.09}
\end{equation}
\end{itemize}
where $p_V=\sin^2(2\theta_V)$ and $p'_V=\sin^2(2\theta'_V)$. The ADC parameter $p_V$ ($p'_V$) is due to the action of an ADC HWP in the path of clockwise propagating (initially) V-polarized beam before (after) the NOT operation.

Upon combining Eqs.~(\ref{eq5.08}) and (\ref{eq5.09}), we get
\begin{equation}
\begin{aligned}
\alpha |HH\rangle+\beta |VV\rangle\rightarrow &  \alpha [(1-p'_H)|VV\rangle |b\rangle+p'_H|HH\rangle |a\rangle] +\\
& \beta[(1-p_V)|HH\rangle|b\rangle+  p_V(1-p'_V)|VV\rangle|a'\rangle + p_V p'_V|HH\rangle|b'\rangle].
\end{aligned}
\label{eq5.10}
\end{equation}

From Eq.~(\ref{eq5.10}), the spatial mode population for different ADC settings can be found. Details are listed in the table~[\ref{tab5.1}]. ESD-manipulation setup has four spatial modes ($|a\rangle$, $|a'\rangle$, $|b\rangle$, and $|b'\rangle$) and there are total five different conditions on population of different spatial modes corresponding to different settings of the ADC values that need to be satisfied for the manipulation experiment to work. For example, for $p_{\scaleto{V}{4pt}}=0,~p'_{\scaleto{V}{4pt}}=0,~\text{and}~p_{\scaleto{H}{4pt}}=0$, input $|H\rangle$- and $|V\rangle$-polarization components get converted to $|V\rangle$- and $|H\rangle$-polarizations, respectively, and go to spatial mode $|b\rangle$. These photons are then coupled to $L_1 R_1$ couplers and give rise to coincidences for $|HH\rangle\langle HH|$ and $|VV\rangle\langle VV|$ projections with equal probability for a maximally entangled state $|\psi\rangle=[|HH\rangle+|VV\rangle]/\sqrt{2}$ as input. Similarly, for other ADC settings, these conditions are listed in the table~[\ref{tab5.2}].

\begin{table} [!htb]
\resizebox{\columnwidth}{!}{%
\renewcommand{\arraystretch}{1.5}
\begin{tabular}{ |p{0.6cm}||p{0.5cm}|p{0.5cm}|p{0.5cm}|p{1.75cm}|p{3cm}| p{2.9cm}|p{2cm}|}
\hline
\multirow{2}{*}{S.N.} & \multicolumn{3}{c}{ADC settings} \vline & \multicolumn{3}{c}{Output mode population}  &   \\
 \cline{2-8}
 & $p_V$ &  $p'_V$ & $p'_H$  & $|a\rangle$ & $|b\rangle$ & $|a'\rangle$ & $|b'\rangle$ \\
 \hline \hline
1 & $p_V$ &  $p'_V$ & $p'_H$ & $\alpha p'_H |HH\rangle$ & $\alpha (1-p'_H)|VV\rangle+\beta (1-p_V)|HH\rangle$ & $\beta p_V(1-~p'_V)|VV\rangle$ & $\beta p_V p'_V|HH\rangle$ \\
 \hline 
2 & 0 & 0 & 0 & 0 & $\alpha |VV\rangle+\beta |HH\rangle$ & 0 & 0 \\
 \hline
3 & 1 & 0 & 0 & 0 & $\alpha |VV\rangle$ & $\beta |HH\rangle$ & 0 \\
 \hline
4 & 1 & 1 & 0 & 0 & $\alpha |VV\rangle$ & 0 &  $\beta |HH\rangle$ \\
 \hline
5 & 1 & 1 & 1 & $\alpha |HH\rangle$ & 0 & 0 &  $\beta |HH\rangle$ \\
 \hline
6 & 1 & 0 & 1 & $\alpha |HH\rangle$ & 0 & $\beta |VV\rangle$ &  0 \\
 \hline
7 & 0 & 0 & 1 & $\alpha |HH\rangle$ & $\beta |HH\rangle$ & 0 &  0 \\
  \hline
8 & 0 & 1 & 1 & $\alpha |HH\rangle$ & $\beta |VV\rangle$ & 0 &  0 \\
 \hline
\end{tabular}}
\caption{\textit{This table summarizes the results of different ADC settings on the output mode population in the ESD-manipulation experiment when NOT operation is applied on both the qubits.}\label{tab5.1}}
\end{table}

\begin{table} [!htb]
\resizebox{\columnwidth}{!}{%
\renewcommand{\arraystretch}{1.5}
\begin{tabular}{ |p{0.6cm}||p{2cm}|p{2.5cm}|p{2.5cm}|p{2cm}|p{1.5cm}| p{2.2cm}| }
 \hline
S.N.&  Input pol. & ADC settings ($p_V$, $p'_V$, $p'_H$)& Projection  & Output pol. & Output mode & Coincidence counts\\
 \hline \hline
\textbf{1}  & $|VV\rangle$ & (0,~0,~0) & $|HH\rangle\langle HH|$ & $|HH\rangle$ & $|b\rangle$ & $L_1 R_1$\\
\textbf{2}  & $|HH\rangle$ &  & $|VV\rangle\langle VV|$ & $|VV\rangle$ & $|b\rangle$ & $L_1 R_1$\\
\hline
3  & $|HH\rangle$ & (1,~0,~0) & $|VV\rangle\langle VV|$ & $|VV\rangle$ & $|b\rangle$ & $L_1 R_1$\\
\textbf{4}  & $|VV\rangle$ &  &  & $|VV\rangle$ & $|a'\rangle$ & $L_2 R_2$\\
\hline
5  & $|HH\rangle$ & (1,~1,~0) & $|VV\rangle\langle VV|$ & $|VV\rangle$ & $|b\rangle$ & $L_1 R_1$\\
6  & $|VV\rangle$ &  & $|HH\rangle\langle HH|$ & $|HH\rangle$ & $|b'\rangle$ & $L_2 R_2$\\
 \hline
\textbf{7}  & $|HH\rangle$ & (1,~1,~1) & $|HH\rangle\langle HH|$ & $|HH\rangle$ & $|a\rangle$ & $L_1 R_1$\\
\textbf{8}  & $|VV\rangle$ &  &  & $|HH\rangle$ & $|b'\rangle$ & $L_2 R_2$\\
 \hline
\end{tabular}}
\caption{\textit{This table summarizes the results of different ADC settings on the input polarization in ESD-manipulation experiment when NOT operation is applied on both the qubits. Out of these only five settings (S.N. 1, 2/3/5, 4, 7, and 8/6 ) give different results.}\label{tab5.2}}
\end{table}

\subsection{Coincidence optimization and QST}

Initially coupling into single mode fibres in $L_2$ and $R_2$ were achieved by methods discussed in section [\ref{S5.5.2}]. Then, coincidence was optimized from different spatial modes for different settings of ADC values in different projection bases as given in Table~[\ref{tab5.2}]. After achieving near optimal coupling from different spatial modes, we performed a QST for maximally entangled state input to DSIs with ADC HWPs at zero-deg and found that the reconstructed state had following properties:
\begin{itemize}
\item Fidelity with the ideal state = 0.7529
\item Concurrence of the reconstructed state = 0.5237
\item Purity of the reconstructed state = 0.6212
\item Relative phase  = $330^\circ$ 
\end{itemize}
The reconstructed state density matrix is given by Eq.~(\ref{eq5.10}) and corresponding 3D plot is shown in Fig.~\ref{fig5.18} below.

\begin{equation}
\rho_\text{exp}=\left(
\begin{array}{cccc}
 0.504 & 0.009 -0.022 i & -0.001-0.033 i & 0.242 +0.120 i \\
 0.009 +0.022 i & 0.007 & 0.008 -0.009 i & 0.014 +0.030 i \\
 -0.001+0.033 i & 0.008 +0.009i & 0.027 & -0.008+0.018 i \\
 0.242 -0.120i & 0.014 -0.030i & -0.008-0.018 i & 0.462 \\
\end{array}
\right)
\label{eq5.11}
\end{equation}

\begin{figure}[H]
\centering
\includegraphics[width=0.8\textwidth]{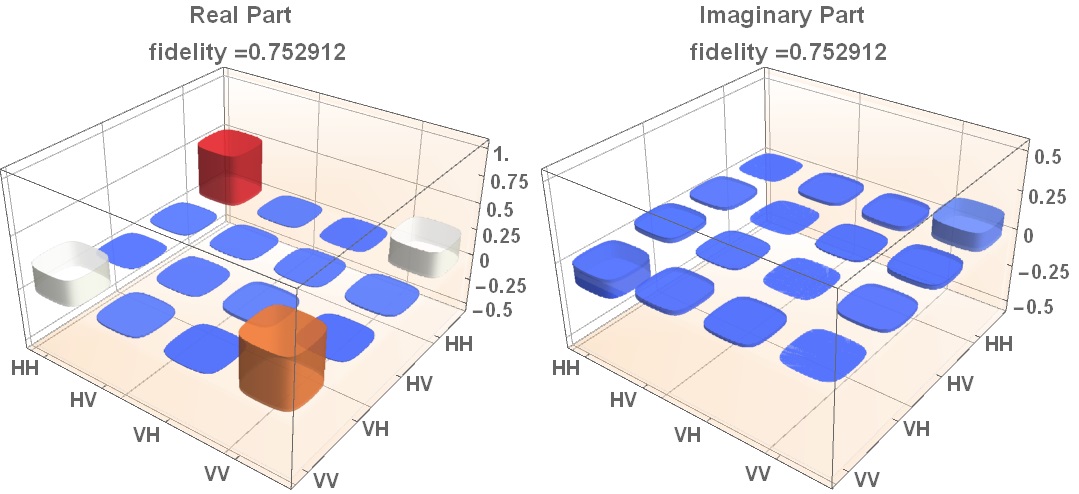}
\caption{\textit{QST result with ESD manipulation setup  for ADC settings (0,0,0).}\label{fig5.18}}
\end{figure}

From the state purity and coherence term of the state in Eq.~(\ref{eq5.11}), it is evident that the reconstructed state has significant phase decoherence. At this point, we were using 810-10 nm bandpass filters before all the couplers. Since we had introduced many optical components in the path-balanced ESD setup to get ESDM setup, some microscopic path delay between clockwise and counter-clockwise propagating components might have been introduced. The effect due to such path differences can be minimized by using a narrow bandpass filter as discussed in the next section.

\subsection{Spectral filtering consideration}
For $\lambda=810$ nm, $\Delta\lambda=10~\text{nm}$ (FWHM of the bandpass filter used for spectral filtering the SPDC photons), the coherence length of the single photons is given by
\begin{equation}
l_c=\frac{\lambda^2}{\Delta\lambda}=66~\mu m
\end{equation}
 And, coherence time of single photons is given by $\tau_c=l_c/c=219~\text{fs}$, where $c$ is the speed of light in vacuum.

We had used 1/2" PBS ($\sim 100~\mu m$) in the H- and V-polarization paths which had thickness tolerance of the order of $100~\mu m$. Since the coherence length of single photons was smaller than the optical path difference due to mismatch in the thickness of 1/2" PBSs used in two paths of the DSI, the polarization amplitudes didn't recombine coherently at PBS, which resulted in partial mixing of the two  outgoing polarization amplitudes in the mode $|b\rangle$. For a given relative delay $\Delta t$ between the two arms of the interferometer, the coherence term will drop as $\exp(-\Delta t/\tau_c)$, where $\tau_c$ is the coherence time of single photons. Time taken by light to travel $100~\mu m$ thick glass plate ($n=1.55$) $\Delta t=517~\text{fs}$. This will lead to drop in the coherence term from $0.5$ to $0.333$ for a maximally entangled state input to DSIs.

Taking a cue from this analysis, we replaced the 810-10 nm BPF with a narrow PBF 810-3nm from Semrock in front of the $L_1R_1$ couplers. After optimizing the coincidence, performed QST for pump HWP at $22.5^\circ$ with ADC settings (0,0,0) in the DSI and got:
\begin{itemize}
\item Fidelity with the ideal state = 0.9058
\item Concurrence of the ideal state = 1
\item Concurrence of the reconstructed state = 0.8183
\item Purity of the state = 0.8319
\item Relative phase $\phi = 102^\circ$
\end{itemize}

Real and imaginary parts of the density matrix are shown in Fig.~\ref{fig5.19} below.
\begin{figure}[H]
\begin{center}
\includegraphics[scale=0.33]{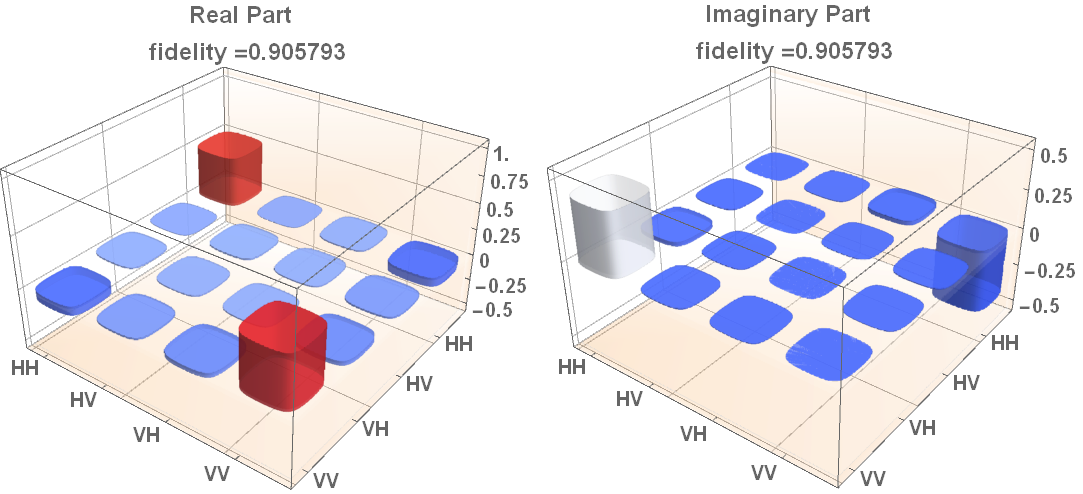}
\caption{\textit{QST results for pump HWP at $22.5^\circ$ with 810-03nm BPF in the $L_1R_1$ couplers for ADC HWPs at (0,0,0) in the DSIs.}\label{fig5.19}}
\end{center}
\end{figure}

Narrow spectral filtering (810-3 nm BPF) lead to an order of magnitude drop in the coincidence counts but state properties significantly improved due to the increase in coherence time $\tau_c$ of the single photons. Therefore, it was decided to use this filter for further coincidence optimization.

\subsection{Current status and way forward}
Currently, state reconstructed at the output ports of the ESDM setup has $90.58\%$ fidelity with respect to maximally entangled state. We aim to improve the fidelity to better than $95\%$ and then we will perform the ESD manipulation experiment so that we can make a meaningful conclusion from such an experiment. For this purpose, we have to ensure that there is no optical path differences between clockwise and counter-clockwise propagating polarization amplitudes only then they would coherently recombine at the PBS and go to mode $|b\rangle$. The drop in fidelity, purity and Concurrence indicate that the coherent recombination of the polarization amplitudes is not taking place perfectly at the PBS in DSI. This can happen either due to (i) imperfect spatial overlap of the two recombining polarization amplitudes at PBS, (ii) losses due to the optical components seen by entangled photons in their path in both the interferomters, or (iii) infinitesimal optical path difference between clockwise and counter-clockwise propagating paths, or (iv) due to cumulative effect of all the above reasons. 

Since single photons are getting coupled into SMF and we get nearly equal coupling (coincidence) for both $|HH\rangle$ and $|VV\rangle$ polarization amplitudes, there is a very little chance that imperfect spatial overlap of the two recombining polarization amplitudes at PBS could be the reason for drop in state purity. Any losses by individual optical components will of course contribute to drop in state purity, which cannot be avoided. However, further coincidence optimization can be done by tweaking different optical mirrors in the DSI as well as outside for zero-deg settings of the ADC HWPs to improve the state properties. Second reasons could be the main contributor towards drop in state purity as we are using $1/2"$ PBS in the two arms, and we know that their thickness have a tolerance of $\sim 100~\mu m$. By using suitable optical path compensation technique in each DSIs, temporal delay can be minimized and therefore drop in state purity due to this reason can be compensated. For this purpose, we can use a thin glass plate (say, 1 mm thickness) in transmitted as well as reflected arms of each DSI, and by slightly tilting one of the glass plates in transmitted or reflected arms of one of the DSI, one can compensate for the microscopic path difference between the two arms of the corresponding DSI.  

These steps are being taken to improve the state fidelity for zero-deg setting of the ADC HWPs in the DSI. We expect to improve the state fidelity at the output modes to better than $95\%$ (in the ESD experiment, for zero-deg setting of ADC, state fidelity was $\sim 97\%$) and then data for ESD-manipulation experiment will be taken in order to make a meaningful conclusion from such an experiment. 

\section{References}
\begin{enumerate}
\itemsep0em 
\item \textit{M. P. Almeida, F. de Melo, M. Hor-Meyll, A. Salles, S. P. Walborn, P. H. Souto Ribeiro, L. Davidovich,``Environment-Induced Sudden Death of Entanglement," \href{https://doi.org/10.1126/science.1139892}{Science \textbf{316}, 555(2007)}}. \label{5.01}
\item \textit{Ashutosh Singh, Siva Pradyumna, A. R. P. Rau, and Urbasi Sinha, ``Manipulation of entanglement sudden death in an all-optical setup," \href{https://doi.org/10.1364/JOSAB.34.000681}{J. Opt. Soc. Am. B \textbf{34}, 681-690 (2017)}}. \label{5.02}
\end{enumerate}

%% file: chapter6.tex
\setcounter{equation}{0}
\chapter{Entanglement protection in higher-dimensional systems}

The inevitable dissipative interaction of an entangled quantum system with its environment causes degradation in quantum correlations present in the system. This can lead to a finite-time disappearance of entanglement, which is known as Entanglement Sudden Death (ESD). Here, we consider an initially entangled qubit-qutrit system and a dissipative noise which leads to ESD, and propose a set of local unitary operations, which when applied on the qubit, qutrit, or both subsystems during the decoherence process, cause ESD to be hastened, delayed, or avoided altogether, depending on its time of application. Delay and avoidance of ESD may find practical application in quantum information processing protocols that would otherwise suffer due to short lifetime of entanglement. The physical implementation of these local unitaries is discussed in the context of an atomic system. The simulation results of such ESD manipulations are presented for two different classes of initially entangled qubit-qutrit systems. A prescription for generalization of this scheme to a qutrit-qutrit system is given. This technique for entanglement protection in the noisy environment is compared with other related techniques such as weak measurement reversal, dynamic decoupling, and quantum Zeno effect.

\section{Introduction}

Quantum decoherence [\ref{1}] is a ubiquitous and unavoidable phenomenon arising because of  entanglement between quantum systems  and their environment. Entanglement, on the other hand, is a fundamentally important phenomenon in the studies of quantum foundations and has great significance in quantum technologies. It is  now seen  as  an  indispensable resource  in Quantum  Information Processing  (QIP) for  various tasks such  as   quantum  computation,  teleportation,   superdense  coding, cryptography, etc., which are either impossible or less efficient using classical correlations [\ref{2}-\ref{4}].

The inevitable dissipative interaction  of an entangled quantum system
with its environment leads to  an irreversible loss of single particle
coherence  as well  as degradation  of entanglement  present in  the
system [\ref{5}]. For  some initial  states, in  the presence  of an
Amplitude Damping Channel (ADC), entanglement degrades asymptotically,
whereas for  others it  can disappear  in finite  time, also  known as
early stage disentanglement or Entanglement Sudden Death (ESD), in
literature [\ref{6}-\ref{8}]. Soon  after its theoretical prediction,
ESD was experimentally demonstrated  in atomic [\ref{9}] and photonic
[\ref{10}]  systems. The  real world  success of several quantum
information,  communication,  and  computation tasks  depends  on  the
resilience of entanglement from noises present in the environment, and
longevity of the entanglement. Thus,  ESD poses a practical limitation
on QIP  tasks.  Therefore,  strategies  which   make entanglement robust against  the detrimental effects of  the noise are
of practical interest in QIP.

Owing to the  simplicity of two-qubit entangled systems and its  usefulness as a resource  in QIP,  there have  been several  theoretical proposals  to combat decoherence  and finite-time  disentanglement in  these systems [\ref{11}-\ref{24}].  Some of these proposals have been experimentally demonstrated   in   atomic,  photonic,  and   solid   state   systems [\ref{25}-\ref{34}]. One of the entanglement protection schemes [\ref{21}] considered a class of two-qubit entangled states which undergo ESD in the presence of an ADC. For such systems, a Local Unitary Operation (LUO), in this case the Pauli $\sigma_x$ operator also known as NOT operation, has been proposed. When LUO is applied  on one or both the subsystems during  the process of decoherence, it can  hasten, delay, or completely avoid the ESD, all depending on the time when NOT operation is applied. This   proposal  was  later  transformed by our group  into  an  all-optical experimental  setup  to study  the  effect  of  LUOs on the disentanglement dynamics in  a photonic system [\ref{22}]. This work proposed  an experimentally  feasible architecture for the implementation of ADC and the local  NOT operation to suitably manipulate the ESD of a two-qubit system in a controlled manner.

Higher-dimensional entangled quantum systems (qudits, $d\geq 3$) can offer practical advantages over the canonical two-qubit entangled systems in QIP protocols. These  systems are more  resilient to errors than their qubit counterparts  in quantum cryptography, and they offer practical advantages; for example, increased channel capacity in quantum communication,  enhanced security in QIP  protocols, efficient
quantum  gates,  and  in  the  tests of  the  foundations  of  quantum
mechanics [\ref{35}]. It  is  therefore  important  to  study  the  effects  of decoherence on these systems.  Entanglement evolution has been studied in higher-dimensional systems present  in the noisy  environment, and despite the  resilience of these  systems to noise [\ref{36}],  ESD is established to be  ubiquitous in all dimensions of  the Hilbert spaces [\ref{37}, \ref{38}]. 

As a first step towards generalization to higher dimensional systems, a qubit-qutrit system is the natural choice as it is intermediate in complexity to a qubit-qubit and qudit-qudit system. Given the availability of full-fledged separability criteria for pure as well as mixed state qubit-qutrit systems, experimental feasibility of state preparation, and the significance of higher-dimensional systems in QIP, they become an important architecture towards understanding and controlling the effects of decoherence in higher-dimensional systems. Some   attempts  have  been  made   to  locally manipulate  the  disentanglement    dynamics    in    qubit-qutrit [\ref{39}, \ref{40}], and qutrit-qutrit [\ref{41}-\ref{43}] systems to retain the entanglement for longer duration. It is found that the quantum interference between two  upper levels  of  the qutrit  can  also be  used  to control  the disentanglement dynamics in higher-dimensional systems [\ref{39}, \ref{44}].

Consider a practical scenario where Charlie prepares a bipartite entangled state for some QIP task and he has to send the entangled particles to Alice and Bob through a quantum channel which is noisy and can potentially cause disentanglement before the particles reach the two parties. In this scenario, we ask the following question: given a higher-dimensional bipartite entangled state which would undergo ESD in the presence of an ADC, can we alter the time of disentanglement by some suitable LUOs during the process of decoherence? As a possible answer to this question, we  explore   the  generalization  of  the   proposal  in  Ref.~[\ref{21}]  for  higher-dimensional  systems,  say qubit-qutrit or qutrit-qutrit system, in  the presence  of an  ADC. If such systems undergo ESD, we propose a set of LUOs, such that when they are applied on the subsystems during  the  process   of  decoherence, they manipulate the disentanglement dynamics, in particular,  delay the  time at which ESD occurs.  Such a  study was  partially done  in Ref.~[\ref{39}], and  here we  generalize their scheme  and propose  a more general class of  LUOs, which  can always suitably manipulate ESD  for an arbitrary initially  entangled state. In the case of qubits, Ref.~[\ref{21}] found it to be NOT operation which is operation corresponding to the Pauli $\sigma_x$ operator. In this work, for qutrits, we propose a set of LUOs, which allow flipping the population between different  levels of the qutrit. Depending on the combination of LUOs, and the time  of their  application, this  method is shown  to be  able to hasten, delay or completely avoid the ESD in qubit-qutrit as well as qutrit-qutrit systems. 

In  the  decoherence process,  pure  states  evolve to  mixed  states.
Therefore, a  computable measure  of entanglement  for mixed  state is
required to quantify  the entanglement. Negativity is  such a measure,
based on the  Positive Partial Transpose (PPT) criterion  due to Peres
and  Horodecki [\ref{45}, \ref{46}]. For $2\otimes2$  and $2\otimes3$ dimensional systems, all PPT  states are separable and  Negative Partial Transpose (NPT) states are entangled.  Negativity is defined as the sum of the absolute values of all the  negative eigenvalues of the partially transposed density matrix  with respect to one of  the subsystems. While the enigmatic features of entanglement were predicted back in 1935, its witnessing and quantification continues to be at the forefront of current research. For a specific class of higher dimensional states $\rho$ which are supported on $d\otimes D$ Hilbert space ($d \leq D$) and with rank $r(\rho) \leq D$, PPT criterion has been found to be necessary and sufficient condition for separability [\ref{47}]. But for more general higher-dimensional systems ($d \otimes D,~\forall~d, D\geq 3~ \&~r(\rho)>D$), PPT criterion is only  a necessary but not sufficient condition for separability.

For these systems, we study Negativity  Sudden Death  (NSD), whose  non-occurrence guarantees Asymptotic Decay of Entanglement (ADE). But for $3\otimes 3$ dimensional PPT states with $r(\rho) > 3$ which have zero Negativity, we cannot comment on the separability and we need some other measure. For this purpose, we have used matrix realignment method for detecting and quantifying entanglement [\ref{48}-\ref{50}] after state is found to have zero negativity using PPT criterion. Realignment criterion states that for any separable state $\rho$, the trace norm of the realignment matrix $\langle m|\otimes\langle\mu|\rho^R|n\rangle\otimes|\nu\rangle=\langle m|\otimes \langle n|\rho|\mu\rangle\otimes|\nu\rangle$ never exceeds one. Therefore, if the realigned negativity given by $R(\rho)= max(0,||\rho^R||-1)$, is non-zero then state $\rho$ is entangled, where $\rho^R_{ij,kl}=\rho_{ik,jl}$. This criterion can detect some of the bound-entangled states which may not be detected by PPT criterion. We use PPT as well as realignment criterion for entanglement detection.

The rest of the paper is  organized as follows: In section  (\ref{S2}), we briefly discuss the physical model in  which qubit-qutrit entangled system can be realized  in an atomic  system and the  natural presence of  ADC in
such  a system. Next, Kraus  operators governing the evolution  of
quantum system  in the presence of ADC  are given. We then propose a set  of LUOs for qubit, and qutrit for manipulating the ESD, and
their physical  implementation for an  atomic system is  discussed. In
section  (\ref{S3})  and  (\ref{S4}), we  present  our  calculations
implementing  the  proposed  LUOs  on  two different class  of qubit-qutrit  system and  their results.  In section (\ref{S5}), we briefly comment on  the generalization of this proposal
to  higher  dimensions, taking an  example of a two-qutrit system. In section (\ref{S6}), the results of ESD manipulations are
compared and contrasted for a given  initial state with respect to the
choice of various LUOs. In the end, we  conclude with  the  advantages  and limitations  of  our proposal for ESD manipulation with other existing schemes.

\section{Physical Model} \label{S2}
\begin{figure}[!htb]
\begin{center}
\includegraphics[clip, trim=5.25cm 12.4cm 8cm 12.6cm, width=0.75\linewidth]{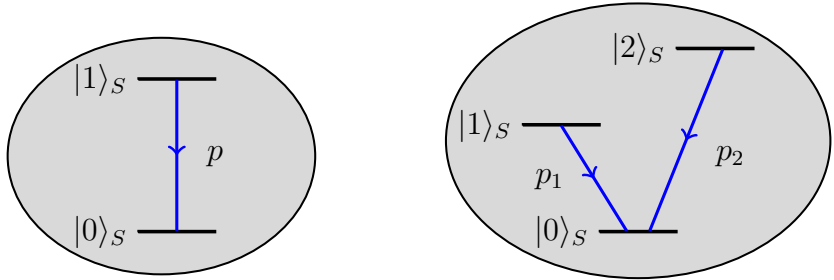}
\caption{\textit{Two-level atom (qubit), and a three-level atom (qutrit) in V-configuration. The parameters $p$, $p_1$, and $p_2$ denote the decay probabilities between the levels $|1\rangle_S\rightarrow|0\rangle_S$ of the qubit, and $|1\rangle_S\rightarrow|0\rangle_S$, and $|2\rangle_S\rightarrow|0\rangle_S$ levels of the qutrit, respectively. The qutrit is taken to be in V-configuration such that the dipole transitions are allowed only between the levels $|1\rangle_S\leftrightarrow|0\rangle_S$ and $|2\rangle_S\leftrightarrow|0\rangle_S$.}}
\label{fig1}
\end{center}
\end{figure}

Consider a hybrid qubit-qutrit entangled system in the presence of an ADC. The qubit and qutrit systems can be realized by a two-level, and a three-level atom in V-configuration (see Fig.~\ref{fig1}), respectively, in an optical trap for instance. The qubit-qutrit entangled state can be prepared by methods discussed in Refs. [\ref{51}-\ref{55}]. Afterwards, the entangled atoms are put into two cavities separated far from each other - by distances larger than the wavelength of the photons emitted from these atoms. The cavities are taken to be at absolute zero temperature and in vacuum state. The spontaneous emission due to atoms interacting locally with their cavities with electromagnetic field in vacuum state, forms identical but independent ADC.

The evolution of a qubit in the presence of an ADC is given by the following quantum map:
\begin{equation}
\begin{aligned}
|0\rangle_S|0\rangle_E & \rightarrow|0\rangle_S|0\rangle_E~,\cr
|1\rangle_S|0\rangle_E & \rightarrow \sqrt{1-p}|1\rangle_S|0\rangle_E+\sqrt{p}|0\rangle_S|1\rangle_E~,
\end{aligned}
\label{eq1}
\end{equation}
where $|0\rangle_S$ and $|1\rangle_S$ are the levels of qubit, and $|0\rangle_E$ and $|1\rangle_E$ are vacuum state and single-photon Fock state of the environment (cavity), respectively. In the Born-Markov approximation: $p=1-\exp(-\Gamma t)$, which is the probability of de-excitation of the qubit from the higher level $|1\rangle_S$ to the lower level $|0\rangle_S$. The subscripts `S' and `E' refer to the system and environment, respectively.

The quantum map of a qutrit (in $V$-configuration ) in the presence of an ADC is given by
\begin{equation}
\begin{aligned}
|0\rangle_S|0\rangle_E & \rightarrow|0\rangle_S|0\rangle_E~,\cr
|1\rangle_S|0\rangle_E & \rightarrow \sqrt{1-p_1}|1\rangle_S|0\rangle_E+\sqrt{p_1}|0\rangle_S|1\rangle_E~,\cr
|2\rangle_S|0\rangle_E & \rightarrow \sqrt{1-p_2}|2\rangle_S|0\rangle_E+\sqrt{p_2}|0\rangle_S|1\rangle_E~,
\end{aligned}
\label{eq2}
\end{equation} 
where $|0\rangle_S$, $|1\rangle_S$, and $|2\rangle_S$ are three levels of the V-type qutrit. There are two decay probabilities for such a qutrit corresponding to the transitions $|1\rangle_S\rightarrow |0\rangle_S$ and $|2\rangle_S\rightarrow |0\rangle_S$, and given by $p_1=1-\exp(-\Gamma_1 t)$ and $p_2=1-\exp(-\Gamma_2 t)$, respectively. Here, $\Gamma_1$ and $\Gamma_2$ represent the decay rate of levels $|1\rangle_S$, and $|2\rangle_S$, respectively.

The Kraus operators governing the evolution of the system in the presence of ADC, for qubit ($M_i$) and qutrit ($\mathbb{M}_i$), are obtained by tracing over the degrees of freedom of the environment from Eqs.~(\ref{eq1}) and (\ref{eq2})~, respectively, and given by
\begin{equation}
\begin{aligned}
M_0=\left( \begin{array}{cccc}
1 & 0  \cr
0 & \sqrt{1-p}\end{array} \right),~~
M_1=\left( \begin{array}{cccc}
0 & \sqrt{p} \cr
0 & 0 \end{array} \right).
\end{aligned}
\label{eq3}
\end{equation}

\begin{equation}
\begin{aligned}
\mathbb{M}_0=\left( \begin{array}{cccc}
1 & 0 & 0  \cr
0 & \sqrt{1-p_1} & 0\cr
0 & 0 & \sqrt{1-p_2}\end{array} \right),~~\cr
\mathbb{M}_1=\left( \begin{array}{cccc}
0 & \sqrt{p_1} & 0  \cr
0 & 0 & 0 \cr
0 & 0 & 0\end{array} \right),~~
\mathbb{M}_2=\left( \begin{array}{cccc}
0 & 0 & \sqrt{p_2}  \cr
0 & 0 & 0\cr
0 & 0 & 0\end{array} \right).
\end{aligned}
\label{eq4}
\end{equation}

The Kraus operators for qubit-qutrit system ($M_{ij}$) are obtained by taking appropriate tensor products of the qubit and qutrit Kraus operators as follows:
\begin{equation}
M_{ij}=M_i \otimes \mathbb{M}_j~~;~~i=0,1,~\&~j=0,1,2.
\label{eq5}
\end{equation}

If initial state of the system is $\rho(0)$ then evolution in the presence of ADC is given by,
\begin{equation}
\rho(p)=\sum_{i,j} M_{ij}~ \rho(0) ~ M_{ij}^\dag.
\label{eq6}
\end{equation}

If such a system undergoes ESD, then manipulation of ESD is achieved by applying LUOs (NOT operations) on qubit and/or qutrit at appropriate time $t=t_n$ as shown in Fig.~\ref{fig2} below. 

\begin{figure}[H]
\begin{center}
\includegraphics[clip, trim=0cm 20cm 1.5cm 1.5cm, width=0.75\linewidth]{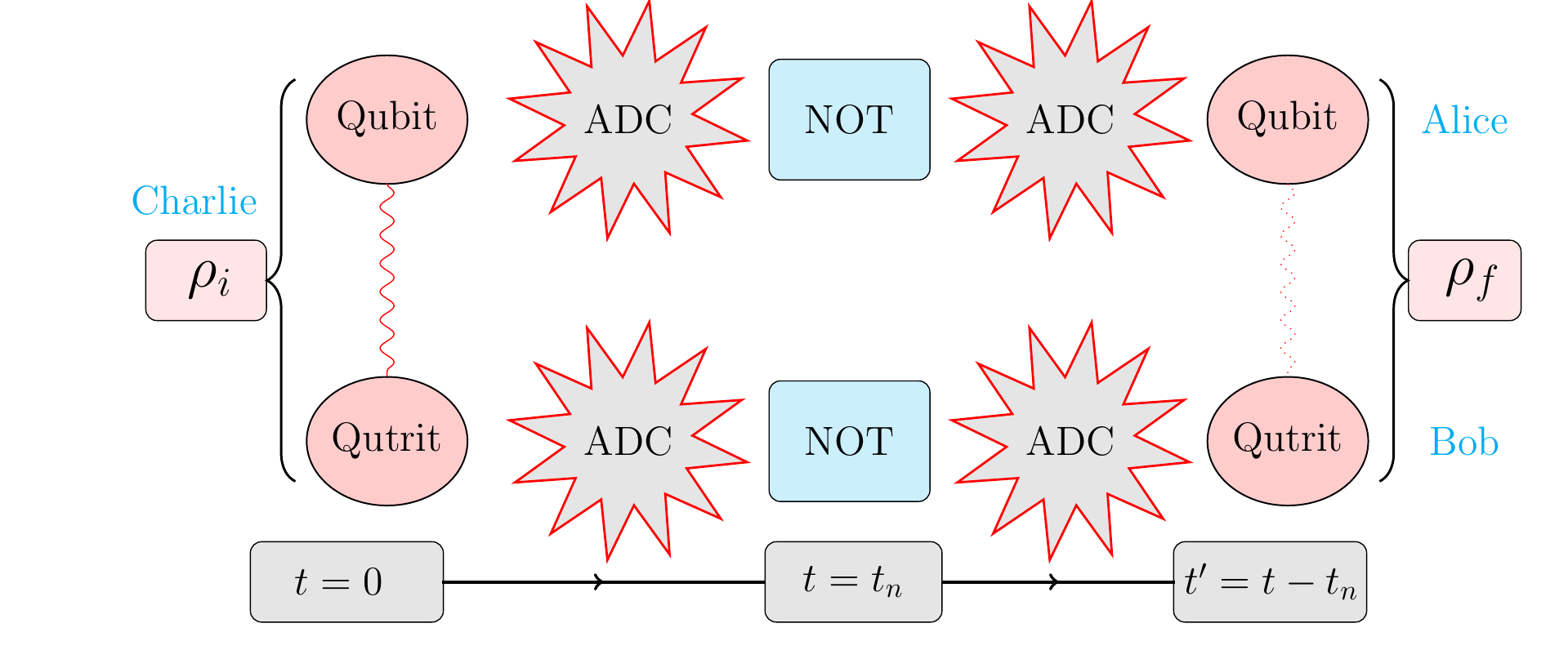}
\caption{\textit{Scheme for prolonging the entanglement in the presence of ADC. Initial state $\rho_i$ undergoes ADC decoherence then NOT operations are performed at $t=t_n$ and system is left to evolve in the ADC. The clock measuring the time is reset at $t=t_n$ and time after the application of NOT operation is measured by $t'=t-t_n$. Such a splitting of time is essential for later discussion.}}
\label{fig2}
\end{center}
\end{figure}

The NOT operation on qubit is Pauli spin operator $\sigma_x$. For qutrit, we propose here a set of operations which swap the populations between the two, or all the three levels of qutrit, namely trit-flip operations: $\mathbb{F}_{01},\mathbb{F}_{02}$, $\mathbb{F}_{102}$, and $\mathbb{F}_{201}$, where
\begin{equation} 
\begin{aligned}
\mathbb{F}_{01}=\left( \begin{array}{cccc}
0 & 1 & 0 \cr
1 & 0 & 0 \cr
0 & 0 & 1\end{array} \right),~~
\mathbb{F}_{02}=\left( \begin{array}{cccc}
0 & 0 & 1 \cr
0 & 1 & 0 \cr
1 & 0 & 0\end{array} \right),~~\cr
\mathbb{F}_{102}=\left( \begin{array}{cccc}
0 & 1 & 0  \cr
0 & 0 & 1  \cr
1 & 0 & 0\end{array} \right),~~
\mathbb{F}_{201}=\left( \begin{array}{cccc}
0 & 0 & 1  \cr
1 & 0 & 0  \cr
0 & 1 & 0\end{array} \right).
\end{aligned}
\label{eq7}
\end{equation}

In an atomic system, the NOT operation on the qubit ($\sigma_x$) can be applied by a $\pi$-pulse on the transition $|0\rangle_S \leftrightarrow |1\rangle_S$, which  interchanges the population between the levels $|0\rangle_S$ and $|1\rangle_S$ of the qubit. The trit-flip operation $\mathbb{F}_{01}$ ( $\mathbb{F}_{02}$) on the qutrit can be applied by a $\pi$-pulse on the transitions $|0\rangle_S \leftrightarrow |1\rangle_S$ ( $|0\rangle_S \leftrightarrow |2\rangle_S$) of the qutrit, which interchanges the population between the respective two levels. The trit-flip operation $\mathbb{F}_{102}$ ($\mathbb{F}_{201}$) on the qutrit can be realized by a $\pi$-pulse applied on the transition $|1\rangle_S \leftrightarrow |0\rangle_S$ ( $|2\rangle_S \leftrightarrow |0\rangle_S$) followed by another $\pi$-pulse to interchange the populations between $|0\rangle_S$ and $|2\rangle_S$ ($|0\rangle_S$ and $|1\rangle_S$). That is, by a series of two $\pi$-pulses $\pi^{|1\rangle_S\leftrightarrow|0\rangle_S}\pi^{|0\rangle_S\leftrightarrow|2\rangle_S}$ ($\pi^{|2\rangle_S\leftrightarrow|0\rangle_S}\pi^{|0\rangle_S\leftrightarrow|1\rangle_S}$).

Mathematically, the application of LUOs at $p=p_n$ can be represented by
\begin{equation}
\rho(p_n)=(U_1\otimes U_2)\rho(p)(U_1\otimes U_2)^\dag~,
\label{eq8}
\end{equation}
where $U_1=\sigma_x$ or $\mathbb{I}_2$, and  $U_2=\mathbb{F}_{01},~\mathbb{F}_{02},~\mathbb{F}_{102},~\mathbb{F}_{201}$ or $\mathbb{I}_3$.

Let us label another set of Kraus operators ($M'_{ij}$) with the parameter $p$ replaced by $p'$ ($t$ replaced by $t'$, $t'=t-t_n$); $p'=1-\exp(-\Gamma t')$, and of the form similar to Eq.~(\ref{eq5}). These Kraus operators are applied to the state (\ref{eq8}) to see the evolution of the system when it undergoes ADC after the application of LUOs as follows:
\begin{equation}
\rho(p',p_n)=\sum_{i,j} M'_{ij}~\rho(p_n)~ M_{ij}^{'\dag}. 
\label{eq9}
\end{equation}

When both the LUOs are identity operations, i.e., $U_1=\mathbb{I}_2$ and $U_2=\mathbb{I}_3$, we have the uninterrupted system evolving in the ADC. The state of the system in this case is given by
\begin{equation}
\rho(p',p)=\sum_{i,j}M'_{ij}\rho(p)M_{ij}^{'\dag}.
\label{eq10}
\end{equation}

Our aim is to investigate whether the phenomenon of hastening, delay and avoidance of ESD also occurs in higher-dimensional entangled systems as seen in $2\otimes 2$ systems [\ref{21}] or not. For this purpose, we compare the Negativity of the system manipulated using LUOs (\ref{eq9}) with that of the uninterrupted system (\ref{eq10}). For the purpose of current analysis, we choose the decay probabilities for the qutrit as $p_1=0.8p$ and $p_2=0.6p$, where $p$ is the decay probability of the qubit.


\section{X-type qubit-qutrit entangled system: State-I}\label{S3}

Let us consider the one-parameter qubit-qutrit entangled state given by,
\begin{equation}
\begin{aligned}
\rho(0) =& \frac{x}{2}[|00\rangle\langle 00|+|01\rangle\langle 01|+|11\rangle\langle 11|+|12\rangle\langle 12|] \cr
& + \frac{1-2x}{2}[|02\rangle\langle 02|+|02\rangle\langle 10|+|10\rangle\langle 02|+|10\rangle\langle 10|]~,
\end{aligned}
\label{eq11}
\end{equation}
where $0\leq x<1/3$.

Assuming that both the subsystems, qubit as well as qutrit, undergo local and independent ADC, evolution of the system is given by Eq.(\ref{eq6}). For state (\ref{eq11}), the mathematical form of Negativity in the presence of ADC and in terms of $x$ and $p$ is given as

\begin{equation}
\begin{adjustbox}{max width=0.8\columnwidth}
$
\begin{aligned}
N(x,p)= & \frac{1}{4} [2 p^2 x-4px+1.6 p+2 x-(0.64 p^4 x^2-1.28 p^3 x^2+10.24 p^2 x^2+2.56 p^3 x \cr
& -12.16 p^2 x+4.96 p^2-25.6 p x^2+25.6 p x-6.4 p+16 x^2-16 x+4)^{1/2}].
\end{aligned}
$
\end{adjustbox}
\label{eq12}
\end{equation}

\begin{figure} [H]
\begin{center}
\includegraphics[width=0.45\linewidth]{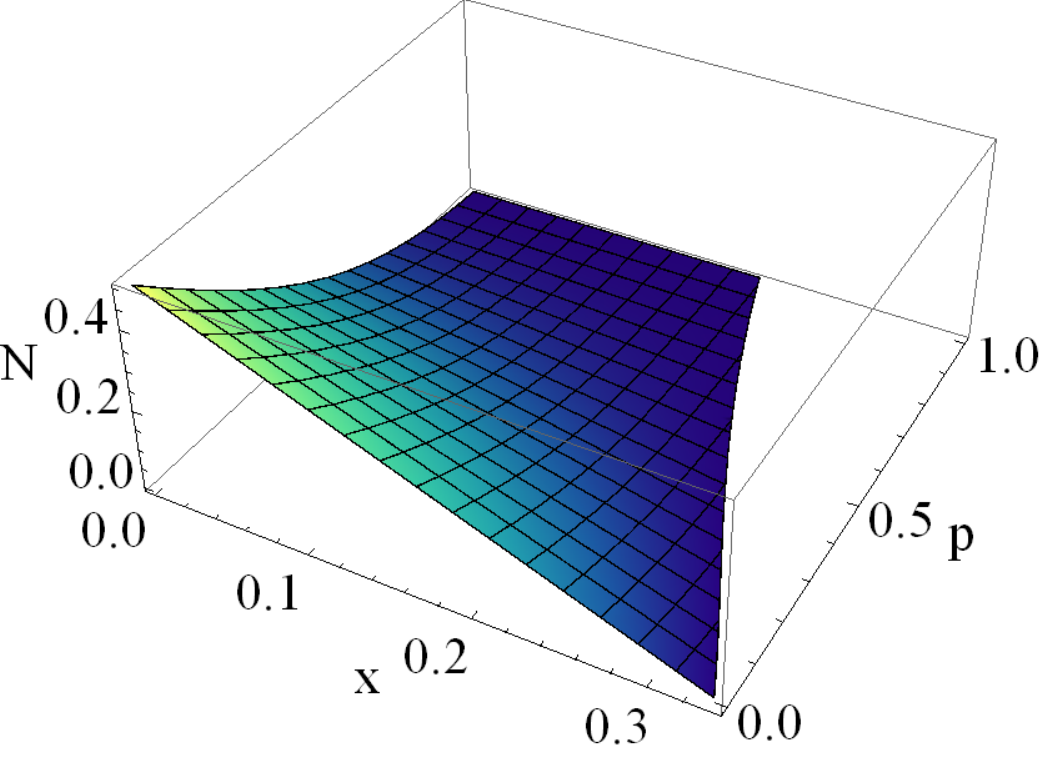}
\caption{\textit{Plot of Negativity vs. ADC probability for  $0\leq x < 1/3$ for the  entangled state (\ref{eq11}). The system undergoes ADE for $0\leq x\leq 0.20$, and ESD for $0.20<x<1/3$.}}
\label{fig3} 
\end{center}
\end{figure}

Evolution of the entanglement vs. ADC parameter ($p$) for $0\leq x<1/3$ is shown in Fig.~\ref{fig3}. The entangled state (\ref{eq11}) undergoes ADE for $0\leq x\leq 0.2$, and ESD for $0.2<x<1/3$. This can be easily verified by Eq. (\ref{eq12}).

\subsection{ESD in the presence of ADC}

We choose $x=0.25$ such that the initial state (\ref{eq11}) undergoes ESD at $p=0.6168$. The mathematical form of Negativity in terms of $p$ and $p'$ is given as

\begin{equation}
\begin{adjustbox}{max width=0.8\columnwidth}
$
\begin{aligned}
N(p,p')=&\frac{1}{4} [0.34 p^2 p'^2-0.84 p^2 p'-0.84 pp'^2+0.5 p p'+0.5 p'^2+0.6 p'+0.5 p^2+0.6 p+0.5\cr
& -\{0.0256 p^4 p'^4-0.1152 p^4 p'^3+0.1936 p^4 p'^2-0.144 p^4 p'-0.1152 p^3 p'^4+0.0096 p^3 p'^3 \cr 
& + 0.8656 p^3 p'^2-1.32 p^3 p'+0.1936 p^2 p'^4+0.8656 p^2 p'^3-0.9676 p^2 p'^2-2.584 p^2 p' \cr 
& -0.144 p p'^4-1.32 p p'^3-2.584 p p'^2+6.48 p p'+0.04 p'^4+0.56 p'^3+2.56 p'^2-1.6 p' \cr
& +0.04 p^4+0.56 p^3+2.56 p^2-1.6 p+1\}^{1/2}].
\end{aligned}
$
\end{adjustbox}
\label{eq13}
\end{equation}

\begin{figure} [H]
\begin{center} 
\includegraphics[width=0.45\linewidth]{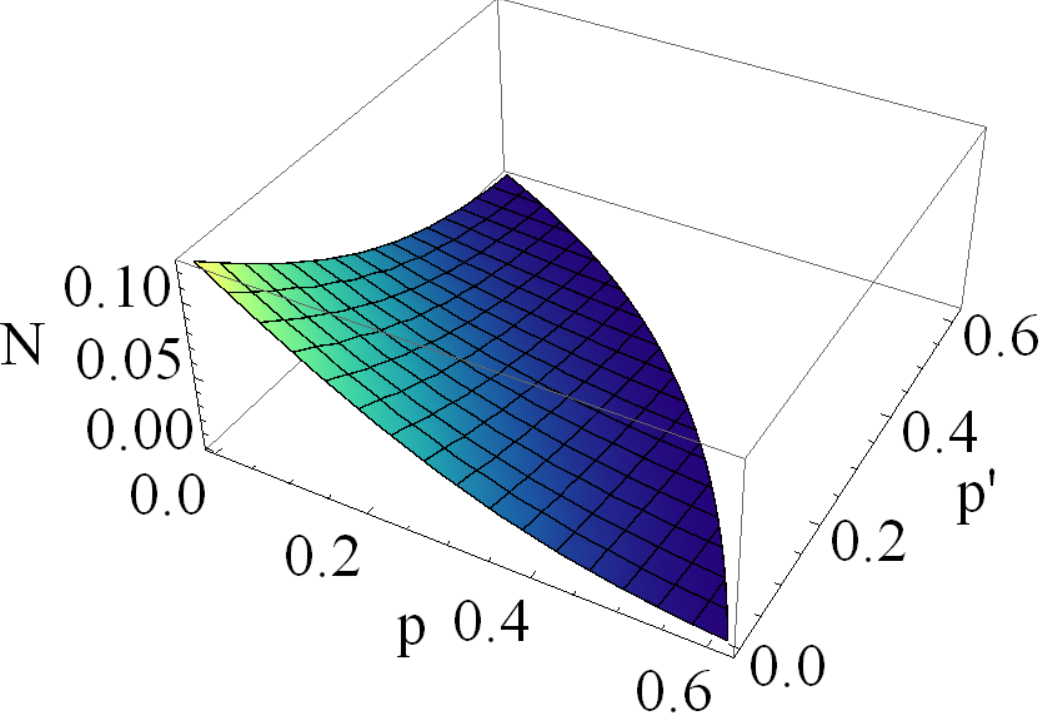}  
\caption{\textit{Plot of Negativity vs. ADC probability ($p,p'$) for $x=0.25$ for the state (\ref{eq11}). It undergoes ESD at $p=0.6168$ for $p'=0$ and vice-versa. For a non-zero $p$, ESD occurs along the curved path in the ($p,p'$) plane.}}
\label{fig4} 
\end{center}
\end{figure}

Using Eq.~(\ref{eq13}), the plot of Negativity vs. ADC probability ($p,p'$) for the state (\ref{eq11}) is shown in Fig.~\ref{fig4}. For $p=0$, ESD occurs at $p'=0.6168$ and for arbitrary values for $p$, ESD occurs along the non-linear curve in $pp'$ plane as shown in Fig.~\ref{fig4}.

For entanglement protection, we apply different combinations of LUOs and their effects are discussed below.

\subsection{$\sigma_x$ applied to qubit and $F_{01}$  applied to qutrit}

A NOT operation ($\sigma_x$) is applied to the qubit and trit-flip operation $\mathbb{F}_{01}$  applied to qutrit part of the state (\ref{eq11}) at $p=p_n$ as in Eq.~(\ref{eq9}).

For uninterrupted system, end of entanglement $p'$ depends on $p$ as follows: 

\begin{equation}
\begin{adjustbox}{max width=0.8\columnwidth}
$
\begin{aligned}
p'=& \frac{1}{2 (0.0625 p^2-0.15 p+0.0875)}[0.15 p^2+0.035 p-0.25 + 
\{(-0.15 p^2-0.035 p+0.25)^2 \cr
& -4 (0.0625 p^2-0.15 p+0.0875)(0.0875 p^2+0.25 p-0.1875)\}^{1/2}].
\end{aligned}
$
\end{adjustbox}
\label{eq14}
\end{equation}

When LUOs are applied at $p = p_n$ , the end of entanglement $p'$ depends on $p_n$ as follows:
\begin{equation}
p'=\frac{\left(p_n-1\right) \left(0.0875 p_n^2+0.25 p_n-0.1875\right)}{\left(0.25 p_n+0.5\right) \left(0.35 p_n^2+p_n+0.25\right)}.
\label{eq15}
\end{equation}

\begin{figure} [H]
\begin{center}
  \includegraphics[width=0.45\linewidth]{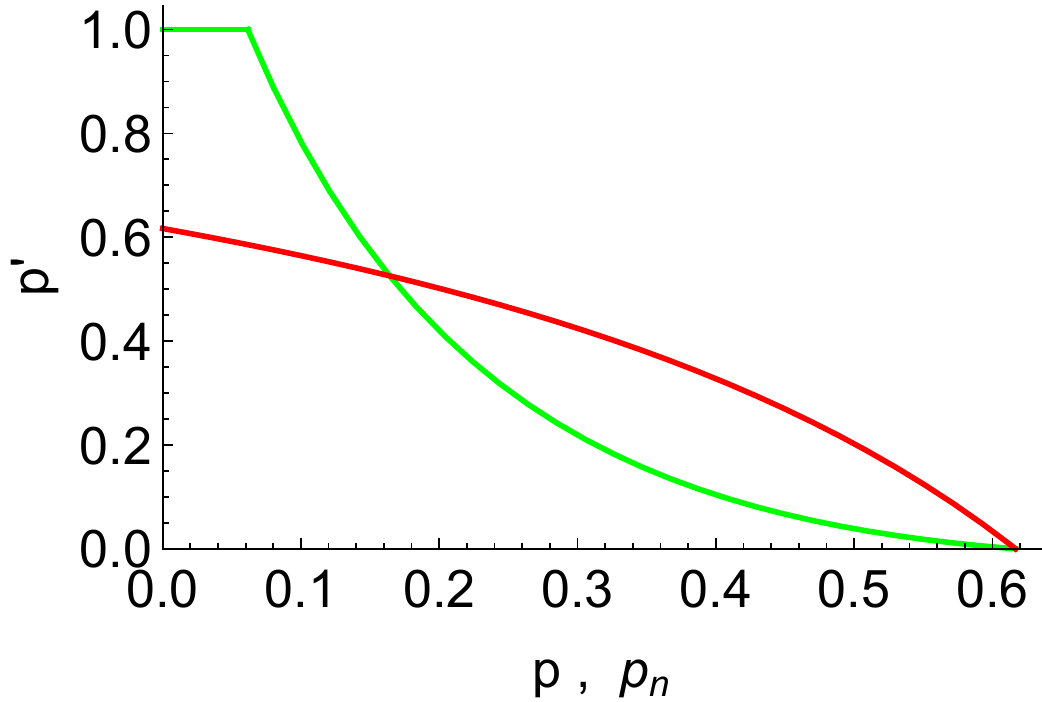}
\end{center}
\caption{\textit{Plot shows the non-linear curvature in  $p'$ vs.  $p ~or~p_n$ for ESD (red curve)  and its manipulation (green curve) such that the NOT operation ($\sigma_x$) is applied on the qubit and trit-flip operation $\mathbb{F}_{01}$ applied on the qutrit at $p=p_n$ for  $x=0.25$. The action of LUOs give rise to avoidance  for $0\leq p_n \leq 0.0615$, delay for $0.0615<p_n<0.1641$, and hastening of ESD for $0.1641<p_n<0.6168$ as the green curve lies below red curve in this range.}}
\label{fig5}
\end{figure}

 The Fig.~\ref{fig5} shows the non-linear curvature in  $p'$ vs.  $p~or~p_n$ for ESD (red curve) and its manipulation (green curve). The manipulation leads to avoidance of ESD for $0\leq p_n \leq 0.0615$, delay for $0.0615<p_n<0.1641$, and hastening of ESD for $0.1641<p_n<0.6168$ as the green curve dips below red curve in this range.  This can be easily verified through Eqs. (\ref{eq14}) and (\ref{eq15}).

\subsection{$F_{01}$ applied to qutrit only}

\begin{figure}[H]
  \centering
  \includegraphics[width=0.45\linewidth]{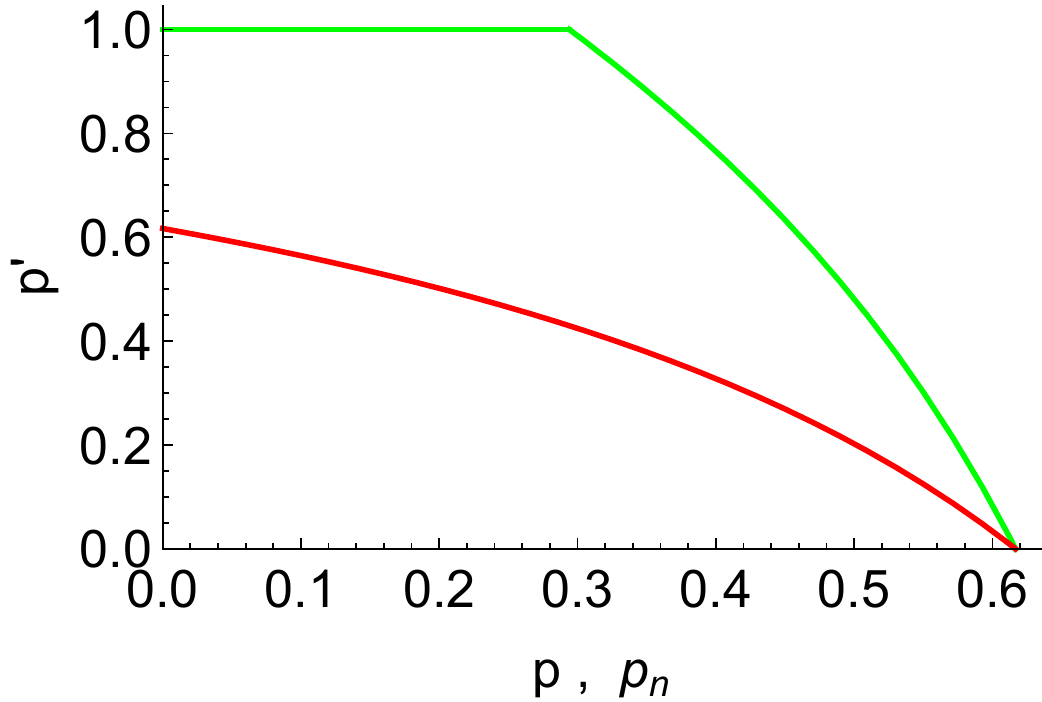}
\caption {\textit{Plot shows the non-linear curvature in  $p'$ vs.  $p ~or~p_n$ in ESD (red curve) and its manipulation (green curve) such that trit-flip operation $\mathbb{F}_{01}$ is applied on the qutrit at $p=p_n$ for $x=0.25$. The action of LUOs give rise to avoidance for $0\leq p_n\leq 0.2941$, delay for $0.2941<p_n<0.6168$, and hastening of ESD  does not occur is this case.}}
\label{fig6}
\end{figure}

The trit-flip operation $\mathbb{F}_{01}$ is applied to the qutrit part of the state (\ref{eq11}) at $p=p_n$ as in Eq.~(\ref{eq9}). The Fig.~\ref{fig6} shows the non-linear curvature in  $p'$ vs.  $p~or~p_n$ in ESD (red curve) and its manipulation (green curve). The manipulation leads to avoidance for $0\leq p_n\leq 0.2941$, delay for $0.2941<p_n<0.6168$, and hastening of ESD  does not occur in this case.

The LUOs $\sigma_x\otimes\mathbb{F}_{102}$ and $\mathbb{I}_2\otimes\mathbb{F}_{102}$ applied on the state (\ref{eq11}) lead to same effect as $\sigma_x\otimes\mathbb{F}_{01}$ and  $\mathbb{I}_2\otimes\mathbb{F}_{01}$, respectively. The other combination of LUOs such as $\sigma_x\otimes\mathbb{F}_{02}$,~$\sigma_x\otimes\mathbb{F}_{201}$,~ $\sigma_x\otimes\mathbb{I}_2$, ~$\mathbb{I}_2\otimes\mathbb{F}_{02}$, and ~$\mathbb{I}_2\otimes\mathbb{F}_{201}$  applied on the state (\ref{eq11}) give rise to only hastening of ESD in the entire range $0< p_n <0.6168$.

Let us now intuitively understand the disentanglement dynamics of the qubit-qutrit system and the occurrence of ESD. The state~(\ref{eq11}) is entangled due to the coherence terms $\rho_{34}$ ($|02\rangle\langle 10|$) and $\rho_{43}$ ($|10\rangle\langle 02|$) of the density matrix. The separability condition depends on the instantaneous value of the quantity $N=\frac{1}{2} \left(\rho_{11}+\rho_{66}-\sqrt{(\rho_{11}- \rho_{66})^2+4 \rho_{34} \rho_{43}}\right)$. If $N$ is negative, the state~(\ref{eq11}) is entangled else separable. In the presence of an ADC, coherence terms $\rho_{34}$ and $\rho_{43}$ decay as $\sim\sqrt{(1-p)(1-p_2)}$ and the term $\rho_{66}$ decays as $\sim(1-p)(1-p_2)$. The population of qubit-qutrit ground state $\rho_{11}$ changes as $\rho_{11} + \rho_{44} p + \rho_{22} p_1 + \rho_{55} p p_1 + \rho_{33} p_2 + \rho_{66} p p_2$. The terms $\rho_{34}$ ($\rho_{43}$) and $\rho_{66}$ decrease with time and $\rho_{11}$ increases. Due to the cumulative evolution of all these terms, the time $\left( t=-\frac{1}{\Gamma} \log_e(1-p)\right)$ at which $N$ becomes zero, is known as the time of sudden death and the state~(\ref{eq11}) becomes separable afterwards. For $p = 1$ or $p_2 = 1$ ($t\to \infty$) system in Eq.~(\ref{eq11}) looses the coherence completely and for $p,~p_1,~p_2 = 1$ qubit-qutrit system is found in the ground state $|00\rangle$.

The physical reason behind the action of different LUOs resulting in hastening, delay, or avoidance of ESD for this class of state can be understood as follows: when we apply the trit-flip operation  $\mathbb{F}_{01}$ (for example) on the system~(\ref{eq11}) after it has evolved in the ADC, it changes the instantaneous population between different levels of the qutrit in such a way that the elements of the density matrix (i) $\rho_{11}$ and $\rho_{22}$, (ii) $\rho_{44}$ and $\rho_{55}$ get swapped, and (iii) the coherence term $\rho_{34}$ ($\rho_{43}$) change their position to $\rho_{35}$ ($\rho_{53}$). When this flipped state evolves in the ADC, new coherence terms $\rho^{(n)}_{35}$ and $\rho^{(n)}_{53}$ (where terms with superscript `(n)' indicate the density matrix elements after the application of LUOs) now decay as $\sim(1-p)(1-p_1)(1-p_2)$, and new $\rho^{(n)}_{66}$ term decays as $\sim (1-p)(1-p_2)$.  The term $\rho^{(n)}_{11}$ evolves as $\rho^{(n)}_{11} + \rho^{(n)}_{44} p + \rho^{(n)}_{22} p_1 + \rho^{(n)}_{55} p p_1 + \rho^{(n)}_{33} p_2 + \rho^{(n)}_{66} p p_2$. After the application of LUOs, condition for ESD now becomes: $N=\frac{1}{2} \left(\rho^{(n)}_{11}+\rho^{(n)}_{66}-\sqrt{\left(\rho^{(n)}_{11}- \rho^{(n)}_{66}\right)^2+4 \rho^{(n)}_{34} \rho^{(n)}_{43}}\right)$, which looks similar to the earlier condition for ESD but it depends on new terms of the density matrix after the LUOs. The instantaneous population of different levels of the qubit-qutrit system depends on the decay rate of the different levels of the qubit-qutrit system, the time when LUO (trit-flip operation) is applied, and time elapsed after the application of LUO. Again, due to cumulative evolution of different terms of the density matrix, when $N$ becomes zero, system becomes separable. The basic idea behind this entanglement protection scheme is to choose correct combination of LUOs depending on the decay rate of different levels of the qubit-qutrit system and the time of application of LUOs such that the state after the application of LUOs results in the delay or avoidance of ESD.


\section{X-type qubit-qutrit entangled system: State-II}\label{S4}

Let us consider another class of one-parameter qubit-qutrit entangled state given by,
\begin{equation}
\begin{aligned}
\rho(0)= & \frac{x}{2}[|00\rangle\langle 00|+|01\rangle\langle 01|+|11\rangle\langle 11|+|12\rangle\langle 12|+|00\rangle\langle 12| 
+|12\rangle\langle 00|] \cr 
& + \frac{1-2x}{2}[|02\rangle\langle 02|+|10\rangle\langle 10|]~,
\end{aligned}
\label{eq16}
\end{equation}
where $1/3<x\leq 1/2$.

Assuming that both the subsystems; qubit as well as qurit, undergo local and independent ADC, evolution of the system is given by Eq.~(\ref{eq6}). For state (\ref{eq16}), evolution of the entanglement vs. ADC parameter ($p$) for $1/3\leq x\leq 1/2$ is computed. The initial entangled state (\ref{eq16}) undergoes ESD in the entire range $1/3<x\leq 1/2$.   For $p=0$, ESD occurs at $p'=0.8452$ and for arbitrary values for $p$, ESD occurs along the non-linear curve in $pp'$ plane similar to Fig.~\ref{fig4}.  We choose $x=0.5$ such that ESD occurs at $p=0.8452$, and study the effect of NOT operation ($\sigma_x$) applied to the qubit, and/or  trit-flip operations $\mathbb{F}_{01}$, $\mathbb{F}_{02}$, $\mathbb{F}_{102}$, $\mathbb{F}_{201}$  applied to the qutrit, in manipulating the ESD. The effect of different combinations of LUOs in entanglement protection are discussed below.

\subsection{$\sigma_x$ applied to qubit and $F_{01}$ applied to qutrit}

\begin{figure} [H]
\centering
\includegraphics[width=0.45\linewidth]{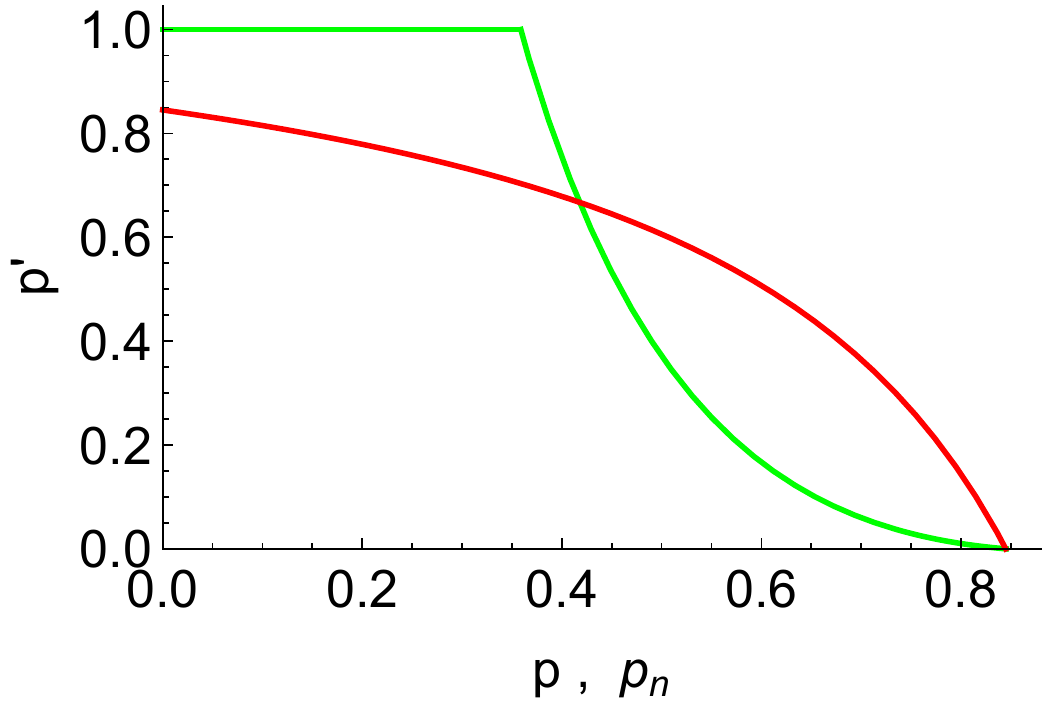}
\caption{\textit{Plot shows the non-linear curvature in  $p'$ vs.  $p ~or~p_n$ for ESD (red curve)  and its manipulation (green curve) such that the NOT operation ($\sigma_x$) is applied on the qubit and trit-flip $\mathbb{F}_{01}$ applied on the qutrit at $p=p_n$ for  $x=0.5$. The action of LUOs give rise to avoidance for $0\leq p_n\leq 0.3586$, delay for $0.3586<p_n<0.4177$, and hastening of ESD for $0.4177<p_n<0.8452$ as green curve lies below the red curve in this range.}}
 \label{fig7}
\end{figure}

The NOT operation ($\sigma_x$) is applied to the qubit and trit-flip operation $\mathbb{F}_{01}$ is applied to the qutrit part of the state (\ref{eq16}) at $p=p_n$ as in Eq.~(\ref{eq9}). The Fig.~\ref{fig7} shows the non-linear curvature in  $p'$ vs.  $p~or~p_n$ in ESD (red curve) and its manipulation (green curve). The manipulation leads to avoidance for $0\leq p_n\leq 0.3586$, delay for $0.3586<p_n<0.4177$, and hastening of ESD for $0.4177<p_n<0.8452$ as green curve lies below the red curve in this range.

\subsection{$\sigma_x$ operation applied to qubit only}

\begin{figure} [H]
\centering
\includegraphics[width=0.45\linewidth]{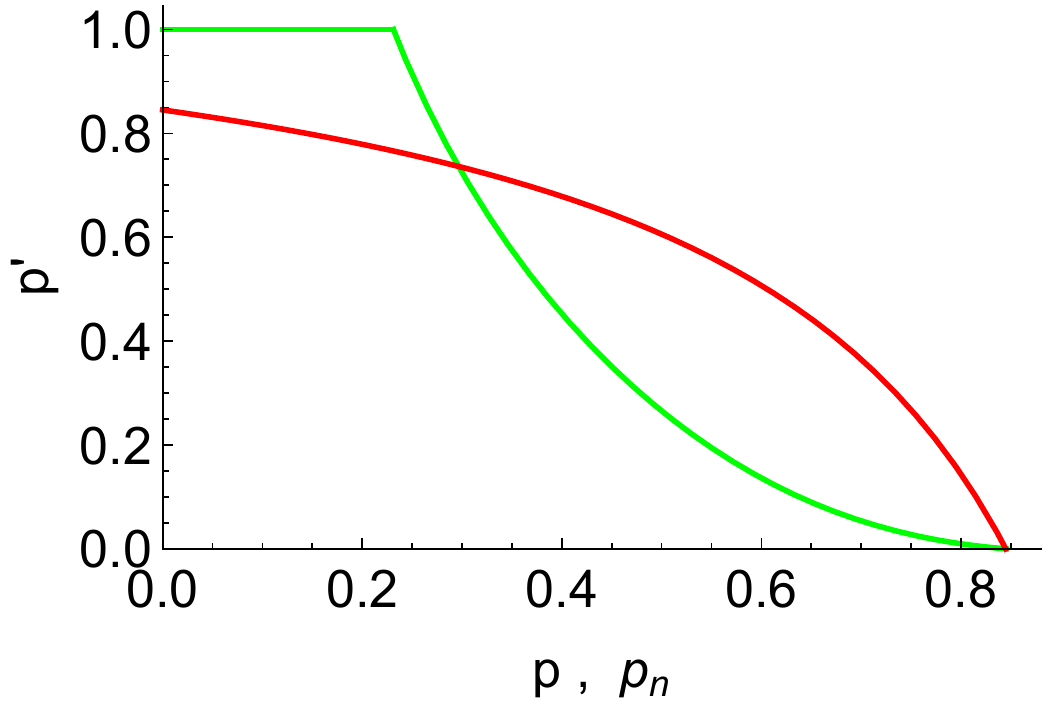}
\caption {\textit{Plot shows the non-linear curvature in  $p'$ vs.  $p ~or~p_n$ in ESD (red curve) and its manipulation (green curve) such that NOT operation is applied only on the qubit at $p=p_n$ for $x=0.5$. The action of LUOs give rise to avoidance for $0\leq p_n\leq 0.2309$, delay for $0.2309<p_n<0.2964$, and hastening of ESD for $0.2964<p_n<0.8452$ as green curve lies below the red curve in this range.}}
  \label{fig8}
\end{figure}

The NOT operation ($\sigma_x$) is applied to the qubit part of the state (\ref{eq16}) at $p=p_n$ as in Eq.~(\ref{eq9}). The Fig.~\ref{fig8} shows the non-linear curvature in  $p'$ vs.  $p~or~p_n$ in ESD (red curve) and its manipulation (green curve). The manipulation leads to avoidance for $0\leq p_n\leq 0.2309$, delay for $0.2309<p_n<0.2964$, and hastening of ESD for $0.2964<p_n<0.8452$ as green curve lies below the red curve in this range.

\subsection{$F_{01}$ applied to qutrit only}

\begin{figure} [H]
\centering
\includegraphics[width=0.45\linewidth]{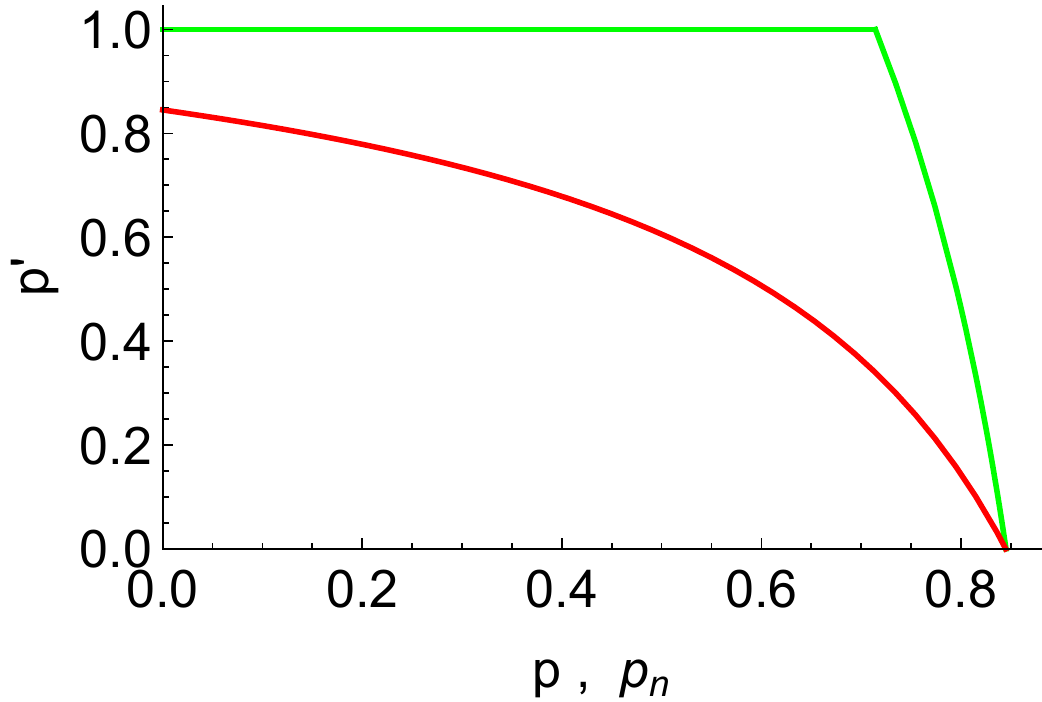}
\caption {\textit{Plot shows the non-linear curvature in  $p'$ vs.  $p ~or~p_n$ in ESD (red curve) and its manipulation (green curve) such that only trit-flip operation $\mathbb{F}_{01}$ is applied on the qutrit at $p=p_n$ for $x=0.5$. The action of LUOs give rise to avoidance for $0\leq p_n\leq 0.7143$, and delay of ESD for $0.7143<p_n<0.8452$ as green curve lies above the red curve but less than one in this range. The hastening of ESD does not occur in this case.}}
 \label{fig9}
\end{figure}

The trit-flip operation ($\mathbb{F}_{01}$) is applied only to the qutrit part of the state (\ref{eq16}) at $p=p_n$ as in Eq.~(\ref{eq9}). The Fig.~\ref{fig9} shows the non-linear curvature in  $p'$ vs.  $p~or~p_n$ in ESD (red curve) and its manipulation (green curve). The manipulation of ESD leads to avoidance for $0\leq p_n\leq 0.7143$, and delay of ESD for $0.7143<p_n<0.8452$ as green curve lies above the red curve but less than one in this range. The hastening of ESD does not occur in this case.

\subsection{$F_{02}$ applied to qutrit only}

\begin{figure} [H]
\centering
\includegraphics[width=0.45\linewidth]{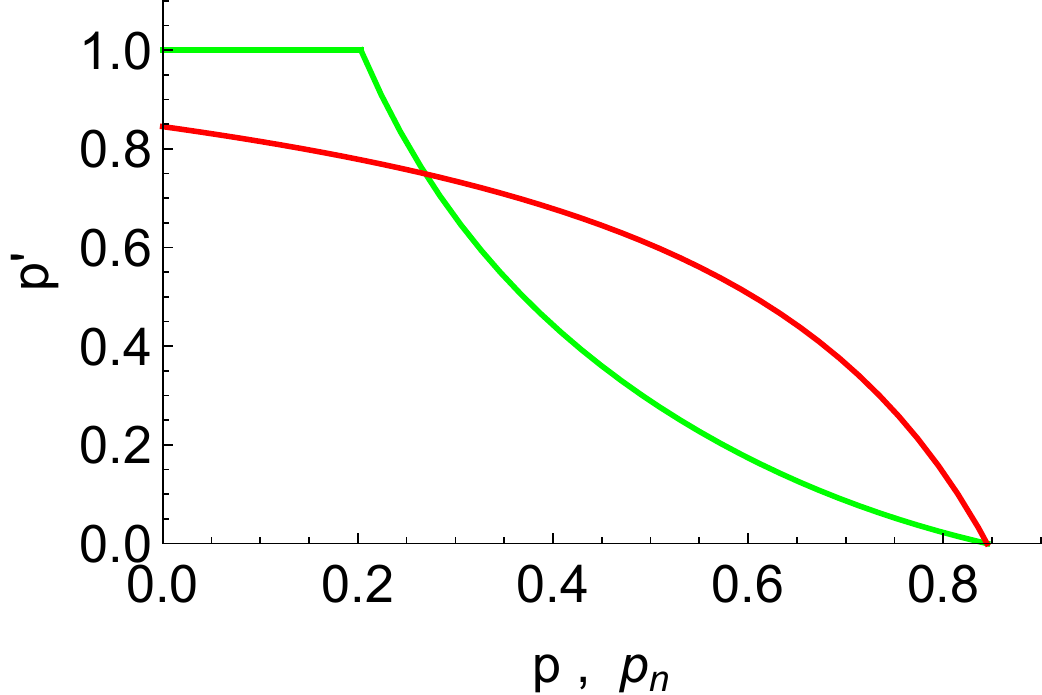}
\caption {\textit{Plot shows the non-linear curvature in  $p'$ vs.  $p ~or~p_n$ in ESD (red curve) and its manipulation (green curve) such that only trit-flip operation $\mathbb{F}_{02}$ is applied on the qutrit at $p=p_n$ for $x=0.5$.  The action of LUOs give rise to avoidance for $0\leq p_n\leq 0.2032$, delay for $0.2032<p_n<0.2693$, and hastening of ESD for $0.2693<p_n<0.8452$ as green curve lies below red curve in this range.}}
 \label{fig10}
\end{figure}

The trit-flip operation $\mathbb{F}_{02}$ is applied to the qutrit part of the state (\ref{eq16}) at $p=p_n$ as in Eq.~(\ref{eq9}). The Fig.~\ref{fig10} shows the non-linear curvature in  $p'$ vs.  $p~or~p_n$ in ESD (red curve) and its manipulation (green curve). The manipulation leads to avoidance for $0\leq p_n\leq 0.2032$, delay for $0.2032<p_n<0.2693$, and hastening of ESD for $0.2693<p_n<0.8452$ as green curve lies below the red curve in this range.

\subsection{$F_{201}$ applied to qutrit only}

\begin{figure} [h!]
  \centering
  \includegraphics[width=0.45\linewidth]{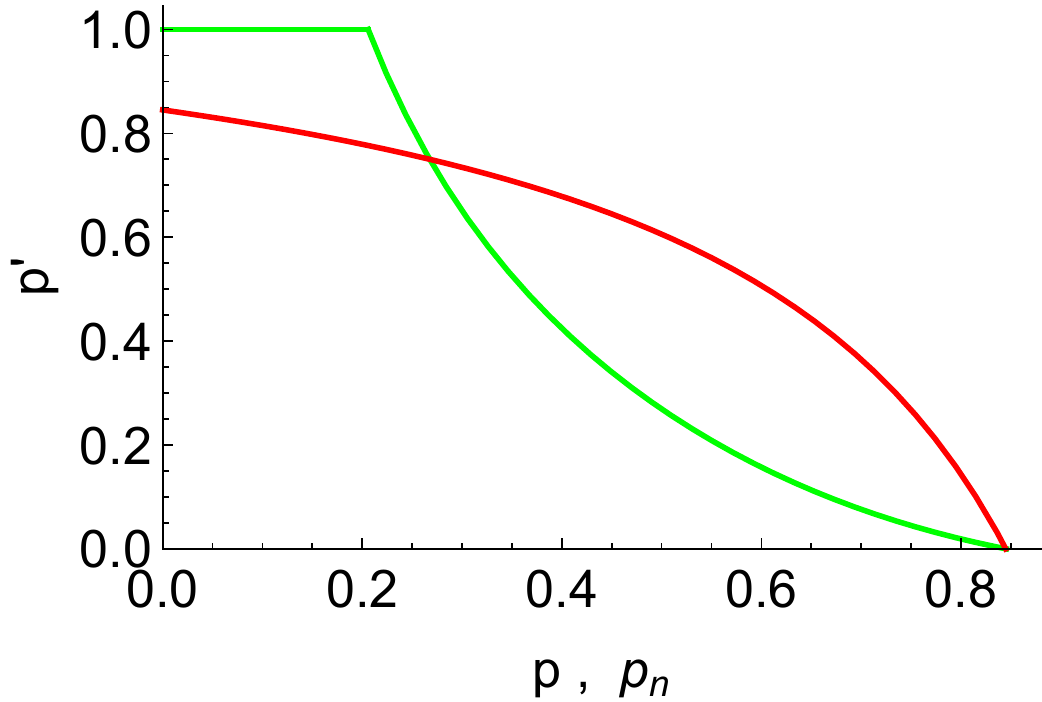}
\caption {\textit{Plot shows the non-linear curvature in  $p'$ vs.  $p ~or~p_n$ in ESD (red curve) and its manipulation (green curve) such that only trit-flip operation ($\mathbb{F}_{201}$) is applied on the qutrit at $p=p_n$ for $x=0.5$.  The action of LUOs give rise to avoidance for $0\leq p_n\leq 0.2059$, delay for $0.2059<p_n<0.2676$, and hastening of ESD for $0.2676<p_n<0.8452$ as green curve lies below the red curve in this range.}}
\label{fig11}
\end{figure}

The trit-flip operation $\mathbb{F}_{201}$ is applied to the qutrit part of the state (\ref{eq16}) at $p=p_n$ as in Eq.~(\ref{eq9}). The Fig.~\ref{fig11} shows the non-linear curvature in  $p'$ vs.  $p~or~p_n$ in ESD (red curve) and its manipulation (green curve). The manipulation leads to avoidance for $0\leq p_n\leq 0.2059$, delay for $0.2059<p_n<0.2676$, and hastening of ESD for $0.2676<p_n<0.8452$ as green curve lies below the red curve in this range.

The LUOs $\sigma_x\otimes\mathbb{F}_{102}$ and $\mathbb{I}_2\otimes\mathbb{F}_{102}$ applied on the state (\ref{eq16}) lead to same effect as $\sigma_x\otimes\mathbb{F}_{01}$ and  $\mathbb{I}_2\otimes\mathbb{F}_{01}$, respectively. The other combination of LUOs such as $\sigma_x\otimes\mathbb{F}_{02}$, and~$\sigma_x\otimes\mathbb{F}_{201}$ applied on the state (\ref{eq16}) only give rise to hastening of ESD in the entire range $0< p_n <0.8452$. 

\section{Generalization to qutrit-qutrit system}\label{S5}

In the  framework of  ADC, the dissipative  interaction of  system and
environment causes the  flow of population of the  system from excited
state to  ground state.  Therefore, any LUO which  reverses this
effect will  change the disentanglement time.  Thus, generalization of
the  above proposal  to  higher-dimensions  is  reasonably straight forward.  For example, the form  of unitary operations for two-qutrits  will be same as in  Eq.~(\ref{eq7}). However, lack of  a well  defined universal entanglement measure  for mixed  entangled states  of dimension  greater than  six, makes  it difficult  to study  the disentanglement  dynamics in  these systems because  even an  initial pure  entangled state  becomes mixed during the evolution.

For  the purpose  of  our study,  we  use Negativity  as  a witness  for entanglement, as  in general, qutrit-qutrit entanglement  is not known
to be  characterized fully, and  that negativity is a  sufficient but
not  necessary  condition  for   entanglement.   Thus,  if  negativity
undergoes  asymptotic  decay  then   this  implies  that  ESD  does not
happen. However, if negativity undergoes  sudden death (NSD), this may
be suggestive of (but does not imply) ESD.  Here, we take $a=1(p_1=ap)$
and $b=0.75~(p_2=bp)$. 

Let us consider an initially entangled two-qutrit system in the presence of ADC as given below.
\begin{equation}
\begin{aligned}
\rho(0)=&\frac{x}{3}(|01\rangle\langle01|+|02\rangle\langle02|+|10\rangle\langle10|+|12\rangle\langle12|+|20\rangle\langle20| 
 + |21\rangle\langle21|) \cr 
& + \frac{1-2x}{3}(|00\rangle\langle00|+ |11\rangle\langle11|+|22\rangle\langle22| 
 + |22\rangle\langle00|+ |00\rangle\langle02|),
 \end{aligned}
\label{eq17}
\end{equation}
where $0\leq x<1/3$.

For two-qutrit system, Kraus operators are given by
\begin{equation}
\mathbb{M}_{ij}=\mathbb{M}_i\otimes\mathbb{M}_j~;~i,j=0,1,2.
\label{eq18}
\end{equation}

Assuming that both the qutrits suffer identical but independent ADC, evolution of the system is given by,
\begin{equation}
\rho(p)=\sum_{i,j} \mathbb{M}_{ij} \rho(0) \mathbb{M}_{ij}^\dag.
\label{eq19}
\end{equation}

Evolution of the entanglement vs. ADC parameter ($p$) for $0\leq x<1/3$ is computed. It is found that the entangled state (\ref{eq17}) undergoes  NSD in the entire range of x-parameter for $0<x<1/3$.

Let us label another set of Kraus operators ($\mathbb{M}'_{ij}$) with the parameter $p$ replaced by $p'$ ($t$ replaced by $t'$, $t'=t-t_n$); $p'=1-\exp(-\Gamma t')$, and of the form similar to (\ref{eq18}), and apply it to the state (\ref{eq17}) to get the state of the uninterrupted system evolving in the ADC. 
\begin{equation}
\rho(p',p)=\sum_{i,j}\mathbb{M}'_{ij} \rho(0)\mathbb{M}_{ij}^{'\dag}.
\label{eq20}
\end{equation}

We choose $x=0.25$ such that the initial state (\ref{eq17}) undergoes NSD at $p=0.3636$. For $p=0$, NSD occurs at $p'=0.3636$ and for arbitrary values for $p$, NSD occurs along the non-linear curve in $pp'$ plane similar to Fig.~\ref{fig4}.

For protecting entanglement, trit-flip operation ($\mathbb{F}_{01}$) is applied to only one of the qutrits at $p=p_n$ as follows,
\begin{equation}
\rho^{(1)}(p_n)=(\mathbb{F}_{01} \otimes \mathbb{I}_3)\rho(p)(\mathbb{F}_{01} \otimes \mathbb{I}_3)^\dag.
\label{eq21}
\end{equation}

Evolution of the system afterwards in ADC is given by,
\begin{equation}
\rho^{(1)}(p',p_n)=\sum_{i,j} \mathbb{M}'_{ij}\rho^{(1)}(p_n)\mathbb{M}_{ij}^{'\dag}.
\label{eq22}
\end{equation}

\begin{figure} [H]
\begin{center}
\includegraphics[width=0.45\columnwidth]{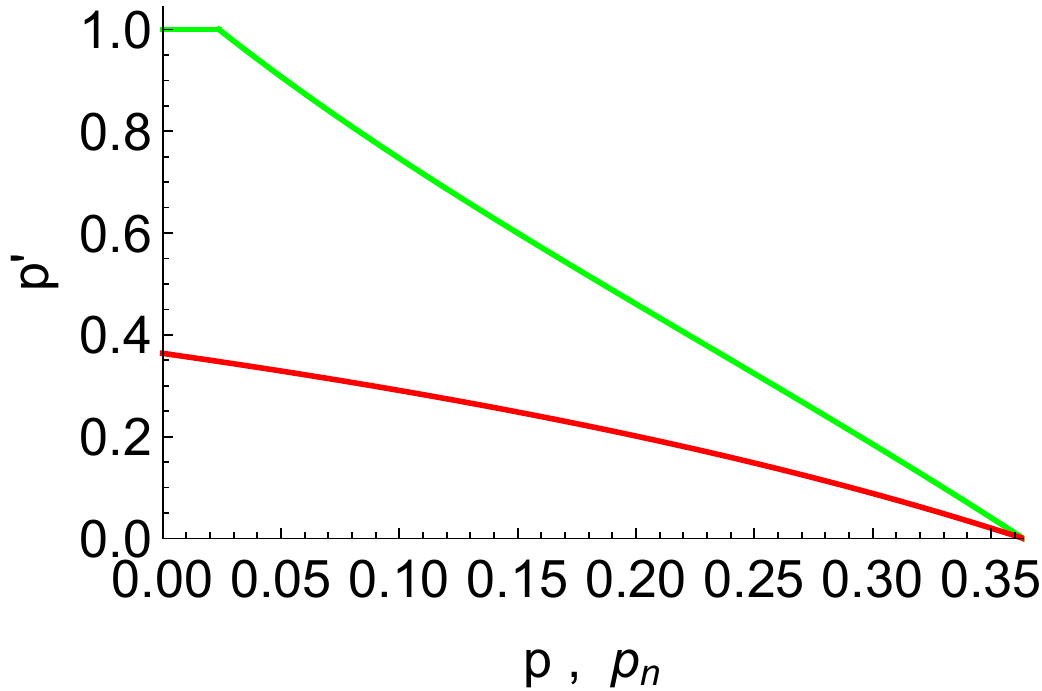}
\caption{\textit{Plot shows the non-linear curvature in  $p'$ vs.  $p ~or~p_n$ in NSD (red curve) and its manipulation (green curve) such that the trit-flip operation ($\mathbb{F}_{01}$) is applied on only one of the qutrits at $p=p_n$ for  $x=0.25$. The action of LUOs give rise to avoidance  for $0\leq p_n \leq 0.0238$, delay for $0.0238<p_n<0.3636$, and hastening of NSD does not occur in this case.}}
\label{fig12} 
\end{center}
\end{figure}

The Fig.~\ref{fig12} shows the non-linear curvature in  $p'$ vs.  $p~or~p_n$ for NSD (red curve) and its manipulation (green curve). The action of LUO ($\mathbb{F}_{01}$) gives rise to avoidance of NSD for $0\leq p_n \leq 0.0238$, delay for $0.0238<p_n<0.3636$, and hastening of NSD does not occur for this choice of parameter.

\section{Summary and discussion}\label{S6}

We have  proposed a set  of Local Unitary Operations (LUOs)  for qubit-qutrit system undergoing Entanglement  Sudden Death (ESD) in  the presence of an amplitude damping channel (ADC), such that when they are applied locally on one or both subsystems, then depending on the initial state, choice of the operation,  and its time  of application, one can  always suitably manipulate  the ESD.   We  have considered  two  different classes  of initially entangled qubit-qutrit systems which undergo ESD, and we find that  for a  given  initial  state, one  can  always  find a  suitable combination  of LUOs,  such that  when applied  at appropriate time  it can always delay  the time of ESD,  and therefore facilitate the tasks which would  not have been possible under shorter entanglement lifetime.

The  results  of   different  combinations  of  LUOs applied on  the qubit-qutrit system on  the manipulation of ESD for two different initially entangled states are summarized in the table~(\ref{tab1}) below.  In some cases, ESD can  be hastened, delayed,  as well  as avoided, whereas  in other cases,  it  can be  only  delayed  and avoided,  or  ESD  can be  only hastened.  Due to  symmetry  in the  population  of initial  entangled state,  the   NOT  operations   $\sigma_x\otimes\mathbb{F}_{102}$  and $\mathbb{I}_2\otimes\mathbb{F}_{102}$  applied  on the  either  states lead   to   same   effect  as   $\sigma_x\otimes\mathbb{F}_{01}$   and $\mathbb{I}_2\otimes\mathbb{F}_{01}$, respectively. Based on the  results in table (\ref{tab1}), the  following LUOs are advisable:  for example,  for state-II,  $\sigma_x$ on  the first qubit   and  no   action   on  the   qutrit   suffices  to   guarantee avoidance of ESD provide it is applied sufficiently early. Since this  is the simplest of all  possible combination of operations,  this  may  be  called   the  optimal  in  terms  of  gate operations.   For state-I,  all operations  allowing avoidance  are two-sided. It is  worth noting here that although the  noise is acting on both the subsystems, still  even LUO applied on only one of the  subsystems can suitably delay or avoid  the ESD as in the case of a two qubit system.

\begin{table}[h]
\begin{center}  
\renewcommand{\arraystretch}{1.25}
\begin{tabular}{ |p{3cm}||p{3cm}||p{3cm}|}
\hline
Operation  & State-I & State-II  \\
\hline \hline
 $\sigma_x \otimes \mathbb{F}_{01}$ & A, D, and H & A, D, and H   \\
  \hline
  $\sigma_x \otimes \mathbb{F}_{02}$ & only H &  only H   \\
  \hline
  $\sigma_x \otimes \mathbb{F}_{102}$ & A, D, and H &  A, D, and H   \\
  \hline
  $\sigma_x \otimes \mathbb{F}_{201}$ & only H &  Only H  \\
  \hline
  $\sigma_x \otimes \mathbb{I}_{3}$ & only H & A, D, and H  \\
  \hline
 $\mathbb{I}_{2} \otimes \mathbb{F}_{01}$ & only A and D  & only A and D  \\
  \hline
  $\mathbb{I}_{2} \otimes \mathbb{F}_{02}$ & only H &  A, D, and H   \\
  \hline
   $\mathbb{I}_{2} \otimes \mathbb{F}_{102}$ & only A and D & only A and D  \\
  \hline
   $\mathbb{I}_{2} \otimes \mathbb{F}_{201}$ & only H &   A, D, and H  \\
  \hline
\end{tabular}
\caption{\textit{Different combination of local  unitary operations applied on the  initial state-I (Eq.~\ref{eq11})  and state-II (Eq.~\ref{eq16}) resulting in avoidance (A), delay (D), and hastening (H) of ESD. Due to symmetry  in the population  of initial entangled state,  LUOs  $\sigma_x\otimes\mathbb{F}_{102}$   and
  $\mathbb{I}_2\otimes\mathbb{F}_{102}$ applied  on the  either states
  lead   to  same   effect  as   $\sigma_x\otimes\mathbb{F}_{01}$  and
  $\mathbb{I}_2\otimes\mathbb{F}_{01}$, respectively}}. \label{tab1}
\end{center}
\end{table}

Such a scheme will find application  where two parties, say Alice and
Bob, share an  entangled pair for some  quantum information processing task  and  they  know  a-priori that  ADC  is  present  in  the
environment, and therefore they can decide whether they are faced
with the prospect  of ESD.  Then, they can locally  apply suitable
LUOs at appropriate time to  delay  or  avoid  the ESD.   Our  proposed  scheme  for preserving  entanglement   longer  will   also  find   application  in entanglement distillation protocols.

 Furthermore, these LUOs can also be effective in protecting higher-dimensional entangled quantum systems in the presence of generalized amplitude damping channel as in the case of two-qubit system [\ref{24}]. Applying LUOs in the presence of generalized ADC towards the entanglement protection of higher dimensional systems could be an interesting work for future study. However, unlike ADC or generalized ADC, LUOs are not effective in protecting entanglement in the presence of bit-flip or phase-flip channel or both and the argument is as follows: under the action of ADC, population of a qubit (or qutrit) system flows from excited state to the ground state. The action of NOT operation (or more general LUO) is to reverse this effect by pumping the population from the ground to an excited state. While LUOs don’t change the amount of entanglement present in the system, they change the subsequent entanglement dynamics leading to hastening, delay or avoidance of ESD. This is the reason for effectiveness of LUOs in controlling the disentanglement dynamics in the presence of ADC. However, same reasoning doesn’t hold true for the bit-flip or phase flip channel as these are non-dissipative noises. 

Other schemes which aim to protect entanglement in a noisy environment include dynamic decoupling [\ref{56}], weak measurement (partial-collapse measurement) and quantum measurement reversal [\ref{57}], and quantum Zeno effect [\ref{58}]. Dynamic decoupling uses a sequence of $\pi$-pulses to protect the quantum states from noise. This scheme can potentially freeze the initial state and thereby preserve the quantum coherence stroboscopically to infinite time. The weak measurement and reversal scheme is based on applying a weak measurement prior to decoherence and then probabilistically reversing this operation on the decohered state. The measurement induced quantum Zeno effect [\ref{58}] for entanglement protection in cavity QED architecture, on the other hand, utilizes a very simple method of monitoring the population of the cavity modes that results in the entanglement protection exceeding its natural life-time. In our proposed method, LUOs alter the state in such a way that the decoherence effect on entanglement is minimized, i.e., for the states which were initially undergoing ESD, either time of ESD is delayed or it is averted altogether. Resource-wise, our scheme is simpler compared to the aforementioned schemes as it requires just one time intervention using LUOs but unlike the weak measurement and reversal protocol it does not restore the initial state after the decoherence.

To conclude, we have presented a scheme based on local unitary operations to protect entanglement from undergoing sudden death in the presence of amplitude damping channel for higher-dimensional systems. We have also compared and contrasted our entanglement protection scheme with other existing schemes. An interesting future line of theoretical study could be the entanglement protection in the presence of generalized amplitude damping channel in higher-dimensional systems. Our scheme will also attain more practical significance through commensurate future experiments.

The published version of the research work reported in this chapter can be found in Ref. [\ref{59}.


\section{References}\label{S8}
\begin{enumerate}
\setlength{\itemsep}{0pt}
\item \textit{W. H. Zurek,  ``Decoherence, einselection, and the quantum origins of the classical," \href{https://doi.org/10.1103/RevModPhys.75.715}{Rev. Mod. Phys. \textbf{75}, 715-775 (2003).}}\label{1}
\item \textit{R. Horodecki, P. Horodecki, M. Horodecki, K. Horodecki, ``Quantum entanglement,"  \href{https://doi.org/10.1103/RevModPhys.81.865}{Rev. Mod. Phys. \textbf{ 81}, 865 (2009).}} \label{2}
\item \textit{C. H. Bennett and D. P. DiVincenzo, ``Quantum information and computation," \href{https://doi.org/10.1038/35005001}{Nature \textbf{404}, 247 (2000).}} \label{3}
\item \textit{M. A. Nielsen  and I. L. Chuang, ``Quantum Computation and Quantum Information," \href{https://doi.org/10.1017/CBO9780511976667}{Cambridge University Press, Cambridge (2000)}}. \label{4}
\item \textit{H. P. Breuer and F. Petruccione, ``The Theory of Open Quantum Systems," \href{https://doi.org/10.1093/acprof:oso/9780199213900.001.0001}{Oxford University Press, Oxford (2002)}}. \label{5}
\item \textit{T. Yu  and J. H. Eberly, ``Finite-Time Disentanglement Via Spontaneous Emission" \href{https://doi.org/10.1103/PhysRevLett.93.140404}{Phys. Rev. Lett. \textbf{93}, 140404 (2004).}} \label{6}
\item \textit{T. Yu  and J. H. Eberly, ``Quantum Open System Theory: Bipartite Aspects," \href{https://doi.org/10.1103/PhysRevLett.97.140403}{Phys. Rev. Lett. \textbf{97}, 140403 (2006).}} \label{7}
\item \textit{T. Yu  and J. H. Eberly, ``Sudden Death of Entanglement," \href{https://doi.org/10.1126/science.1167343}{Science \textbf{ 323} , 598 (2009).}}\label{8}
\item \textit{J. Laurat, K. S. Choi, H. Deng, C. W. Chou, H. J. Kimble, ``Heralded Entanglement between Atomic Ensembles: Preparation, Decoherence, and Scaling," \href{https://doi.org/10.1103/PhysRevLett.99.180504}{Phys. Rev. Lett. \textbf{99}, 180504 (2007).}} \label{9}
\item \textit{M. P. Almeida, F. de Melo, M. Hor-Meyll, A. Salles, S. P. Walborn, P. H. Souto Ribeiro, L. Davidovich, ``Environment-Induced Sudden Death of Entanglement," \href{https://doi.org/10.1126/science.1139892}{ Science \textbf{316}, 555 (2007).}} \label{10}
\item \textit{D. A. Lidar, I. L. Chuang, and K. B. Whaley,  ``Decoherence-free subspaces for quantum computation,"  \href{https://doi.org/10.1103/PhysRevLett.81.2594}{ Phys. Rev. Lett. \textbf{ 81}, 2594 (1998).}} \label{11}
\item \textit{Lorenza Viola, Emanuel Knill and Seth Lloyd, ``Dynamical Decoupling of Open Quantum Systems," \href{https://doi.org/10.1103/PhysRevLett.82.2417}{Phys. Rev. Lett. \textbf{82}, 2417 (1999).}} \label{12}
\item \textit{P.W. Shor, ``Scheme for reducing decoherence in quantum computer memory,"  \href{https://doi.org/10.1103/PhysRevA.52.R2493}{Phys. Rev. A \textbf{52}, R2493 (1995).}} \label{13}
\item \textit{A.M. Steane, ``Error Correcting Codes in Quantum Theory,"  \href{https://doi.org/10.1103/PhysRevLett.77.793}{Phys. Rev. Lett. \textbf{77}, 793 (1996).}} \label{14}
\item \textit{P. Facchi, D. A. Lidar, and S. Pascazio, ``Unification of dynamical decoupling and the quantum Zeno effect,"  \href{https://doi.org/10.1103/PhysRevA.69.032314}{Phys. Rev. A \textbf{69}, 032314 (2004).}} \label{15}
\item\textit{Naoki Yamamoto, Hendra I. Nurdin, Matthew R. James, and Ian R. Petersen, ``Avoiding entanglement sudden death via measurement feedback control in a quantum network," \href{https://doi.org/10.1103/PhysRevA.78.042339}{Phys. Rev. A \textbf{78}, 042339 (2008).}} \label{16}
\item \textit{S. Maniscalco, F. Francica, R. L. Zaffino, N. L. Gullo, and F. Plastina, ``Protecting entanglement via the quantum Zeno effect," \href{https://doi.org/10.1103/PhysRevLett.100.090503}{Phys. Rev. Lett. \textbf{100}, 090503 (2008).}} \label{17}
\item \textit{J. G. Oliveira, Jr., R. Rossi, Jr., and M. C. Nemes 2008 Protecting, enhancing, and reviving entanglement Phys. Rev. A  \href{https://doi.org/10.1103/PhysRevA.78.044301}{\textbf{78}, 044301.}} \label{18}
\item \textit{Q. Sun, M. Al-Amri, L. Davidovich, and M. S. Zubairy 2010 Reversing entanglement change by a weak measurement Phys. Rev. A\href{https://doi.org/10.1103/PhysRevA.82.052323}{\textbf{ 82}, 052323.}}\label{19}
\item \textit{A. N. Korotkov and K. Keane 2010 Decoherence suppression by quantum measurement reversal Phys. Rev. A \href{https://doi.org/10.1103/PhysRevA.81.040103}{\textbf{81},040103(R).}}\label{20}
\item \textit{A.R.P. Rau, M. Ali, and G. Alber 2008 Hastening, delaying or averting sudden death of quantum entanglement EPL \href{https://doi.org/10.1209/0295-5075/82/40002}{\textbf{82}, 40002.}} \label{21}
\item \textit{Ashutosh Singh, Siva Pradyumna, A R P Rau, and Urbasi Sinha 2017 Manipulation of entanglement sudden death in an all-optical setup J. Opt. Soc. Am. B \href{https://doi.org/10.1364/JOSAB.34.000681}{\textbf{34}, 681-690.}} \label{22}
\item\textit{Mahmood Irtiza Hussain, Rabia Tahira and Manzoor Ikram, ``Manipulating the Sudden Death of Entanglement in Two-qubit Atomic Systems," \href{https://doi.org/10.3938/jkps.59.2901}{J. Korean Phys. Soc. \textbf{59}, 4, pp. 2901-2904 (2011).}} \label{23}
\item \textit{M. Ali, A. R. P. Rau, and G. Alber, ``Manipulating entanglement sudden death of two-qubit X-states in zero- and finite-temperature reservoirs," \href{https://doi.org/10.1088/0953-4075/42/2/025501} {J. Phys. B: At. Mol. Opt. Phys. \textbf{ 42}, 025501(8) (2009).}}\label{24}
\item \textit{P. G. Kwiat, A. J. Berglund, J. B. Altepeter, and A. G. White, ``Experimental verification of decoherence-free subspaces"  \href{https://doi.org/10.1126/science.290.5491.498} {Science \textbf{ 290}, 498 (2000).}} \label{25}
\item \textit{D. Kielpinski, V. Meyer, M. A. Rowe, C. A. Sackett, W. M. Itano, C. Monroe, and D. J. Wineland, ``A decoherence-free quantum memory using trapped ions," \href{https://doi.org/10.1126/science.1057357}{Science \textbf{ 291}, 1013 (2001).}}\label{26}
\item \textit{L. Viola, E. M. Fortunato, M. A. Pravia, E. Knill, R. Laflamme, and D. G. Cory, ``Experimental realization of noiseless subsystems for quantum information processing," \href{https://doi.org/10.1126/science.1064460}{Science \textbf{293}, 2059 (2001).}}\label{27}
\item \textit{Michael J. Biercuk, Hermann Uys, Aaron P. VanDevender, Nobuyasu Shiga, Wayne M. Itano \& John J. Bollinger,  ``Optimized dynamical decoupling in a model quantum memory," \href{https://doi.org/10.1038/nature07951}{Nature \textbf{458}, 996 (2009).}} \label{28}
\item\textit{Jiangfeng Du, Xing Rong, Nan Zhao, Ya Wang, Jiahui Yang \& R. B. Liu, ``Preserving electron spin coherence in solids by optimal dynamical decoupling,"  \href{https://doi.org/10.1038/nature08470}{Nature \textbf{461}, 1265 (2009).}} \label{29}
\item \textit{ J.C. Lee, Y.C. Jeong, Y.S. Kim, and Y.H. Kim, ``Experimental demonstration of decoherence suppression via quantum measurement reversal," \href{https://doi.org/10.1364/OE.19.016309}{Opt. Express \textbf{19}, 16309 - 16316 (2011).}} \label{30}
\item \textit{Y.S. Kim, J.C. Lee, O. Kwon, and Y.-H. Kim, ``Protecting entanglement from decoherence using weak measurement and quantum measurement reversal," \href{https://doi.org/10.1038/nphys2178}{Nature Phys. \textbf{8}, 117 (2012).}} \label{31}
\item \textit{H.T. Lim, J.C. Lee, K.H. Hong, and Y.H. Kim, ``Avoiding entanglement sudden death using single-qubit quantum measurement reversal," \href{https://doi.org/10.1364/OE.22.019055}{Opt. Express \textbf{22}, 19055 (2014).}} \label{32}
\item \textit{Jong-Chan Lee, Hyang-Tag Lim, Kang-Hee Hong, Youn-Chang Jeong, M.S. Kim \& Yoon-Ho Kim, ``Experimental demonstration of delayed-choice decoherence suppression," \href{https://doi.org/10.1038/ncomms5522}{Nature Communications \textbf{5}, 4522 (2014).}} \label{33}
\item \textit{Jin-Shi Xu, Chuan-Feng Li, Ming Gong, Xu-Bo Zou, Cheng-Hao Shi, Geng Chen, and Guang-Can Guo, ``Experimental demonstration of photonic entanglement collapse and revival",  \href{https://doi.org/10.1103/PhysRevLett.104.100502}{Phys. Rev. Lett. \textbf{ 104}, 100502 (2010).}} \label{34}
\item \textit{Erhard M., Krenn M. \& Zeilinger A., ``Advances in high-dimensional quantum entanglement," \href{https://doi.org/10.1038/s42254-020-0193-5}{Nat Rev Phys \textbf{2}, 365–381 (2020).}}\label{35}
\item \textit{Daniel Collins, Nicolas Gisin, Noah Linden, Serge Massar, and Sandu Popescu, ``Bell inequalities for arbitrarily high-dimensional systems," \href{https://doi.org/10.1103/PhysRevLett.88.040404}{Phys. Rev. Lett. \textbf{88}, 040404 (2002).}}\label{36}
\item \textit{Ann Kevin, and Jaeger Gregg, ``Entanglement sudden death in qubit-qutrit systems," \href{https://doi.org/10.1016/j.physleta.2007.07.070} {Phys. Lett. A \textbf{372}, 579-583 (2008).}} \label{37}
\item \textit{ M. Ali, A R P Rau, and Kedar Ranade, ``Disentanglement in qubit-qutrit systems," arXiv \href{https://arxiv.org/abs/0710.2238}{0710.2238 (2007).}} \label{38}
\item \textit{Mazhar Ali, ``Quantum Control of Finite-time Disentanglement in Qubit-Qubit and Qubit-Qutrit Systems," \href{http://tuprints.ulb.tu-darmstadt.de/id/eprint/1895}{Ph.D. Thesis (2009).}}\label{39}
\item \textit{Xing Xiao, ``Protecting qubit-qutrit entanglement from amplitude damping decoherence via weak measurement and reversal," \href{http://dx.doi.org/10.1088/0031-8949/89/6/065102}{Phys. Scr. \textbf{89}, 065102 (2008).}}\label{40}
\item\textit{X. Xiao and Y. L. Li, ``Protecting qutrit-qutrit entanglement by weak measurement and reversal," \href{http://dx.doi.org/10.1140/epjd/e2013-40036-3}{Eur. Phys. J. D \textbf{67}, 204 (2013).}}\label{41} 
\item\textit{Xiang-Ping Liao and Mao-Fa Fang and Man-Sheng Rong and Xin Zho, ``Protecting free-entangled and bound-entangled states in a two-qutrit system under decoherence using weak measurements," \href{https://doi.org/10.1080/09500340.2016.1271149}  {Journal of Modern Optics \textbf{64}, 12, pp. 1184-1191 (2017).}}\label{42}
\item \textit{Mazhar Ali, ``Distillability sudden death in qutrit-qutrit systems under amplitude damping," \href{https:/doi.org/10.1088/0953-4075/43/4/045504}{J. Phys. B \textbf{43}, 045504 (2010).}}\label{43}
\item \textit{L. Derkacz and L. Jakobczyk, ``Quantum interference and evolution of entanglement in a system of three-level atoms," \href{https://doi.org/10.1103/PhysRevA.74.032313} {Phys. Rev. A \textbf{74}, 032313 (2006).}}\label{44}
\item \textit{A. Peres, ``Separability Criterion for Density Matrices," \href{https://doi.org/10.1103/PhysRevLett.77.1413}{Phys. Rev. Lett. \textbf{77}, 1413 (1996).}}\label{45} 
\item \textit{M.Horodecki, P. Horodecki  and R. Horodecki, ``Separability of mixed states: necessary and sufficient conditions," \href{https://doi.org/10.1016/S0375-9601(96)00706-2}{Phys. Lett. A \textbf{223}, 1 (1996).}} \label{46}
\item \textit{Paweł Horodecki, Maciej Lewenstein, Guifré Vidal, and Ignacio Cirac, ``Operational criterion and constructive checks for the separability of low-rank density matrices," \href{https://link.aps.org/doi/10.1103/PhysRevA.62.032310}{Phys. Rev. A \textbf{62}, 032310 (2000).}} \label{47}
\item \textit{O. Rudolph, ``Computable Cross-norm Criterion for Separability," \href{https://doi.org/10.1007/s11005-004-0767-7}{Letters in Mathematical Physics \textbf{70}, 57–64 (2004)}}.\label{48}
\item \textit{O. Rudolph, ``Further results on the cross norm criterion for separability," \href{https://10.1007/s11128-005-5664-1}{Quant. Inf. Proc. \textbf{4}, 3, pp. 219 - 239 (2005)}}.\label{49}
\item \textit{Kai Chen, Ling-An Wu, ``A matrix realignment method for recognizing entanglement,"  \href{http://dl.acm.org/citation.cfm?id=2011534.2011535}{Quantum Inf. Comput. \textbf{3}, 3, pp.193-202 (2003).}} \label{50}
\item \textit{E. Hagley, X. Ma$\hat{\iota}$tre, G. Nogues, C. Wunderlich, M. Brune, J. M. Raimond, and S. Haroche, ``Generation of Einstein-Podolsky-Rosen Pairs of Atoms,"  \href{https://doi.org/10.1103/PhysRevLett.79.1}{Phys. Rev. Lett. \textbf{79}, 1 (1997).}}  \label{51}
\item \textit{B. P. Lanyon, T. J. Weinhold, N. K. Langford, J. L. O’Brien, K. J. Resch, A. Gilchrist, and A. G. White, ``Manipulating Biphotonic Qutrits," \href{https://doi.org/10.1103/PhysRevLett.100.060504}{Phys. Rev. Lett. \textbf{100}, 060504 (2008).}}\label{52}
\item \textit{Robert Fickler, Radek Lapkiewicz, William N. Plick, Mario Krenn, Christoph Schaeff, Sven Ramelow, Anton Zeilinger, ``Quantum Entanglement of High Angular Momenta," \href{https://doi.org/10.1126/science.1227193}{Science \textbf{338}, 640-643 (2012).}} \label{53}
\item \textit{E. J. Galvez, S. M. Nomoto, W. H. Schubert, and M. D. Novenstern, ``Polarization-Spatial-Mode Entanglement of Photon Pairs," International Conference on Quantum Information, OSA Technical Digest (CD) (Optical Society of America), paper QMI18 (2011).} \label{54}
\item \textit{Zheng Shi-Biao, ``Production of Entanglement of Multiple Three-Level Atoms with a Two-Mode Cavity," \href{https://doi.org/10.1088/0253-6102/45/3/031} {Commun. Theor. Phys. \textbf{45}, 539-541 (2006).}} \label{55}
\item\textit{ L. Viola, E. Knill, and S. Lloyd,  ``Dynamical Decoupling of Open Quantum Systems," \href{https://doi.org/10.1103/PhysRevLett.82.2417}{Phys. Rev. Lett. \textbf{82}, 2417 (1999).}}\label{56}
\item \textit{Yong-Su Kim, Jong-Chan Lee, Osung Kwon and Yoon-Ho Kim, ``Protecting entanglement from decoherence using weak measurement and quantum measurement reversal," \href{https://doi.org/10.1038/NPHYS2178}{Nature Physics \textbf{8}, 117–120 (2012).}} \label{57}
\item \textit{S. Maniscalco, F. Francica, R. L. Zaffino, N. L. Gullo, and F. Plastina, ``Protecting entanglement via the quantum Zeno effect", \href{https://doi.org/10.1103/PhysRevLett.100.090503}{Phys. Rev. Lett. \textbf{100}, 090503  (2008)}}. \label{58}
\item \textit{Ashutosh Singh and Urbasi Sinha, ``Entanglement protection in higher-dimensional systems,"  \href{https://doi.org/10.1088/1402-4896/ac8200}{Phys. Scr. \textbf{97}, 085104 (2022).}} \label{59}
\end{enumerate}

%% file: chapter7.tex
\setcounter{equation}{0}
\chapter{Summary and future scope}

We have reported our theoretical and experimental investigations towards Creation, characterization, and manipulation of quantum entanglement in a photonic system. We have studied two different aspects of quantum entanglement: (i) A review study on the comparison between different Entanglement Measures (EMs) for non-maximally entangled two-qubit pure states is done towards quantification of entanglement in such states and then this study is extended to higher-dimensional systems, (ii) Entanglement dynamics of a two-qubit system is studied in the presence of an Amplitude Damping Channel  (ADC) and a scheme based on local unitary operations is presented to protect entanglement from undergoing Entanglement Sudden Death (ESD). This decoherence study is then extended to qubit-qutrit and qutrit-qutrit entangled systems and entanglement protection scheme is proposed for higher dimensional systems.

For the experimental study, a high-fidelity ($\sim 98\%$) polarization-entangled photon source based on type-I Spontaneous Parametric Down-Conversion (SPDC) process is prepared and characterized using quantum state tomography. It is observed that when thickness of each non-linear crystal in the paired-BBO crystal geometry of type-I SPDC source was comparable to, or greater than, the coherence length of the pump laser, it gave rise to the partial distinguishability of the photons generated in the first and second crystal by their arrival time statistics measurement. This lead to decoherence and thus drop in the amount of entanglement in the two-qubit system. Theoretical estimation of such a temporal delay is done. For pre-compensating the delay and erasing the temporal distinguishability, a birefringent temporal compensator is used to improve the quality of entanglement in the generated bi-photon state. We have reported our experimental observations on the effect of spectral filtering conditions in the SPDC process on the quality of two-qubit entanglement. 

Regarding the comparison between different EMs for the purpose of quantification of the amount entanglement in non-maximally entangled two-qubit pure states, we have defined suitable parameters which quantify the deviation of a given non-maximally entangled state from maximally entangled states. In this process, we found that different EMs give different deviations of a given non-maximally entangled two-qubit pure state from maximally entangled state and they can differ as much as $23.5\%$ when Entanglement of Formation and Log-Negativity are used as EMs. Then we briefly commented on such comparison between EMs for higher-dimensional systems. While in the case of a bipartite two-qubit entangled states, although different EMs show different deviations of a given state from maximally entangled state, the EMs remain monotonic with respect to each other. But in the bipartite qutrit case, different EMs not only provide different estimations of the deviation of any non-maximally entangled state from the maximally entangled state, but the EMs can also be non-monotonic with respect to each other. In sum, the results of studies in this work bring out the need for exploring different ideas for quantifying how close (far) a given entangled state is to (from) the maximally entangled state and comparing the results obtained by using such quantifiers with those based on different EMs. This leads to the following question: Is there any fundamental criterion for assessing which quantifier is the appropriate one to be used for addressing questions such as the one posed in this work, or would such a criterion have to be operationally defined essentially dependent on the specific context in which the entangled state is used as a resource? A comprehensive study is required for shedding further light on this issue as well as for gaining a deeper understanding of the comparison between different EMs, taking into account the studies probing their respective physical significance.

In our study of the evolution of entanglement in two-qubit systems in the presence of an ADC, we found that there are two class of states: one which undergo Asymptotic Decay of Entanglement (ADE), and the other one where states disentangle in finite time known as Entanglement Sudden Death (ESD). Then, we proposed a scheme for manipulation of ESD using local NOT operation on one or both the qubits of a bipartite entangled state to protect entanglement for a longer time. It is worth noting that the action of local unitary operations on individual subsystems cannot change the amount of entanglement in the system but subsequent dynamics can be altered. We have theoretically shown that the phenomenon of hastening, delay and avoidance of ESD occur depending on the time of application of NOT operation. The next step was to set up the experiment for the manipulation of ESD. In this direction, we  have experimentally demonstrated the phenomenon of ESD and ADE for two different class of initially entangled states. Finally, we built the experimental setup to demonstrate the hastening, delay and avoidance of ESD using NOT operation on both the qubits. We have discussed different constraints that need to be simultaneously taken care of in such an experiment. We have outlined our endeavour in this direction and reported preliminary results for zero-degree settings  of the ESD-manipulation experiment. An experimental realization of our proposal will be important for practical noise engineering in quantum information processing and is underway.

Our proposal for manipulation of ESD has an advantage over decoherence suppression using weak measurement and quantum measurement reversal and delayed-choice decoherence suppression. There, as the strength of weak interaction increases, the success probability of decoherence suppression decreases. On the other hand, using our scheme we can manipulate the ESD, in principle, with unit success probability as long as we perform the NOT operation at the appropriate ADC parameter value. Delay and avoidance of ESD, in particular, will find application in the practical realization of quantum information and computation protocols that might otherwise suffer due to a short lifetime of entanglement. Also, it will have implications toward such control over other physical systems. The advantage of the manipulation of ESD in a photonic system is that one has complete control over the damping parameters, unlike in most atomic systems.  Resource-wise, our scheme is simpler compared to the other two aforementioned schemes but it does not restore the state after the decoherence back to initial state.

Regarding the theoretical extension of the idea of manipulation of ESD using local unitary operations to higher dimensional systems, we have considered qubit-qutrit and qutrit-qutrit entangled systems evolving in the ADC. For the states which undergo ESD, we have proposed a more general class of local population-flip operators in the context of a bipartite atomic system. We have shown that even in higher dimensional systems, ESD can be suitably manipulated and phenomenon of hastening, delay, and avoidance of ESD is observed. We have compared and contrasted pros and cons of entanglement protection in higher-dimensional systems using our scheme with other existing schemes in the literature.

Future scope of this work may include looking into the entanglement protection in the presence of other noise models and more general noises that occur in real-world physical systems such as a combination of many of the noises acting together. Further attempts can be made towards transforming the current state-dependent entanglement protection scheme into a more general state-independent scheme which would further benefit the real-world quantum information processing. One can also explore the efficacy of such local unitary operations on other quantum correlations such as quantum discord and geometric discord evolving in the noisy environment.
